\documentclass[twoside,fleqn]{ActaStyle}
\usepackage[colorlinks=true, pdfstartview=FitV, linkcolor=black,
            citecolor=black, urlcolor=black]{hyperref}
\usepackage{graphicx}
\usepackage{dcolumn}
\usepackage{rotating}
\usepackage{times,cite}
\usepackage{amsmath}
\usepackage{amssymb}
\usepackage{amsfonts}
\usepackage{amsthm}
\usepackage{multicol}
\usepackage{dsfont}
\usepackage{float}
\usepackage{color}
\usepackage{wrapfig}

\usepackage{fancyhdr}
\renewcommand{\subsectionmark}[1]{}
\def\softd{{\leavevmode\setbox1=\hbox{d}%
\hbox to 1.05\wd1{d\kern-0.4ex{\char039}\hss}}}%cstocs
\def\softt{{\leavevmode\setbox1=\hbox{t}%
\hbox to \wd1{t\kern-0.6ex{\char039}\hss}}}%cstocs
\def\softl{l\kern-0.45ex\raise0.1ex\hbox{'}\kern-0.10ex}%cstocs
\def\softL{L\kern-0.8ex\raise0.1ex\hbox{'}\kern0.1ex}%cstocs

\newcommand\curl{\text{\rm curl\,}}
\newcommand\Div{{\text{\rm div\,}}}
\newcommand\grad{\text{\rm grad\,}}
\newcommand\hor{\text{\rm hor\,}}

\renewcommand{\thesection}{\arabic{section}}
\renewcommand{\thesubsection}{\thesection.\arabic{subsection}}
\renewcommand{\theequation}{\thesubsection.\arabic{equation}}
\renewcommand{\thefigure}{\thesection.\arabic{figure}}
\renewcommand{\thetable}{\thesection.\arabic{table}}

\def\authorheading{M. Fecko}
\def\shorttitle{Modern geometry in not-so-high echelons of physics: Case studies}

\begin{document}
\pagerange{261}{359}
 \title{MODERN GEOMETRY IN NOT-SO-HIGH ECHELONS OF PHYSICS:\\ CASE STUDIES}
\author{M.~Fecko\email{fecko@fmph.uniba.sk}}
{Department of Theoretical Physics and Didactics of Physics,\\ Comenius University, Bratislava, Slovakia}

\abstract{In this mostly pedagogical tutorial article a brief introduction to modern geometrical treatment
     of fluid dynamics and electrodynamics is provided.
The main technical tool is standard theory of differential forms.
In fluid dynamics, the approach is based on general theory of integral invariants
     (due to Poincar\'e and Cartan).
Since this stuff is still not considered common knowledge,
the first chapter is devoted to an introductory and self-contained exposition
of both Poincar\'e version as well as Cartan's extension of the theory.
The main emphasis in fluid dynamics part of the text is on explaining basic classical results
on vorticity phenomenon (vortex lines, vortex filaments etc.) in ideal fluid.
In electrodynamics part, we stress the aspect of how different (in particular, rotating) observers
perceive the same space-time situation.
Suitable $3+1$ decomposition technique of differential forms proves to be useful for that.
As a representative (an simple) example we analyze Faraday's law of induction
(and explicitly compute the induced voltage) from this point of view.}

\vspace{0.3cm}

\begin{minipage}{2.5cm}
\quad{\small {\sf KEYWORDS:}}
\end{minipage}
\begin{minipage}{10cm}
Ideal fluid, barotropic flow, vortex lines, transport theorem, Helmholtz theorem,
lines of solenoidal field, integral invariant, 3+1 decomposition, rotating frame, Faraday's law
\end{minipage}

\tableofcontents
\setcounter{equation}{0} \setcounter{figure}{0} \setcounter{table}{0}\newpage

\section{Introduction}

Among theoretical physicists, modern differential geometry is typically associated with its ``higher echelons'',
like advanced general relativity, string theory, topologically non-trivial solutions in gauge theories,
Kaluza-Klein theories and so on.

However, geometrical methods also proved to be highly effective in several other branches of physics
    which are usually treated as more ``mundane'' or, put it differently,
    as ``not-so-high echelons'' of theoretical physics.
Good old fluid dynamics (or, more generally, dynamics of continuous media) and electrodynamics
may serve as prominent examples.

Nowadays, some education in modern differential geometry (manifolds, differential forms, Lie derivatives, ...)
   becomes a standard part of theoretical physics curriculum.
After learning those things, however, the potential strength of this mathematics is rarely demonstrated in real physics courses.

Although I definitely do not advocate entering of modern geometry into ``first round'' physics courses
    (of, say, above mentioned fluid dynamics and electrodynamics), it seems to me that to show how it is really used
     in some ``second round'' courses might be quite a good idea.
First, in this way some more advanced material in the particular subject may be explained
     in a simple and lucid way so typical for modern geometry.
Second, from the opposite perspective, this exposition is the best way to show
     how differential geometry itself really works.

If, on the contrary, this is not done so, modern geometry is segregated from real life and forced out to the above mentioned ``higher echelons'',
   with the natural consequence that for majority of students
   who put considerable energy into grasping this stuff in mathematics courses
   all their work is completely in vain.

Now, a few words about the structure of this tutorial article.

In the fluid dynamics part, we restrict to \emph{ideal} (inviscid) fluid and, in addition, are only interested
     in the \emph{barotropic} case (except for Ertel's theorem, which is more ge\-ne\-ral).
Our exposition rests on theory of \emph{integral invariants} due to Poincar\'e and Cartan.
I~think this approach is well suited for treating classical material concerning \emph{vorticity}
(like Helmholtz and Kelvin's theorems).
Perhaps it is worth noting that we treat fluid dynamics in terms of ``extended'' space
as the underlying manifold
(i.e. the space where time is a full-fledged dimension rather than just a parameter).

In electrodynamics part, we first derive Maxwell equations in terms of differential forms in 4-dimensional space-time
   (this is achieved by learning the structure of general forms in Minkowski space and checking it versus the standard
    3-dimensional version of the equations).
Then, in the second step, we introduce the concept of \emph{observer field} in space-time, intimately connected
to the concept of reference frame.
Using appropriate technique of 3+1 (space + time) decomposition of forms (and operations on them) with respect to the observer field
we can easily compute what various observers ``see'' when they ``look at invariant space-time electromagnetic reality''.
As an elementary example of this approach, we explicitly compute relativistically induced electric field seen in the rotating frame of a wire rim
as well as its line integral along the rim (the induced voltage) in the Faraday's law demonstration setting.

The reader is supposed to have basic working knowledge of modern geometry mentioned above
(manifolds, differential forms, Lie derivatives, ...;
if not, perhaps the crash course \cite{fecko2007} might help as a first aid treatment).
More tricky material is explained in the text
(and mostly a detailed computation is given when needed).

In Appendix \ref{app:vectoranalysis} we collect useful formulas which relate expressions in the language of differential forms
in 3-dimensional Euclidean space (as well as 4-dimensional Minkowski space) to their counterparts from usual vector analysis.
Ability to translate various expressions from one language to another (go back and forth at any moment)
is essential for effective use of forms in both fluid dynamics and electrodynamics.

Appendices \ref{app:cannotend}  and \ref{app:linesandtubes} are devoted to answering the question
whether field lines of solenoidal (\emph{divergence-free}) vector field indeed cannot end (or start) inside the
domain where there are defined (they \emph{can} :-).

\newpage

\section{Integral invariants - what Poincar\'e and what Cartan}
\label{euler_poincare_cartan}

\setcounter{equation}{0}
\subsection{Motivation - why the topic appears here}
\label{whyishere}

Theory of integral invariants is a well-established part of geometry
with various applications.
It is probably best-known from Hamiltonian mechanics.

Integral invariants were first formally introduced and studied by Poincar\'{e} in his
celebrated memoir \cite{Poincare1}.
He explained them in more detail in the book \cite{Poincare2}.
Then the concept was  developed by Cartan and summarized in his monograph \cite{cartan1922}.

What was Cartan's contribution?
Roughly speaking, while Poincar\'{e} considered invariants in phase space,
Cartan studied these objects in \emph{extended} phase space.
This led him to a true generalization:
one can associate, with each Poincar\'{e} invariant, corresponding Cartan's invariant.
The latter proves to be invariant with respect to a ``wider class of operations'' (see more below).
In addition, and this point of view will be of particular interest for us,
going from Poincar\'e version to Cartan's one may be regarded, in a sense,
as going from time-independent situation to time-dependent one.

Since the theory of integral invariants is both instructive in its own right
and used in Chapter \ref{euler}, we placed it in the very beginning of the paper.

In Chapter \ref{euler}, we first use its original Poincar\'e version in Section \ref{conseq_euler_stationary_barotropic}
and then, already the Cartan's extension, as a tool providing non-stationary
fluid dynamics equations from the known form of the stationary case in Sec \ref{euler_non_stationary}
and for gaining useful information from it in Sec. \ref{conseq_euler_non_stationary_barotropic}.

Remarkably, if we do it in this way, the resulting more general non-stationary equation looks
\emph{simpler} than the stationary one.
In addition, its consequences, like Helmholtz theorem on behavior of vortex lines in inviscid fluid,
look very naturally in this picture.

\setcounter{equation}{0}
\subsection{Poincar\'e}
\label{poincare}

Let's start with Poincar\'e invariants.

Consider a manifold $M$ endowed with dynamics given by a \emph{vector field} $v$
\begin{equation} \label{dynamicsgivenbyv}
       \dot \gamma = v
       \hskip 1.5cm
       {\dot x}^i = v^i(x)
\end{equation}
The field $v$ generates the dynamics (time evolution) via its flow $\Phi_t \leftrightarrow v$.
We will call the structure \emph{phase space}
\begin{equation} \label{defphasespace}
      (M,\Phi_t \leftrightarrow v) \hskip 1cm \text{{\color{red}\emph{phase space}}}
\end{equation}
In this situation, let's have a $k$-form $\alpha$ and consider its integrals over various
$k$-chains ($k$-dimensional surfaces) $c$ on $M$.
Due to the flow $\Phi_t$ corresponding to $v$, the $k$-chains flow away, $c \mapsto \Phi_t (c)$.
Compare the value of integral of $\alpha$ over the original $c$ and integral over $\Phi_t (c)$.
If, {\it for any chain} $c$, the two integrals are equal, it clearly reflects a remarkable
property of the form $\alpha$ with respect to the field $v$. We call it integral invariant:
\begin{equation} \label{integrinvariant}
       \int_{\Phi_t (c)} \alpha = \int_c \alpha
       \hskip 1cm \Leftrightarrow \hskip 1cm
       \int_c \alpha \  \  \  \text{is {\color{red}\emph{integral invariant}}}
\end{equation}
Let's see what this says for {\it infinitesimal} $t\equiv \epsilon$. Then
\begin{equation}
      \label{odtec1}
       \int_{\Phi_\epsilon (c)} \alpha
       = \int_c \alpha + \epsilon \int_c \mathcal L_v\alpha
\end{equation}
(plus, of course, higher order terms in $\epsilon$; here $\mathcal L_v$ is \emph{Lie derivative} along $v$).
So, in the case of integral invariant, the condition
\begin{equation} \label{podmienka1}
    \int_c \mathcal L_v\alpha = 0
\end{equation}
is to be fulfilled.
Since this is to be true {\it for each} $c$, the form (under the integral sign) itself
in (\ref{podmienka1}) is to vanish
\begin{equation} \label{podmienka2}
       \mathcal L_v\alpha = 0
\end{equation}
This is the \emph{differential version} of the statement (\ref{integrinvariant}).

There is, however, an important subclass of $k$-chains, namely $k$-\emph{cycles}.
These are chains whose boundary vanish:
\begin{equation} \label{defcyklu}
    \partial c = 0 \hskip 2cm c = \ \text{\emph{cycle}}
\end{equation}
In specific situations, it may be enough that some integral only behaves invariantly when restricted to cycles.
If this is the case, the condition (\ref{podmienka2}) is overly strong. It can be weakened to the requirement
that the form under the integral sign in (\ref{podmienka1}) is \emph{exact}, i.e.
\begin{equation} \label{podmienka3}
       \mathcal L_v\alpha = d\tilde \beta
\end{equation}
for some form $\tilde \beta$.
\vskip .5cm
\noindent $\blacktriangledown \hskip 0.5cm$
Indeed, in one direction, Eqs. (\ref{defcyklu}) and (\ref{podmienka3}) then give
\begin{equation} \label{podmienka33}
       \int_c \mathcal L_v\alpha  = \int_c d\tilde \beta = \int_{\partial c} \tilde \beta = 0
\end{equation}
so that (\ref{podmienka1}) \emph{is} fulfilled.
In the opposite direction, if the integral (\ref{podmienka1}) is to vanish for \emph{each cycle},
the form under the integral sign is to be \emph{exact}
(due to \emph{de Rham theorem}), so (\ref{podmienka3}) holds.
\hfill $\blacktriangle$ \par
\vskip .5cm

According to whether the integrals of forms are invariant for arbitrary $k$-chains or just for $k$-cycles,
integral invariants are divided into \emph{absolute} invariants (for any $k$-chains) and \emph{relative} ones
(just for $k$-cycles).
We can summarize what we learned yet as follows:
\begin{eqnarray} %{rcl}
      \label{odtec2}
      \{ \mathcal L_v\alpha = 0, \   c \ \text{arbitrary} \}
       &\Leftrightarrow& \int_c \alpha \  \  \  \text{is {\it absolute} integral invariant}     \\
       \label{odtec3}
      \{ \mathcal L_v\alpha = d\tilde \beta, \  c = \ \text{cycle} \}
       &\Leftrightarrow& \oint_c \alpha \  \  \  \text{is {\it relative} integral invariant}
\end{eqnarray}
Let's see, now, what else we can say about \emph{relative} integral invariants.
The condition (\ref{podmienka3}) may be rewritten (using $\mathcal L_v = i_vd+di_v$) as
\begin{equation} \label{ivdalphajeexaktna}
        i_vd\alpha = d\beta
\end{equation}
(where $\beta = \tilde \beta - i_v\alpha$).
Therefore it holds, trivially,
\begin{equation} \label{jetotoiste1}
        i_vd\alpha = d\beta
        \hskip 1cm \Leftrightarrow \hskip 1cm
        \mathcal L_v\alpha = d \tilde \beta
\end{equation}
and so also the following main statement on relative invariants (reformulation of (\ref{odtec3}))
is true:
\begin{equation} \label{jetotoiste2}
        \boxed{i_vd\alpha = d\beta
        \hskip .5cm \Leftrightarrow \hskip .5cm
        \mathcal L_v\alpha = d\tilde \beta
        \hskip .5cm \Leftrightarrow \hskip .5cm
        \oint_c\alpha = \ \ \text{\emph{relative} invariant}}
\end{equation}
So we can identify the presence of relative integral invariant \emph{in differential version}:
on phase space $(M,v)$, we see a form $\alpha$ fulfilling \emph{any} of the two equations mentioned in
Eq. (\ref{jetotoiste1}).

\vskip .5cm
\noindent
[Perhaps we should stress how the second equivalence sign is to be interpreted.
 There is no $\beta$ under the integral sign.
 Therefore, from the rightmost statement of Eq. (\ref{jetotoiste2}), it is not possible to reconstruct
 any \emph{particular} $\beta$, present in the leftmost statement.
 So one should read the second equivalence sign, in particular its right-to-left direction,
 as the assertion that, provided the rightmost statement holds, there \emph{exists} a form $\beta$
 such that the leftmost statement is true.
 (And, of course, one should adopt the same attitude with respect to the middle statement and $\tilde \beta$.)]
\vskip .5cm

Notice that, as a consequence of Eq. (\ref{podmienka3}), we also get the equation
\begin{equation} \label{dalphajeinvar}
        \mathcal L_v(d\alpha) = 0
\end{equation}
This says, however (see Eq. (\ref{odtec2})), that integral of $d\alpha$ is absolute integral invariant.
So, if we find a relative invariant given by $\alpha$,
then $d\alpha$ provides an absolute invariant:
\begin{equation} \label{pravidlorelabs}
       \boxed{\oint_c \alpha  \  \ \text{is relative invariant}
       \hskip .6cm \Rightarrow \hskip .6cm
       \int_D d\alpha  \  \ \text{is absolute invariant}}
\end{equation}
(here $\partial c=0$, whereas $\partial D$ may not vanish).

Conversely, if we find an absolute invariant then it is, clearly, also a relative one
(if something is true for \emph{all} chains then it is, in particular, true for closed chains, i.e. for cycles).
Absolute invariants thus present a part (subset) of relative invariants and the exterior derivative $d$
maps relative invariants into absolute invariants.
\vskip .5cm
\noindent
[Notice that whenever we find a ''good'' triple $(v,\alpha ,\beta)$ (i.e. $i_vd\alpha = d\beta$ holds),
 we can generate, for the same dynamics $(M,v)$, a series of additional ''good'' triples
\begin{eqnarray} %{rcl}
      \label{good1}
       (v,\alpha ,\beta )_{\text{new}}
       &\leftrightarrow& (v, \alpha \wedge d\alpha , 2\beta \wedge d\alpha)                           \\
       \label{good2}
       &\leftrightarrow&  (v, \alpha \wedge d\alpha \wedge d\alpha, 3\beta \wedge d\alpha \wedge d\alpha)  \\
       &\dots & \\
       \label{good3}
       &\leftrightarrow&  (v, \alpha \wedge (d\alpha)^k , (k+1)\beta \wedge (d\alpha)^k
                          \hskip 1cm k=0,1,2,\dots
\end{eqnarray}
(check) so that we get a series of relative invariants
\begin{equation}\label{newrelinv}
                      \oint_{c_{(0)}} \alpha
                \hskip .7cm
                      \oint_{c_{(1)}} \alpha \wedge d\alpha
                \hskip .7cm
                      \dots
                \hskip .71cm
                      \oint_{c_{(k)}} \alpha \wedge (d\alpha)^k
                      \hskip .7cm k=0,1,2,\dots
\end{equation}
(Here $c_{(k)}$ are \emph{cycles} of appropriate dimensions.
 For deg $d\alpha$ = odd and $k\ge 2$ we get, clearly, vacuous statements, since $(d\alpha)^k=0$.)]
\vskip .5cm
\noindent
{\bf Example \ref{poincare}.1}:
Consider the \emph{Hamiltonian} mechanics (the \emph{autonomous} case, yet, i.e. with the Hamiltonian $H$
independent of time).
Here the dynamical field $v$ is the \emph{Hamiltonian} field $\zeta_H$ given by the equation
\begin{equation} \label{hamiltonova1}
                 i_{\zeta_H}d\theta = -dH
                 \hskip 1cm \theta = p_adq^a
\end{equation}
(see Ch.14 in (\cite{fecko2006})). Comparison with Eq. (\ref{ivdalphajeexaktna})
\begin{equation} \label{porovnaniePoincHam}
                  i_{\zeta_H}d\theta
                = -dH
    \hskip 1cm  \leftrightarrow \hskip 1cm
                  i_v d\alpha
                = d\beta
\end{equation}
reveals that
a good $\alpha$ is the 1-form $\theta$.
The role of the corresponding form $\beta$ (potential) is played by the (minus) Hamiltonian $H$.
\vskip .5cm
\noindent
[Notice that this property of $\theta$ is actually true w.r.t. the field $v=\zeta_H$ for \emph{arbitrary} $H$,
 i.e. w.r.t. a whole \emph{family} of dynamical fields on $M$.
 So, in this particular realization  of the triple $(M,v,\alpha)$,
 a \emph{single} $\alpha$ is good for a whole \emph{family} of dynamical vector fields $v$
 (namely, for all \emph{Hamiltonian} fields).]
\vskip .5cm

According to Eqs. (\ref{good1}) - (\ref{good3}), we have also additional triples $(v,\alpha, \beta)$,
given as
\begin{eqnarray} %{rcl}
      \label{hampr1}
       (v,\alpha ,\beta )
       &\leftrightarrow& (\zeta_H, \theta , -H)                           \\
       \label{hampr2}
       &\leftrightarrow&  (\zeta_H, \theta \wedge \omega , -2H\omega)  \\
       \label{hampr3}
       &\leftrightarrow&  (\zeta_H, \theta \wedge \omega \wedge \omega,
                                                          -3H\omega \wedge \omega)   \\
       &\text{etc.}&
\end{eqnarray}
(where $\omega = d\theta$) and, consequently, relative integral invariants
\begin{equation}\label{newrelinvham}
                      \oint_{c_{(0)}} \theta
                \hskip .7cm
                      \oint_{c_{(1)}} \theta \wedge \omega
                \hskip .7cm
                      \dots
                \hskip .71cm
                      \oint_{c_{(k)}} \theta \wedge \omega^k
                      \hskip .7cm k=0,1,2,\dots
\end{equation}
Because of Eq. (\ref{pravidlorelabs}), we can also deduce that
\begin{equation}\label{newabsinvham}
                      \int_{D_{(0)}} \omega
                \hskip .7cm
                      \int_{D_{(1)}} \omega \wedge \omega
                \hskip .7cm
                      \dots
                \hskip .71cm
                      \int_{D_{(k)}} \omega \wedge \omega^k
                      \hskip .7cm k=0,1,2,\dots
\end{equation}
are absolute integral invariants.
The end of Example \ref{poincare}.1.

\setcounter{equation}{0}
\subsection{Life on $M\times \mathbb R$}
\label{formsonmtimesr}

In order to clearly understand Cartan's contribution to the field of integral invariants
(i.e. sections \ref{poincareincartans} and \ref{cartanincartans}),
a small technical digression might be useful.
What we need to understand is how differential forms
(as well as vector fields)
on $M$ and $M\times \mathbb R$ are related.

It is useful to interpret the $\mathbb R$-factor as \emph{time} axis added to $M$.
Then, if $M$ is phase space (see Eq. (\ref{defphasespace})), we call $M\times \mathbb R$
\emph{extended} phase space
\begin{equation} \label{defextphasespace}
      M\times \mathbb R \hskip 1cm \text{{\color{red}\emph{extended phase space}}}
\end{equation}

On $M \times \mathbb R$, a $p$-form $\alpha$ may be uniquely decomposed as
\begin{equation}
      \boxed{\alpha = dt \wedge \hat s + \hat r}
      \label{decomposition1}
\end{equation}
where both $\hat s$ and $\hat r$ are {\it spatial}, i.e. they do not contain the factor $dt$
in its coordinate presentation (the property of being spatial is denoted by hat symbol, here).
Simply, after writing the form in adapted coordinates $(x^i,t)$, i.e. in
those where $x^i$ come from $M$ and $t$ comes from $\mathbb R$,
one groups together all terms which do contain $dt$ once and, similarly,
all terms which do not contain $dt$ at all  (there is no other possibility :-).

Since spatial forms $\hat s$ and $\hat r$ do not contain $dt$, they look at first sight
(when written in coordinates), as if they lived on $M$ (rather than on $M \times \mathbb R$,
 where they actually live).

Notice, however, that $t$ still can enter {\it components} of \emph{any} form.
(And spatial forms are no exception.) We say that $\hat s$ and $\hat r$ are, in general,
\emph{time-dependent}.

 Therefore, when performing exterior derivative $d$ of a {\it spatial} form, say $\hat r$,
 there is a part, $\hat d \hat r$, which does not take into account the $t$-dependence
 of the components (if any; as if $d$ was performed just on $M$), plus a part which,
 on the contrary, sees the $t$ variable alone.
 (In Sec. \ref{faraday}, we encounter a more complicated version of $\hat d$.)
 Putting to\-gether, we have
 \begin{equation}
         \label{donspatial}
        d\hat r = dt \wedge \mathcal L_{\partial_t} \hat r + \hat d \hat r
\end{equation}
Then, for a \emph{general} form (\ref{decomposition1}), we get
\begin{equation}
       \boxed{d\alpha = dt \wedge (-\hat d \hat s + \mathcal L_{\partial_t} \hat r ) + \hat d \hat r}
       \label{dongeneral}
\end{equation}
Notice that the resulting form also has the general structure given in Eq. (\ref{decomposition1}).

Consider now an important particular case. There is a natural projection
\begin{equation}
       \pi:M \times \Bbb R \to M
       \hskip 1cm (m,t)\mapsto m
       \hskip 1cm (x^i,t)\mapsto x^i
       \label{projectiononm}
\end{equation}
We can use it to pull-back forms from $M$ onto $M \times \Bbb R$
\begin{equation}
       \pi^* : \Omega (M) \to \Omega (M \times \Bbb R)
       \label{pullbackform}
\end{equation}
From the coordinate presentation of $\pi: (x^i,t)\mapsto x^i$ we see, that
\emph{any} form $\rho$ on $M \times \Bbb R$, which results from such pull-back from $M$, is

1. spatial

2. time-independent

\noindent
And also the converse is clearly true: if a form on $M \times \Bbb R$ is \emph{both} spatial \emph{and}
 time-independent, then there is a \emph{unique} form on $M$ such that the form under consideration
 can be obtained as pull-back of the form on $M$ (just think in coordinates; the coordinate presentation
 of the two forms, in \emph{adapted} coordinates $(x^i,t)$ and $x^i$, \emph{coincides}).

\vskip .5cm
\noindent
[The two properties may also be expressed more invariantly:
\begin{eqnarray} %{rcl}
      \label{priestoroveformy}
    i_{\partial_t} \rho &=& 0  \hskip .6cm \text{{\color{red}\emph{spatial}}}
                               \hskip 2.2cm (\text{no $dt$ in coordinate presentation})          \\
       \label{tnezavisleformy}
      {\mathcal L}_{\partial_t} \rho &=& 0 \hskip .6cm \text{{\color{red}\emph{time-independent}}}
        \hskip .71cm (\text{no $t$ in components})
\end{eqnarray}
(notice that the vector field $\partial_t$ as well as the $1$-form $dt$ are \emph{canonical}
on $M\times \mathbb R$).]
\vskip .5cm

Take two such forms. Since they are spatial, we can denote them by
$\hat s :=\pi^* s$ and $\hat r :=\pi^* r$
(let their degrees be $p-1$ and $p$, respectively; the un-hatted forms $s$ and $r$ live on $M$)
and compose a form $\alpha$ on $M \times \Bbb R$ according to Eq. (\ref{decomposition1}).

Is the resulting $p$-form $\alpha$, for the most general choice of $s$ and $r$ on $M$, i.e. the form
\begin{equation}
      \alpha = dt \wedge \pi^* s + \pi^* r
      \label{decomposition2}
\end{equation}
the most general $p$-form on $M \times \Bbb R$? No, it is not, because of the \emph{property} 2.
of the forms $\hat s :=\pi^* s$ and $\hat r :=\pi^* r$. The forms
$\hat s$ and $\hat r$ obtained in \emph{this} particular way (as pull-backs of some $s$
and $r$ on $M$) are necessarily \emph{time-independent}, whereas in general
the two forms which figure in the decomposition (\ref{decomposition1})
need not be necessarily such; what is only strictly needed is the \emph{property} 1.,
they \emph{are} to be \emph{spatial}.

We can summarize the message of this part of the section by the following statements:
\vskip .2cm
\noindent
St.\ref{formsonmtimesr}.1.: Any form on $M \times \Bbb R$ decomposes according to Eq.
                            (\ref{decomposition1})

\noindent
St.\ref{formsonmtimesr}.2.: The forms $\hat s$ and $\hat r$, resulting from Eq.
                            (\ref{decomposition1}), are spatial

\noindent
St.\ref{formsonmtimesr}.3.: The forms $\hat s$ and $\hat r$ are not necessarily time-independent

\noindent
St.\ref{formsonmtimesr}.4.: A form on $M \times \Bbb R$ is both spatial and time-independent
\newline \indent \indent \indent iff it is pull-back from $M$
\vskip .2cm

If $\hat s$ and $\hat r$ (in the decomposition  (\ref{decomposition1}) of a general form $\alpha$
on $M \times \Bbb R$ are time-dependent,
a useful point of view (especially in physics) is to regard them as \emph{time-dependent} objects
living \emph{on} $M$.
\vskip .5cm
\noindent
[In this case, however, $t$ is no longer a coordinate, it becomes ``just a'' parameter.
  The point of Sec. \ref{formsonmtimesr} is, on the contrary, that going from $M$ to $M \times \Bbb R$
  may simplify the life in that we get \emph{standard} forms on $M \times \Bbb R$ rather than
  forms on $M$ carrying ``parameter-like'' labels.]
\vskip .5cm

And what about \emph{vector fields} on $M \times \Bbb R$? The situation is similar to forms:
a ge\-ne\-ral vector field $W$ may be uniquely decomposed into temporal and spatial parts
\begin{equation}
       W = a\partial_t +v = a(x,t)\partial_t +v^i(x,t)\partial_i
       \label{vectorfielddecomp}
\end{equation}
If $a(x,t)$ and $v^i(x,t)$ do not depend on time, the field $W$ on $M \times \Bbb R$ corresponds to
a pair of a scalar and a vector field \emph{on} $M$, otherwise a useful point of view
is to regard $W$ as a pair of \emph{time-dependent} scalar and vector field,
respectively, \emph{on} $M$.

In particular, consider a vector field of the structure
\begin{equation} \label{vectorfieldlikeonM}
      \xi = \partial_t +v
\end{equation}
with time-\emph{independent} components $v^i(x)$.
Its flow, taking place on the extended phase space $M \times \Bbb R$,
combines a trivial flow $t_0 \mapsto t_0+t$ along the temporal factor $\mathbb R$
with an independent flow on the phase-space factor $M$,
given by the vector field $v=v^i(x)\partial_i$ living on $M$.
This can be used from the opposite side: the dynamics on $M$ given by a vector field $v$
on $M$ (the situation considered in Sec. \ref{poincare}) may be \emph{equivalently} replaced by dynamics
on $M \times \Bbb R$, governed by the vector field (\ref{vectorfieldlikeonM}).
(If $(m_0,t_0)\mapsto (m(t),t_0+t)$ is the solution on $M \times \Bbb R$,
the solution of the original dynamics on $M$ is simply given by the projection of the result
onto the $M$ factor, i.e. as $m_0\mapsto m(t)$.)

\subsubsection{Digression: Reynolds transport theorem(s)}
\label{reynoldstransport}

Let's use the formalism introduced in Section \ref{formsonmtimesr} for a proof of
a classical theorem (see \cite{reynolds1903}), which is still widely used in applications.

Consider a spatial (possibly time-dependent) $k$-form on $M\times \mathbb R$
(i.e. a $k$-form $\alpha$ in Eq. (\ref{decomposition1}) with $\hat s =0$).
Fix a spatial $k$-chain $D_0$ in hyper-plane $t=t_0$ ($k$-dimensional surface whose points lie in the hyper-plane)
and let $D(t)=\Phi_t(D_0)$ be its image w.r.t. the flow $\Phi_t \leftrightarrow \xi$, where $\xi = \partial_t+v$
with spatial (possibly time-dependent) $v$. (Notice that $D(t)$ is spatial as well, it lies
in the hyper-plane with time coordinate fixed to $t_0+t$.)
Then integral of $\hat r$ over $D(t)$
is a function of time (because of time-dependence of both $D(t)$ and $\hat r$) and one
may be interested in its time derivative.
Using standard computation (for the last but one equation sign, see 4.4.2 in \cite{fecko2006}) we get
\begin{equation} \label{timederivative2a}
                  \frac{d}{dt}\int_{\Phi_t(D_0)} \hat r
                  =
                  \frac{d}{dt}\int_{D_0} \Phi_t^*\hat r
                  =
                  \int_{D_0} \frac{d}{dt}\Phi_t^*\hat r
                  =
                  \int_{D_0} \Phi_t^*{\mathcal L}_\xi \hat r
                  =
                  \int_{D(t)} {\mathcal L}_\xi \hat r
\end{equation}
Now
$$
\begin{array} {rcl}
  {\mathcal L}_\xi \hat r
  &=& i_\xi d\hat r  + di_\xi \hat r                                                   \\
  &=& i_\xi (\hat d \hat r +dt\wedge {\mathcal L}_{\partial_t} \hat r) + di_v \hat r   \\
  &=& (i_v \hat d \hat r + {\mathcal L}_{\partial_t} \hat r) + dt \wedge (\dots) + di_v \hat r   \\
\end{array}
$$
The details of (\dots) are of no interest since the term does not survive
(because of the presence of the factor $dt$) integration over \emph{spatial} surface $S(t)$.
Therefore, when this expression is plugged into Eq. (\ref{timederivative2a}) and
Stokes theorem is applied to the last term, we immediately get the desired general
``transport theorem'' in the form
\begin{equation} \label{timederivative2b}
            \boxed{\frac{d}{dt}\int_{D(t)} \hat r
                  =
                  \int_{D(t)} ({\mathcal L}_{\partial_t} \hat r + i_v \hat d \hat r)
                  +
                  \int_{\partial D(t)} i_v\hat r}
                  \hskip 1cm \text{{\color{red}\emph{transport theorem}}}
\end{equation}
Let us specify the result for the usual $3$-dimensional Euclidean space, $M=E^3$.
Here, we have the following expressions representing general spatial $k$-forms
\begin{equation} \label{allformsinE3}
           f(\bold r,t)
           \hskip 1cm
           \bold A(\bold r,t) \cdot d\bold r
           \hskip 1cm
           \bold B(\bold r,t) \cdot d\bold S
           \hskip 1cm
           h(\bold r,t) dV
\end{equation}
for $k=0,1,2$ and $3$, respectively.
Therefore, we have \emph{as many as four} versions of the transport theorem, here
(separate version for each $k$).
Namely, using well-known formulas from vector analysis in the language of differential forms
in $E^3$ (see Tab. \ref{tab1}), Eq. (\ref{timederivative2b}) takes the following four
appearances (so we get {\color{red}\emph{classical Reynolds transport theorems}}):
\begin{eqnarray} %{rcl}
      \label{transport0}
      &k=0& \hskip .1cm \frac{d}{dt}f(\bold r(t),t) \hskip .45cm
                  =
                   \ \partial_tf  + (\bold v \cdot \boldsymbol \nabla)f  \\
  \label{transport1}
      &k=1& \hskip .1cm \frac{d}{dt}\int_{c(t)} \bold A \cdot d\bold r \hskip .1cm
                  =
                  \int_{c(t)} (\partial_t\bold A + \curl \bold A \times \bold v) \cdot d\bold r
                  +
                  (\bold v \cdot \bold A)|_{P_1(t)}^{P_2(t)} \\
  \label{transport2}
      &k=2& \hskip .1cm \frac{d}{dt}\int_{S(t)} \bold A \cdot d\bold S
                  =
                  \int_{S(t)} (\partial_t\bold A + (\boldsymbol \nabla \cdot \bold A) \bold v) \cdot d\bold S
                  +
                  \oint_{\partial S(t)} (\bold A \times \bold v) \cdot d\bold r \\
  \label{transport3}
      &k=3& \hskip .1cm \frac{d}{dt}\int_{V(t)} fdV \hskip .25cm
                  =
                  \int_{V(t)} (\partial_tf)dV
                  +
                  \oint_{\partial V(t)} f\bold v \cdot d\bold S
\end{eqnarray}

\renewcommand{\arraystretch}{1.5}
\tabcolsep=10pt
\begin{table}[t]
\begin{center}
\begin{tabular}{|c||c|c|c|}
\hline
$\hat r$
&
$\hat d \hat r$
&
 $i_v\hat r$
&
 ${\mathcal L}_{\partial_t} \hat r$
 \\
\hline \hline
$f$ &
$\boldsymbol \nabla f\cdot d\bold r$
&
$0$
&
$\partial_tf$
\\
\hline
$\bold A \cdot d\bold r$ &
$(\curl \bold A) \cdot d\bold S$
&
$\bold v \cdot \bold A$
&
$(\partial_t\bold A) \cdot d\bold r$
\\
\hline
$\bold A \cdot d\bold S$ &
$(\Div \bold A) dV$
&
$(\bold A \times \bold v) \cdot d\bold r$
&
$(\partial_t\bold A) \cdot d\bold S$
\\
\hline
$fdV$  &
$0$
&
$f\bold v \cdot d\bold S$
&
$(\partial_tf)dV$
\qquad \\
\hline
\end{tabular}
\caption[whythis]{Relevant operations on (possibly time dependent) differential forms in $E^3$
                  (see Appendix \ref{app:vectoranalysis} or, in more detail, Sections 8.5 and 16.1 in \cite{fecko2006}).}
\label{tab1}
\end{center}
\end{table}
\renewcommand{\arraystretch}{1.0}
\tabcolsep=6pt

\noindent
Comments:

For Eq. (\ref{transport0}), recall that integral of a $0$-form $f$ over a point $P$
           is defined as $f(P)$ (evaluation of $f$ at $P$).
           So, the integral at the l.h.s. of Eq. (\ref{timederivative2b}) reduces to evaluation
of $f$ at $(\bold r(t),t)$.

In (\ref{transport1}), $c(t)$ is a (spatial) curve (at time $t$)
connecting $P_1(t)$ and $P_2(t)$, so that $\partial c(t)=P_2(t) - P_1(t)$.

In Eq. (\ref{transport2}), $S(t)$ is a (spatial) surface (at time $t$) with boundary $\partial S(t)$;
        see e.g. \$13.5 in \cite{nearing} and the end of Sec. \ref{inducedvoltage} here.

In Eq. (\ref{transport3}), $V(t)$ is a volume (at time $t$) with boundary $\partial V(t)$;
        in fluid dynamics, it is often referred to as \emph{material volume}
       (no mass is transported across the surface that encloses the volume).

\setcounter{equation}{0}
\subsection{Poincar\'e from Cartan's perspective}
\label{poincareincartans}

In this section, we present Cartan's point of view on (\ref{ivdalphajeexaktna})
and (\ref{jetotoiste2}).

First, we switch to \emph{extended} phase space $M\times \mathbb R$
and just retell, \emph{there}, the story considered in Sec. \ref{poincare}.
At the end, surprisingly, even at this stage of the game, we get \emph{more} than
we learned in Sec. \ref{poincare}.

We know from Eq. (\ref{vectorfieldlikeonM}) that rather than to study the
(dynamics given by the) vector field $v$ on $M$, we may (equivalently) study, on $M\times \mathbb R$,
the (dynamics given by the) field
\begin{equation} \label{definiciaxi1}
      \xi = \partial_t +v
\end{equation}
Now, our $v$ on $M$ satisfies $i_vd\alpha = d\beta$, i.e. Eq. (\ref{ivdalphajeexaktna}).
Cartan succeeded to find an equation \emph{on} $M\times \mathbb R$ which, in terms of the field
$\xi$, says the same. Construction of the resulting equation is as follows:

First, pull-back the forms $\alpha$ and $\beta$
(w.r.t. the natural projection $\pi: M\times \mathbb R \to M$) and get \emph{spatial}
and \emph{time-independent} forms $\hat \alpha = \pi^*\alpha$ and $\hat \beta = \pi^*\beta$
on $M\times \mathbb R$ (see Eq. (\ref{pullbackform}) and the text following the equation).

Second, combine them to produce the $k$-form $\sigma$ (\`{a} la Eq. (\ref{decomposition2})):
\begin{equation} \label{definiciasigma1}
       \sigma = \hat \alpha + dt \wedge \hat \beta
\end{equation}

Third, check that
\begin{equation} \label{zakladnarovnica1}
      i_\xi d\sigma = 0
\end{equation}
holds on $M\times \mathbb R$ if and only if $i_vd\alpha = d\beta$ is true on $M$.
\vskip .5cm
\noindent $\blacktriangledown \hskip 0.5cm$
Recall that $\mathcal L_{\partial_t}$ \emph{vanishes} on
$\hat \alpha = \pi^*\alpha$ and $\hat \beta = \pi^*\beta$.
Then, using Eq. (\ref{dongeneral}),
$$
\begin{array} {rcl}
  d\sigma
  &=& d(\hat \alpha + dt \wedge \hat \beta) \\
  &=& \hat d \hat \alpha + dt \wedge (\mathcal L_{\partial_t} \hat \alpha -\hat d \hat \beta) \\
  &=& \hat d \hat \alpha + dt \wedge ( -\hat d \hat \beta)
\end{array}
$$
and, due to Eq. (\ref{ivdalphajeexaktna}),
$$
\begin{array} {rcl}
  i_\xi d\sigma
  &=& i_{\partial_t +v}[\hat d \hat \alpha + dt \wedge (-\hat d \hat \beta)]           \\
  &=& (i_v\hat d \hat \alpha -\hat d \hat \beta) + dt \wedge (i_v \hat d \hat \beta)   \\
  &=& 0 + dt \wedge 0       \\
  &=& 0
\end{array}
$$
since we get from (\ref{ivdalphajeexaktna})
$$i_v \hat d \hat \beta = i_v i_v \hat d \hat \alpha = 0
$$
\hfill $\blacktriangle$ \par
\vskip .5cm
So indeed
\begin{eqnarray} %{rcl}
      \label{poincarecartan1}
      \boxed{i_v \hat d \hat \alpha = \hat d \hat \beta
  \hskip .5cm \Leftrightarrow \hskip .5cm
  i_\xi d\sigma =0}
  \hskip 1cm \text{for} \hskip 1cm
  \sigma &=& \hat \alpha + dt \wedge \hat \beta              \\
       \label{poincarecartan2}
       \xi    &=& \partial_t +v
\end{eqnarray}
holds.

Yet, we have just rewritten Eq. (\ref{ivdalphajeexaktna}),
which is a statement about something happening on phase space, into the form given
in (\ref{zakladnarovnica1}),
which is a statement about something happening on \emph{extended} phase space.

And what is it good for to switch from phase space to extended phase space?

In the first step, it reveals (as early as here, in Sec. \ref{poincareincartans})
that already using Poincar\'e's assumptions alone, a \emph{more general} statement about invariance,
in comparison with (\ref{jetotoiste2}), holds.

And in addition, in the second step (which we study in detail in Sec. \ref{cartanincartans}),
the structure of Eq. (\ref{zakladnarovnica1}) provides a hint to \emph{further}
generalization of Eq. (\ref{ivdalphajeexaktna}), such that the new, more general,
statement \emph{still} will be true.

So let us proceed to the first step.
In extended phase space $M\times \Bbb R$, consider integral curves of the field
$\xi = \partial_t +v$, i.e.~the~\emph{time development} curves.
\vskip .5cm
\noindent
[Formally, time development of \emph{points} in extended phase space is meant, here.
 In applications, the points may have various concrete interpretations.
 In fluid dynamics, as an example, the points correspond to positions of infinitesimal amounts
 of mass $dm$ of the fluid, so the curves correspond to the
 \emph{``real'' motion} of the fluid,
 whereas in Hamiltonian mechanics the points correspond to (abstract, pure) \emph{states}
 of the Hamiltonian system.]
\vskip .5cm

\begin{figure}[tb]
\begin{center}
\includegraphics[clip,scale=0.40]{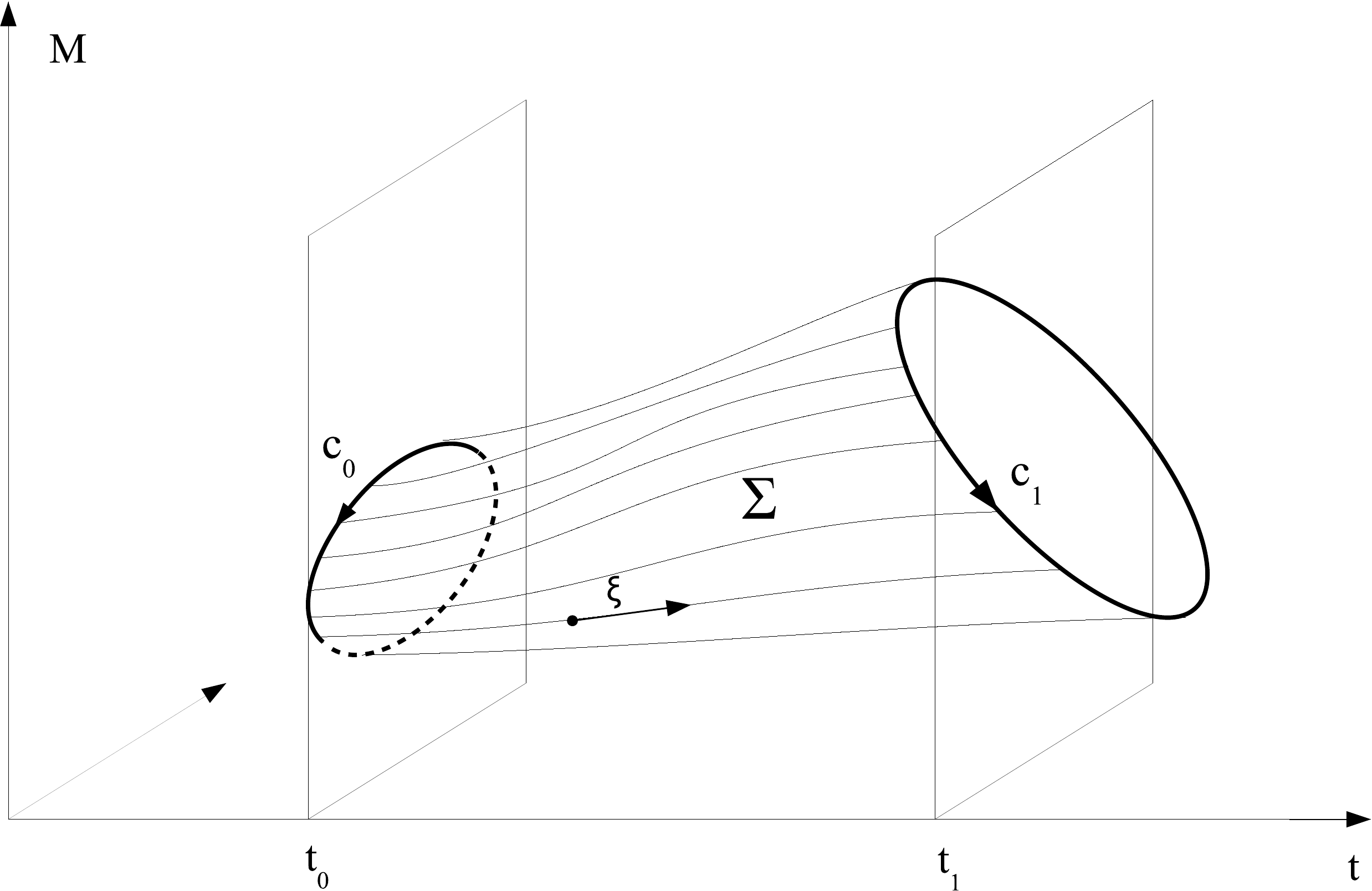}
\caption{The cycles $c_0$ and $c_1$ encircle the same tube
         of integral curves of the vector field $\xi = \partial_t +v$
         on extended phase space $M\times \Bbb R$;
         in general, they do not lie in hyper-planes of constant time.}
\label{c0c1encirclesigma1}
\end{center}
\end{figure}

Concentrate on a family of such integral curves given as follows:
Let their ``left ends'' ema\-nate from a $k$-cycle $c_0$ on $M\times \Bbb R$
(i.e. the points of the $k$-cycle $c_0$ serve as \emph{initial values} needed for the first-order
dynamics given by $\xi$) and ``right ends''
terminate at a $k$-cycle $c_1$ on $M\times \Bbb R$.
The family of such curves forms a $(k+1)$-chain (surface) $\Sigma$, whose boundary consists
of precisely the two cycles (closed surfaces) $c_0$ and $c_1$
\begin{equation} \label{hranicaretazca}
       \partial \Sigma = c_0 - c_1
\end{equation}
We say that the integral curves ``connecting'' the cycles $c_0$ and $c_1$ form a \emph{tube},
 and the cycle $c_0$ \emph{encircles} the tube. Then, clearly, the cycle $c_1$
 encircles \emph{the same} tube that $c_0$ does (see Fig.~\ref{c0c1encirclesigma1}).

\begin{figure}[tb]
\begin{center}
\includegraphics[clip,scale=0.40]{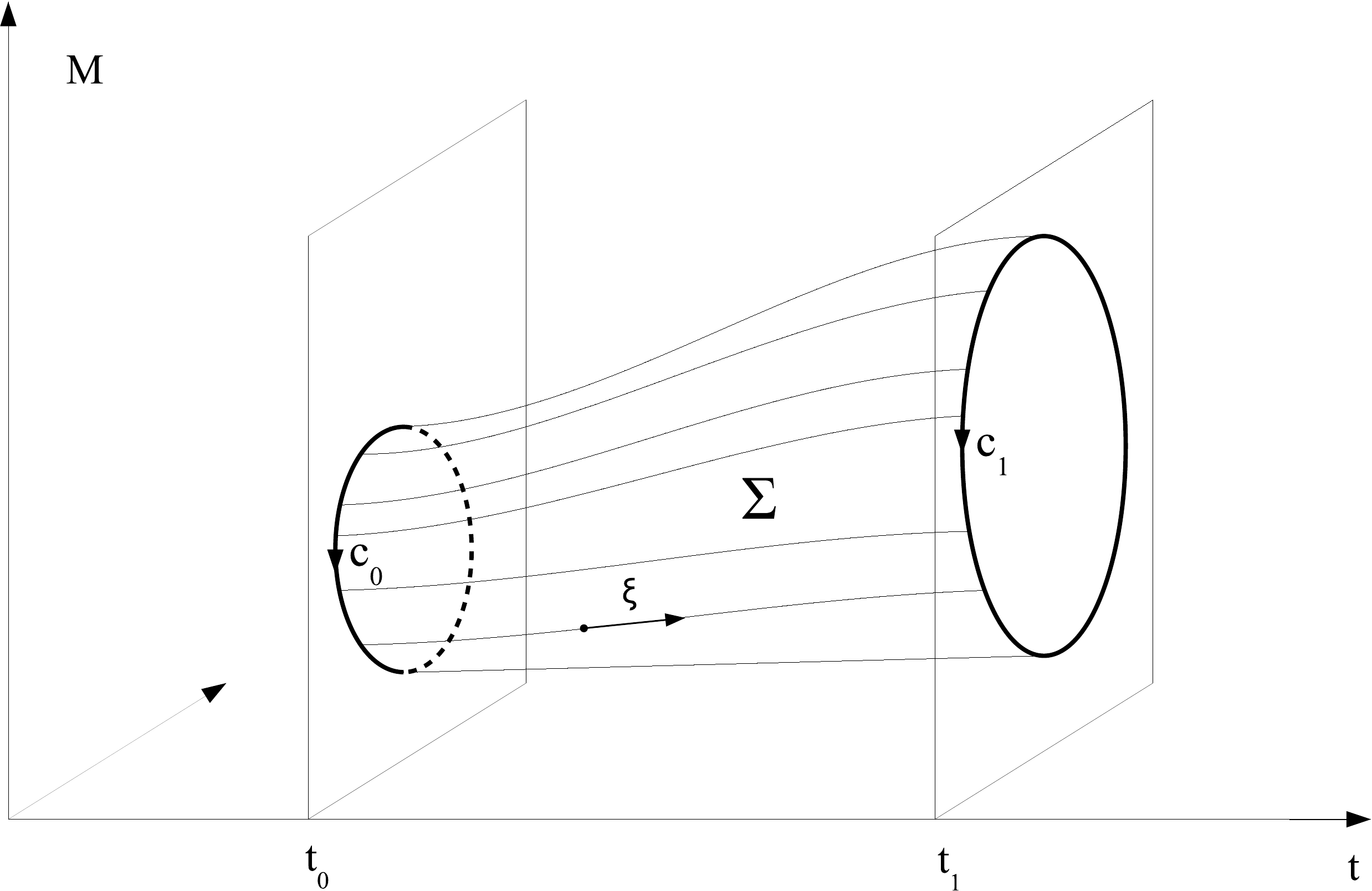}
\caption{The cycles $c_0$ and $c_1$ lie in hyper-planes of constant time and encircle the same tube
         of integral curves of the vector field $\xi = \partial_t +v$
         on extended phase space $M\times \Bbb R$.}
\label{c0c1encirclesigma2}
\end{center}
\end{figure}

\vskip .5cm
\noindent
[Here is an example of how such surface $\Sigma$ may be constructed
 (first very special, then its reshaping to a general one):
 take, in time $t_0$, a $k$-cycle in phase space $M$.
 We regard it as a $k$-cycle in the extended phase space $M\times \Bbb R$, which, by accident, completely
 lies in the hyperplane $t=t_0$.
 Now we let evolve all its points in time (according to the dynamics given by $\partial_t +v$).
 At time $t_1$ the family of curves produces, clearly, a new $k$-cycle in extended phase space $M\times \Bbb R$,
 lying completely in the hyperplane $t=t_1$, now. The points of the curves of time evolution
 between times $t_0$ and $t_1$ form together a $(k+1)$-dimensional surface $\Sigma$
 (rather special, yet; see Fig.~\ref{c0c1encirclesigma2}).

 If we proceed along the lines above, the two boundary cycles do lie in the hyper-surfaces of constant time.
 In general, it is not required, however, the boundary cycle $c_0$ (as well as $c_1$)
 is \emph{any} cycle in $M\times \Bbb R$, i.e. it may contain points
 \emph{at different times}.
 Such, more general, surface may be produced from the particular one described above as follows.
 We let flow the points of the particular $c_0$ along integral curves of the field $\xi$,
 with the parameter of the flow, however, being (smoothly) dependent of the point on $c_0$.
 What we get in this way still remains to be a cycle; its points, however, do not have, in general,
 the same value of the time coordinate (see Fig.~\ref{c0c1encirclesigma1}).]
\vskip .5cm

And the statement (already due to Cartan) is that the integral of the form $\sigma$ is
\emph{relative integral invariant}, which means, now, the following:
\begin{equation} \label{relatcartaninv}
       \oint_{c_0}\sigma = \oint_{c_1}\sigma
\end{equation}
where $c_0$ and $c_1$ are \emph{any two} cycles \emph{encircling a common tube}.

\vskip .5cm
\noindent $\blacktriangledown \hskip 0.5cm$
The proof is \emph{amazingly} simple:
\begin{eqnarray} %{rcl}
        \label{argumentarnold1}
  \int_{\Sigma} d\sigma
       &\overset{1.} {=}& \int_{\partial \Sigma} \sigma = \oint_{c_0} \sigma - \oint_{c_1} \sigma \\
               \label{argumentarnold2}
       &\overset{2.} {=}& 0
\end{eqnarray}
The second equality (saying that the surface integral actually vanishes) results from clever
observation how an \emph{elementary} contribution to the integral looks like:
In each point, $\Sigma$ locally spans on two vectors tangent to the surface
and one of them may be chosen to be the vector $\xi$.
So, in the process of integration of $d\sigma$ over $\Sigma$, one sums terms
of the structure
\begin{equation} \label{ofthestructure}
       d\sigma (\xi,\dots )\equiv i_{\xi}d\sigma (\dots)
\end{equation}
\emph{Any} such term, however, vanishes
due to the key equation (\ref{zakladnarovnica1}).
\hfill $\blacktriangle$ \par
\vskip .5cm

Therefore, the analogue of Eq. (\ref{jetotoiste2}) is the statement:
\begin{equation} \label{jetotoiste3}
         \boxed{i_\xi d\sigma = 0
        \hskip 1cm \Leftrightarrow \hskip 1cm
        \oint_c\sigma = \ \ \text{\emph{relative} invariant}}
\end{equation}
If we, already at this stage, make a comparison of the statement of Poincar\'e (\ref{jetotoiste2})
versus the corresponding one due to Cartan, (\ref{jetotoiste3}) and (\ref{relatcartaninv}),
we see that the Cartan's  one is \emph{stronger}.

For, if both cycles in Cartan's statement are \emph{special}, namely such that they lie in hyper-surfaces of
\emph{constant time}, we simply return to the Poincar\'e statement
(from the form $\sigma \equiv \hat \alpha + dt  \wedge \hat \beta$,
 it is enough to take seriously the part $\hat \alpha$, since the factor $dt$ \emph{vanishes} on
 special integration domains under consideration).
 If we use, however, general cycles allowed by Cartan, we get a \emph{brand new} statement,
 not mentioned at all by Poincar\'e.

Actually, in Sec. \ref{cartanincartans} we will see that the statement encoded in Eq. (\ref{jetotoiste3})
can be given \emph{even stronger} meaning.
\newline \newline \noindent
{\bf Example \ref{poincareincartans}.1}:
Let's return to \emph{Hamiltonian  mechanics} once again
(still the \emph{autonomous} case, i.e. with the Hamiltonian $H$ independent of time).
Putting together concrete objects from (\ref{hampr1}) and the general receipt from
(\ref{definiciasigma1}), we get the form $\sigma$ as follows
\begin{equation} \label{hamilsigma}
       \sigma = p_adq^a - Hdt
\end{equation}
The dynamical field $\xi$ becomes
\begin{equation} \label{hamilxi}
       \xi = \partial_t +\zeta_H
\end{equation}
and Hamilton equations take the form
\begin{equation} \label{hamilequations}
      \boxed{i_\xi d\sigma =0} \hskip 1cm \text{{\color{red}\emph{Hamilton equations}}}
\end{equation}
The general Cartan's statement (\ref{relatcartaninv}) is realized as follows:
\begin{equation} \label{relatcartaninvham1}
       \oint_{c_0}(p_adq^a - Hdt) = \oint_{c_1}(p_adq^a - Hdt)
\end{equation}
(where $c_0$ and $c_1$ encircle the same tube of solutions,
 so the situation is represented by Fig.~\ref{c0c1encirclesigma1})

If we choose the cycles $c_0$ and $c_1$ in constant time hyperplanes
(then $c_1$ results from time development of the cycle $c_0$), we get the \emph{original Poincar\'e} statement
\begin{equation} \label{relatcartaninvham2}
       \oint_{c_0}p_adq^a = \oint_{c_1}p_adq^a
       \hskip 1cm
       c_0 \ \text{at} \ t_0, \ c_1 \ \text{at} \ t_1
\end{equation}
(here, Fig.~\ref{c0c1encirclesigma2} is appropriate).
The end of Example \ref{poincareincartans}.1.

\setcounter{equation}{0}
\subsection{Cartan from Cartan's perspective}
\label{cartanincartans}

At the end of Sec. \ref{poincareincartans} we learned that the \emph{first} Cartan's generalization
of the statement of Poincar\'e consisted in observation that switching from phase space to extended phase space
and, at the same time, augmenting differential form under the integral sign
\begin{equation} \label{MnaMRaalphanasigmu}
       M \mapsto M \times \mathbb R
       \hskip 1cm
       \alpha \mapsto \sigma = \hat \alpha +dt \wedge \hat \beta
\end{equation}
(where $\beta$ is from $i_vd\alpha = d\beta$) enables one to extend the class of cycles, for which the integral is invariant
(namely from cycles which completely reside in hyper-planes of constant time,
 \`{a} la Fig. \ref{c0c1encirclesigma2}, to cycles whose points may have different values
 of time coordinate, \`{a} la Fig. \ref{c0c1encirclesigma1};
 what remains compulsory is just to encircle, by both cycles, common tube of trajectories in extended phase space).

However, according to Cartan, there is a \emph{still further} possibility how the situation may be generalized.

Recall that the forms $\hat \alpha$ and $\hat \beta$ on $M\times \Bbb R$, occurring in the formula
(\ref{definiciasigma1}),
were just the forms $\alpha$ and $\beta$ (defined in Eq. (\ref{jetotoiste2})) \emph{pulled-back} from $M$
\begin{equation} \label{hatalphahatbeta}
       \hat \alpha := \pi^* \alpha \hskip 1cm \hat \beta := \pi^* \beta
\end{equation}
w.r.t. the natural projection
\begin{equation} \label{projekcianaM}
       \pi:M\times \mathbb R \to M \hskip 1cm (m,t)\mapsto m \hskip 1cm (x^i,t)\mapsto x^i
\end{equation}
(So, no new input was added in comparison with the situation in Sec. \ref{poincare}
 considered by Poincar\'{e}.)
Because of this fact, the forms $\hat \alpha$ and $\hat \beta$ are both \emph{spatial} and
\emph{time-independent} (see the discussion near Eq. (\ref{decomposition2})).

Let us focus our attention, now, on the role of time-independence of the forms.
Imagine that the forms $\hat \alpha$ and $\hat \beta$ in the decomposition (\ref{definiciasigma1})
were time-\emph{dependent} (i.e., according to Eq. (\ref{decomposition1}),
that $\sigma$ was a \emph{general $k$-form} on the extended phase space $M\times \mathbb R$).
Does it mean that integrals of the form $\sigma$ over cycles encircling common tube of solutions
cease to be equal?

When we return to the (``amazingly simple'') proof given in Eqs. (\ref{argumentarnold1}) and
(\ref{argumentarnold2}) we see that the \emph{only} fact used was validity
of Eq. (\ref{zakladnarovnica1}), i.e. $i_\xi d\sigma = 0$ (see the l.h.s. of (\ref{jetotoiste3})).
Therefore, the Cartan's variant of the statement concerning integral invariants
\emph{still holds}.

The ``decomposed version'' of the equation $i_\xi d\sigma = 0$, however, gets a bit more complex
than $i_vd\alpha = d\beta$, now.
          Namely, if we (re)compute expression $i_\xi d\sigma$ (not assuming
\footnote{Contrary to the computation between (\ref{definiciasigma1}) and (\ref{poincarecartan1}),
           where time-independence \emph{was} used!}
           time-independence) and equate it to zero, we get
\begin{equation} \label{newequation}
       {\mathcal L}_{\partial_t} \hat \alpha +i_v \hat d \hat \alpha = \hat d \hat \beta
\end{equation}
So, we can say that
\begin{eqnarray} %{rcl}
      \label{poincarecartan3}
      \boxed{{\mathcal L}_{\partial_t} \hat \alpha +i_v \hat d \hat \alpha = \hat d \hat \beta
  \hskip .5cm \Leftrightarrow \hskip .5cm
  i_\xi d\sigma =0}
  \hskip 1cm \text{for} \hskip 1cm
  \sigma &=& \hat \alpha + dt \wedge \hat \beta              \\
       \label{poincarecartan4}
       \xi    &=& \partial_t +v
\end{eqnarray}
Notice that \emph{a new term},
\begin{equation} \label{newterm}
       {\mathcal L}_{\partial_t} \hat \alpha
\end{equation}
emerges the equation, in comparison with the time-independent case
(\ref{poincarecartan1}), (\ref{poincarecartan2}).
It is also worth noticing that time-derivative of \emph{the other} form, $\hat \beta$,
is \emph{absent} in the resulting equation.
\vskip .5cm
\noindent $\blacktriangledown \hskip 0.5cm$
Repeating once more the computation between (\ref{definiciasigma1}) and (\ref{poincarecartan1})
\emph{not assuming}, however, validity of (\ref{tnezavisleformy}), we get:

\begin{eqnarray} %{rcl}
  \label{desigma1}
  d\sigma
  &=& d(\hat \alpha + dt \wedge \hat \beta) \\
  \label{desigma2}
  &=& \hat d \hat \alpha + dt \wedge ({\mathcal L}_{\partial_t} \hat \alpha -\hat d \hat \beta)
\end{eqnarray}
\begin{eqnarray} %{rcl}
  \label{ixidesigma1}
  i_\xi d\sigma
  &=& i_{\partial_t +v}[\hat d \hat \alpha + dt \wedge ({\mathcal L}_{\partial_t} \hat \alpha - \hat d \hat \beta)]  \\
  \label{ixidesigma2}
  &=& ({\mathcal L}_{\partial_t} \hat \alpha -\hat d \hat \beta + i_v\hat d \hat \alpha)
                                                  - dt \wedge i_v ({\mathcal L}_{\partial_t} \hat \alpha -\hat d \hat \beta)
\end{eqnarray}
Equating this to zero is equivalent to writing down as many as \emph{two spatial} equations
\begin{eqnarray} %{rcl}
  \label{rovnica1}
  {\mathcal L}_{\partial_t} \hat \alpha +i_v\hat d \hat \alpha   &=& \hat d\hat \beta   \\
  \label{rovnica2}
  i_v ({\mathcal L}_{\partial_t} \hat \alpha -\hat d \hat \beta) &=& 0
\end{eqnarray}
The second equation is, however, a simple \emph{consequence} of the first one (just apply $i_v$ on the first),
so it is enough to consider the first equation alone.
\hfill $\blacktriangle$ \par
\vskip .5cm

Thus what Cartan added (as the \emph{second generalization} of Poincar\'e) was the possible
dependence of spatial forms on time.
Then, however, one must not forget, when writing the spatial version of the elegant equation
$i_\xi d\sigma =0$, to add the time-derivative term ${\mathcal L}_{\partial_t} \hat \alpha$.

So we conclude the section by stating the \emph{final Cartan's result}:

\begin{equation} \label{jetotoiste4}
        \boxed{{\mathcal L}_{\partial_t} \hat \alpha +i_v \hat d \hat \alpha = \hat d \hat \beta
        \hskip .5cm \Leftrightarrow \hskip .5cm
        i_\xi d\sigma =0
        \hskip .5cm \Leftrightarrow \hskip .5cm
        \oint_c\sigma = \ \ \text{\color{red}\emph{relative invariant}}}
\end{equation}
where the last statement means, in detail,
\begin{equation} \label{jetotoiste5}
        \boxed{\oint_{c_0}\sigma = \oint_{c_1}\sigma
               \hskip 1.5cm
               \text{if $c_0$ and $c_1$ {\color{red}\emph{encircle a common tube}} of solutions}}
\end{equation}

\begin{figure}[tb]
\begin{center}
\includegraphics[clip,scale=0.30]{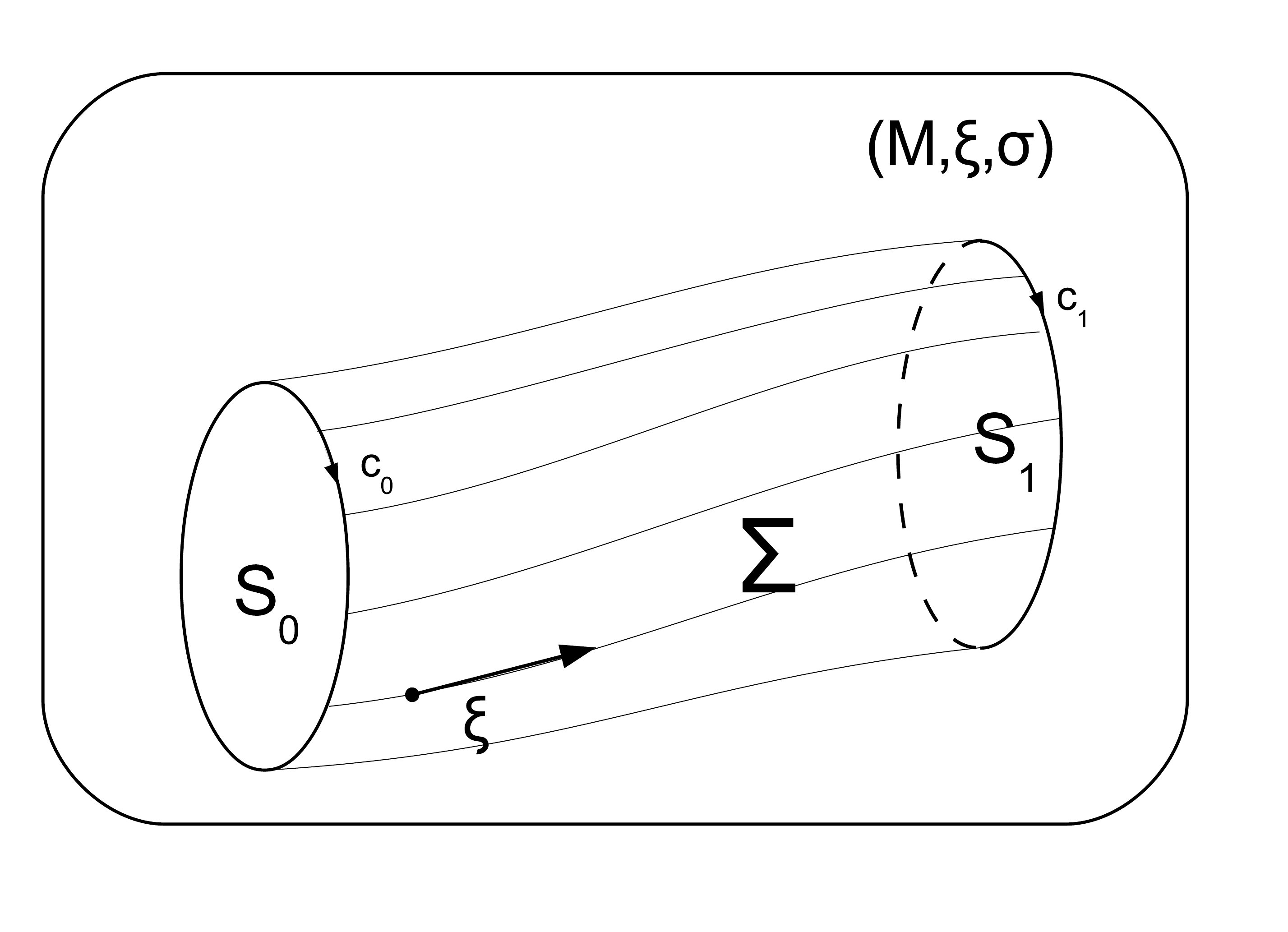}
\caption{$V$ is solid cylinder (the solid tube \emph{inside}) made of solutions emanating from the left cap $S_0$
         and entering the right cap $S_1$.
         Boundary $\partial V$ of the solid cylinder $V$ consists of 3 parts, hollow cylinder $\Sigma$ (``side'' of the solid cylinder),
         and the two caps, $S_0$ and $S_1$.
         The cycles $c_0$ and $c_1$ are boundaries of the caps, $c_0 = \partial S_0$ and $c_1= \partial S_1$.
         They encircle the same tube
         of integral curves of the vector field $\xi$
         on a manifold $(M,\xi,\sigma)$.}
\label{c0c1encirclesigma3}
\end{center}
\end{figure}

Similarly, one can write down a corresponding statement concerning \emph{absolute} invariant
obtained by integration of the \emph{exterior derivative} $d\sigma$ of $\sigma$:
\begin{equation} \label{jetotoiste6}
        \boxed{\int_{S_0}d\sigma = \int_{S_1}d\sigma
               \hskip .5cm
               \text{if $S_0$ and $S_1$ {\color{red}\emph{cut a common (solid) tube}} of solutions}}
\end{equation}
\vskip .2cm
\noindent $\blacktriangledown \hskip 0.5cm$
\newline \noindent
Proof 1.: Plug $c_0=\partial S_0$, $c_1=\partial S_1$ into Eq. (\ref{jetotoiste5})
          and use Stokes theorem.
\newline \noindent
Proof 2.: Start from scratch: consider a dynamical vector field $\xi$ on a manifold $\mathcal M$.
(So integral curves of $\xi$ are ``solutions'' and they define the dynamics on $\mathcal M$.)
Let $\xi$ satisfy $i_\xi d\sigma =0$ where $\sigma$ is a $k$-form on $\mathcal M$.
Now, consider $V$, the \emph{solid} tube of solutions. By this we mean the $(k+2)$-dimensional domain enclosed
          by the \emph{hollow} $(k+1)$-dimensional tube of solutions $\Sigma$ and two $(k+1)$-dimensional ``cross section''
          surfaces $S_0$ and $S_1$, see Fig. \ref{c0c1encirclesigma3}.
          So $\partial V = \Sigma + S_1 - S_0$
          (and $0=\partial \partial V = \partial \Sigma + c_1 - c_0$).
Then
\begin{equation} \label{solidtube}
                 0 = \int_{V}dd\sigma = \int_{\partial V}d\sigma = \int_{\Sigma}d\sigma +\int_{S_1}d\sigma -\int_{S_0}d\sigma
\end{equation}
But the integral over $\Sigma$ vanishes (due to the argument mentioned in Eq. (\ref{ofthestructure}))
and we get Eq. (\ref{jetotoiste6}).
\hfill $\blacktriangle$ \par
\vskip .5cm
\noindent
{\bf Example \ref{cartanincartans}.1}:
Third time is the charm - let's return again to {\it Hamiltonian} mechanics.
But now, for the first time, let's \emph{allow} condescendingly time-\emph{dependent} Hamiltonian $H$,
i.e. let's consider the general, {\it non-autonomous} case.

From the identification (cf. (\ref{hampr1}))
$$
   (v,\hat \alpha ,\hat \beta ) \leftrightarrow (\zeta_H, \theta , -H)
$$
we see, in spite of our generous offer, complete \emph{lack of interest}, in the case of the form
$\hat \alpha \equiv p_adq^a$, to depend on time.
This is \emph{not} the case, however, for $\hat \beta \equiv -H(q,p,t)$:
there we see a sincere interest to firmly grasp the chance of a lifetime.
But since time dependence of $\hat \alpha$ alone matters for the resulting equation (\ref{newequation}),
the spatial version of Hamiltonian equations
\begin{equation} \label{hamiltonequations}
                 i_\xi d \sigma  = 0
\end{equation}
remains, formally, completely \emph{intact},
\begin{equation} \label{hamiltonova4}
                 i_{\zeta_H}\hat \omega = -\hat dH
\end{equation}
(Its actual time \emph{dependence} is unobtrusively hidden inside $H$ and it penetrates, via equation
(\ref{hamiltonova4}), to the vector field $\zeta_H$ and, in the upshot, to the dynamics itself.)

\vskip .5cm
\noindent
[We know that if we write down Hamilton equations ``normally'', as
\begin{equation} \label{hamiltonove}
              {\dot q}^a = \frac{\partial H}{\partial p_a}
              \hskip 1cm
              {\dot p}_a = -\frac{\partial H}{\partial q^a}
\end{equation}
there is no visible formal difference, in the time-dependent Hamiltonian case, with respect to the case
when the Hamiltonian does not depend on time.
Of course, after unwinding the equations (performing explicitly the partial derivatives)
the equations get more complicated (since they are non-autonomous),
but \emph{prior to} the unwinding there is \emph{no extra term} because of time-dependent Hamiltonian.]
\vskip .5cm

The general Cartan's statement (\ref{relatcartaninv}) is still (also in non-autonomous case)
realized as follows:
\begin{equation} \label{relatcartaninvham}
       \boxed{\oint_{c_0}(p_adq^a - Hdt) = \oint_{c_1}(p_adq^a - Hdt)}
\end{equation}
if $c_0$ and $c_1$ encircle a common tube of solutions.
And Eq. (\ref{jetotoiste6}) adds that
\begin{equation} \label{absolcartaninvham}
       \boxed{\int_{S_0}(dp_a \wedge dq^a - dH\wedge dt) = \int_{S_1}(dp_a\wedge dq^a - dH\wedge dt)}
\end{equation}
if $S_0$ and $S_1$ cut (enclose) a common solid tube of solutions.
The end of Example \ref{cartanincartans}.1.
\newline \newline
And finally, let us make the following remark concerning \emph{absolute} integral invariants.
Recall that, still at the level of Poincar\'e (i.e. of Sec. \ref{poincare}),
absolute and relative invariants differ in that
the Lie derivative $\mathcal L_v\alpha$ vanishes (for absolute invariants, Eq. (\ref{podmienka2}))
or it is just exact, $d\tilde \beta$ (for relative ones, Eq. (\ref{podmienka3})).
The relative case was then rewritten into the form $i_vd\alpha = d\beta$ using
the identity $\mathcal L_v\alpha = i_vd\alpha +di_v\alpha$.
Notice, however, that the same identity enables one to write the ``absolute'' condition
$\mathcal L_v\alpha =0$ in the form of the ``relative'' one $i_vd\alpha = d\beta$;
one just needs to put
\begin{equation} \label{absoluteasrelative}
                 \beta = -i_v\alpha
\end{equation}
Then, when switching to Cartan's approach (including time-dependence of spatial forms),
we are to make corresponding changes in all formulas of interest.
We get, in this way, the following ``absolute invariant'' version of the original
``relative invariant'' statement given in Eqs. (\ref{poincarecartan3}) and (\ref{poincarecartan4}):
\begin{eqnarray} %{rcl}
      \label{abspoincarecartan3}
      \boxed{{\mathcal L}_{\partial_t}\hat \alpha  + {\hat {\mathcal L}}_v\hat \alpha = 0
  \hskip .5cm \Leftrightarrow \hskip .5cm
  i_\xi d\sigma =0}
  \hskip .5cm \text{for} \hskip .5cm
  \sigma &=& \hat \alpha - dt \wedge i_v\hat \alpha              \\
       \label{abspoincarecartan4}
       \xi    &=& \partial_t +v
\end{eqnarray}
where the following abbreviation
\begin{equation} \label{spatialLiederivative}
       {\hat {\mathcal L}}_v := i_v \hat d +\hat d i_v
       \hskip 2cm
       \text{\emph{spatial Lie derivative}}
\end{equation}
was introduced.

\vskip .5cm
\noindent $\blacktriangledown \hskip 0.5cm$
For new definition of $\sigma$ one just replaces $\hat \beta \mapsto -i_v\hat \alpha$;
$\xi$ remains intact. For the new spatial version of $i_\xi d\sigma =0$ we get
$$
\begin{array} {rcl}
  {\mathcal L}_{\partial_t} \hat \alpha +i_v\hat d \hat \alpha   &=& \hat d(-i_v\hat \alpha) \\
  {\mathcal L}_{\partial_t} \hat \alpha +(i_v\hat d \hat \alpha + \hat di_v\hat \alpha)  &=& 0 \\
  {\mathcal L}_{\partial_t} \hat \alpha + {\hat {\mathcal L}}_v \hat \alpha &=& 0
\end{array}
$$
Warning: notice that
$$ {\hat {\mathcal L}}_v
   \neq
   {\mathcal L}_v
$$
(since ${\hat {\mathcal L}}_v := i_v \hat d +\hat d i_v$ whereas ${\mathcal L}_v := i_v d + d i_v$;
 the hat matters :-).
Therefore
$$ {\mathcal L}_{\partial_t} + {\hat {\mathcal L}}_v
   \neq
   {\mathcal L}_{\partial_t} + {\mathcal L}_v
$$
i.e. the operator ${\mathcal L}_{\partial_t} + {\hat {\mathcal L}}_v$ acting on $\hat \alpha$
in Eq. (\ref{abspoincarecartan3}) \emph{should not} be confused with
${\mathcal L}_{\partial_t} + {\mathcal L}_v \equiv {\mathcal L}_{\partial_t +v} \equiv {\mathcal L}_{\xi}$.
\hfill $\blacktriangle$ \par
\vskip .5cm
\noindent
[Like in computation of \emph{spatial exterior} derivative $\hat d$ (see Eq. (\ref{donspatial})),
 the \emph{spatial Lie} derivative (of a spatial form $\hat \alpha$) simply does not take into account
 $t$-dependence of components (if any; as if it was performed just on $M$).
 Here, however, we speak of the $t$-dependence of components of \emph{both} $\hat \alpha$ \emph{and} $v$.]
\vskip .5cm

\setcounter{equation}{0}
\subsection{Continuity equation}
\label{continuity}

Let's start with\emph{ time-independent} case.

On $(M,v)$ one often encounters \emph{volume form} $\Omega$, i.e. a maximum degree, everywhere
non-vanishing differential form. Then we define the volume of a domain $D$ as
\begin{equation} \label{defvolume}
                 \text{vol} \ D := \int_{D} \Omega
                 \hskip 1.5cm
                 \text{\emph{volume of $D$}}
\end{equation}
Let $\rho$ be \emph{density} of some scalar quantity on $M$.
For concretness, let's speak of \emph{mass density}. Then
\begin{equation} \label{defmass}
                 m(D) := \int_D \rho \Omega
                 \hskip 1.5cm
                 \text{\emph{total mass in $D$}}
\end{equation}
(Clearly, we can treat in the same way other scalar quantities like, say, electric charge, entropy,
 number of states etc.)

Now what we mean by the statement that mass (or the scalar quantity in question) is \emph{conserved}?
Well, precisely that the integral in Eq. (\ref{defmass}) is to be promoted, in particular theory under discussion,
to be \emph{absolute integral invariant}:
\begin{equation} \label{massisconserved}
                 \boxed{
                 \int_D \rho \Omega
                 = \
                 \text{\emph{absolute integral invariant}}
                       }
\end{equation}
\vskip .5cm
\noindent
[Notice that it is integral Eq. (\ref{defmass}) rather than  Eq. (\ref{defvolume}) which
 is to be treated as integral invariant.
 The volume of some particular domain $D$ may change in time
 (except for very special cases, see Eq. (\ref{withOmega})),
 but the mass encompassed by the domain is to be constant since the velocity $v$
 is assumed to be identified with motion of the ``mass particles'',
 so the domain moves together with these ``particles'':
\begin{eqnarray} %{rcl}
      \label{volumenotconserved}
      \text{vol} \ D(t) &\neq& \text{vol} \ D(0) \hskip .5cm \text{in general}              \\
       \label{massconserved}
       m(D(t))          &=&    m(D(0)) \hskip .5cm \text{assumed}
\end{eqnarray}
(Here $D(t):= \Phi_t(D(0))$, $\Phi_t \leftrightarrow v$.
 Keep in mind, however, that ``mass'' is not to be interpreted literally, here.
 As an example it may be, as we already mentioned above, a quantity like appropriate probability
 or number of particles in Hamiltonian phase space, see Example \ref{continuity}.1).]
\vskip .5cm
As we know from Sec. \ref{poincare} (see Eq. (\ref{podmienka2})),
the differential version of the statement that Eq. (\ref{massisconserved}) represents
absolute integral invariant, reads
\begin{equation} \label{contstat1}
                 \mathcal L_v (\rho \Omega) = 0
\end{equation}
This is nothing but the \emph{continuity equation} for the \emph{time-independent case}.
It can also be expressed in several alternative (and more familiar) ways.

First, recall that \emph{divergence} of a vector field $u$ is defined by
\begin{equation} \label{divergencedef}
                 \mathcal L_u \Omega = : (\Div u)\Omega
\end{equation}
(see 8.2.1 and 14.3.7 in (\cite{fecko2006})). Then Eq. (\ref{contstat1}) is equivalent to
\begin{equation} \label{contstat2}
                  \boxed{
                  {\Div} (\rho v)  = 0
                       }
                 \hskip 1cm
                 \text{{\color{red}\emph{continuity equation}}} \ (\text{\emph{time-independent}})
\end{equation}
or, in a bit longer form, to
\begin{equation} \label{contstat3}
                 v\rho +\rho \ \Div v = 0
\end{equation}
\vskip .5cm
\noindent $\blacktriangledown \hskip 0.5cm$
First notice that
$$
\begin{array} {rcl}
  \mathcal L_v (\rho \Omega)
  &=& (i_vd+di_v) (\rho \Omega) \\
  &=& di_v (\rho \Omega) \\
  &=& d(i_{\rho v} \Omega) \\
  &=& \mathcal L_{\rho v} \Omega
\end{array}
$$
So, combining Eq. (\ref{contstat1}) with Eq. (\ref{divergencedef}) we get Eq. (\ref{contstat2}).
On the other hand,
$$
\begin{array} {rcl}
  \mathcal L_v (\rho \Omega)
  &=& (\mathcal L_v \rho) \Omega + \rho \mathcal L_v \Omega \\
  &=& (v \rho) \Omega + \rho (\Div v) \Omega \\
  &=& (v \rho + \rho \ \Div v) \Omega \\
\end{array}
$$
So vanishing of $\mathcal L_v (\rho \Omega)$ also leads to  Eq. (\ref{contstat3}).
\hfill $\blacktriangle$ \par
\vskip .5cm
Thus we can write \emph{continuity equation} (in the time-independent case) in \emph{any}
of the following four versions:
\begin{equation} \label{contstat4}
                 \mathcal L_v (\rho \Omega) = 0
                 \hskip .8cm
                 \mathcal L_{\rho v} \Omega = 0
                 \hskip .8cm
                 \Div (\rho v) = 0
                 \hskip .8cm
                 v\rho +\rho \ \Div v = 0
\end{equation}
This reduces, for \emph{incompressible} case (when the \emph{volume} itself is conserved), to
\emph{any} of the two versions:
\begin{equation} \label{contstatincompr}
                 \mathcal L_v \Omega = 0
                 \hskip 1.8cm
                 \Div  v = 0
                  \hskip 1.8cm
                 \text{\emph{incompressible}}
\end{equation}

Now we proceed to \emph{general}, possibly time-dependent, case.
In order to achieve this goal we can simply use the general procedure described
in Sec. \ref{cartanincartans}.
In particular, since our integral invariant is \emph{absolute},
we are to use the version based on Eqs. (\ref{abspoincarecartan3}) and (\ref{abspoincarecartan4}).

Namely, on $M\times \mathbb R$, we define
\begin{equation} \label{sigmaforconteq}
                 \sigma := \rho \hat \Omega - dt \wedge i_v \rho \hat \Omega
                 \hskip 1.8cm
                 \xi := \partial_t +v
\end{equation}
Then, according to Eq. (\ref{abspoincarecartan3}) the full, time-dependent version
of \emph{continuity equation} reads
\begin{equation} \label{conttimedepend1}
                 i_\xi d\sigma = 0
                 \hskip 1cm \text{or, equivalently} \hskip 1cm
                 {\mathcal L}_{\partial_t} (\rho \hat \Omega)  +
                     {\hat {\mathcal L}}_v (\rho \hat \Omega) = 0
\end{equation}
The spatial version can be further rewritten to the following, more standardly looking form:
\begin{equation} \label{conttimedepend3}
                 \boxed{
                 \partial_t \rho  + \hat {\Div} (\rho v)  = 0
                       }
                 \hskip 1cm
                 \text{{\color{red}\emph{continuity equation}}} \ (\text{\emph{general case}})
\end{equation}
where, for any \emph{spatial} vector field $u$, the following operation
\begin{equation} \label{spatialdivergence}
       {\hat {\mathcal L}}_u \hat \Omega = : (\hat {\Div} u) \ \hat \Omega
       \hskip 2cm
       \text{\emph{spatial divergence}}
\end{equation}
was introduced.
\vskip .5cm
\noindent $\blacktriangledown \hskip 0.5cm$
First notice that the volume form $\Omega$ on $M$ typically does not depend on time, so
${\mathcal L}_{\partial_t} \hat \Omega = 0$. Therefore
$$
\begin{array} {rcl}
   {\mathcal L}_{\partial_t} (\rho \hat \Omega)
  &=&  ({\mathcal L}_{\partial_t} \rho) \hat \Omega  \\
  &=&  (\partial_t \rho )\hat \Omega
\end{array}
$$
Now
$$
\begin{array} {rcl}
  {\hat {\mathcal L}}_v (\rho \hat \Omega)
  &=& \hat di_v (\rho \hat \Omega) + i_v\hat d (\rho \hat \Omega) \\
  &=& \hat di_{\rho v} \hat \Omega \\
  &=& \hat di_{\rho v} \hat \Omega + i_{\rho v} \hat d \hat \Omega \\
  &=& {\hat {\mathcal L}}_{\rho v} \hat \Omega \\
  &=& (\hat {\Div} \rho v) \ \hat \Omega
\end{array}
$$
(we used $\hat d (\rho \hat \Omega) = 0 = \hat d \hat \Omega$, since $\hat \Omega$
 already has maximum \emph{spatial} degree).
So, combining both results we get
$${\mathcal L}_{\partial_t} (\rho \hat \Omega)  +
                     {\hat {\mathcal L}}_v (\rho \hat \Omega) =
  (\partial_t \rho + \hat {\Div} \rho v) \ \hat \Omega
$$
from which Eq. (\ref{conttimedepend3}) follows.

Like in computation of \emph{spatial exterior} derivative $\hat d$ (see Eq. (\ref{donspatial}))
 and the \emph{spatial Lie} derivative ${\hat {\mathcal L}}_u$ (see Eq. (\ref{spatialLiederivative})),
 the \emph{spatial divergence} $\hat {\Div} u$ (of a \emph{spatial} vector field $u$)
 simply does not take into account
 $t$-dependence of its components (if any; as if it was performed just on $M$).
\hfill $\blacktriangle$ \par
\vskip .5cm
An important special case represents the situation when \emph{both} $\text{vol} \ D$ \emph{and} $m(D)$
are absolute integral invariants, i.e. both \emph{volume and mass} are conserved.
(See Example \ref{continuity}.1 illustrating this phenomenon in \emph{Hamiltonian mechanics}
 and Section \ref{massandentropy_stat}, where we encounter it in ideal fluid dynamics.)
Here, rather than just Eq. (\ref{conttimedepend1}), as many as \emph{two} similar equations hold,
one for $\sigma$ containing $\rho \hat \Omega$ and one for $\sigma$ with just $\hat \Omega$:
\begin{eqnarray} %{rcl}
      \label{withOmega}
      i_\xi d\sigma_1    = 0 \hskip 1cm \sigma_1    &:=&  \hat \Omega - dt \wedge i_v \hat \Omega
                             \hskip 1cm \text{(volume conserved)}     \\
       \label{withrhoOmega}
      i_\xi d\sigma_\rho = 0 \hskip 1cm \sigma_\rho &:=&  \rho \hat \Omega - dt \wedge i_v \rho \hat \Omega
                             \hskip 1cm \text{(mass conserved)}
\end{eqnarray}
Clearly, the general continuity equation, Eq. (\ref{conttimedepend3}), is still true
(because of Eq. (\ref{withrhoOmega})).
But the additional piece of wisdom contained in Eq. (\ref{withOmega})
also enables one to write a brand new, much \emph{simpler}
equation, namely
\begin{equation} \label{liouville}
                 \boxed{
                 \xi \rho = 0
                       }
                 \hskip .5cm \text{i.e.} \hskip .5cm
                 \boxed{
                 \partial_t \rho + v\rho = 0
                       }
                 \hskip 1cm \text{{\color{red}\emph{Liouville equation}}}
\end{equation}
\vskip .5cm
\noindent
[This may also be grasped intuitively: if volume is conserved and, in addition, the ``weighted'' volume
 is conserved as well, the ``weight'' itself (the scalar multiple of the volume form) is to be conserved.
 Conserved here means constant along dynamical curves, so application of $\xi$ on the scalar function,
 i.e. differentiation along dynamical curves, is to vanish.]
\vskip .5cm
\noindent $\blacktriangledown \hskip 0.5cm$
First notice that $\sigma_\rho = \rho \sigma_1$. Therefore
$$
\begin{array} {rcl}
   i_\xi d\sigma_\rho
  &=&  i_\xi d(\rho \sigma_1)   \\
  &=&  i_\xi (d\rho \wedge \sigma_1) +\rho i_\xi d\sigma_1 \\
  &=&  (\xi\rho) \sigma_1 -d\rho \wedge i_\xi \sigma_1 +\rho i_\xi d\sigma_1
\end{array}
$$
Now $i_\xi d\sigma_1$ vanishes due to Eq. (\ref{withOmega}) and
$$
\begin{array} {rcl}
  i_\xi \sigma_1
  &=& i_\xi (\hat \Omega - dt \wedge i_v \hat \Omega) \\
  &=& i_v \hat \Omega - i_v \hat \Omega + dt \wedge i_\xi i_v \hat \Omega \\
  &=& dt \wedge i_v i_v \hat \Omega  \\
  &=& 0
\end{array}
$$
(we used that both $\hat \Omega$ and $i_v\hat \Omega$ are \emph{spatial} and $\xi = \partial_t +v$).
So, combining all results we get
$$
  i_\xi d\sigma_\rho = (\xi\rho) \sigma_1
$$
from which (together with Eq. (\ref{withrhoOmega})), finally, Eq. (\ref{liouville}) follows.
\hfill $\blacktriangle$ \par
\vskip .5cm
\noindent
{\bf Example \ref{continuity}.1}:
Fourth time is the charm - let's return again to general,
time-\emph{dependent Hamiltonian} mechanics.

The role of $\Omega$ on $M$ is played by (a \emph{constant} multiple of) the $n$-th power of $\hat \omega$
(present in Hamilton equations (\ref{hamiltonova4}), see 14.3.6 and 14.3.7 in \cite{fecko2006})
\begin{equation} \label{volumehamilton}
              \hat \Omega \propto \hat \omega \wedge \dots \wedge \hat \omega
              \hskip 1cm
              \text{\emph{Liouville form}}
\end{equation}
(see the last integral in Eq. (\ref{newabsinvham})).
Then, using the philosophy of Eq. (\ref{abspoincarecartan3}), we can switch to time-dependent case
by constructing
\begin{equation} \label{volumehamiltontrick}
              \hat \Omega - dt \wedge i_{\zeta_H} \hat \Omega
\end{equation}
Integral of this form is absolute invariant in the (broader) sense of Cartan
(i.e. with general solid tube of solutions, \`{a} la Eq. (\ref{jetotoiste6})).
Standardly only (narrower) ``Poincar\'e version'' is used (with the integrals restricted to two
fixed-time hypersurfaces)
and it is then nothing but the celebrated Liouville theorem
on conservation of the phase space volume
\begin{equation} \label{Liouvilletheorem}
              \boxed{\int_{\hat D}\hat \Omega = \int_{\Phi_t (\hat D)}\hat \Omega}
              \hskip 1cm
              \text{{\color{red}\emph{Liouville theorem}}}
\end{equation}
(where $\hat D$ is any \emph{spatial} $2n$-dimensional domain). Notice that the theorem
is still true in time-dependent case.

In classical \emph{statistical mechanics} a state, say at time $t=0$, is given
in terms of \emph{distribution function} $\rho$ on $M$.
By definition, \emph{probability} of finding a particle within $D\subset M$ is given by
the very expression Eq. (\ref{defmass})
\begin{equation} \label{distributionfunc}
                 m(D) := \int_D \rho \Omega
                 \hskip 1.5cm
                 \text{\emph{probability to find particle in $D\subset M$}}
\end{equation}
(Note that the ``total mass'' is equal to unity, here. If total number of particles is $N$,
 $Nm(D)$ is number of particles in $D\subset M$.)
This integral is, however, \emph{also con\-ser\-ved}.
\vskip .5cm
\noindent $\blacktriangledown \hskip 0.5cm$
Indeed, since $M=D(0) \cup (M\backslash D(0))$,
the probability $p$ to find particle within $D(t)\equiv \Phi_t(D)$ at time $t$
is equal to $p_1p_2+p_3p_4$, where

- $p_1$ is probability to find it within $D(0)\equiv D$ at time $t=0$

- $p_2$ is probability to find it within $D(t)$ at $t$ provided it was in $D(0)$ at $t=0$

- $p_3$ is probability to find it outside $D(0)$ at time $t=0$

- $p_4$ is probability to find it within $D(t)$ at $t$ provided it was outside $D(0)$ at $t=0$.

Now $p_2=1$ (trivially, by definition of $D(t)$ as image of $D(0)$ w.r.t. the dynamics),
$p_4=0$ since trajectories \emph{do not intersect}
(no points from outside can penetrate inside). So, $p=p_1$, i.e. $m(D(t))=m(D(0))$.
\hfill $\blacktriangle$ \par
\vskip .5cm

Therefore, $m(D)$ is indeed an absolute integral invariant, too. This means,
for the distribution function $\rho$
(already in Cartan's language, as a function on \emph{extended} phase space $M\times \mathbb R$)
that it fulfills Liouville equation (\ref{liouville}). Since $v=\zeta_H$, here, it reads
\begin{equation} \label{liouvilleinhamil1}
                 \xi \rho = 0
                 \hskip 1cm \text{i.e.} \hskip 1cm
                 \partial_t \rho + \zeta_H \rho = 0
\end{equation}
In canonical coordinates $(x^a,p_a)$ on $M$, we have
\begin{equation} \label{hamfield}
                 \zeta_H = \frac{\partial H}{\partial p_a}
                           \frac{\partial}{\partial x^a}
                         -
                           \frac{\partial H}{\partial x^a}
                           \frac{\partial}{\partial p_a}
                 \hskip 1cm \text{i.e.} \hskip 1cm
                 \zeta_H \rho = \{ H,\rho   \}
\end{equation}
where
\begin{equation} \label{poisbracket}
                 \{ f,h \} :=
                           \frac{\partial f}{\partial p_a }
                           \frac{\partial h}{\partial x^a}
                         -
                           \frac{\partial f}{\partial q^a}
                           \frac{\partial h}{\partial p_a}
                 \hskip 1cm \text{\emph{Poisson bracket}}
\end{equation}
In terms of Poisson bracket, Eq. (\ref{liouvilleinhamil1})
may be written, at last, in its well-known form
\begin{equation} \label{liouvilleinhamil2}
                 \boxed{\partial_t \rho + \{ H,\rho   \} = 0}
                 \hskip .5cm \text{{\color{red}\emph{Liouville equation}}}
                 \hskip .3cm \text{(in \emph{Hamiltonian mechanics})}
\end{equation}
The end of Example \ref{continuity}.1.

\setcounter{equation}{0}
\subsection{Remarkable integral surfaces}
\label{remarkable_surface}

It turns out that, under general conditions studied by Cartan, one can find a family of surfaces,
whose behavior is truly remarkable.
Namely, the family is \emph{invariant} w.r.t. the flow $\Phi_t$ of vector field $\xi$.
Put it differently, if one takes such a surface $\mathcal S$ and lets it evolve in time
($\mathcal S \mapsto \Phi_t(\mathcal S) \equiv \mathcal S(t)$), the resulting surface
         $\mathcal S(t)$ is again a member of the family.

As we will see, first in Sec. \ref{euler_stationary_helmholtz}
and then in Sec.
\ref{euler_non_stationary_helmholtz}, vortex lines in fluid dynamics are just particular
(one-dimensional) cases of the surfaces.
In this sense, the surfaces may be regarded as generalization of vortex lines
\footnote{Btw. I am not aware of whether this material
           is known in the literature.}
and their property mentioned above is a generalization of Helmholtz's celebrated result
on vortex lines ``frozen into fluid''.

Let us see how (simply) this comes about.

Consider the general situation in Cartan's approach to relative integral invariants
(described in Section \ref{cartanincartans}),
i.e. a $k$-form $\sigma$ and a dynamical vector field $\xi$ given by Eqs.
(\ref{poincarecartan3}) and (\ref{poincarecartan4}) respectively and related by equation $i_\xi d\sigma =0$.

Now, consider two \emph{distributions} on $M\times \mathbb R$, given by
(those vectors which annihilate) the forms $d\sigma$ and $dt$, respectively:
\begin{eqnarray} %{rcl}
      \label{distribution1}
      \mathcal D^{(1)}
         &:=& \{ \text{vectors} \ \ w \ \ \text{such that} \ \  i_w d\sigma = 0 \ \ \text{holds}  \} \\
  \label{distribution2}
      \mathcal D^{(2)}
         &:=& \{ \text{vectors} \ \ w \ \ \text{such that} \ \  i_w dt = 0 \ \ \ \text{holds}  \}
\end{eqnarray}
Their intersection is the distribution
\begin{equation} \label{distribution3}
       \mathcal D \equiv \mathcal D^{(1)} \cap \mathcal D^{(2)}
         := \{ \text{vectors} \ w \ \text{such that} \  i_w d\sigma = 0 \ \text{and} \ i_w dt = 0 \ \text{holds}  \}
\end{equation}
Both distributions $\mathcal D^{(1)}$ and $\mathcal D^{(2)}$ are \emph{integrable}, so that we can, locally,
consider their \emph{integral surfaces}.
It is clear that intersections of integral surfaces of distributions $\mathcal D^{(1)}$ and $\mathcal D^{(2)}$
are integral surfaces in its own right, namely of the distribution $\mathcal D$.

\vskip .5cm
\noindent $\blacktriangledown \hskip 0.5cm$
Recall a version of the integrability criterion due to Frobenius:
a distribution is integrable if, along with any two vector fields belonging to the distribution,
the same holds for their commutator
(see Sec. \ref{integrabilityandspatial} here and Sec. 19.3 in \cite{fecko2006}).

So, let $u,w$ $\in \mathcal D^{(1)}$, i.e. $i_ud\sigma = i_wd\sigma =0$.

Then, using identity $[\mathcal L_u,i_w]=i_{[u,w]}$ (see 6.2.9 in \cite{fecko2006}), we have
$$i_{[u,w]}d\sigma = \mathcal L_ui_w d\sigma - i_w\mathcal L_u d\sigma = - i_w(i_ud+di_u) d\sigma = 0
$$
Therefore, the commutator belongs to the distribution as well. The same holds for $dt$.

Btw. integrability of $\mathcal D^{(2)}$ is clear from the outset - integral submanifolds are simply fixed-time
 hyper-surfaces $t=$ const. Vectors belonging to $\mathcal D^{(2)}$ are just \emph{spatial} vectors
 introduced in (\ref{vectorfielddecomp}).

One should notice that the issue of \emph{dimension} of the distribution $\mathcal D$
needs more careful examination of \emph{ranks} of the forms involved, in particular of
rank of the form $d\sigma$. (In general, the rank of a form may not be constant and, consequently,
the dimension may vary from point to point.)
\hfill $\blacktriangle$ \par
\vskip .5cm
Now, both distributions $\mathcal D^{(1)}$ and $\mathcal D^{(2)}$ happen to be, in addition to their integrability,
\emph{invariant} w.r.t. the time development, i.e. w.r.t. the flow $\Phi_t \leftrightarrow \xi$
\begin{equation} \label{distrinvariant1}
               \Phi_t(\mathcal D^{(1)}) = \mathcal D^{(1)}
               \hskip 1cm
               \Phi_t(\mathcal D^{(2)}) = \mathcal D^{(2)}
\end{equation}

\vskip .5cm
\noindent $\blacktriangledown \hskip 0.5cm$
We have
$$
\begin{array} {rcl}
       \mathcal L_\xi(d\sigma)
  &=&  (i_\xi d+di_\xi)(d\sigma) = d(i_\xi d\sigma) = 0 \\
       \mathcal L_\xi(dt)
  &=&  d(\xi t) = d (\partial_t t) = d1 =0
\end{array}
$$
But
\begin{equation} \label{distrinvariant3}
       \mathcal L_\xi(d\sigma) = 0
       \hskip 1cm
       \mathcal L_\xi(dt) = 0
\end{equation}
is just infinitesimal version of
\begin{equation} \label{distrinvariant2}
       \Phi_t^*(d\sigma) = d\sigma
       \hskip 1cm
       \Phi_t^*(dt) = dt
\end{equation}
Invariance of generating differential forms, however, results in invariance of the corresponding
distributions.
\hfill $\blacktriangle$ \par
\vskip .5cm
Therefore, also the ``combined'' distribution $\mathcal D$ \emph{is invariant} w.r.t. the time development
\begin{equation} \label{distrinvariant4}
               \Phi_t(\mathcal D) = \mathcal D
\end{equation}
And, consequently, any integral surface $\mathcal S$ of the distribution $\mathcal D$
evolves to the surface $\Phi_t(\mathcal S)\equiv \mathcal S(t)$ which is \emph{again} integral surface of the (same !)
distribution $\mathcal D$:
\begin{figure}[tb]
\begin{center}
\includegraphics[clip,scale=0.40]{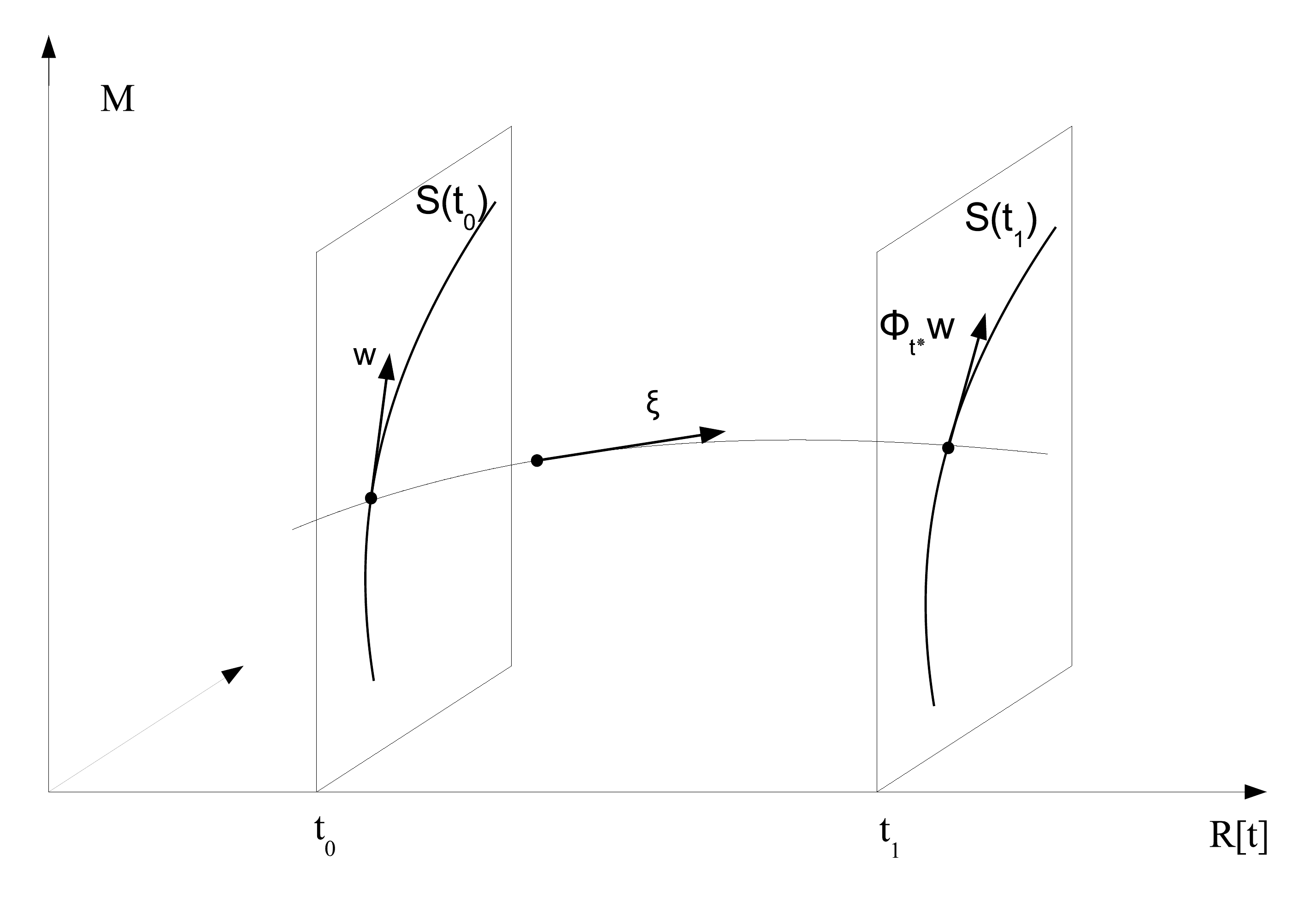}
\caption{$S(t_0)$ is the integral surface of $\mathcal D$ passing through a point on $M$ at time $t_0$
         and $w$ is a vector in this point tangent to $S(t_0)$.
         The flow maps $w$ to $\Phi_{t*}w$. $S(t_1)$ is the integral surface of $\mathcal D$ passing through the point of tangency of $\Phi_{t*}w$.
         It turns out that (locally) the surface $S(t_1)$ is nothing but the image of the surface $S(t_0)$ w.r.t. the flow of $\xi$.
         So, in fluid dynamics parlance, integral surfaces of $\mathcal D$ are \emph{frozen} into ``fluid''.}
\label{remarkablesurfaces}
\end{center}
\end{figure}
\begin{equation} \label{surfaceinvariant1}
       \boxed{
       \{ \mathcal S \ \text{is integral surface of} \ \mathcal D \}
       \hskip .5cm \Rightarrow \hskip .5cm
       \{ \mathcal S(t) \ \text{is integral surface of} \ \mathcal D \}
             }
\end{equation}
\vskip .5cm
\noindent $\blacktriangledown \hskip 0.5cm$
Let $w$ be tangent to $\mathcal S$ (see Fig. \ref{remarkablesurfaces}).
So, at some point of $\mathcal S$, it annihilates both $d\sigma$ and $dt$.
As $\mathcal S$ evolves to $\Phi_t(\mathcal S)\equiv \mathcal S(t)$, the tangent vector $w$
evolves to $\Phi_{t*}w$ (it follows from the \emph{definition} of push-forward operation).
The issue is whether $\Phi_{t*}w$ is tangent to $\mathcal S(t)$ or, put another way, whether
\begin{equation} \label{issuew1}
                 i_{w}d\sigma = 0
                 \hskip 1cm \Rightarrow \hskip 1cm
                 i_{\Phi_{t*} w}d\sigma = 0
\end{equation}
\begin{equation} \label{issuew2}
                 i_{w}dt = 0
                 \hskip 1.1cm \Rightarrow \hskip 1cm
                 i_{\Phi_{t*} w}dt = 0
\end{equation}
holds (i.e. whether also $\Phi_{t*}w$ annihilates both $d\sigma$ and $dt$).

For $dt$, it is straightforward (since it is just a 1-form):
\begin{equation} \label{fordtyes}
                 i_{\Phi_{t*} w}dt = (dt)(\Phi_{t*} w) = (\Phi^*_t dt)(w) = dt (w) = i_wdt =0
\end{equation}

For $d\sigma$ we have to take care of more arguments (since $\sigma$ is a $k$-form):
the issue is whether
\begin{equation} \label{issuew3}
                 (d\sigma)( \Phi_{t*} w,u,\dots )= 0
\end{equation}
\emph{for any} $u,\dots$ sitting at the same point as $\Phi_{t*} w$.
But \emph{any} $u,\dots$ may be regarded as $\Phi_{t*} \tilde u, \dots$ for some $\tilde u,\dots$
(sitting at the same point as $w$; namely $\tilde u=\Phi_{-t*} u$).
So, we get for the l.h.s. of Eq. (\ref{issuew3})
\begin{equation} \label{issue0}
                 (d\sigma)( \Phi_{t*} w,\Phi_{t*} \tilde u,\dots)
              =  (\Phi_t^* d\sigma)(w, u,\dots)
              =  (d\sigma)(w,u,\dots)
              =  (i_wd\sigma)(u,\dots)
              = 0
\end{equation}
Therefore, in general, whenever a form is invariant w.r.t. the flow $\Phi_t$,
vectors which annihilate the form at some time flow to vectors
which also do annihilate the form at later times. And this means that
integral surfaces given by the form always flow to integral surfaces of the form again.
\hfill $\blacktriangle$ \par
\vskip .5cm
Now, when applied to fluid dynamics (vortex lines),
it turns out to be fairly useful to understand the matter also from the perspective of $M$ alone
(rather than only on $M\times \mathbb R$).

Well, first recall that the whole theory about integral invariants only holds when the equation
$i_{\xi}d\sigma = 0$ is true.
Therefore, we can use it whenever we need.
And we could need it, for example, to \emph{rewrite the form} $d\sigma$ \emph{itself}.
Namely, we have:
\begin{eqnarray} %{rcl}
      \label{dsigmavzdyvseob}
      d\sigma &=& \hat d \hat \alpha + dt \wedge ({\mathcal L}_{\partial_t} \hat \alpha -\hat d \hat \beta)
                  \hskip .8cm \text{always} \\
  \label{dsigmanariesenivseob}
        &=& \hat d \hat \alpha +dt\wedge (-i_v\hat d\hat \alpha)
                       \hskip 1.4cm \text{\emph{on solutions} of} \hskip .8cm
                       i_{\xi}d\sigma = 0
\end{eqnarray}
The first line is simply the result of straightforward computation (see Eq. (\ref{desigma2})).
The second line arises when \emph{spatial} version of the equation $i_{\xi}d\sigma = 0$,
i.e. the leftmost equation in (\ref{jetotoiste4}), is used in the first line.

Then note that the key distribution $\mathcal D$ may also be characterized as being generated
by the forms $\hat d \hat \alpha$ and $dt$ instead of $d\sigma$ and $dt$:
\begin{eqnarray} %{rcl}
      \label{distribution3old}
      \mathcal D
         &:=& \{ \text{vectors} \ w \ \ \text{such that} \ \ i_w d\sigma = 0 \ \ \text{and} \ \ i_w dt = 0 \ \ \text{holds}  \} \\
  \label{distribution3new}
         &:=& \{ \text{vectors} \ w \ \ \text{such that} \ \ i_w \hat d \hat \alpha = 0 \ \ \text{and} \ \ i_w dt = 0 \ \ \text{holds}  \}
\end{eqnarray}
\vskip .5cm
\noindent $\blacktriangledown \hskip 0.5cm$
Indeed, because of Eq. (\ref{dsigmanariesenivseob}), we have
$$
\begin{array} {rcl}
  i_wd\sigma
  &=& i_w\hat d \hat \alpha +i_wdt \wedge (-i_v\hat d\hat \alpha) +dt\wedge (i_wi_v\hat d\hat \alpha) \\
  &=& i_w\hat d \hat \alpha +i_wdt \wedge (-i_v\hat d\hat \alpha) -dt\wedge (i_v(i_w\hat d\hat \alpha)) \\
\end{array}
$$
If we denote $i_w\hat d \hat \alpha \equiv \hat b$ (it is a \emph{spatial} $1$-form) and $i_wdt \equiv c$
(a function), we get
$$
  i_wd\sigma
  = \hat b -c(i_v\hat d\hat \alpha) -dt\wedge i_v\hat b
$$
from which immediately results
$$
             \{
             i_w d\sigma = 0
             \hskip .3cm \text{and} \hskip .3cm
             c =0
             \}
             \hskip 1cm \Leftrightarrow \hskip 1cm
             \{
             \hat b  = 0
             \hskip .3cm \text{and} \hskip .3cm
             c =0
             \}
$$
\hfill $\blacktriangle$ \par
\vskip .3cm \indent
The very concept of (a single) surface $\mathcal S$ is actually naturally tied to $M$ itself:
since $\mathcal S$ always lies in a \emph{fixed-time}
hyper-surface of $M\times \mathbb R$, it may be regarded, at any time $t$, as lying just on $M$.

Then a possible point of view of the situation is that, \emph{on} $M$ \emph{alone}, we have
\emph{time-dependent} form $\hat d \hat \alpha$ generating, consequently,
\emph{time-dependent distribution}
\begin{equation} \label{distributiononM}
       \hat {\mathcal D}
         := \{ \text{vectors} \ w \ \text{on $M$ such that} \  i_w \hat d \hat \alpha = 0 \ \text{holds}  \}
\end{equation}
Since the form $\hat d \hat \alpha$ depends on time, the family of all integral surfaces depends
on time as well.
And the (nontrivial) statement is that if we pick up, at time $t=0$, particular integral surface
$\mathcal S$ and let it evolve in time, i.e. compute $\mathcal S \mapsto \mathcal S(t)$
using \emph{time-dependent} vector field $v$ on $M$, then, at time $t$,
\begin{equation} \label{statementonM1}
 \text{the \emph{evolved surface} $\mathcal S(t)$ is \emph{integral surface} of the \emph{evolved distribution},}
\end{equation}
i.e. $\mathcal S(t)$ is an element of the evolved family of integral surfaces.
\vskip .5cm
\noindent
[This is indeed non-trivial. Time evolution of the surface, $\mathcal S \mapsto \mathcal S(t)$,
 is based on properties of (time-dependent) vector field $v$. On the other hand,
 time evolution of the distribution, $\hat {\mathcal D} \mapsto \hat {\mathcal D}(t)$,
 is based on properties of the (time-dependent) differential form $\hat d \hat \alpha$.
 Therefore, the statement Eq. (\ref{statementonM1}) assumes, at each time, a precise well-tuned relation between
 $v$ and $\hat d \hat \alpha$.
 Validity of the statement shows that Eq. (\ref{newequation}), the spatial version of $i_\xi d\sigma =0$,
 provides exactly the needed relation.]
\vskip .5cm
This can also be restated in more figurative ``fluid dynamic'' parlance, regarding $v$ as
the \emph{velocity} field of an abstract ``fluid'' on ($n$-dimensional!) $M$:
\begin{equation} \label{statementonM2}
      \boxed{\text{The surfaces $\mathcal S(t)$ are \emph{frozen} into the ``fluid''.}}
\end{equation}

\setcounter{equation}{0} \setcounter{figure}{0} \setcounter{table}{0}\newpage
\section{Ideal fluid dynamics and vorticity}
\label{euler}

In this section a geometric formulation of the dynamics of ideal fluid, is presented.
The dynamics is described, as is well known, by
\begin{equation} \label{eulernorm}
                 \boxed{\rho \left(\partial_t \bold v + (\bold v \cdot \boldsymbol \nabla) \bold v \right)
                        =
                        - \boldsymbol \nabla p -\rho \boldsymbol \nabla \Phi}
                        \hskip 1cm
                        \text{{\color{red}\emph{Euler equation}}}
\end{equation}
Here the mass density $\rho$, the velocity field $\bold v$, the pressure $p$ and the potential
$\Phi$ of the volume force field
($g z$ for the usual gravitational field) are functions of $\bold r$ and $t$.
In what follows we rewrite the equation into the form
(see Eq. (\ref{eulerstatform3}) in Sec. (\ref{euler_stationary_barotropic})
 and Eq. (\ref{eulernestac2}) in Sec. (\ref{eulerintegrinvarnonstat})),
from which classical theorems of Kelvin (on conservation of velocity circulation or vortex flux),
Helmholtz (on conservation of vortex lines) and Ertel
(on conservation of a more complicated quantity), result with remarkable ease.

The exposition goes as follows.

First, in Sec. \ref{euler_stationary},
we present a formalism appropriate for stationary flow and we show, in Sec. \ref{euler_stationary_helmholtz},
how Helmholtz statement may be extracted in this particular case.

Then, in Sec. \ref{euler_non_stationary}, we use our knowledge of integral invariants stuff
developed in Sec. \ref{euler_poincare_cartan}
(in particular, \emph{Cartan}'s contribution to Poincar\'{e} picture from Sec. \ref{cartanincartans})
to obtain Euler equation for general, not necessarily stationary, flow in a remarkably succinct form.
It turns out (see Sec. \ref{euler_non_stationary_helmholtz}) that this presentation
of Euler equation enables one to understand the Helmholtz statement in more or less
the same way as it was the case for the stationary situation in Sec. \ref{euler_stationary_helmholtz}.

As we already mentioned in the Introduction,
our treatment of fluid dynamics is based on systematic use of ``extended'' space
(with coordinates $(\bold r,t)$) as the underlying manifold,
i.e. we work on space where time is a \emph{full-fledged dimension} rather than just a parameter.
In our opinion this approach makes the topic simpler.

\setcounter{equation}{0}
\subsection{Stationary flow}
\label{euler_stationary}

The aim of this subsection is to formulate the Euler equation for stationary flow
in the language of differential forms.
Later on, when discussing vortex lines (in Sec. \ref{euler_stationary_helmholtz}),
this proves to be very convenient.

\subsubsection{Acceleration term and covariant derivative}
\label{euler_stationary_covariant}

For stationary flow, all time derivatives vanish, so we get from Eq. (\ref{eulernorm})
\begin{equation} \label{eulerstatnorm1}
                     \rho (\bold v \cdot \boldsymbol \nabla) \bold v
                   =
                   - \boldsymbol \nabla p -\rho \boldsymbol \nabla \Phi
\end{equation}
Here the mass density $\rho$, the velocity field $\bold v$, the pressure $p$ and the potential
$\Phi$ of the volume force field are only functions of $\bold r$.
So the underlying manifold, where everything takes place, is the common Euclidean space $E^3$.

Recall, where the \emph{acceleration} term
\begin{equation} \label{accelstat1}
                 \bold a  = (\bold v \cdot \boldsymbol \nabla) \bold v
\end{equation}
in the (stationary) Euler equation comes from:
One compares the velocity of a mass element in a slightly later moment $t +\epsilon$
with the velocity  of the same mass element just now (at time $t$).
And, of course, one should not forget about the fact that the element
which is at $t$ situated at $\bold r$, is at $t +\epsilon$ situated at a slightly shifted place,
namely at $\bold r +\epsilon \bold v(\bold r)$:
\begin{equation} \label{accelstat2}
                 \bold a (\bold r) = \frac{\bold v(\bold r+\epsilon \bold v(\bold r))
                                     - \bold v(\bold r)}{\epsilon}
\end{equation}
This leads directly to (\ref{accelstat1}).

It also reveals, that we actually encounter \emph{covariant} derivative
(of $\bold v$ along $\bold v$), here:
\begin{equation} \label{accelstatcovar}
                 \bold a = \nabla_{\bold v}\bold v
\end{equation}
This is because, in order to make the comparison legal,
we are to \emph{translate} the ``later'' velocity vector into the point of the ``sooner'' one
(both velocities are to sit at a common point in order they may be subtracted
 one from another)
and the path of the element alone (rather than trajectories of neighboring elements, too)
is enough for gaining the resulting vector
(so the translation is \emph{parallel} and, consequently,
 the derivative is covariant rather than Lie).
So, equation (\ref{eulerstatnorm1}) becomes
\begin{equation} \label{eulerstatgeom1}
                 \nabla_{\bold v}\bold v
                   = - \frac{1}{\rho} \ \boldsymbol \nabla p - \boldsymbol \nabla \Phi
\end{equation}
Ignoring bold-face (i.e. using standard geometrical notation), we write it as
\begin{equation} \label{eulerstatgeom2}
                 \nabla_v v
                   = - \frac{1}{\rho} \ \grad p - \grad \Phi
\end{equation}

\subsubsection{Vorticity two-form and vortex lines}
\label{euler_stationary_vorticity}

In what follows (see Sec. \ref{euler_stationary_helmholtz}; then also \ref{euler_non_stationary_helmholtz}),
we will be interested in behavior, under the flow of ideal fluid (given by Eq. (\ref{eulerstatgeom2})),
of \emph{vortex lines}, i.e. the lines tangent, at each point, to \emph{vorticity} vector field
\begin{equation} \label{defvorticityvector}
                 \boldsymbol \omega := \curl \bold v
\end{equation}
If the lines are (arbitrarily) parametrized by some $\lambda$, the corresponding curves become
$\bold r(\lambda)$ and the tangent vector is ${\bold r}'$.
(The prime means differentiation w.r.t. $\lambda$, here.)

By definition, this is to be parallel to $\boldsymbol \omega$.
Therefore the (differential) equation for computing vortex lines may be written as
\begin{equation} \label{eqforvortelines1}
                 {\bold r}' \times \curl \bold v = \bold 0
\end{equation}
Equation (\ref{eqforvortelines1}) can be rewritten in terms of \emph{differential forms}.
Why we should do this?
The point is that, in a while, we succeed to do the same with Euler equation (\ref{eulerstatgeom2}).
And it turns out then
that the vortex lines stuff is handled with remarkable ease
in the language of differential forms.

Velocity field, $v$, is a \emph{vector} field.
Information stored in it may equally well be expressed via the corresponding \emph{covector}
field $\tilde v$ (\emph{velocity $1$-form}) .
This is simply defined through ``lowering index'' procedure
(with respect to the standard metric tensor in $E^3$)
\begin{equation} \label{lowering_index}
                 v \mapsto \tilde v \equiv g(v, \ . \ )
                 \hskip 1cm \text{i.e.} \hskip 1cm
                 v^i \mapsto  v_i \equiv g_{ij}v^j
\end{equation}
Now, the theory of differential forms, when applied to standard 3-dimensional vector analysis,
teaches us (see Appendix \ref{app:vectoranalysis} or, in more detail, \$8.5 in \cite{fecko2006}) that
\begin{eqnarray} %{rcl}
      \label{velocityform}
            \tilde v  &=& \bold v \cdot d\bold r \\
      \label{vorticityform}
           d\tilde v  &=& (\curl \bold v) \cdot d\bold S \ \equiv \ \boldsymbol \omega \cdot d\bold S \\
      \label{vortexlinesexpr}
           i_{\gamma'}d\tilde v  &=& ({\bold r}' \times \curl \bold v) \cdot d\bold r
\end{eqnarray}
Here, $\gamma'$ is just an abstract notation for the tangent vector
${\bold r}'$ to the curve $\gamma (\lambda) \leftrightarrow \bold r(\lambda)$.
\vskip .5cm
\noindent
[It is more common to denote the abstract tangent vector to a curve $\gamma$ as $\dot \gamma$.
 Here, however, it might cause a confusion: there is a \emph{time} development of points, here, too
 (each point of the fluid flows along the vector field $v$) and dot also standardly denotes
 the \emph{time} derivative, i.e., here, it might also denote the directional derivative
 along the streamlines of the flow. Our prime, on the other hand, denotes the directional derivative,
 \emph{at a fixed time}, along the \emph{vortex} line $\gamma$.]
\vskip .5cm
\begin{figure}[tb]
\begin{center}
\includegraphics[clip,scale=0.30]{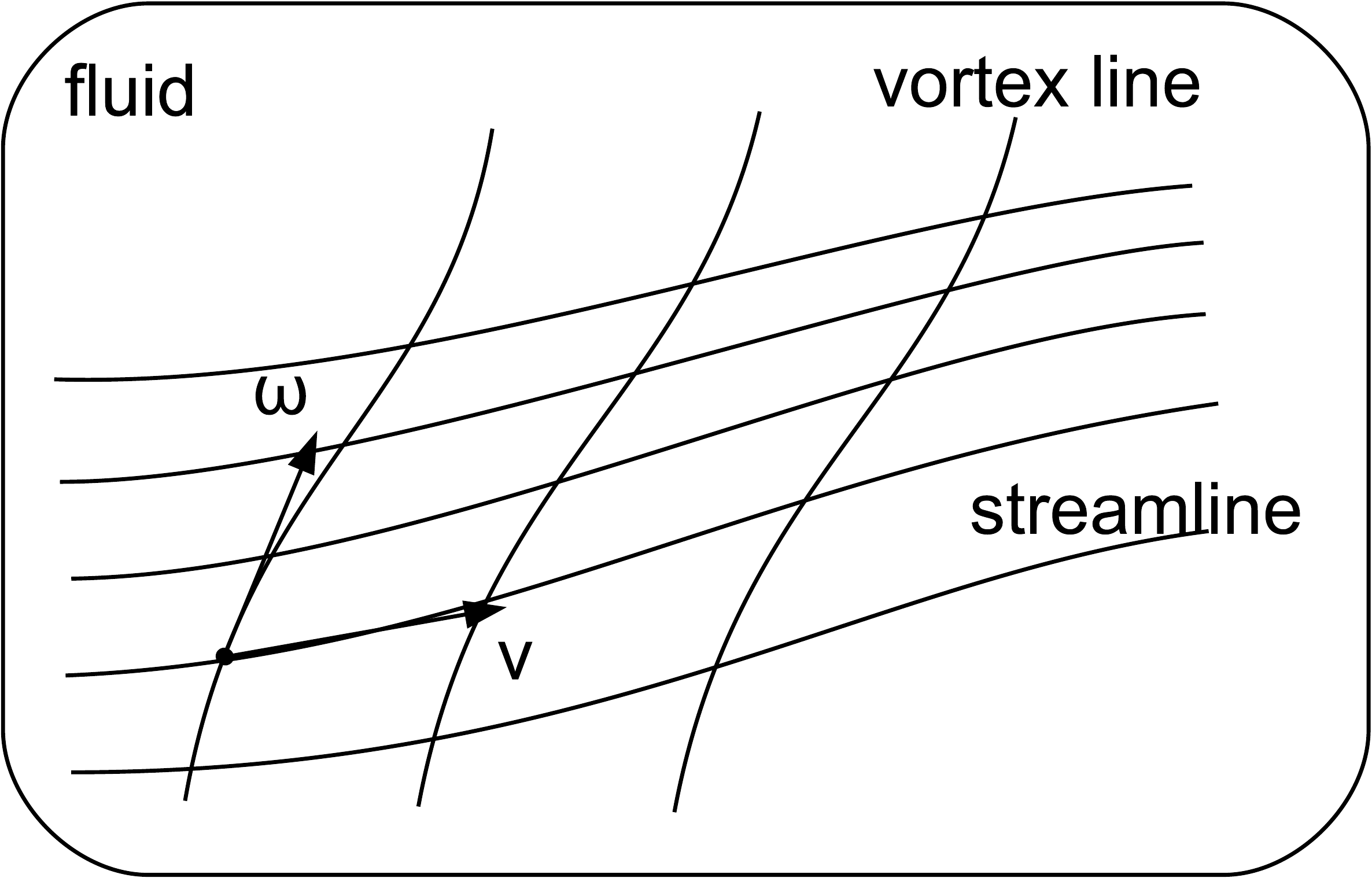}
\caption{Streamlines are lines within the fluid which are tangent, at each point, to velocity vector field $\bold v$.
         Vortex lines realize the same idea with the replacement of velocity field by \emph{vorticity}
         vector field $\boldsymbol \omega$.}
\label{fig:streamandvortexline}
\end{center}
\end{figure}

So we see that we can also express Eq. (\ref{eqforvortelines1}) as
\begin{equation} \label{eqforvortelines2}
                 \boxed{i_{\gamma'}d\tilde v = 0}
                 \hskip 1cm
                 \text{{\color{red}\emph{vortex lines equation}}}
\end{equation}
The exterior derivative of the velocity 1-form $\tilde v$, i.e. the 2-form $d\tilde v$
introduced in (\ref{vorticityform}), is called \emph{vorticity $2$-form}.
We see from Eq. (\ref{eqforvortelines2}) that, geometrically speaking, vortex lines direction
is the direction which \emph{annihilates} the vorticity $2$-form.
\vskip .5cm
\noindent
[The concept of a \emph{vortex line} is not to be confused with a different concept of \emph{line vortex}.
 The latter denotes the situation (particular flow) when the magnitude of vorticity vector
 is negligible outside a small vicinity of a line (\emph{tornado} providing a well-known example).
 So, in the case of line vortex, there is a single particular line within the fluid volume
 (defined as the line where vorticity is sufficiently large) whereas, usually,
 there is a lot of vortex lines, one passing through each point of the volume.]
\vskip .5cm

\subsubsection{Acceleration term reexpressed}
\label{euler_stationary_reexpressed}

In Eq. (\ref{eulerstatgeom2}) two \emph{vector} fields are equated.
One can easily express the same content by equating two \emph{covector} fields
(i.e. 1-forms; forms turn out to be fairly convenient for treating vor\-tices,
 as we already mentioned in the last paragraph).

On the l.h.s. of Eq. (\ref{eulerstatgeom2}), notice that the (standard = Riemann/Levi-Civita)
covariant de\-ri\-vative \emph{commutes} with raising and lowering of indices
(since the connection is \emph{metric}). Therefore
\begin{equation} \label{commutes}
                 \tilde a \equiv g(a, \ . \ ) \equiv  g(\nabla_v v, \ . \ )
               = \nabla_v (g(v, \ . \ )) \equiv \nabla_v \tilde v
\end{equation}
On the r.h.s. of Eq. (\ref{eulerstatgeom2}), the relation between gradient as a vector field and gradient
as a~co\-vec\-tor field may be used:
\begin{equation} \label{grad_vect_covect}
                 g (\grad f, \ . \ ) := df
                  \hskip 1cm \text{i.e.} \hskip 1cm
                 (\grad f)^i = g^{ij} (df)_j \equiv g^{ij} \partial_j f
\end{equation}
Putting all this together we get
\begin{equation} \label{eulerstatform1}
                 \nabla_v\tilde v
                   = - \frac{1}{\rho} \ dp - d \Phi
\end{equation}
Now, we can reexpress the covariant derivative in terms of Lie derivative and, finally,
in terms of the exterior and interior derivatives:
\begin{eqnarray} %{rcl}
      \label{covtolie1}
            \nabla_v\tilde v  &=& \mathcal L_v \tilde v - (\nabla v) (\tilde v) \\
      \label{covtolie2}
                              &=& i_vd\tilde v + d(v^2/2)
                                  \hskip 1cm
                                  v^2 \equiv g(v,v)
\end{eqnarray}
\vskip .5cm
\noindent $\blacktriangledown \hskip 0.5cm$
A proof: First, in general we have
$$
\begin{array} {rcl}
       \mathcal L_W \langle \alpha , V \rangle
  &=&  \nabla_W \langle \alpha , V \rangle \\
       \langle \mathcal L_W \alpha , V \rangle + \langle \alpha , \mathcal L_W V \rangle
  &=&  \langle \nabla_W \alpha , V \rangle + \langle \alpha , \nabla_W V \rangle\\
       \langle \mathcal L_W \alpha , V \rangle + \langle \alpha , [W, V] \rangle
  &=&  \langle \nabla_W \alpha , V \rangle + \langle \alpha , \nabla_V W + [W,V] \rangle\\
       \langle \mathcal L_W \alpha , V \rangle
  &=&  \langle \nabla_W \alpha , V \rangle + (\nabla W)(V,\alpha) \\
       \mathcal L_W \alpha
  &=&  \nabla_W \alpha  + (\nabla W)(\alpha)
\end{array}
$$
Here, $\nabla W$ is a tensor of type $\binom 11$ and $(\nabla W)(\alpha)$
looks in components
$$((\nabla W)(\alpha))_i= (\nabla W)_i^j\alpha_j = W^j_{ \ ; i}\alpha_j
$$
Now, in our particular case in \emph{Cartesian} coordinates, where $i)$ all $\Gamma$'s vanish
and $ii)$ $v^j = v_j$, we have
$$((\nabla v)(\tilde v))_i=  v^j_{ \ ; i}v_j =  v_{j , i}v_j =
                           (v_jv_j/2)_{  , i}  = (dv^2/2)_i
$$
This is, however, tensorial equation (i.e. independent of coordinates), so
$$\nabla_v\tilde v = \mathcal L_v \tilde v - d(v^2/2)
$$
Finally, due to Cartan's identity $\mathcal L_v = i_vd+di_v$, we see that Eq. (\ref{covtolie2}) holds.
\hfill $\blacktriangle$ \par
\vskip .5cm
Plugging Eq. (\ref{covtolie2}) into Eq. (\ref{eulerstatform1}) we get
\begin{equation} \label{eulerstatform2}
                 i_v d\tilde v
                   = - \frac{1}{\rho} \ dp - d (\Phi + v^2/2)
\end{equation}
This can be, of course, expressed back in the language of standard vector analysis.
Using formulas collected in Appendix \ref{app:vectoranalysis}, we get
\begin{equation} \label{eulerstatform2vecanal}
                 \bold v \times (\boldsymbol \nabla \times \bold v)
                   = \frac{1}{\rho} \ \boldsymbol \nabla p + \boldsymbol \nabla (\Phi + v^2/2)
\end{equation}

\subsubsection{Conservation of mass and entropy}
\label{massandentropy_stat}

Let $\Omega$ be the standard \emph{volume form} and $D\subset M$. Consider the following three integrals:
\begin{eqnarray} %{rcl}
      \label{volumefluid}
            \text{vol} \ D &:=& \int_{D} \Omega
                                \hskip 1.5cm
                                \text{\emph{volume of $D$}} \\
      \label{massfluid}
             m(D) &:=& \int_D \rho \Omega
                 \hskip 1.35cm
                 \text{\emph{total mass in $D$}}            \\
      \label{entropyfluid}
             S(D) &:=& \int_D s\rho \Omega
                 \hskip 1.2cm
                 \text{\emph{total entropy in $D$}}
\end{eqnarray}
\vskip .5cm
\noindent
[One could wonder why the product $s\rho$, and not just $s$, enters the expression
 in Eq. (\ref{entropyfluid}) of total entropy in domain $D\subset M\equiv E^3$.
 The reason is that $s$ denotes entropy \emph{per unit mass}.
 So, by definition, $s\delta m$ is entropy of infinitesimal mass $\delta m$ of the fluid.
 Since $\delta m = \rho \delta V$ (i.e. $\rho$ denotes mass \emph{per unit volume}),
 the amount $\delta S$ of entropy in volume $\delta V$ is
 $\delta S = s\delta m = s\rho \delta V$.]
\vskip .5cm
The total mass of the fluid in $D$, Eq. (\ref{massfluid}), is \emph{conserved}.
Simply $m(D(t))$ is the same amount of mass as $m(D(0))$, it just traveled
(formally via the flow $\Phi_t$ generated by $v$) to some other place.

The total entropy of the fluid in $D$, Eq. (\ref{entropyfluid}), is \emph{conserved as well}.
This is because of the fact, that the fluid is \emph{ideal}.
There is no heat exchange between different parts of the fluid,
no energy dissipation caused by internal friction, viscosity, in the fluid.
So the motion of ideal fluid is to be treated as \emph{adiabatic}.
Entropy $S(D(t))$ is the same as $S(D(0))$.

We can also express the two facts differently:
$$\text{\emph{both} $m(D)$ and $S(D)$ are \emph{absolute integral invariants}.}
$$
Such a situation was already discussed in detail in general context of integral invariants
in Section \ref{continuity}.
So here, first, we have as many as \emph{two} continuity equations
\`{a} la Eq. (\ref{contstat2}),
playing role of differential versions of Eqs. (\ref{massfluid}) and (\ref{entropyfluid}):
\begin{equation}
      \label{massconteq1}
            \boxed{\Div (\rho v)  = 0}
                             \hskip 1.1cm \text{(\emph{continuity equation for mass})}
\end{equation}
\begin{equation}
       \label{entropyconteq1}
           \boxed{\Div (s\rho v)  = 0}
                             \hskip 1cm \text{(\emph{continuity equation for entropy})}
\end{equation}
This just corresponds to the existence of two ``isolated'' integral invariants.

But the two integral invariants are actually related in a specific way described
in Sec. \ref{continuity}, see Eq. (\ref{withOmega}) and below.
Notice, however, that the correspondence is a bit tricky:
$$\text{
        $\Omega \ , \ \rho \Omega$ \ \ there
        \hskip 1cm $\leftrightarrow$ \hskip 1cm
        $\rho \Omega \ , \ s(\rho \Omega)$ \ \ here
       }
$$
From this fact we can deduce, according to (time-independent version of) Eq. (\ref{liouville}),
that
\begin{equation} \label{vsvanishes}
                 \boxed{vs=0}
\end{equation}
This says that $s$ is constant \emph{along streamlines}
(integral lines of the velocity field $v$).
The constant may, in general, take different values on different streamlines.
\vskip .5cm
\noindent $\blacktriangledown \hskip 0.5cm$
A direct proof: Eq. (\ref{entropyconteq1}) says
$$\Div (s\rho v) = \rho (vs) +s\Div (\rho v) =0
$$
Then Eq. (\ref{massconteq1}) leads to Eq. (\ref{vsvanishes}).
We used $\Div (fu) = uf + f\Div u$, which follows from
$$0=i_u(df \wedge \Omega) = (uf)\Omega - df\wedge i_u\Omega
$$
and
$$
\begin{array} {rcl}
       0
  &=&  \Div (fu)\Omega \\
  &=&  (di_{fu}+i_{fu}d)\Omega \\
  &=&   d(fi_u\Omega) \\
  &=& df \wedge i_u\Omega + (f\Div u)\Omega \\
  &=& (uf + f\Div u)\Omega
\end{array}
$$
\hfill $\blacktriangle$ \par
\vskip .5cm

\subsubsection{Barotropic fluid}
\label{barotropic}

In general, equation of state of the fluid may be written as $p=p(V, s)$,
where $s$ is (specific) entropy (i.e. entropy per unit mass) and $V$
is (specific) volume (i.e. volume per unit mass).
Or, alternatively, as
\begin{equation} \label{generalfluid}
                 p=p(\rho, s) \hskip 1cm \text{general fluid}
\end{equation}
since $\rho = 1/V$.
In this case, Eq. (\ref{eulerstatform2}) is the final form of \emph{Euler equation}
for \emph{stationary} flow.

However, one can think of an important model, when the pressure depends \emph{on} $\rho$ \emph{alone}:
\begin{equation} \label{barotropicfluid}
                 p=p(\rho)
                 \hskip 1.3cm \text{\emph{barotropic} fluid}
\end{equation}
(In particular, we speak of \emph{polytropic} case, if $p=K\rho^\gamma$, $K=$ const.).
It is a useful model for fluid behavior in a wide variety of situations.
Then, clearly
\begin{equation} \label{barotropicclosed}
                 d\left( \frac{1}{\rho} \ dp \right) = 0
\end{equation}
So, there is a function, $P$, such that
\begin{equation} \label{barotropic_P}
                 \frac{1}{\rho} \ dp = dP
\end{equation}

In order to interpret $P$ physically, consider the general
ther\-mo\-dynamic relation
\begin{equation} \label{enthalpygeneral}
                 dw = Tds +Vdp \equiv Tds +\frac{1}{\rho} \ dp
\end{equation}
where $w$ is \emph{enthalpy} (heat function) \emph{per unit mass}.
So $dp/\rho$ is \emph{not} exact, \emph{in general}. From Eq. (\ref{vsvanishes}) we see that
$ds(v)=0$, This means that, at each point of the fluid, the 1-form $ds$ vanishes
when restricted onto a particular
1-dimensional \emph{subspace} (spanned by $v$) but it, in general, does not vanish
\emph{as a 1-form}, i.e. on \emph{arbitrary} vector at that point.
(There are good reasons, mentioned above, for $s$ to be constant along a single streamline,
 however, there is no convincing reason for the constant to have the same value
 on neighboring streamlines.)

So, if we search for a situation, when $dp/\rho$ \emph{is} exact
(i.e. Eq. (\ref{barotropic_P}) holds) we have to have, according to Eq. (\ref{enthalpygeneral}),
$ds=0$ as a 1-form (on \emph{any} argument, no just on $v$). This means, however
\begin{equation} \label{sequalsconst}
                 s = \ \text{const.}
\end{equation}
throughout the volume of the fluid, not just along streamlines
(or, put it differently, the constants the function $s$ takes on different streamlines
 \emph{should be equal}, now).
So, barotropic case is the one when Eq. (\ref{vsvanishes}) is fulfilled ``trivially''
as Eq. (\ref{sequalsconst}). Although it may seem to be a rare situation,
one can read in \$2 of \cite{lanlif87} that, on the contrary, ``it usually happens''
(\emph{barotropic} is called \emph{isentropic}, there).
Denoting $w=P$, we get (\ref{barotropic_P}).

\subsubsection{Final form of stationary Euler equation}
\label{euler_stationary_barotropic}

Using Eq. (\ref{barotropic_P}) in Eq. (\ref{eulerstatform2}), we get our final form of
the equation of motion governing stationary and barotropic flow of ideal fluid:
\begin{equation} \label{eulerstatform3}
                 \boxed{i_v d\tilde v = - d\mathcal E}
                 \hskip 1cm
                 \text{{\color{red}\emph{Euler equation}}}
                 \hskip .5cm
                 (\text{{\color{red}\emph{stationary}, \emph{barotropic}}})
\end{equation}
Here
\begin{equation} \label{deffunctionE}
                 \boxed{\mathcal E := v^2/2 +P+ \Phi}
                  \hskip 1cm
                 \text{\emph{Bernoulli function}}
\end{equation}
Remember, however, that for more general, non-barotropic flow we have to use
more general equation (\ref{eulerstatform2}).
In what follows we mainly concentrate on Eq. (\ref{eulerstatform3}).

Equation (\ref{eulerstatform3}) can also be expressed in terms of good old vector analysis.
Using formulas collected in Appendix \ref{app:vectoranalysis}, we get
(see Eq. (\ref{eulerstatform2vecanal}))
\begin{equation} \label{eulerstatform3vecanal}
                 \bold v \times (\boldsymbol \nabla \times \bold v)
                   = \boldsymbol \nabla \mathcal E
\end{equation}

\setcounter{equation}{0}
\subsection{Simple consequences of stationary Euler equation}
\label{conseq_euler_stationary_barotropic}

\subsubsection{Fluid statics (hydrostatics)}
\label{hydrostatics}

Statics means that fluid does not flow at all, i.e. $v=0$.
(So what still remains unknown is $p$ and $\rho$ as functions of $\bold r$.)

Plugging this into Eqs. (\ref{eulerstatform3}) and (\ref{deffunctionE}) we get
\begin{equation} \label{eulervis0}
                 dP+ d\Phi =0
\end{equation}
Or, more generally (relaxing barotropic assumption, see Eq. (\ref{barotropic_P})),
\begin{equation} \label{hydrostatics1}
                 \boxed{\frac{dp}{\rho} + \ d\Phi =0}
                 \hskip 1cm
                 \text{{\color{red}\emph{fluid statics equation}}}
                  \hskip .3cm
                 \text{(\emph{hydrostatic equilibrium})}
\end{equation}
\vskip .5cm
\noindent
{\bf Example \ref{hydrostatics}.1}: Water in a swimming pool.

Water is incompressible, so $\rho = \rho_0 =$ const.
(So the only unknown quantity is $p(\bold r)$.)
For gravitational field we have $\Phi (x,y,z) = gz$. Therefore
\begin{equation} \label{water1}
                 dp = - \rho_0 gdz = d(-\rho_0 gz)
\end{equation}
and
\begin{equation} \label{water2}
                 p(x,y,z) = p_0 - \rho_0 gz
\end{equation}
So the pressure linearly increases with depth in water.
The end of Example \ref{hydrostatics}.1.
\vskip .5cm
\noindent
{\bf Example \ref{hydrostatics}.2}: Polytropic atmosphere.

The atmosphere (air) is compressible, so both $p$ and $\rho$ are unknown.
Since we are still in gravitational field, $\Phi (x,y,z) = gz$ again.
From symmetry we can assume both quantities depend on $z$ alone, $p(z)$ and $\rho (z)$.

The main assumption is that
\begin{equation} \label{polytropicfluid}
                 p(\rho) = K\rho^{\gamma}
                 \hskip 1cm
                 \text{\emph{polytropic fluid}}
                 \hskip 1cm
                 K,\gamma = \ \text{const.}
\end{equation}
(This is a particular instance of \emph{baro}tropic assumption $p= p(\rho)$ from Eq. (\ref{barotropicfluid}).)

Let us start with \emph{isothermal} case, i.e. $\gamma = 1$.
\vskip .5cm
\noindent
[Equation of state for $\nu$ moles of \emph{ideal gas} is
 \begin{equation} \label{idealgas}
                 pV = \nu RT
                 \hskip 1cm
                 \text{i.e.}
                 \hskip 1cm
                 p = p(\rho,T)=\frac{RT}{M_1} \ \rho
\end{equation}
where $V=\nu V_1$ is volume of $\nu$ moles and $M=\rho V = \rho \nu V_1 = \nu M_1$ is mass of $\nu$ moles.
For $T=T_0=$ const. we get
Eq. (\ref{polytropicfluid}) for $K=RT_0/M_1$ and $\gamma =1$.]
\vskip .5cm
Then Eq. (\ref{hydrostatics1}) says
\begin{equation} \label{isothermalfluidequation}
                 \frac{d\rho}{\rho} = - \left( \frac{M_1g}{RT_0} \right) dz \equiv - \frac{dz}{z_0}
                 \hskip 1cm
                 \rho = \rho (z)
\end{equation}
where the \emph{isothermal scale-height} of the atmosphere is
\begin{equation} \label{isothermalfluidz0}
                 z_0 := \frac{RT_0}{M_1g} = \frac{kT_0}{m_1g}
\end{equation}
\vskip .3cm
\noindent
[Recall that $R=N_Ak$, where $N_A$ is Avogadro number (number of constituent particles, usually atoms or molecules, per mole)
 and $k$ is Boltzmann constant.
 Moreover, $M_1=N_Am_1$, where $m_1$ is mass of a single particle.
 So $z_0 = (N_AkT_0)/(N_Am_1g)$, or
\begin{equation} \label{meaningofz0}
                 m_1gz_0 = kT_0
\end{equation}
Therefore $z_0$ is the height in which gravitational potential energy of the particle equals energy $kT_0$.]
\vskip .5cm
From Eqs. (\ref{isothermalfluidequation}) and (\ref{idealgas}) we get
that both mass density and pressure decrease \emph{exponentially} with altitude
\begin{eqnarray} %{rcl}
      \label{isothermalfluidrho}
            \rho (z) &=& \rho_0 e^{- z/z_0}   \\
      \label{isothermalfluidp}
               p (z) &=& p_0 e^{- z/z_0}
\end{eqnarray}
where $p_0=p(0)$, $\rho_0=\rho (0)$.
Notice that
\begin{equation} \label{boltzmannfactor}
                 e^{- z/z_0} = e^{- m_1gz/kT_0}
\end{equation}
is nothing but the \emph{Boltzmann factor} $e^{-U/kT}$.

\emph{Adiabatic} case corresponds to $1 \neq \gamma = c_p/c_V$ and $K= p_0/\rho_0^\gamma$
in the polytropic formula Eq. (\ref{polytropicfluid}).
We get, instead of Eq. (\ref{isothermalfluidequation}),
\begin{equation} \label{polytropicfluidequation}
                 K^{1/\gamma}p^{-1/\gamma}dp = - gdz
                 \hskip .9cm \text{i.e.} \hskip .9cm
                 d\left( \frac{K^{1-\Gamma}}{\Gamma}p^\Gamma \right) = d(-gz)
\end{equation}
where $\Gamma \equiv 1-1/\gamma$. This results in
\begin{eqnarray} %{rcl}
      \label{adiabaticfluidrho}
            \rho (z) &=& \rho_0
                         \left(1 - \frac{\gamma - 1}{\gamma} \frac{z}{z_0} \right)^{\frac{1}{\gamma -1}}   \\
      \label{adiabaticfluidp}
               p (z) &=& p_0
                          \left(1 - \frac{\gamma - 1}{\gamma} \frac{z}{z_0} \right)^{\frac{\gamma}{\gamma -1}} \\
      \label{adiabaticfluidT}
                T(z) &=& T_0  \left(1 - \frac{\gamma - 1}{\gamma} \frac{z}{z_0} \right)
\end{eqnarray}
\vskip .3cm
\noindent
[For adiabatic case $pV^{\gamma}=p_0V_0^{\gamma}$ with $\gamma = c_p/c_V$. With the help of
 $\rho V = \nu M_1$ this can be rewritten as $p/\rho^{\gamma} = p_0/\rho_0^{\gamma}$, or
 $p= (p_0/\rho_0^{\gamma})\rho^{\gamma}$, so that $p= K\rho^{\gamma}$ with $K= p_0/\rho_0^\gamma$.
 Integration of Eq. (\ref{polytropicfluidequation}) gives Eq. (\ref{adiabaticfluidp}),
 $p= K\rho^{\gamma}$ leads to Eq. (\ref{adiabaticfluidrho}) and finally equation of state
 Eq. (\ref{idealgas}) provides us with Eq. (\ref{adiabaticfluidT}).]
\vskip .5cm
As we see from Eq. (\ref{adiabaticfluidT}), the temperature of the atmosphere falls off \emph{linearly}
          with increasing height above ground level.
\footnote{It turns out - see e.g. a nice account in \$ 6.6-6.8 in \cite{fitzpatrick2006} - that
          inserting realistic numbers leads to $z_0 \doteq 8,4$ km and the ``1 degree colder per 100 meters higher
          rule of thumb used in ski resorts''.}

Also notice that for $\gamma \to 1$ we return to isothermal case treated before,
i.e. we get $T(z)=T_0$ and exponential decrease of mass density and pressure given
by Eqs. (\ref{isothermalfluidrho}) and (\ref{isothermalfluidp}).
The end of Example \ref{hydrostatics}.2.

\subsubsection{Bernoulli equation}
\label{euler_stationary_bernoulli}

\begin{figure} %[h]
\begin{center}
\hskip -1.60cm
\scalebox{0.15}{\includegraphics[clip,scale=1.70]{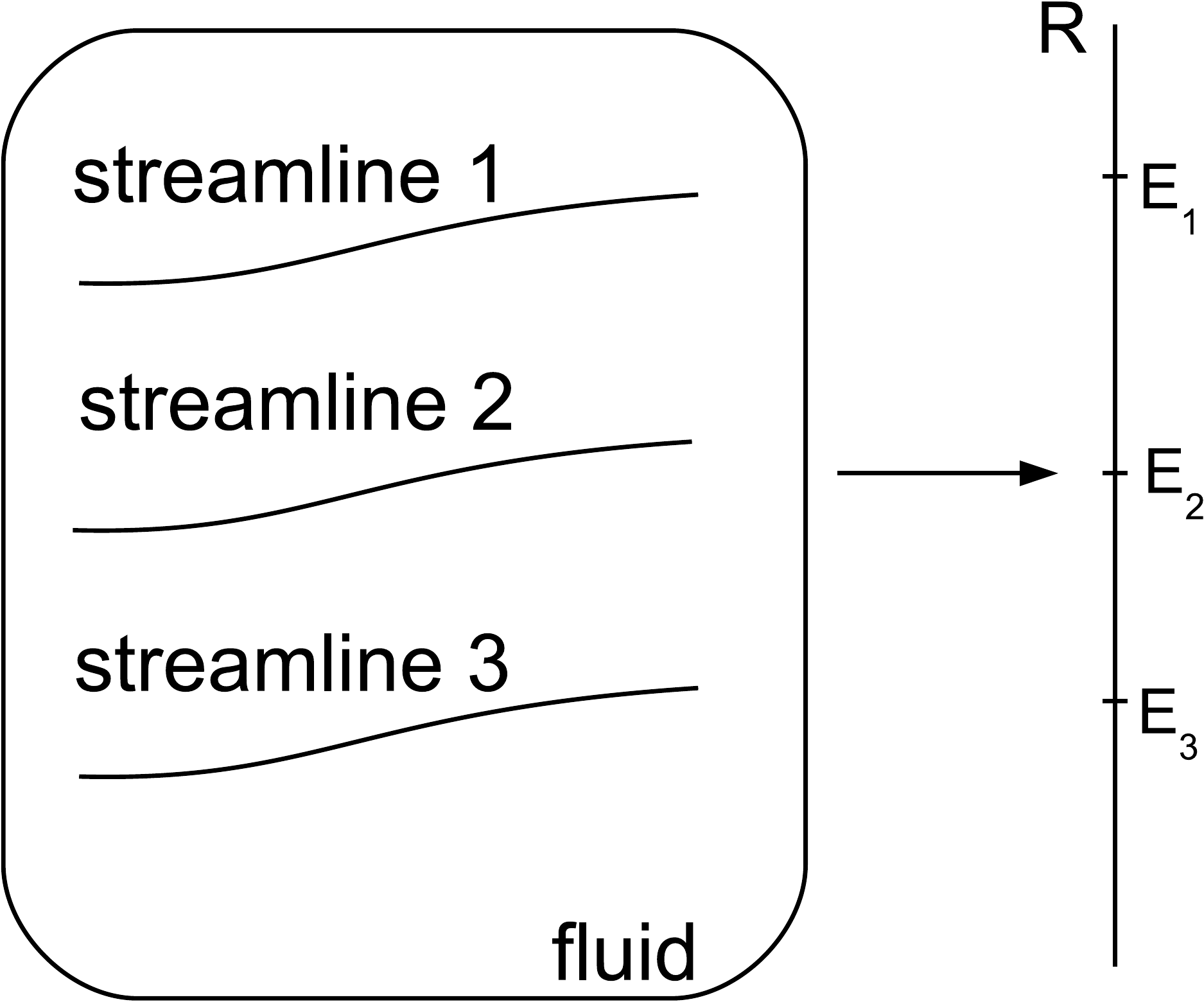}}
\hskip 1.00cm
\scalebox{0.15}{\includegraphics[clip,scale=1.70]{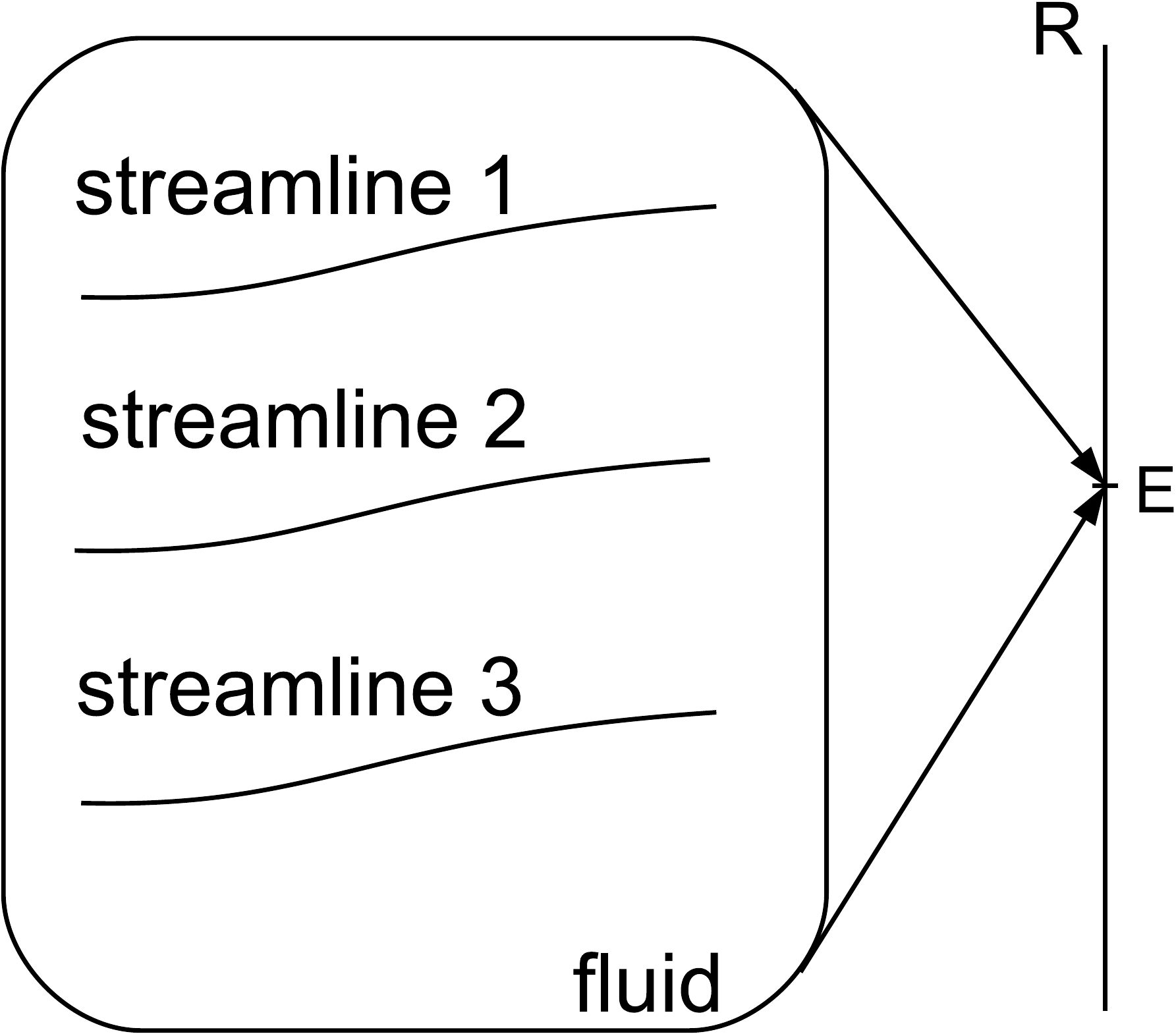}}
\caption{Left: according to Bernoulli theorem Eq. (\ref{bernoulli3}), the function $\mathcal E$
         is constant on any streamline. However, the constant may take different values on different stream\-lines.
         Right: if the flow is vorticity-free, the constant is the same even on different streamlines,
         see Eq. (\ref{bernoulli1}).}
\label{fig:bernoulli}
\end{center}
\end{figure}

Let us apply $i_v$ on both sides of Eq. (\ref{eulerstatform3}). We get
\begin{equation} \label{bernoulli3}
                 \boxed{v\mathcal E =0}
                 \hskip 1cm
                 \text{{\color{red}\emph{Bernoulli equation}}}
\end{equation}
(D.Bernoulli 1738).
This says that $\mathcal E$ is constant \emph{along streamlines}
(integral lines of the velocity field $v$).
The constant may take different values on different streamlines
(see Fig. \ref{fig:bernoulli}).

For \emph{incompressible} fluid (i.e. when $\rho =\rho_0=$ const., e.g. for water) in
\emph{gravitational} field (i.e. when $\Phi (\bold r)=gz$), equation (\ref{bernoulli3}) takes
its more familiar form
\begin{equation} \label{bernoulli2}
                 \rho_0 v^2/2 +p+ \rho_0 gz = \ \text{const.}
\end{equation}

Another way to regard Eq. (\ref{bernoulli3}) comes from comparison with the statement $vs=0$,
Eq. (\ref{vsvanishes}).
We know that the latter is just differential way to express the fact that entropy is conserved
in the sense that Eq. (\ref{entropyfluid}) is absolute integral invariant.
So, Bernoulli equation also says that ``moving particle''
(in hydrodynamic terminology, a fixed small amount of mass of the fluid;
 equivalent expressions are ``fluid particle'', ``fluid element'', ``fluid parcel'' or ``material element'')
carries with it a constant value of $\mathcal E$.

Now, consider a stationary, barotropic and, in addition, \emph{vorticity-free} flow.
In this particular case, the vorticity 2-form vanishes in Eq. (\ref{eulerstatform3}), $d\tilde v = 0$
(cf. (\ref{defvorticityvector}) and (\ref{vorticityform})).
We get $d\mathcal E = 0$, or, consequently
\begin{equation} \label{bernoulli1}
                 \boxed{\mathcal E = \ \text{const.}}
\end{equation}

So, $\mathcal E$ is constant along streamlines irrespective of whether vorticity vanishes or not
(stationary and barotropic flow is enough),
but in order that $\mathcal E$ is constant \emph{in the bulk} of the fluid,
the flow is to be vorticity-free.

\subsubsection{Kelvin's theorem on circulation - stationary case}
\label{euler_stationary_kelvin}

Let us display, side by side in one line,
Eq. (\ref{eulerstatform3}) versus Eq. (\ref{ivdalphajeexaktna}),
i.e.
the \emph{stationary} Euler equation versus
the core equation of the general theory of \emph{Poincar\'e version} of integral invariants:
\begin{equation} \label{porovnanie}
              i_v (d\tilde v)
                   = - d\mathcal E
    \hskip 1cm  \leftrightarrow \hskip 1cm
                 i_v (d\alpha)
                   = d\beta
\end{equation}
We see that stationary Euler equation \emph{may} be regarded as a particular example of the equation
which is central in general scheme of integral invariants, provided that we make the identification
\begin{equation} \label{stotoznenie}
             \alpha \leftrightarrow \tilde v
             \hskip 1cm
             \beta  \leftrightarrow -\mathcal E
             \hskip 1cm
             v  \leftrightarrow v
\end{equation}
Then, however, all consequences are (almost) automatic.
Here we simply rewrite the general result on relative integral invariant (\ref{jetotoiste2})
using the vocabulary from Eq. (\ref{stotoznenie}) and obtain
\begin{equation} \label{Kelvin_stat1}
        \boxed{i_vd\tilde v = - d\mathcal E
        \hskip .5cm \Rightarrow \hskip .5cm
        \oint_c\tilde v = \ \ \text{\emph{relative} invariant}}
\end{equation}
\vskip .5cm
\noindent
[Notice the implication sign here where equivalence sign is present in Eq. (\ref{jetotoiste2}).
 The reason is actually explained, in general context, in the note following Eq. (\ref{jetotoiste2}).
 Here, the fact that the integral on the r.h.s. of Eq. (\ref{Kelvin_stat1}) behaves invariantly
 would be true for \emph{any} function $\mathcal E$ in $i_vd\tilde v = - d\mathcal E$.
 In Euler equation, however, $\mathcal E$ is a concrete function, namely
 $\mathcal E := v^2/2 +P+\Phi$, see Eq. (\ref{deffunctionE}). Clearly, this particular $\mathcal E$
 cannot be deduced from the invariance property of the integral.]
\vskip .5cm
In traditional notation (see (\ref{velocityform})), the integral in Eq. (\ref{Kelvin_stat1})
is called
\begin{equation} \label{circulation}
                 \oint_c\tilde v
                 \equiv
                 \oint_c \bold v \cdot d\bold r
                 \hskip 1.5cm \text{\emph{velocity circulation}}
\end{equation}
and the statement is that it represents a relative integral invariant
\begin{equation} \label{Kelvin_stat2}
                 \boxed{
                 \oint_c \bold v \cdot d\bold r = \text{const.}}
                 \hskip 1.5cm \text{{\color{red}\emph{Kelvin's theorem}}
                       }
\end{equation}
(\cite{kelvin1869}; see e.g.
\$8 in \cite{lanlif87},
\$1.2 in \cite{chormarsden1993} or
\$5.3 in \cite{batchelor}).
In words: in an ideal fluid, the velocity circulation round a closed fluid contour is constant in time.
(Here ``fluid contour'' is a closed path moving together with the fluid. That is, a contour
 moving by the flow $c\mapsto \Phi_t(c)$, where $\Phi_t \leftrightarrow v$.)
\vskip .5cm
\noindent
[Notice that $c$ in Eq. (\ref{Kelvin_stat2}) is a general 1-cycle (loop). That is, not necesarilly
 a \emph{boundary} of a 2-chain (i.e. of 2-dimensional surface).
 (So $\partial c =0$ is compulsory, $c=\partial S$ is not.)
 As an example, it may be a tangly knot.]
\vskip .5cm

By applying the simple general rule (\ref{pravidlorelabs}) we also get an \emph{absolute} invariant
given as a \emph{surface} integral of the corresponding two-form:
\begin{equation} \label{Kelvin_statabs1}
        \int_S d\tilde v = \ \ \text{\emph{absolute} invariant}
\end{equation}
In traditional notation (see (\ref{vorticityform})), the integral in Eq. (\ref{Kelvin_statabs1})
is called
\begin{equation} \label{vorticityflux}
                 \int_S d\tilde v
                 \equiv
                 \int_S (\curl \bold v) \cdot d\bold S
                 \equiv
                 \int_S \boldsymbol \omega \cdot d\bold S
                 \hskip 1.5cm \text{\emph{vorticity flux}}
\end{equation}
and the statement is that it represents an absolute integral invariant
\begin{equation} \label{Kelvin_statabs2}
                 \boxed{\int_S \boldsymbol \omega \cdot d\bold S = \text{const.}}
\end{equation}
In words: in an ideal fluid, the vorticity flux through fluid surface is constant in time.
(Here ``fluid surface'' is a surface moving together with the fluid. That is, a surface
 moving by the flow $S\mapsto \Phi_t(S)$, where $\Phi_t \leftrightarrow v$.)
\vskip .5cm
\noindent
[Let $\boldsymbol \omega$ be, at some fixed time $t$, (non-vanishing) vorticity vector in
 a given point $P$ of the fluid.
 At this point consider surface elements $d\bold S = \bold n dS$ with fixed $dS$ and all possible
 directions $\bold n$. Then clearly $\boldsymbol \omega \cdot d\bold S$ is \emph{maximum} for $d\bold S$
 directed \emph{along }$\boldsymbol \omega$. Due to Stokes theorem, however, $\boldsymbol \omega \cdot d\bold S$
 is integral of $\bold v$ around the boundary of the (infinitesimal) surface $d\bold S$,
 i.e. the \emph{velocity circulation round the boundary} of $d\bold S$.
 So, the direction of the vorticity vector $\boldsymbol \omega$ is the direction of
 \emph{maximum velocity circulation}. Therefore, going along a \emph{vortex line} is actually
 going, in each point of the line, in the direction with maximum velocity circulation
 round a small circle drawn in the plane perpendicular to the direction. Naively, this is then
 the direction about which, at the point $P$ at time $t$, the fluid particle \emph{rotates}
 (in addition to its translational motion).]
\vskip .5cm

\subsubsection{Helmholtz theorem on vortex lines - stationary case}
\label{euler_stationary_helmholtz}

In order to obtain Bernoulli equation (\ref{bernoulli3}) we just applied the \emph{interior} derivative $i_v$
on both sides of Euler equation (\ref{eulerstatform3}).

Another obvious thing to do with Euler equation (\ref{eulerstatform3}) to get something
interesting is to apply the \emph{exterior} derivative $d$ on both sides.
What we obtain now is
\begin{equation} \label{vorticityisinv1}
                 \mathcal L_v (d\tilde v) = 0
\end{equation}
This is, however, nothing but infinitesimal version of the statement
\begin{equation} \label{vorticityisinv2}
                 \Phi_t^* (d\tilde v) = d\tilde v
                 \hskip 2cm
                 \Phi_t \leftrightarrow v
\end{equation}
So the vorticity 2-form $d\tilde v$ is \emph{invariant}
w.r.t. the flow $\Phi_t$ generated by $v$, i.e. w.r.t. ``real flow'' of the fluid.

Consider, at time $t_0$, an arbitrarily parametrized vortex line $\gamma (\lambda)$.
(So it is, actually, a ``vortex curve''; the parametrization is, however, arbitrary).
Due to the flow $\Phi_t$, its point $\gamma (\lambda)$ moves along the streamline (integral curve)
of $v$ and, at time $t_0+t$, it is at the point
\begin{equation} \label{newvortexline}
                 \Phi_t(\gamma (\lambda)) \equiv (\Phi_t \circ \gamma) (\lambda) =: \gamma_t(\lambda)
\end{equation}
Since this happens with each point of the original vortex line $\gamma$, the line itself moves,
within the time interval from $t_0$ to $t_0+t$, to a new line, $\gamma_t$.
And it is not clear at all, apriori, whether or not the new line, $\gamma_t$, is a \emph{vortex}
line, too.

More than one and a half century ago (in 1858), Helmholtz gave positive answer to the question
(\cite{helmholtz1858}):
\begin{equation} \label{Helmholtztheorem}
                 \{ \gamma = \text{vortex line} \}
                 \hskip .5cm \Rightarrow \hskip .5cm
                 \{ \gamma_t = \text{vortex line} \}
                 \hskip .5cm \text{{\color{red}\emph{Helmholtz theorem}}}
\end{equation}
His theorem asserts that, under the flow of ideal barotropic fluid,
vortex lines always move to vortex lines again. It is often expressed by saying that
vortex lines are ``\emph{frozen}'' into a perfectly inviscid (= ideal) barotropic fluid
(in the sense that they ``move with the fluid'') or that they ``move as material lines''.
\begin{figure}[tb]
\begin{center}
\includegraphics[clip,scale=0.30]{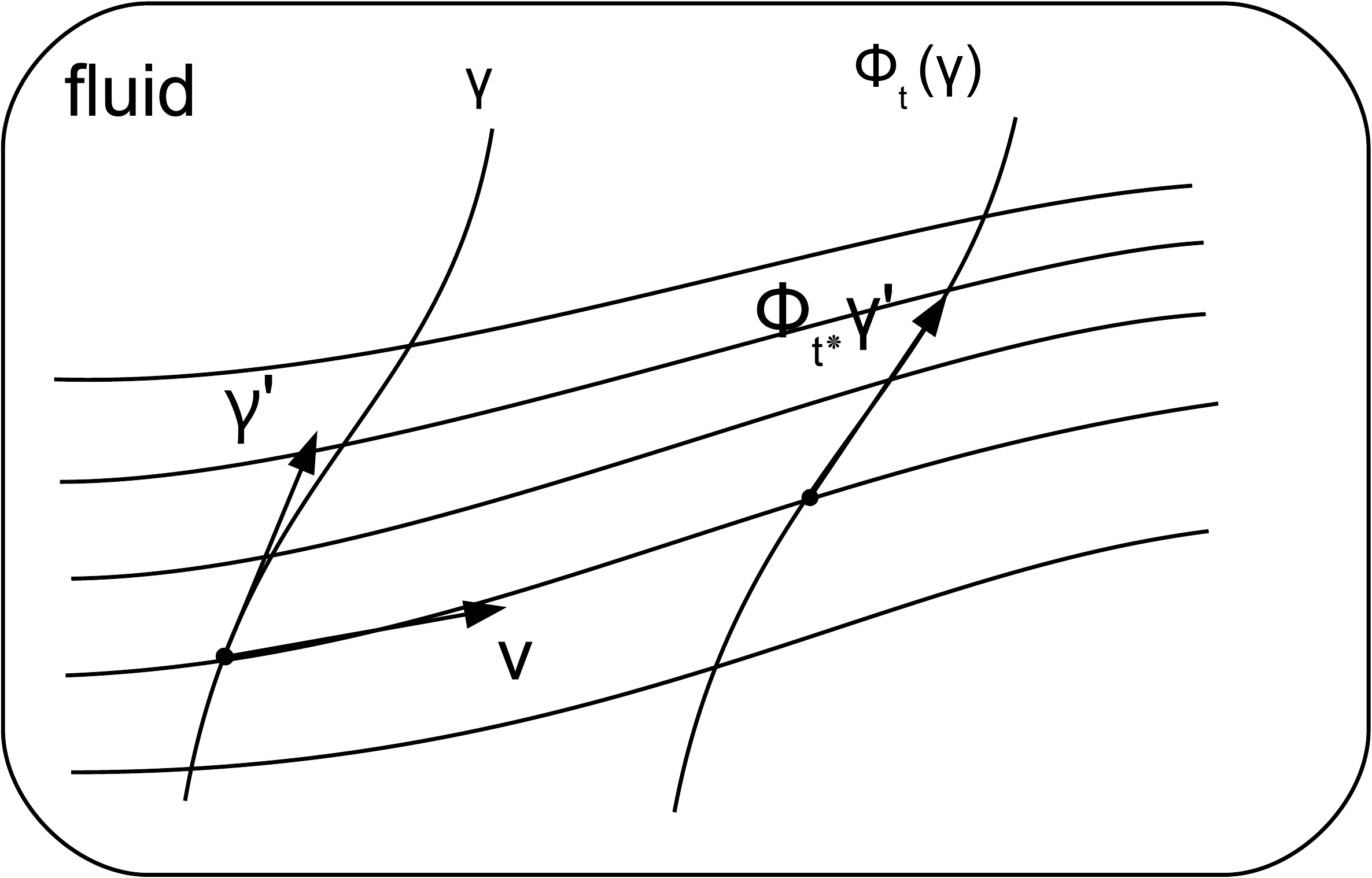}
\caption{Helmholtz theorem Eq. (\ref{Helmholtztheorem}) says that,
         under the flow of ideal barotropic fluid, vortex lines always move to vortex lines again
         (so vortex lines are ``\emph{frozen}'' into the fluid).}
\label{fig:helmholtztheorem}
\end{center}
\end{figure}

\vskip .5cm
\noindent
[It turns out that the physics behind the theorem is closely related to the ubiquitous
 \emph{angular momentum} conservation.
 \footnote{Recall that angular momentum conservation is also an eminence grise behind
           the ``purely geometrical'' Kepler's second law of planetary motion
           (the same area is swept out in a given time).}
The angular momentum of the fluid in a small piece
 of a vortex tube cannot change, since the fluid is ideal (i.e. inviscous) and therefore there are no
 \emph{tangential} forces on the surface of the piece of the tube.
 See a nice exposition in Sec. 40-5 of \cite{feynman1964}.

 On the other hand, one is to proceed with due caution,
 as we can read in Sec. 5.3 of \cite{batchelor}:

``There is an obvious temptation to interpret some of these results as a
consequence of conservation of angular momentum of the fluid contained in
the vortex-tube. Such an interpretation is possible when the cross-section of
the vortex-tube is small and circular, and remains circular, since the (inviscid)
stress at the boundary of the vortex-tube then exerts no couple on the fluid
in the vortex-tube. In more general circumstances the strength of a vortex-
tube is not simply proportional to the angular momentum of the fluid in unit
length of the tube, and more is involved in the above results than a simple
conservation of angular momentum.'']
\vskip .5cm

Let's see how (easily) this follows from Eq. (\ref{vorticityisinv2}).
At time $t_0$, the tangent vector $\gamma '$ to the original ``vortex curve'' $\gamma (\lambda)$
annihilates, according to Eq. (\ref{eqforvortelines2}),
the vorticity $2$-form $d\tilde v$. At time $t_0+t$, the curve becomes
$(\Phi_t \circ \gamma) (\lambda)$. Its tangent vector, therefore, becomes $\Phi_{t*} \gamma '$
(it follows from the \emph{definition} of push-forward operation).
So, the issue is whether
\begin{equation} \label{issue1}
                 i_{\gamma'}d\tilde v = 0
                 \hskip 1cm \Rightarrow \hskip 1cm
                 i_{\Phi_{t*} \gamma '}d\tilde v = 0
\end{equation}
holds or, equivalently, whether
\begin{equation} \label{issue2}
                 (d\tilde v)( \Phi_{t*} \gamma ',w)= 0
\end{equation}
\emph{for any} $w$ (sitting at the same point as $\Phi_{t*} \gamma '$).
But \emph{any} $w$ may be regarded as $\Phi_{t*} u$ for some $u$
(sitting at the same point as $\gamma '$; namely $u=\Phi_{-t*} w$).
So, we get for the l.h.s. of Eq. (\ref{issue2})
\begin{equation} \label{issue3}
                 (d\tilde v)( \Phi_{t*} \gamma ',\Phi_{t*} u)
              =  (\Phi_t^* d\tilde v)(\gamma ', u)
              =  (d\tilde v)(\gamma ', u)
              =  (i_{\gamma'}d\tilde v)(u)
              = 0
\end{equation}
Therefore, because of invariance of $d\tilde v$ w.r.t. the flow $\Phi_t$,
tangent vectors which annihilate the $2$-form at some time flow to tangent vectors
which also do annihilate the $2$-form at later times and this means that
vortex lines always flow to vortex lines again.

The reader who grasped material discussed in the Chapter \ref{euler_poincare_cartan}
(devoted to integral invariants in general)
probably noticed that the main statement presented here may be regarded as
a particular case of the result presented in Sec. \ref{remarkable_surface}
(and also the proof is very similar).
We return to Helmholtz theorem again in Sec. \ref{euler_non_stationary_helmholtz}.
There we learn that it holds even for \emph{non-stationary} flow.
(Sec. \ref{remarkable_surface} treats non-stationary theory from the very beginning.)

\subsubsection{Related Helmholtz theorems - stationary case}
\label{euler_stationary_helmholtz_related}

In literature one often finds three ``classical'' Helmholtz theorems closely related to
our statement Eq. (\ref{Helmholtztheorem}).
They follow easily from what was said before,
including Kelvin's theorem on circulation, Eq. (\ref{Kelvin_stat2}).
Recall, however, that the Helmholtz paper was published
11 years sooner than the Kelvin's one - in 1858 versus 1869, respectively).

Let us first introduce a few new concepts.

\emph{Vortex sheet} is a (2-dimensional) surface composed from vortex lines
(i.e. it is a 1-parameter family of vortex lines, passing through the same curve in space).
In particular it reduces to \emph{vortex tube} if the curve becomes a loop,
the boundary of a surface in space
(so the vortex tube is a vortex sheet rolled so as to form a tube).
And finally, \emph{vortex filament} is a vortex tube whose ``cross-section'' is of infinitesimal dimensions.
(A common name for all these objects is vortex \emph{structures}.)

It is obvious that Helmholtz theorem concerning vortex \emph{lines}, Eq. (\ref{Helmholtztheorem}),
immediately leads to corresponding statement concerning vortex \emph{tubes} (or structures, in general):
\begin{equation} \label{vortextubesmove}
                 \text{\emph{Vortex tubes always move to vortex tubes again.}}
\end{equation}

At time $t=0$, fix a vortex tube. Consider a cross-section $S$ of the tube (2-dimensional surface cutting the tube).
Then its boundary, $c=\partial S$, is a loop which encircles the tube once.
Stokes theorem says that
\begin{equation} \label{stokescircflux}
                 \int_S  d\tilde v = \int_c \tilde v
\end{equation}
So, velocity circulation round the boundary $c$ of the cross-section $S$ equals the vorticity
flux through $S$ itself.

Now, take another cross-section, $S'$, with the \emph{same boundary} $c$ (so $c=\partial S = \partial S'$).
Then the two fluxes (through $S$ and $S'$, respectively) are equal,
\begin{equation} \label{anotherS}
                 \int_S  d\tilde v =  \int_{S'}  d\tilde v
\end{equation}
since the r.h.s. of Eq. (\ref{stokescircflux}) does not contain any particular choice
of the cross-section, it feels its boundary alone.

What is more interesting, however, is that the flux even does not depend on the choice of the \emph{loop} $c$
encircling the tube! So, it is a number which characterizes the \emph{tube itself}:
\begin{equation} \label{strengthofthetube}
                 \int_S d\tilde v
                 \equiv
                 \int_S \boldsymbol \omega \cdot d\bold S
                 \hskip 1cm \text{\emph{stregth of the tube}}
\end{equation}
(It is the \emph{vorticity flux} through \emph{any} cross-section $S$ cutting the tube.)
\vskip .5cm
\noindent $\blacktriangledown \hskip 0.5cm$
We provide as many as two proofs, since each one might be instructive,
albeit from different point of view.

\begin{figure}[tb]
\begin{center}
\includegraphics[clip,scale=0.30]{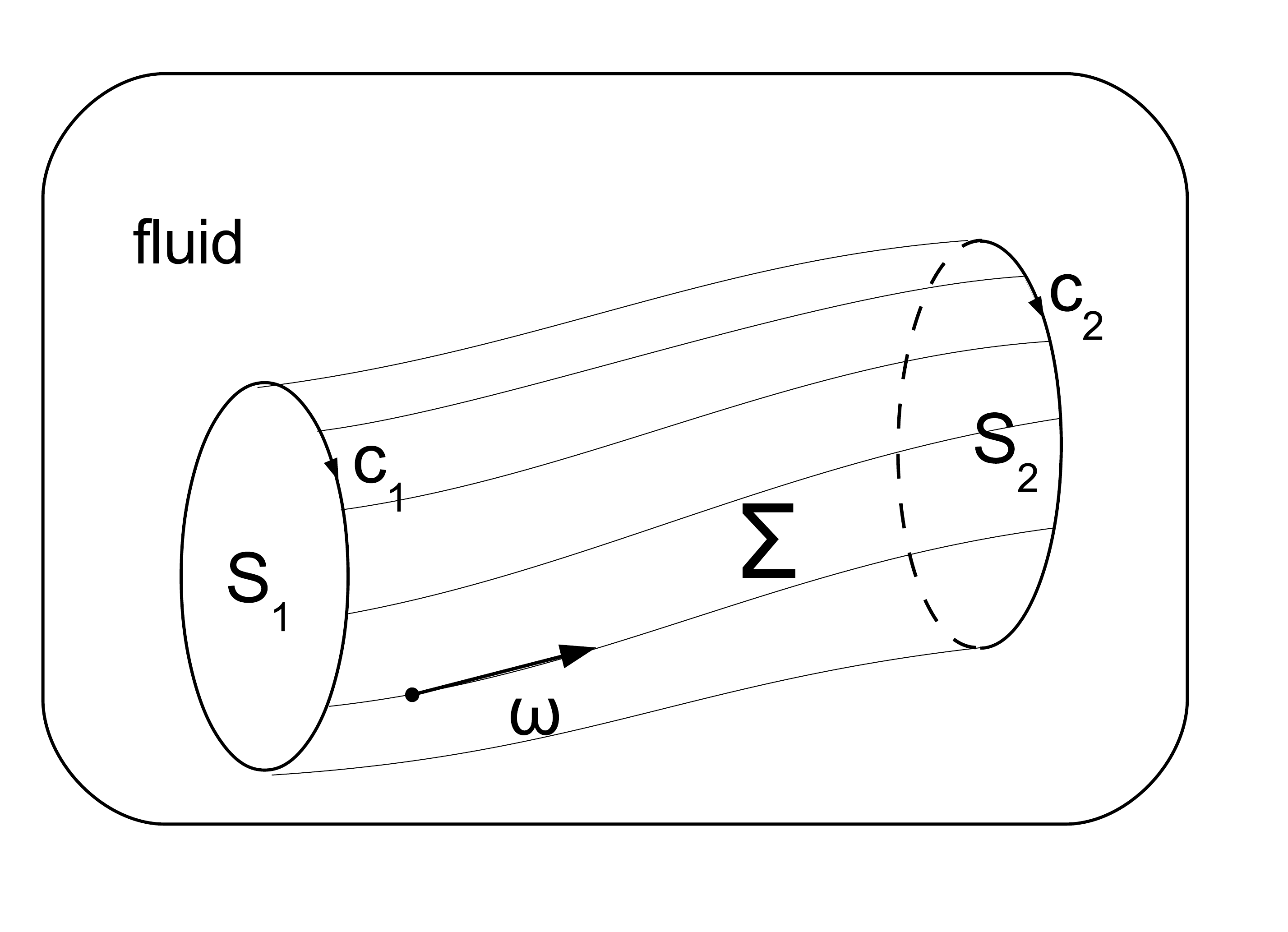}
\caption{A solid vortex tube $V$ (the tube \emph{inside}) is made of vortex lines emanating from the left cap $S_1$
         and entering the right cap $S_2$.
         Boundary $\partial V$ of the solid cylinder $V$ consists of 3 parts, $\partial V = \Sigma +S_2-S_1$,
         hollow cylinder $\Sigma$ (``side'' of the solid cylinder) and the two caps, $S_1$ and $S_2$.
         We have $0=\partial \partial V = \partial \Sigma +\partial S_2- \partial S_1$,
         so $\partial \Sigma = \partial S_1- \partial S_2 = c_1 - c_2$.
         The cycles $c_1$ and $c_2$ encircle the same tube of vortex lines.}
\label{fig:vortextube}
\end{center}
\end{figure}

\noindent
Proof 1.: This is just a version of the ``amazingly simple'' proof from Eqs.
(\ref{argumentarnold1}) and (\ref{argumentarnold2}): let $c_1$ and $c_2$ be the boundaries
of $S_1$ and $S_2$ respectively (see Fig. \ref{fig:vortextube}) and let $\Sigma$ be the (2-dimensional) part of the tube
enclosed by $c_1$ and $c_2$ (so that $\partial \Sigma = c_1 - c_2$). Then
\begin{eqnarray} %{rcl}
        \label{argumentarnold11}
  \int_{\Sigma} d\tilde v
       &\overset{1.} {=}& \oint_{\partial \Sigma} \tilde v
                      =   \oint_{c_1} \tilde v - \oint_{c_2} \tilde v
                      = \int_{S_1} d\tilde v - \int_{S_2} d\tilde v \\
               \label{argumentarnold22}
       &\overset{2.} {=}& 0
\end{eqnarray}
Justification of the second equality (saying that the surface integral actually vanishes)
repeats the argument used in the text following Eqs. (\ref{argumentarnold1}) and (\ref{argumentarnold2}),
replacing $\xi$ by $u$ (a vector along the \emph{vortex line}, i.e. obeying $i_ud\tilde v = 0$).

\noindent
Proof 2.: Let $u$ be a vector \emph{field} defined by $i_ud\tilde v = 0$, i.e.
a field tangent, at each point, to the vortex line passing through the point;
there is a freedom $u\mapsto fu$ in the choice of $u$, $f$ being a function.
So, the flow $\Phi_s$ of $u$ leaves each vortex line invariant (as a set).
Therefore, the same holds for the vortex \emph{tube}.

Consider the (artificial!) ``dynamics'' given by $u$. Then the equation $i_ud\tilde v = 0$
may be regarded as a particular case of the basic equation (\ref{jetotoiste2}), $i_vd\alpha = d\beta$,
from the general theory of integral invariants
(with $v\mapsto u$, $\alpha \mapsto \tilde v$ and $\beta \mapsto 0$;
 notice that \emph{solutions} from Sec.\ref{poincareincartans} become \emph{vortex lines}, here,
 and the \emph{tube of solutions} becomes the \emph{vortex tube}). So,
\begin{equation} \label{jerelativny}
                 \oint_c \tilde v = \text{relative invariant}
\end{equation}
and, consequently,
\begin{equation} \label{jeabsolutny}
                 \int_S d\tilde v = \text{absolute invariant}
\end{equation}
both of them for our ``artificial dynamics'' generated by $u$
(as opposed to the \emph{real dynamics} generated by the fluid velocity field $v$).
However, Eq. (\ref{jeabsolutny}) exactly says that the vorticity flux does not depend
on the particular choice of cross-section $S$ cutting the tube.
(Just realize that, starting from a fixed cross-section $S_1$, one can produce any other $S_2$
 by means of the flow $\Phi_s$ with appropriate $s$ and $u\sim fu$.)
\hfill $\blacktriangle$ \par
\vskip .5cm
We can write the proved statement as
\begin{equation} \label{indepofsection}
                 \boxed{\int_S \boldsymbol \omega \cdot d\bold S = \text{const.}}
\end{equation}
which looks exactly as Eq. (\ref{Kelvin_statabs2}), although it has completely different meaning, now.
While Eq. (\ref{Kelvin_statabs2}) says that the integral is constant \emph{in time}
(i.e. when the surface flows away as a ``material surface'' - it changes according to
$S\mapsto \Phi_t (S)$, $\Phi_t \leftrightarrow v$), Eq. (\ref{indepofsection}), on the contrary,
says that the integral is constant ``\emph{along the vortex tube}''
(i.e. when $S$ moves - \emph{at fixed time} - along the tube, so
 $S\mapsto \Phi_s (S)$, $\Phi_s \leftrightarrow u$, $i_ud\tilde v =0$)).
 These two meanings of the equally looking formula may also be expressed differentially:
 the vorticity two form $d\tilde v$ is \emph{Lie-invariant} w.r.t. as many as \emph{two} vector fields:
\begin{eqnarray} %{rcl}
        \label{Liederivwrtfieldv}
          \mathcal L_v d\tilde v &=& 0
             \hskip 1cm \text{invariancy w.r.t. motion \emph{in time}}       \\
        \label{Liederivwrtfieldu}
          \mathcal L_u d\tilde v &=& 0
             \hskip 1cm \text{invariancy w.r.t. motion \emph{along vortex tube}}
\end{eqnarray}
\vskip .5cm
\noindent $\blacktriangledown \hskip 0.5cm$
The first one immediately results (see Eq. (\ref{vorticityisinv1})) from applying $d$ on $i_vd\tilde v = -d\mathcal E$
(Euler equation), the second one from doing the same on $i_ud\tilde v=0$
(definition of the field $u$ along vortex lines).
\hfill $\blacktriangle$ \par
\vskip .5cm
Let us summarize the following properties of vortex tubes, which we already know:

- the strength is \emph{intrinsic invariant} associated with the tube

- it is \emph{constant in time} as the tube moves

- the tube moves together with the fluid (it is ``frozen'' in it)

\vskip .5cm
\noindent $\blacktriangledown \hskip 0.5cm$
The first statement follows immediately from Eq. (\ref{indepofsection}), the second one
from Eq. (\ref{Kelvin_statabs2}). The third property expresses the content
of Eq. (\ref{Helmholtztheorem}).
\hfill $\blacktriangle$ \par
\vskip .5cm
\vskip .5cm
\noindent
[As the fluid (and consequently a vortex tube) moves, it sometimes happens that the cross section
of the tube \emph{decreases} its area.
(The tube becomes thinner as time goes on.
 A possible reason may be stretching of the tube in incompressible fluid - like when water
 flows down a bathtub drain).
Since the strength of the tube is to remain the same,
the magnitude of the vorticity (i.e. the length of the vorticity vector)
necessarily \emph{increases} to keep the strength. This very phenomenon occurs in \emph{tornado}.
Stretching of vortex tube caused by updrafts of hot air in the atmosphere
can result in dramatic magnitude of vorticity.]
\vskip .5cm

Well, and now let us mention what the three Helmholtz celebrated theorems say:
\newline \newline
\noindent
\emph{First theorem}:
            The strength of a vortex filament is constant along its length.
\newline \newline
\noindent
\emph{Second theorem}:
A vortex filament cannot end in a fluid;
\newline
\indent
\hskip 2.1cm
it must extend to the boundaries of the fluid or form a closed path.
\newline \newline
\noindent
\emph{Third theorem}:
A fluid that is initially irrotational remains irrotational.
\newline
\indent
\hskip 2cm
(In the sense that \emph{fluid elements} initially free of vorticity
\newline
\indent
\hskip 2cm
remain free of vorticity.)
\vskip .5cm
\noindent $\blacktriangledown \hskip 0.5cm$
The first theorem is nothing but the statement given in Eq. (\ref{indepofsection}).

The second one is discussed at some length (deservedly) in Appendix \ref{app:cannotend} and \ref{app:linesandtubes}.

The third one: consider two points, $P$ and $Q=\Phi_t(P)$. If some ``fluid element'' sits at $P$ at time $t=0$,
it is at $Q$ at time $t$.
Assume the contrary: let the ``fluid element'' be ``rotational'' at $Q$, i.e. vorticity $2$-form
$d\tilde v$ is non-zero
at $Q$ at time $t$. This means that there exists an infinitesimal surface $S$ around $Q$ such that
$\int_Sd\tilde v \neq 0$. The surface $S$ is $\Phi_t$-image of some $S_0$ around $P$
(namely of $S_0 = \Phi_{-t}S$). Therefore
$$0\neq \int_Sd\tilde v = \int_{\Phi_t(S_0)}d\tilde v = \int_{S_0}\Phi_t^*d\tilde v = \int_{S_0}d\tilde v
$$
and this means that vorticity $2$-form $d\tilde v$ is non-zero at $P$ and, consequently, that
the ``fluid element'' is ``rotational'' at $P$.
\hfill $\blacktriangle$ \par
\vskip .5cm

\subsubsection{Ertel's theorem}
\label{euler_stationary_Ertel}

All the statements of the last three sections
(Bernoulli equation in Sec. \ref{euler_stationary_bernoulli},
 Kelvin theorem in Sec. \ref{euler_stationary_kelvin}
 and Helmholtz theorem in Sec. \ref{euler_stationary_helmholtz})
rest heavily on \emph{barotropic} assumption $p=p(\rho)$, see Eq. (\ref{barotropicfluid}),
added to Euler equation in its more general form Eq. (\ref{eulerstatform2}).
So they assume that Euler equation has its simpler form $i_vd\tilde v = -d\mathcal E$
(see Eq. (\ref{eulerstatform3})).

Here we mention a weaker statement due to Ertel (1942) which is, however, more general
in that it \emph{does not} assume barotropic regime of the fluid.

Recall that the crucial point of derivation of Helmholtz theorem from Euler equation
consisted in application of $d$ on Eq. (\ref{eulerstatform3}).
Let us mimic the procedure replacing, however, Eq. (\ref{eulerstatform3}) by the more general
one, Eq. (\ref{eulerstatform2}).

We get, instead of Eq. (\ref{vorticityisinv1}), the following equation:
\begin{equation} \label{vorticityisnotinv1}
                 \mathcal L_v (d\tilde v) = \frac{d\rho \wedge dp}{\rho^2}
\end{equation}
So, the vorticity $2$-form $d\tilde v \equiv \boldsymbol \omega \cdot d\bold S$ is \emph{no longer} invariant
w.r.t. the flow $\Phi_t \leftrightarrow v$.

\vskip .5cm
\noindent
[Therefore surface \emph{integral} of the vorticity $2$-form $d\tilde v \equiv \boldsymbol \omega \cdot d\bold S$ is
no longer integral invariant. Since the integral may be transformed to velocity circulation
round the boundary of the surface, \emph{Kelvin}'s theorem no longer holds. Looking at Eq. (\ref{vorticityisinv1})
we see, that neither \emph{Helmholtz} theorem holds for non-barotropic flow.
And, finally, applying $i_v$ on Eq. (\ref{eulerstatform2}), rather than Eq. (\ref{eulerstatform3}),
we convince ourselves that \emph{Bernoulli} equation cease to hold as well.]
\vskip .5cm

There is, however, a braver \emph{$3$-form} which, in spite of truly hard times in this merciless
non-barotropic world still \emph{remains} to be devoted to the noble idea of the concept of
$\Phi_t \leftrightarrow v$-invariancy.

In order to find this admirable $3$-form, notice that the equation $vs=0$ (\ref{vsvanishes}),
which is \emph{always} true for ideal fluid (i.e. it does not need the fluid to be barotropic),
may be written in a slightly more scientific form as
\begin{equation} \label{sisinvariant1}
                 \mathcal L_v s = 0
\end{equation}
Applying $d$ we get
\begin{equation} \label{dsisinvariant1}
                 \mathcal L_v ds = 0
\end{equation}
This means, however, that
\begin{equation} \label{dsdvisinvariant1}
                 \mathcal L_v (ds \wedge d\tilde v) = 0
\end{equation}
\vskip .5cm
\noindent $\blacktriangledown \hskip 0.5cm$
Indeed, we get, first
$$\mathcal L_v (ds \wedge d\tilde v) = (\mathcal L_v ds) \wedge d\tilde v
                                    + ds \wedge \frac{d\rho \wedge dp}{\rho^2}
                                    = \frac{ds \wedge d\rho \wedge dp}{\rho^2}
$$
But the (general) equation of state Eq. (\ref{generalfluid}) enables one to write $s=s(\rho ,p)$,
so that $ds = (\dots)d\rho +(\dots)dp$ and the expression above vanishes.
\hfill $\blacktriangle$ \par
\vskip .5cm
Therefore
\begin{equation} \label{invariantertel1}
                 \int_Dds \wedge d\tilde v = \ \ \text{\emph{absolute} integral invariant}
\end{equation}
In traditional notation, the statement in Eq. (\ref{invariantertel1}) is known as
\begin{equation} \label{invariantertel2}
                 \int_D ds \wedge d\tilde v
                 \equiv
                 \boxed{\int_D (\boldsymbol \omega \cdot \boldsymbol \nabla s) dV
                 = \ \text{const.}}
                 \hskip 1cm \text{{\color{red}\emph{Ertel's theorem}}}
\end{equation}
(\cite{ertel1942}; see also exposition in \$5.18 of \cite{schutz}
or \$8 in \cite{lanlif87}).
\vskip .5cm
\noindent $\blacktriangledown \hskip 0.5cm$
Using formulas from Appendix \ref{app:vectoranalysis} we have
$$
\begin{array} {rcl}
       ds \wedge d\tilde v
  &=&  (\boldsymbol \nabla s \cdot d\bold r) \wedge d(\bold v \cdot d\bold r) \\
  &=&  (\boldsymbol \nabla s \cdot d\bold r) \wedge (\boldsymbol \omega \cdot d\bold S) \\
  &=&  (\boldsymbol \omega \cdot \boldsymbol \nabla s) dV
\end{array}
$$
\hfill $\blacktriangle$ \par
\vskip .5cm
There is an alternative way to express the message of Eq. (\ref{invariantertel2}).
Notice that each $3$-form na $E^3$ is proportional to the volume form $\Omega$ or,
also, to the $3$-form $\rho \Omega$. So we can \emph{define} function $f$ such that
\begin{equation} \label{defertelfunction}
                 ds \wedge d\tilde v =: f \rho \Omega
\end{equation}
Explicitly, in vector analysis notation,
\begin{equation} \label{ertelfunctionexplic}
                 f = \frac{\boldsymbol \omega \cdot \boldsymbol \nabla s}{\rho}
\end{equation}
\vskip .5cm
\noindent $\blacktriangledown \hskip 0.5cm$
Indeed,
$$
\begin{array} {rcl}
       f\rho dV
  &=& f\rho \Omega                                            \\
  &=& ds \wedge d\tilde v                                     \\
  &=& (\boldsymbol \omega \cdot \boldsymbol \nabla s) dV
\end{array}
$$
so that $f\rho = \boldsymbol \omega \cdot \boldsymbol \nabla s$ as claimed.
\hfill $\blacktriangle$ \par
\vskip .5cm
From the invariance of the l.h.s. of Eq. (\ref{defertelfunction}), see Eq. (\ref{dsdvisinvariant1}),
we can immediately infer that
\begin{equation} \label{fisconstant}
                 \boxed{vf = 0}
\end{equation}
This says that, in addition to the function $s$, see Eq. (\ref{vsvanishes}), also the new function
$f$ is constant \emph{along streamlines}
(integral lines of the velocity field $v$).
The constant may take different values on different streamlines.
\vskip .5cm
\noindent
[Here we see why exactly the choice of $\rho \Omega$ in the definition of function $f$
 in Eq. (\ref{defertelfunction}) is much better than to use just $\Omega$.
 Namely the proportionality coefficient (function) is only $\mathcal L_v$-invariant
 (i.e. Eq. (\ref{fisconstant}) holds) if it ``stands between'' two differential forms
 which are $\mathcal L_v$-\emph{invariant themselves}.

 Since $\mathcal L_v(\rho \Omega) =0$ (i.e. mass is conserved)
 whereas $\mathcal L_v\Omega \neq 0$ (i.e. volume is not conserved),
 the right choice is $\rho \Omega$.]
\vskip .5cm

So, for the most general flow of \emph{ideal} fluid we encountered as many as \emph{three}
absolute integral invariants, all of them based on 3-forms
(i.e. given by \emph{volume} integrals):
\begin{equation} \label{3invariants1}
                 \int_D \rho \Omega
                 \hskip 1cm
                 \int_D s\rho \Omega
                 \hskip 1cm
                 \int_D f\rho \Omega
\end{equation}
The same objects, when written using notation more frequent in fluid dynamics:
\begin{equation} \label{3invariants2}
                 \int_D dm
                 \hskip 1cm
                 \int_D sdm
                 \hskip 1cm
                 \int_D fdm
\end{equation}
In words: by any ``moving particle''
(in hydrodynamic terminology, a fixed small amount of mass of the fluid)
constant values of as many as three quantities are carried:
$\delta m$ (its mass), $s\delta m$ (its entropy) and $f\delta m$
(the new vorticity-based quantity introduced by Ertel).

What happens with the three integral invariants if we return back to the more specific,
\emph{barotropic} situation?
Since $s=$ const. (in the sense of Eq. (\ref{sequalsconst}), not just along $v$),
we deduce that

- the second integral is just a constant multiple of the first one

- the constancy of the third integral is the statement $0=$ const.

\noindent
(The 1-form $ds$ vanishes and both Eqs. (\ref{invariantertel2}) and (\ref{fisconstant})
  become vacuous. Well, the statement $0=$ const. \emph{is} true, we have to admit,
  but it seldom helps very much.)

On the other hand, do not forget about \emph{Bernoulli} equation $v\mathcal E=0$ (\ref{bernoulli3}),
which also may be regarded in terms of $\mathcal E \delta m$ carried by a ``moving particle''
(as we discussed earlier in Sec. \ref{euler_stationary_bernoulli}). So, in the barotropic case,
 we can safely state that we have \emph{two} non-trivial absolute integral invariants:
\begin{equation} \label{2invariantsbarotrop1}
                 \int_D \rho \Omega
                 \hskip 1cm
                 \int_D \mathcal E\rho \Omega
\end{equation}
or, in the more usual notation,
\begin{equation} \label{2invariantsbarotrop2}
                 \int_D dm
                 \hskip 1cm
                 \int_D \mathcal E dm
\end{equation}

\setcounter{equation}{0}
\subsection{Non-stationary flow}
\label{euler_non_stationary}

In this section we learn how the general scheme of integral invariants,
studied in some detail in Sec. \ref{euler_poincare_cartan}, directly offers to express
Euler non-stationary equations in a remarkable form from which, among other things,
Helmholtz vortex-lines statement results as easily as it was demonstrated, for stationary case,
in Section \ref{euler_stationary_helmholtz}.

\subsubsection{Euler equation again - stationary case}
\label{eulerintegrinvarstat}

Recall that already in Eq. (\ref{porovnanie}) we displayed, side by side in one line,
the stationary Euler equation versus the core equation of the general theory of Poincar\'e version
of integral invariants
\begin{equation} \label{porovnanie2}
              i_v (d\tilde v)
                   = - d\mathcal E
    \hskip 1cm  \leftrightarrow \hskip 1cm
                 i_v (d\alpha)
                   = d\beta
\end{equation}
This resulted in the following identification of key objects of the two theories
\begin{equation} \label{stotoznenie2}
             \alpha \leftrightarrow \tilde v
             \hskip 1cm
             \beta  \leftrightarrow -\mathcal E
             \hskip 1cm
             v  \leftrightarrow v
\end{equation}
and led to understanding of the Euler stationary equation from Poincar\'e perspective.

Now the goal is to retell \emph{the same} physics \emph{\`{a} la Cartan}
(i.e. with \emph{no change} of the content, \emph{yet}).
It is easy, since we know the general algorithm:

First, in the spirit of Eqs. (\ref{definiciaxi1}) - (\ref{definiciasigma1}),
we introduce the $1$-form
\begin{equation} \label{defsigma1}
             \sigma := \tilde v - \mathcal E dt
\end{equation}
(here, both objects only depend on $\bold r$, they are stationary) and the vector field
\begin{equation} \label{defxi1}
             \xi := \partial_t + v
\end{equation}
(the field $v$ only depends on $\bold r$ as well, it is stationary, too).
Then, due to Eq. (\ref{poincarecartan1}), the stationary Euler equation (\ref{eulerstatform3})
is equivalent to
\begin{equation} \label{ixidesigmajenula1}
             i_{\xi}d\sigma = 0
\end{equation}
It is that simple.

\subsubsection{Euler equation again - non-stationary case}
\label{eulerintegrinvarnonstat}

Now we switch to \emph{non-stationary} case.
This is also pretty straightforward, we just follow the logic of general
Eqs. (\ref{poincarecartan3}) and (\ref{poincarecartan4})).
So, the idea is that the value above all is to keep valid the structure of
Eq. (\ref{ixidesigmajenula1}) even for non-stationary case, i.e. when objects,
from which $\sigma$ and $\xi$ are composed, \emph{do} depend on \emph{time}
(in addition, of course, to their dependence on $\bold r$):
\begin{equation} \label{defsigma2}
             \sigma := \hat v - \mathcal E dt
\end{equation}
\begin{equation} \label{defxi2}
             \xi := \partial_t + v
\end{equation}
where, now,
\begin{equation} \label{zavisiaodt}
             v          = v(\bold r,t)      \hskip 1cm
             \hat v     = \hat v(\bold r,t) \hskip 1cm
             \mathcal E = \mathcal E(\bold r,t)
\end{equation}

\vskip .5cm
\noindent
[Notation of the (field of the) $1$-form $\tilde v$ was changed, in the process of switching
 from space to space-time,
 namely from $\tilde v$ to $\hat v$ (in order to make it clear that we speak of a \emph{spatial} $1$-form);
 notation of the fields $v$ and $\mathcal E$ remains unchanged,
 one has to keep in mind, however, that $v$ is a \emph{spatial} vector field
 (its time component vanishes) and that, of course, all three objects live in space-time,
 i.e. they \emph{do depend on} $t$ as well.]
\vskip .5cm

And what about the ,,decomposed'' version of the equation (\ref{ixidesigmajenula1})?
We already know the answer in general - it is given in Eq. (\ref{poincarecartan3}).
So, using the identification of $\hat \alpha$ and $\hat \beta$ according to
(\ref{defsigma2}) and (\ref{defxi2}), we get
\begin{equation} \label{eulernestac1}
       {\mathcal L}_{\partial_t}\hat v + i_v \hat d \hat v = - \hat d \mathcal E
\end{equation}
Therefore, (\emph{non-stationary}, but still \emph{barotropic})
           \emph{Euler equation}
may be written in any of the two versions:
\begin{equation}
      \label{eulernestac2}
      \boxed{i_\xi d\sigma =0}
      \hskip 1.8cm
      \text{{\color{red} \emph{Euler equation}}}
\end{equation}
or
\begin{equation}
      \label{eulernestac3}
      \boxed{{\mathcal L}_{\partial_t}\hat v + i_v \hat d \hat v = - \hat d \mathcal E}
      \hskip .5cm
      \text{{\color{red} \emph{Euler equation (spatial version)}}}
\end{equation}
where
\begin{equation}
      \label{eulernestac4}
  \sigma = \hat v - \mathcal E dt
           \hskip 1cm
  \xi    = \partial_t +v
           \hskip 1cm
  \mathcal E
         = v^2/2+P+\Phi
\end{equation}
\vskip .5cm
\noindent $\blacktriangledown \hskip 0.5cm$
Let's check explicitly that this indeed exactly matches Euler equation written in the traditional
form (\ref{eulernorm}) by unpacking in Cartesian coordinates:
Work pieces:
$$
\begin{array} {rcl}
      \hat v
  &=& v_idx_i                          \\
      \mathcal L_{\partial_t}\hat v
  &=& v_{i,t} dx_i                     \\
      \hat d \hat v
  &=& v_{i,j}dx_j \wedge dx_i
\end{array}
$$
$$
\begin{array} {rcl}
      i_v \hat d \hat v
  &=& v_{i,j}(v_j dx_i -v_idx_j ) \\
  &=& v_j(v_{i,j} -v_{j,i})dx_i \\
  &=& [(\bold v \cdot \boldsymbol \nabla)v_i - ({\bold v}^2/2)_{,i} ]dx_i
\end{array}
$$
$$
\begin{array} {rcl}
      \hat d \mathcal E
  &=&  {\mathcal E}_{,i}dx_i \\
  &=& [(P+\Phi)_{,i} + ({\bold v}^2/2)_{,i} ]dx_i
\end{array}
$$
Plugging this into Eq. (\ref{eulernestac1}) and equating expressions combined with $dx_i$
on both sides we find:
\begin{equation} \label{vrajeuler2}
       v_{i,t} + (\bold v \cdot \boldsymbol \nabla)v_i - ({\bold v}^2/2)_{,i}
    = -(P+\Phi)_{,i} - ({\bold v}^2/2)_{,i}
\end{equation}
so we see the $i$-th component if vector equation
\begin{equation} \label{vrajeuler3}
       \partial_t \bold v + (\bold v \cdot \boldsymbol \nabla)\bold v = - \boldsymbol \nabla (P+\Phi)
\end{equation}
And that's exactly Euler equation (\ref{eulernorm}) in the barotropic case (i.e. when Eq. (\ref{barotropic_P}) holds).
\hfill $\blacktriangle$ \par
\vskip .5cm
\noindent
[Still another way in which Euler equation (\ref{eulernestac2}) may be written is
\begin{equation} \label{eulernonstatform3vecanal}
                 \partial_t \bold v + \bold v \times (\boldsymbol \nabla \times \bold v)
                   = \boldsymbol \nabla \mathcal E
\end{equation}
In order to see this is true, just compare
Eqs. (\ref{eulerstatform3}), (\ref{eulernestac3}) and (\ref{eulerstatform3vecanal}).]
\vskip .5cm

Although Euler equation looks extremely simply when written in the form (\ref{eulernestac2}),
remember that the three unknown quantities, $v$, $p$ and $\rho$, which are to be determined from it
as functions of $\bold r,t$, enter it in a fairly complicated way.
This is best visible when one writes down the equation explicitly:
\begin{equation} \label{ixidesigmajenulavdrobnych}
             i_\xi d\sigma \equiv
             i_{\underbrace{(\partial_t + v)}_\xi}d(\underbrace{\hat v - (\overbrace{v^2/2 +P+\Phi}^{\mathcal E}) dt}_{\sigma}) = 0
\end{equation}
In particular notice that the velocity field $v$, which is unknown, sits \emph{both} in
vector field $\xi$ \emph{and} in $1$-form $\sigma$.
(It is even \emph{twice} in $\sigma$; both linear and quadratic terms in $v$ are there).

So one looks for such functions $p(\bold r,t)$ and $\rho(\bold r,t)$ as well as
spatial vector field $v(\bold r,t)$ which, when plugged into (\ref{defsigma2}) and (\ref{defxi2}),
result in $\xi$ and $\sigma$, which satisfy elegant equation (\ref{eulernestac2}).

\subsubsection{Conservation of mass and entropy again}
\label{massandentropy}

In Sec. \ref{massandentropy_stat} we came to conclusion that, for \emph{stationary} flow,
the following expressions
\begin{eqnarray} %{rcl}
      \label{massfluidnonstat}
             m(D) &:=& \int_D \rho \Omega
                 \hskip 1.35cm
                 \text{\emph{total mass in $D$}}            \\
      \label{entropyfluidnonstat}
             S(D) &:=& \int_D s\rho \Omega
                 \hskip 1.2cm
                 \text{\emph{total entropy in $D$}}
\end{eqnarray}
should be \emph{absolute integral invariants}, i.e. that both mass and entropy should be \emph{conserved}.
Is this statement to be modified, when switching to \emph{time-dependent} case?

No.

First, still holds that $m(D(t))$ is the same amount of mass as $m(D(0))$, it just traveled
(formally via the flow $\Phi_t$ generated by $\xi$) to some other place.

Second, still holds that the fluid is \emph{ideal}.
There is still no heat exchange between different parts of the fluid,
neither energy dissipation caused by internal friction, viscosity, in the fluid.
So the motion of ideal fluid is to be still treated as \emph{adiabatic}.
Entropy $S(D(t))$ is still to be equal to $S(D(0))$.

The only thing which is to be modified is mathematical expression of mass and entropy conservation.
It becomes a bit more complex, now.

Recall, however, that conservation laws of the type encountered here were already
discussed in detail in general context of integral invariants in Section \ref{continuity}
and, in particular, the time-dependent ``solution'' was given there.

So, we already know how our \emph{two} continuity equations look like, now:
\begin{equation}
      \label{massconteq1non}
            \boxed{\partial_t \rho  + \hat {\Div} (\rho v)  = 0}
                             \hskip 1.6cm \text{(\emph{continuity equation for mass})}
\end{equation}
\begin{equation}
       \label{entropyconteq1non}
           \boxed{\partial_t (s\rho)  + \hat {\Div} (s\rho v)  = 0}
                             \hskip 1cm \text{(\emph{continuity equation for entropy})}
\end{equation}
where the \emph{spatial divergence} was introduced in Eq. (\ref{spatialdivergence}).
They play the role of differential versions of Eqs. (\ref{massfluidnonstat})
and (\ref{entropyfluidnonstat})):

The two integral invariants are still related in a specific way described
in Sec. \ref{continuity}, see Eq. (\ref{withOmega}) and below.
Recall, that the correspondence is a bit tricky:
$$\text{
        $\Omega \ , \ \rho \Omega$ \ \ there
        \hskip 1cm $\leftrightarrow$ \hskip 1cm
        $\rho \Omega \ , \ s(\rho \Omega)$ \ \ here
       }
$$
From this fact we can deduce, according to Eq. (\ref{liouville}), that
\begin{equation} \label{vsvanishesnonstat}
                 \boxed{\xi s=0}
                 \hskip .5cm \text{or, equivalently} \hskip .5cm
                 \boxed{\partial_t s +vs =0}
\end{equation}
This says that ``total time derivative'' of $s$ vanishes.
(Total derivative consists of partial time derivative plus
 derivative along ``momentary streamline'').
Geometrically it means that $s$ is constant along
``streamlines in the extended space''.
The constant may, in general, take different values on different ``streamlines in the extended space''.
\vskip .5cm
\noindent $\blacktriangledown \hskip 0.5cm$
A proof of the general claim Eq. (\ref{liouville}) was given at the place it was stated.
A direct proof of Eq. (\ref{vsvanishesnonstat}) is also easy combining
Eqs. (\ref{massconteq1non}) and (\ref{entropyconteq1non}) in a way similar to what was done
in direct proof of Eq. (\ref{vsvanishes}).
\hfill $\blacktriangle$ \par
\vskip .5cm

\setcounter{equation}{0}
\subsection{Simple consequences of non-stationary Euler equation}
\label{conseq_euler_non_stationary_barotropic}

In Sec. \ref{conseq_euler_stationary_barotropic} we discussed, in the context of \emph{stationary}
Euler equation, its four simple consequences:

- Bernoulli result on conservation of the quantity $\mathcal E = v^2/2 +P+\Phi$

- Kelvin's result on conservation of velocity circulation (or vorticity flux)

- Helmoltz result on ``conservation of vortex lines''

- Ertel's result on conservation of the quantity $f = (\boldsymbol \omega \cdot \boldsymbol \nabla s)/\rho$

The first three laws were only proved under \emph{additional} assumption that the flow is
\emph{barotropic} (isentropic).
Moreover, in Sec. \ref{euler_stationary_Ertel} we came to conclusion that the barotropic assumption
is necessary, the theorems \emph{do not} hold for non-barotropic situation.
(Ertel's theorem remains, on the contrary, to be true.)

And what about the first assumption?
Are the first three results still true when we consider generalization to \emph{non-stationary}
(but, at the same time, barotropic) flow discussed here? Well, let's have a look.

\subsubsection{Bernoulli statement - no longer true}
\label{euler_non_stationary_bernoulli}

When $i_v$ is applied on Eq. (\ref{eulernestac3}), we get
\begin{equation} \label{bernoulli_non_stat}
        \partial_t \left( \frac{v^2}{2} \right) + v\mathcal E = 0
\end{equation}
\vskip .5cm
\noindent $\blacktriangledown \hskip 0.5cm$
What we actually obtain is
$$i_v{\mathcal L}_{\partial_t}\hat v + v \mathcal E =0
$$
But, in Cartesian coordinates,
$$
\begin{array} {rcl}
       i_v{\mathcal L}_{\partial_t}\hat v
  &=&  i_v (v_{i,t}dx^i)\\
  &=&   v_iv_{i,t} \\
  &=&  \partial_t(v_iv_i/2)
\end{array}
$$
\hfill $\blacktriangle$ \par
\vskip .5cm
And this \emph{differs} from $\xi \mathcal E=0$. So, $\mathcal E$ is no longer conserved
for time-dependent flow.

\subsubsection{Kelvin's theorem on circulation - still true}
\label{euler_non_stationary_kelvin}

\begin{figure}[tb]
\begin{center}
\includegraphics[clip,scale=0.40]{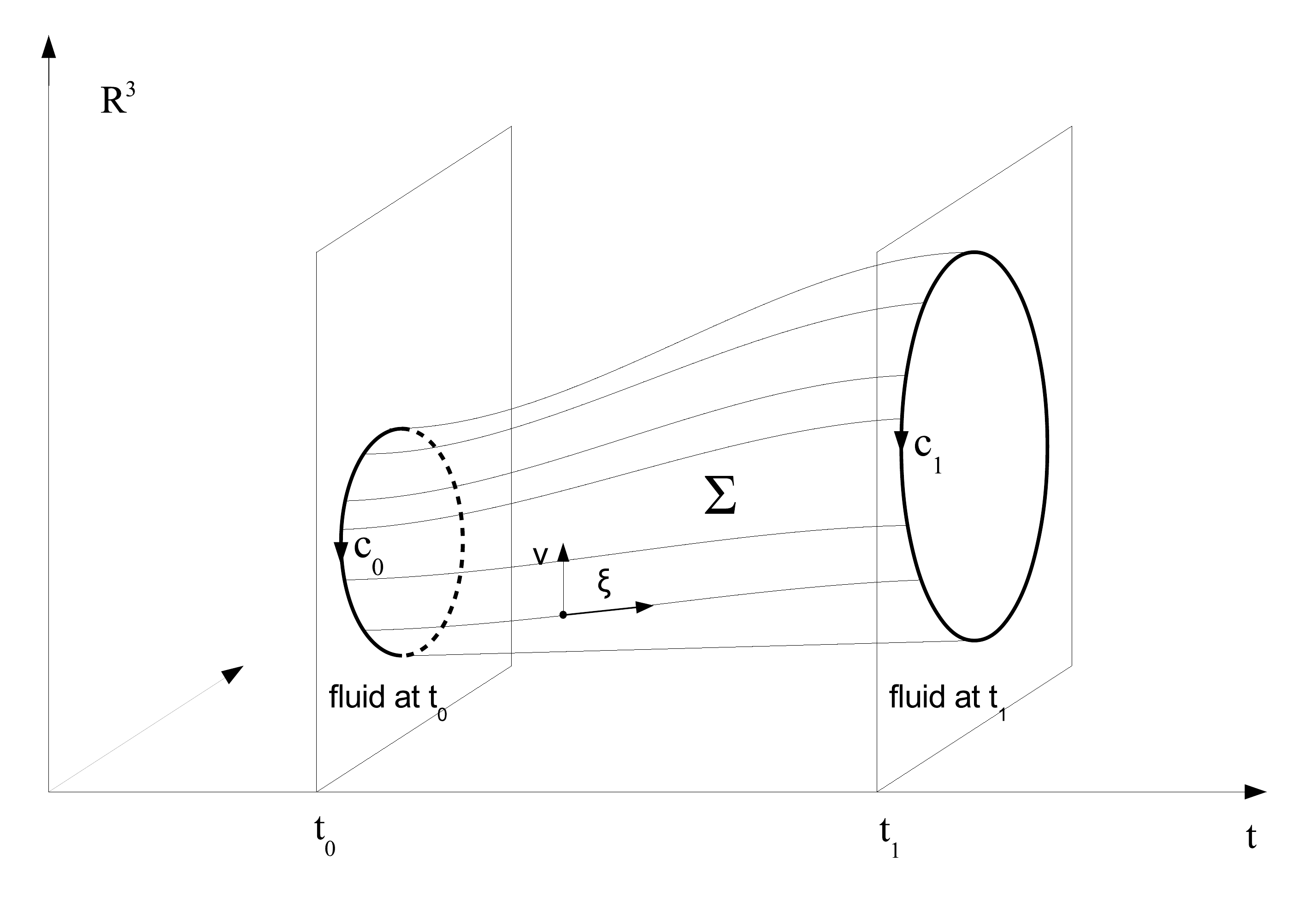}
\caption{The cycles $c_0$ and $c_1$ lie in hyper-planes of constant time
         on extended space $\mathbb R^3 \times \Bbb R$,
         so they correspond to cycles in the fluid at time $t_0$ and $t_1$ respectively.
         They encircle the same tube of integral curves of the vector field $\xi = \partial_t +v$,
         so if $c_0$ is regarded as an arbitrary \emph{material} cycle in the fluid at time $t_0$,
         then $c_1$ contains the same fluid particles at time $t_1$.}
\label{kelvinnonstationary}
\end{center}
\end{figure}

The positive answer results simply from straightforward application of the general result
from the theory of integral invariants displayed in Eqs. (\ref{jetotoiste4}) and (\ref{jetotoiste5}):
since (the time-dependent) Euler equation has the form $i_\xi d\sigma = 0$, integration
of $\sigma$ over a cycle results necessarily in (relative) integral invariant:
\begin{equation} \label{abc1}
        \text{Euler equation reads} \ \ i_\xi d\sigma =0
        \hskip .7cm \Rightarrow \hskip .7cm
        \oint_c\sigma = \ \ \text{\emph{relative invariant}}
\end{equation}
where the last statement means, in detail,
\begin{equation} \label{abc2}
        \oint_{c_0}\sigma = \oint_{c_1}\sigma \hskip 1.5cm \text{if $c_0$ and $c_1$ {\emph{encircle a common tube}} of solutions}
\end{equation}
After inserting the explicit form of $\sigma$ from Eq. (\ref{eulernestac4}), we get
\begin{equation} \label{abc3}
        \oint_{c_0}(\hat v -\mathcal E dt) = \oint_{c_1}(\hat v -\mathcal E dt)
        \hskip .5cm
        \text{if $c_0$ and $c_1$ {\emph{encircle a common tube}} of solutions}
\end{equation}
Restricting the (1-dimensional) cycles $c_0$ and $c_1$ to fixed-time hyper-surfaces
$t=t_0$ and $t_1$, respectively (i.e. \`{a} la Poincar\'e rather than \`{a} la Cartan), we get
(see Fig. \ref{kelvinnonstationary})
\begin{equation} \label{Kelvin_non_stat}
                 \boxed{
                 \oint_c \bold v \cdot d\bold r = \text{const.}}
                 \hskip 1.5cm \text{{\color{red}\emph{Kelvin's theorem}}
                       }
\end{equation}
\vskip .5cm
\noindent
[A proof in terms of vector analysis may start from Reynolds transport theorem
Eq. (\ref{transport1}) for $k=1$: if we put $\bold A = \bold v$
 and take into account that $\bold v$ \emph{there} is just our $\bold v$ \emph{here}
 and that $c(t)$ is \emph{closed}, here ($\partial c=0$, so that the last term drops out), we get
\begin{equation} \label{KelvinviaReynolds}
\frac{d}{dt}\oint_{c(t)} \bold v \cdot d\bold r
                  =
                  \int_{c(t)} (\partial_t\bold v + \curl \bold v \times \bold v) \cdot d\bold r
                  =
                  \int_{c(t)} \boldsymbol \nabla \mathcal E \cdot d\bold r
                  =
                  \int_{\partial c(t)} \mathcal E
                  = 0
\end{equation}
(In the second equation sign we used Euler equation in the form Eq. (\ref{eulernonstatform3vecanal}).)]

\subsubsection{Helmholtz theorem on vortex lines - still true}
\label{euler_non_stationary_helmholtz}

Here we show that the possibility to write Euler equation (not restricted to stationary flow) in the form
of Eq. (\ref{eulernestac2}) enables one to prove the Helmholtz theorem
(stating that vortex lines are ``frozen'' into perfectly inviscid barotropic fluid)
equally easily as it was done (for the case of stationary flow) in
Section \ref{euler_stationary_helmholtz}
(it is quite instructive to check our steps here versus the corresponding steps there).

First, realize that from Eq. (\ref{eulernestac2}) we immediately get
\begin{equation} \label{dsigmajeinvar1}
             \mathcal L_{\xi}(d\sigma) = 0
\end{equation}
(just apply $d$ and use $di_\xi +i_\xi d=\mathcal L_\xi$),
i.e. the invariance of the $2$-form $d\sigma$ w.r.t. the flow $\Phi_t$ of the field $\xi$
\begin{equation} \label{dsigmajeinvar2}
             \Phi_t^*(d\sigma) = d\sigma
             \hskip 2cm
             \Phi_t \leftrightarrow \xi
\end{equation}
We already know (from Eq. (\ref{issue1}) and the text after it) that this implies
that if a vector $w$ \emph{annihilates} the $2$-form $d\sigma$ at some time $t_0$,
then the vector $\Phi_{t*}w$ annihilates the $2$-form $d\sigma$ at time $t_0+t$.
That is, if $w$ annihilates $d\sigma$ at some time,
then, being dragged by the flow $\Phi_t$ (so \emph{Lie-dragged}; technically $w\mapsto \Phi_{t*}w$),
it keeps this property ``forever'':
\begin{equation} \label{iwdsigma0isLiedragge}
             i_wd\sigma =0
             \hskip 1cm \Rightarrow \hskip 1cm
             i_{\Phi_{t*}w}(d\sigma) = 0
\end{equation}
\begin{figure}[tb]
\begin{center}
\includegraphics[clip,scale=0.40]{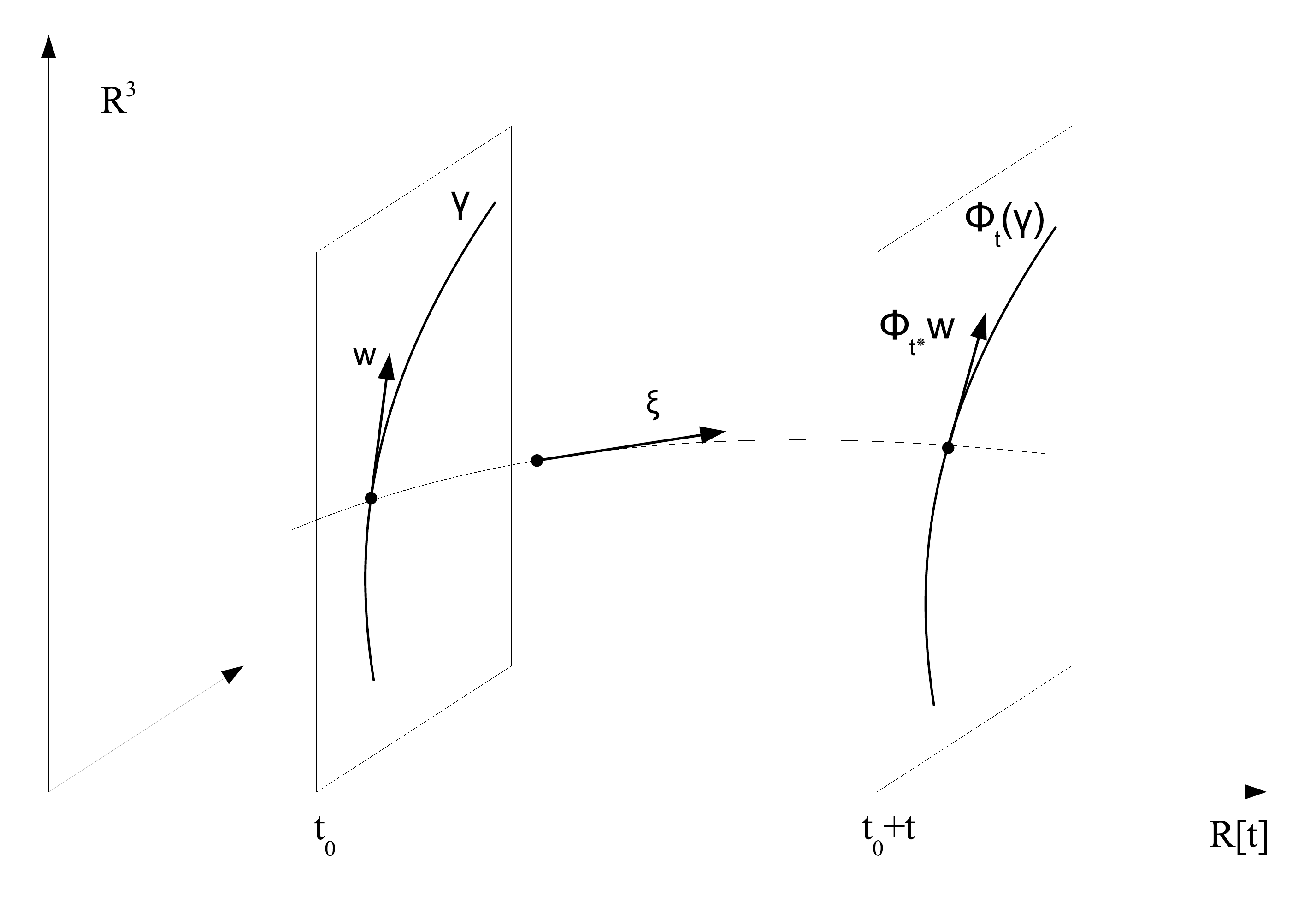}
\caption{$\gamma$ is a vortex line at time $t_0$ and $w$ is a vector tangent to $\gamma$ (i.e. a multiple of the vorticity vector at that point).
         The flow maps $w$ to $\Phi_{t*}w$.
         It turns out that the vortex line passing through the point of tangency of $\Phi_{t*}w$
         is nothing but the image $\Phi_t(\gamma)$ of the vortex line $\gamma$ w.r.t. the flow of $\xi$.
         So vortex lines are \emph{frozen} into fluid also in non-stationary case.}
\label{fig:helmholtznonstationary}
\end{center}
\end{figure}
Now consider, at time $t_0$, a vortex line $\gamma$ (see Fig. \ref{fig:helmholtznonstationary}). Parametrize it arbitrarily by $\lambda$.
Its tangent vector $\gamma'(\lambda)$ is a spatial vector $w$ which, by definition,
annihilates $2$-form $\hat d \hat v$
\begin{equation} \label{wannihilateshatdhatv}
             w= \gamma'(\lambda) = \ \text{spatial}
             \hskip 1cm
             i_w\hat d \hat v = 0
\end{equation}
Use the general wisdom learned in Eqs. (\ref{dsigmavzdyvseob}), (\ref{dsigmanariesenivseob}).
Here, it says
\begin{eqnarray} %{rcl}
      \label{dsigmavzdyvseob2}
      d\sigma &=& \hat d \hat v + dt \wedge ({\mathcal L}_{\partial_t} \hat v -\hat d \mathcal E)
                  \hskip .8cm \text{always} \\
  \label{dsigmanariesenivseob2}
        &=& \hat d \hat v +dt\wedge (-i_v\hat d\hat v)
                       \hskip 1.4cm \text{\emph{on solutions} of Euler equation}
\end{eqnarray}
But then it should be clear that, \emph{assuming that Euler equation is true},
\begin{equation} \label{drobnyfakt}
             i_w(d\sigma) = 0
             \hskip 1cm \Leftrightarrow \hskip 1cm
             i_w(\hat d \hat v) = 0
\end{equation}
\noindent $\blacktriangledown \hskip 0.5cm$
Indeed,
$$
\begin{array} {rcl}
  i_wd\sigma
  &=& i_w\hat d \hat v +dt\wedge (i_wi_v\hat d\hat v) \\
  &=& i_w\hat d \hat v -dt\wedge (i_v(i_w\hat d\hat v)) \\
\end{array}
$$
If we denote $i_w\hat d \hat v \equiv \hat b$ (it is a \emph{spatial} $1$-form), we get
$$
  i_wd\sigma
  = \hat b -dt\wedge i_v\hat b
$$
from which immediately
$$
             i_w(d\sigma) = 0
             \hskip 1cm \Leftrightarrow \hskip 1cm
             \hat b  = 0
$$
\hfill $\blacktriangle$ \par
\vskip .2cm \indent
Let us collect together once more all relevant equations,
Eqs. (\ref{iwdsigma0isLiedragge}), (\ref{wannihilateshatdhatv}) and (\ref{drobnyfakt}),
in as suggestive a way as possible:
\begin{eqnarray} %{rcl}
      \label{gammaisvortexline}
            i_{\gamma'}(\hat d \hat v) &=& 0 \hskip 3cm \text{$\gamma$ is a vortex line}              \\
      \label{anihdviffanihdsigma}
            i_{\gamma'}(\hat d \hat v) = 0
            \hskip .3cm  &\Leftrightarrow& \hskip .3cm i_{\gamma'}(d \sigma) = 0                      \\
       \label{gamaanihajfitgammaanih}
            i_{\gamma'}(d \sigma) = 0
      \hskip .3cm  &\Rightarrow& \hskip .3cm i_{\Phi_{t*}\gamma'}(d \sigma) = 0
\end{eqnarray}
Then, clearly, we can deduce that
\begin{equation} \label{wecandeduce}
             i_{\Phi_{t*}\gamma'}(\hat d \hat v) = 0
             \hskip 1cm \text{i.e.} \hskip 1.95cm
             \text{$\Phi_t \circ \gamma$ is a vortex line as well}
\end{equation}
Expressed in plain English: Helmholtz theorem \emph{holds} for non-stationary flow, too.

As already mentioned at the end of Sec. \ref{euler_stationary_helmholtz},
the reader who grasped material discussed in the Chapter \ref{euler_poincare_cartan}
(devoted to integral invariants in general)
probably noticed that the main statement presented here may be regarded as
a particular case of the result presented in Sec. \ref{remarkable_surface}
(and also the proof is very similar). Note that the section \ref{remarkable_surface}
treats non-stationary situation from the very beginning.

\vskip .5cm
\noindent
[Usually analysis of vortex dynamics is based on equation
\begin{equation} \label{vortexdynamics}
             \partial_t \boldsymbol \omega = \boldsymbol \nabla \times (\bold v \times \boldsymbol \omega )
             \hskip 1cm
             \text{\emph{vorticity equation}}
\end{equation}
In is obtained as follows: we apply $\hat d$ on the spatial version of Euler equation,
Eq. (\ref{eulernestac3}). Since $\hat d$ commutes with $\mathcal L_{\partial_t}$
(check), we get
$$\mathcal L_{\partial_t}\hat \omega + \hat d i_v \hat \omega = 0
  \hskip 2cm
  \hat \omega := \hat d \hat v \equiv \boldsymbol \omega \cdot d\bold S
$$
Using formulas from Appendix \ref{app:vectoranalysis} we immediately get Eq. (\ref{vortexdynamics}).]
\vskip .5cm

\subsubsection{Digression: Helmholtz theorem in Hamiltonian mechanics?}
\label{Helmholtz_Hamiltonian}

Recall that we found, in general theory of integral invariants,
``remarkable integral surfaces'' whose behavior may be concisely characterized
as that they are \emph{frozen into the fluid}
(see Eq. (\ref{statementonM2}) in Sec. \ref{remarkable_surface}).

When studying ideal fluid dynamics we learned that the fundamental dynamical equation
has the form of equation typical for the theory of integral invariants.
So, ``remarkable integral surfaces'' have to have some meaning in fluid dynamics, too.
They become \emph{1-dimensional}, in this particular case, and a closer inspection
reveals that they are nothing but \emph{vortex lines} well-known from ideal fluid dynamics.
(Their property of being ``frozen into the fluid'' is the content of Helmholtz theorem.)

Now, Hamiltonian mechanics is a paradigmatic example of the theory of integral invariants, too.
(Actually people mostly encounter the topics in Hamiltonian mechanics alone.)
What role then ``remarkable integral surfaces'' play in Hamiltonian mechanics?
What is the analogue of vortex lines in Hamiltonian mechanics?

In order to clearly see how the concept we are seeking is defined, in Table \ref{tab2}
the correspondence of relevant objects in each of the three theories is displayed.

\begin{table}[t]
\begin{center}
\begin{tabular}{|p{4cm}|p{4cm}|p{4cm}|}
\hline
\vspace{-2em}
\begin{eqnarray*}
&\text{general theory}&
\end{eqnarray*}
\vspace{-2em}
&
\vspace{-2em}
\begin{eqnarray*}
 &\text{ideal fluid dynamics}&
\end{eqnarray*}
\vspace{-2em}
&
\vspace{-2em}
\begin{eqnarray*}
  &\text{Hamiltonian mechanics}&
\end{eqnarray*}
\vspace{-2em}                                    \\
\hline \hline
\vspace{-1.8em}
\begin{eqnarray*}
&i_\xi d\sigma = 0&
\end{eqnarray*}
\vspace{-2em}
&
\vspace{-2em}
\begin{eqnarray*}
&i_\xi d\sigma = 0&
\end{eqnarray*}
\vspace{-2em}
&
\vspace{-2em}
\begin{eqnarray*}
&i_\xi d\sigma = 0&
\end{eqnarray*}
\vspace{-2em}                                    \\
\hline
\vspace{-1.8em}
\begin{eqnarray*}
&\sigma = \hat \alpha +dt \wedge \hat \beta&
\end{eqnarray*}
\vspace{-2em}
&
\vspace{-2em}
\begin{eqnarray*}
&\sigma = \hat v -\mathcal E dt&
\end{eqnarray*}
\vspace{-2em}
&
\vspace{-2em}
\begin{eqnarray*}
&\sigma = \hat \theta -Hdt&
\end{eqnarray*}
\vspace{-2em}                                    \\
\hline
\vspace{-1.8em}
\begin{eqnarray*}
&\xi = \partial_t +v&
\end{eqnarray*}
\vspace{-2em}
&
\vspace{-2em}
\begin{eqnarray*}
&\xi = \partial_t +v&
\end{eqnarray*}
\vspace{-2em}
&
\vspace{-2em}
\begin{eqnarray*}
&\xi = \partial_t +\zeta_H&
\end{eqnarray*}
\vspace{-2em}                                    \\
\hline
\vspace{-1.8em}
\begin{eqnarray*}
&d\sigma = \hat d \hat \alpha -dt \wedge i_v \hat d \hat \alpha&
\end{eqnarray*}
\vspace{-2em}
&
\vspace{-2em}
\begin{eqnarray*}
&d\sigma = \hat d \hat v -dt \wedge i_v \hat d \hat v&
\end{eqnarray*}
\vspace{-2em}
&
\vspace{-2em}
\begin{eqnarray*}
&d\sigma = \hat \omega -dt \wedge i_{\zeta_H} \hat \omega&
\end{eqnarray*}
\vspace{-2em}                                    \\
\hline
\vspace{-2em}
\begin{eqnarray*}
&i_w\hat d \hat \alpha = 0&
\end{eqnarray*}
\vspace{-2em}
&
\vspace{-2em}
\begin{eqnarray*}
&i_{\gamma'} \hat d \hat v = 0&
\end{eqnarray*}
\vspace{-2em}
&
\vspace{-2em}
\begin{eqnarray*}
&i_w \hat \omega = 0&
\end{eqnarray*}
\vspace{-2em} \qquad                             \\
\hline
\vspace{-1.8em}
\begin{eqnarray*}
&\text{``remarkable surfaces''}&
\end{eqnarray*}
\vspace{-2em}
&
\vspace{-2em}
\begin{eqnarray*}
&\text{vortex lines}&
\end{eqnarray*}
\vspace{-2em}
&
\vspace{-2em}
\begin{eqnarray*}
&\text{?}&
\end{eqnarray*}
\vspace{-2em}                                    \\
\hline
\end{tabular}
\caption[comparison]{Correspondence of relevant objects in general theory of integral invariants,
                                                           dynamics of ideal (and barotropic) fluid
                                                           and Hamiltonian mechanics.
                     The question is: what plays the role of vortex lines in Hamiltonian mechanics?
                     What exactly the question mark in the bottom right corner of the table denotes?}
\label{tab2}
\end{center}
\end{table}
So what we are seeking is a spatial surface $\mathcal S$ whose tangent vectors annihilate
the 2-form $\hat \omega$
\begin{equation} \label{annihomega}
             i_w \hat \omega = 0
\end{equation}
And here the reader enters the less optimistic part of the story.
Since $\hat \omega \equiv dp_a \wedge dq^a$ is \emph{non-degenerate}, the only vector satisfying
Eq. (\ref{annihomega}) is (by definition of non-degeneracy) the null-vector ($w=0$).
And this means that the ``surface'' is actually 0-dimensional, i.e. it reduces,
in this particular case, to a \emph{point}. The otherwise potentially interesting theorem about
surfaces frozen into the fluid becomes a ``trivial'' fact saying that \emph{phase points} evolve
according to Hamilton's equations.

\vskip .5cm
\noindent
[In fluid dynamics $\hat d \hat v$, regarded as a closed $2$-form on $M$ has, according to Darboux theorem,
 rank $2$ (if it is non-zero) or $0$ (if it vanishes). So, for non-vanishing $\hat d \hat v$
 (i.e. for non-vanishing $\boldsymbol \omega \equiv \curl \bold v$), we have $2$-form of rank $2$
 in $3$-dimensional space and, therefore, a \emph{unique} direction which annihilates it,
 so a unique vortex line starting at a given point.]
\vskip .5cm
\noindent
\textbf{Remark \ref{Helmholtz_Hamiltonian}.1}: There is a completely different, and much more interesting, understanding of the concept
        of ``vortex lines'' in the context of Hamiltonian mechanics. Namely, comparison of
\begin{equation} \label{hamiltoncompare}
                 i_\xi d\sigma = 0
                 \hskip 1cm
                 \text{\emph{Hamilton equations}}
\end{equation}
versus
\begin{equation} \label{vortexlinescompare}
                 i_{\gamma'}d\tilde v = 0
                 \hskip 1cm
                 \text{\emph{vortex lines equation}}
\end{equation}
(see Eqs. (\ref{hamilequations}) and (\ref{eqforvortelines2}))
reveals their identical formal structure. This leads to the claim that
``the vortex lines of the form $pdq -Hdt$ are the integral curves of the canonical equations''
(see \cite{arnold1989}, p.236).

Vortex lines (in this particular sense) for appropriately modified \emph{two-form} turn out to
play formally the same role in \emph{Nambu mechanics} (see \cite{fecko2013a} and \cite{fecko2013b}).

\subsubsection{Related Helmholtz theorems - still true}
\label{euler_non_stationary_helmholtz_related}

Recall, what the three Helmholtz celebrated theorems say
(see the end of Sec. \ref{euler_stationary_helmholtz_related}):
\newline \newline
\noindent
\emph{First theorem}:
            The strength of a vortex filament is constant along its length.
\newline \newline
\noindent
\emph{Second theorem}:
A vortex filament cannot end in a fluid;
\newline
\indent
\hskip 2.1cm
it must extend to the boundaries of the fluid or form a closed path.
\newline \newline
\noindent
\emph{Third theorem}:
A fluid that is initially irrotational remains irrotational.
\newline
\indent
\hskip 2cm
(In the sense that \emph{fluid elements} initially free of vorticity
\newline
\indent
\hskip 2cm
remain free of vorticity.)
\newline \newline
\indent
Concerning the first two statements, notice that vortex filament is a genuinely spatial concept,
it is defined, from the point of view of extended space, in a fixed-time hyper-surface $t=$ const.
No change of dynamics can influence any claim which only refers to something
happening in a fixed time $t=$ const. So, if the statements were true before,
they remain to be true now.

The situation is different for the third statement. The words ``initially''
and ``remains'' signalize that details of time development can matter, now.

Well, in non-stationary case, the analog of $\mathcal L_vd\tilde v = 0$ and  $\Phi_t^*d\tilde v = d\tilde v$
(i.e. Eqs. (\ref{vorticityisinv1}) and (\ref{vorticityisinv2}))
are the equations $\mathcal L_\xi d\sigma = 0$ and  $\Phi_t^*d\sigma = d\sigma$
(i.e. (\ref{dsigmajeinvar1}) and (\ref{dsigmajeinvar2})).
Since $\xi$ generates \emph{time-evolution} of ``fluid elements'', the flow-invariance of $d\sigma$
w.r.t. $\Phi_t \leftrightarrow \xi$ says that $d\sigma$ is \emph{Lie-constant} along the trajectory
(in extended space) of the fluid element.

Now, we know from Eq. (\ref{dsigmanariesenivseob2}) that we can write ``\emph{on shell}''
(i.e. when Euler equation $i_\xi d\sigma =0$ is satisfied or, put it differently,
 for a \emph{physically possible} flow of the fluid)
\begin{equation}
      \label{onshelldsigma}
      d\sigma = \hat d \hat v +dt\wedge (-i_v\hat d\hat v)
\end{equation}
This means we can actually repeat the proof from the very end of
Sec. \ref{euler_stationary_helmholtz_related}:
assume the contrary, i.e. let the ``fluid element'' be ``rotational'' at $(Q,t)$, so vorticity $2$-form
$d\hat v$ is non-zero at $Q$ at time $t$. Then there exists a~\emph{spatial}
infinitesimal surface $S$ around $(Q,t)$ such that $\int_Sd\hat v \neq 0$.
The surface $S$ is $\Phi_t$-image of some \emph{spatial} $S_0$ around $(P,t=0)$
(namely of $S_0 = \Phi_{-t}S$). Therefore
$$0\neq \int_Sd\hat v = \int_S            d\sigma
                        = \int_{\Phi_t(S_0)}d\sigma
                        = \int_{S_0}\Phi_t^*d\sigma
                        = \int_{S_0}d\sigma
                        = \int_{S_0}d\hat v
$$
and this means that vorticity $2$-form $d\hat v$ is non-zero at $(P,t=0)$ and, consequently,
that the ``fluid element'' is ``rotational'' already at $P$ at $t=0$.

\setcounter{equation}{0} \setcounter{figure}{0} \setcounter{table}{0}\newpage
\section{Faraday's law and rotating reference frames}
\label{faraday}

When a wire rim rotates in a homogeneous and static magnetic field, a non-zero voltage is induced
at the ends of the rim. This is a manifestation of the celebrated Faraday's law.

The phenomenon can be easily qualitatively explained, as is well known, in both ``laboratory'' frame
and the frame rotating with the rim.

In the former case, there is no electric field. But an electron (in the wire) moves
(because of the rotation of the wire) and, consequently, it feels \emph{magnetic part} of the Lorentz force,
$q\bold v \times \bold B$.

In the latter case, the electron does not move. However, it feels the electric field
``induced'' by \emph{relativistic transformation} of the original static magnetic field
to the local co-moving rotating reference frame (i.e. the frame moving with the electron).
So, it feels \emph{electric part} of the Lorentz force, $q\bold E$.

\vskip .5cm
\noindent
[Recall that, in general, what is a strictly magnetic (or strictly electric) field in one reference frame
 may be a ``mixture'' of both electric and magnetic fields, when observed in another reference frame.]
\vskip .5cm

In either case, the electron feels a force, therefore it \emph{moves along the wire} and
if a light bulb is connected across the terminals of the rim, the bulb \emph{shines}.

By definition, the voltage at the ends of the rim is given by the line integral of the electric field
along the rim, $\int \bold E \cdot d\bold r$. Here $\bold E$ is the field obtained by transformation
of the ``original'' homogeneous static magnetic field to the instantaneous local frame
in which the particular piece of the rim
(needed for the infinitesimal contribution $\bold E \cdot d\bold r$) is still.
Since, however, the new frame is rather complicated
(e.g. its velocity $\bold v$, needed for the relativistic factor $\sqrt{1-v^2}$,
depends on particular position of the element $d\bold r$ within the rim),
explicit voltage calculation might
be a bit involved.

In what follows we present an approach in terms of differential forms.
It is pretty general and well adapted to a lot of otherwise intricate situations.
Clearly, the computation of the voltage itself (in the situation of the rim discussed above)
may be performed by elementary means.
(Say, as the time derivative of the magnetic flux through the surface enclosed by the rotating rim,
 which is a high-school algebra.)

So, the aim of our exposition is not to show how an easy subject may be discussed in a complicated way :-).
Rather, we aim to introduce the reader to a fairly powerful and general formalism enabling one
to treat electromagnetic fields in various, possibly complicated, reference frames.
And, just for fun, to show how to compute explicitly the induced electric field and its line integral
along the rim. So, the computation of the voltage only serves as a pretext for introduction
to the general formalism and testing it in a well-known elementary situation.

\vspace*{0.5cm}
\setcounter{equation}{0}
\subsection{From (very special) 3+1 to 4}
\label{3plus1to4}

           In this section we show how {\color{red}\emph{Maxwell equations}} in vacuum
\footnote{This holds in {\it Heaviside's} system of units and setting $c=1$; otherwise
          various constants may enter the equations which are, however, irrelevant for us.}
$$\hskip -2cm
  \text{\emph{inhomogeneous}}
  \hskip 4cm
  \text{\emph{homogeneous}}
$$
\begin{eqnarray} %{rcl}
            \label{maxw1}
            \Div \bold E
      &=& \rho  \hskip 3cm \curl \bold E +\partial_t \bold B \ = \ 0 \\
            \label{maxw2}
            \curl \bold B -\partial_t \bold E
      &=& \bold j \hskip 4.2cm  \Div \bold B   \ = \ 0
\end{eqnarray}
may be written in terms of differential forms in Minkowski space.
In order to do that, we have to learn, first, how a general differential form in Minkowski space
looks like and what is the effect of applying standard operations on forms
(namely exterior derivative and Hodge star operator).

\subsubsection{Differential forms in Minkowski space}
\label{formsinminkowski}

Consider Minkowski space $E^{1,3}$, i.e. the manifold $\mathbb R^4$ endowed with the \emph{metric tensor}
of the form
\begin{equation}
       \label{minkowskimetric}
      g\equiv \eta = dt \otimes dt - (dx \otimes dx + dy \otimes dy + dz \otimes dz)
                   =:
                   dt \otimes dt - \hat g
\end{equation}
when expressed in standard (global, Cartesian) coordinate system $(t,x,y,z) \equiv (t, \bold r)$.
So, the matrix of components (in standard notation $(t,x,y,z) \equiv (x^0, x^1,x^2,x^3)$) is
\begin{equation} \label{matrixofmetric}
  g_{\mu \nu}=
  \left( \begin{matrix}
                        1 &  0 &  0 &  0\\
                        1 & -1 &  0 &  0\\
                        1 &  0 & -1 &  0\\
                        1 &  0 &  0 & -1
                        \end{matrix} \right)
\end{equation}
Then, as explained in Sec. \ref{formsonmtimesr} (and also in Appendix \ref{app:vectoranalysis}),
each differential form may be decomposed as follows
\begin{equation}
      \label{decompositionminkowski}
      \boxed{\alpha = dt \wedge \hat s + \hat r}
\end{equation}
Here \emph{spatial} forms $\hat s$ and $\hat r$ do not contain factor $dt$ and are \emph{time-dependent}
versions of expressions displayed in Eq. (\ref{possibleformsine3}). So, explicitly,
possible $0,1,2,3$ and $4$-forms are
\begin{equation}
        \label{possibleformsminkowski}
        f
        \hskip .7cm
        fdt + \bold a \cdot d\bold r
        \hskip .7cm
        dt \wedge \bold a \cdot d\bold r + \bold b \cdot d\bold S
        \hskip .7cm
        dt\wedge \bold a \cdot d\bold S + f dV
        \hskip .7cm
        fdt\wedge dV
\end{equation}
with both scalar and vector fields $f,\bold a, \bold b$ depending, in general,
on $\bold r$ \emph{and} $t$.

We already learned in Sec. \ref{formsonmtimesr} how exterior derivative of such forms
is to be expressed in terms of \emph{spatial} exterior derivative $\hat d$ (see Eq. (\ref{dongeneral})):
\begin{equation}
       \label{dongeneralmink}
       \boxed{d\alpha \equiv d(dt \wedge \hat s + \hat r) = dt \wedge (-\hat d \hat s + \mathcal L_{\partial_t} \hat r ) + \hat d \hat r}
\end{equation}
Another important operation on forms, on manifolds where metric tensor (and orientation) is available,
is \emph{Hodge star} operator $*$ (see, e.g., \$5.8 in \cite{fecko2006}).

Just as we have ``two $d$ operators'', the ``full'' $d$ and the spatial $\hat d$
(and Eq. (\ref{dongeneralmink}) shows how the two are related), there are also two star operators,
the ``full'' $*$ and the spatial $\hat *$. Why? There are two metric tensors (and orientations)
available. First $\eta$, the metric tensor in Minkowski space (of type + - - - in our convention).
But also $\hat g$ (see Eq. (\ref{minkowskimetric})), the metric tensor in the Euclidean subspace
(of type + + +).

Consider, as an example, 2-form $dx\wedge dy$. When the two Hodge star operators are applied on it,
the two results are completely different: it should be a 2-form when regarded in $E^{1,3}$ and a 1-form
when regarded in $E^3$ (recall that in general it is a $(n-p)$-form in $n$-dimensional space).
Explicitly, using standard formulas, one gets
\begin{eqnarray}
      \label{fullstaronit}
                 *(dx \wedge dy)
             &=& dt \wedge dz               \\
      \label{spatialstaronit}
                \hat *(dx \wedge dy)
             &=& dz
\end{eqnarray}
So, we can also write the action of the full star in terms of the action of the spatial star
\begin{equation}
       \label{fullintermsof}
       *(dx \wedge dy) = dt \wedge \hat *(dx \wedge dy)
\end{equation}
If one plays a bit with various other possibilities, general formula
\begin{equation}
       \label{starongeneralmink}
       \boxed{*\alpha \equiv *(dt \wedge \hat s + \hat r) = dt \wedge (\hat * \hat r) + \hat * \hat \eta \hat s}
\end{equation}
may be derived quite easily (see, e.g., \$16.1 in \cite{fecko2006}).

\subsubsection{Maxwell equations}
\label{maxwelleq}

Now, let us use our new piece of wisdom in electrodynamics.
Let us apply exterior derivative on a general \emph{two-form} expressed in the ``decomposed'' way
displayed in Eq. (\ref{possibleformsminkowski}). We get
\begin{equation}
       \label{dontwoform}
       d(dt \wedge \bold a \cdot d\bold r + \bold b \cdot d\bold S) =
       dt\wedge (\partial_t \bold b-\curl \bold a ) \cdot d\bold S +
                               (\Div \bold b)dV
\end{equation}
\noindent $\blacktriangledown \hskip 0.5cm$
Indeed, using Eqs. (\ref{curl}) and (\ref{divergence}) from Appendix \ref{app:vectoranalysis}, we have
$$
\begin{array} {rcl}
  d(dt \wedge \bold a \cdot d\bold r + \bold b \cdot d\bold S)
  &=& dt\wedge (\mathcal L_{\partial_t} (\bold b \cdot d\bold S) -\hat d (\bold a \cdot d\bold r))
                                                                 + \hat d (\bold b \cdot d\bold S)   \\
  &=& dt\wedge ((\partial_t \bold b) \cdot d\bold S) -(\curl \bold a) \cdot d\bold S))
                                                     + (\Div \bold b ) dV                            \\
  &=& dt\wedge (\partial_t \bold b-\curl \bold a ) \cdot d\bold S + (\Div \bold b)dV
\end{array}
$$
\hfill $\blacktriangle$ \par
\vskip .2cm \indent
Put in this formula, just for fun, $\bold a = \bold E$, $\bold b = - \bold B$
(so, electric and minus magnetic fields $(\bold E, - \bold B)$ rather than arbitrary
 vector fields $(\bold a,\bold b)$). What we get now is
\begin{equation}
       \label{dontwoformF}
       d(dt \wedge \bold E \cdot d\bold r - \bold B \cdot d\bold S) =
       dt\wedge (-\partial_t \bold B-\curl \bold E ) \cdot d\bold S -
                               (\Div \bold B)dV
\end{equation}
A quick glance on Eqs. (\ref{maxw1}) and (\ref{maxw2}) shows that to fulfill the homogeneous part
of Maxwell equations is the same thing as to make the r.h.s. of Eq. (\ref{dontwoformF}) vanish.

\vskip .5cm
\noindent
[Remember that, in general, vanishing of $\alpha \equiv dt \wedge \hat s + \hat r$ is the same thing
 as \emph{simultaneous} vanishing of \emph{both} $\hat s$ \emph{and} $\hat r$.
 In particular, here the three-forms $dt\wedge d\bold S$ and $dV$ comprise a basis for three-forms
 in Minkowski space $E^{1,3}$ and, consequently, vanishing of the expression on the r.h.s.
 of Eq. (\ref{dontwoformF}) is the same thing as vanishing coefficients standing by these
 particular three-forms. This gives, however, exactly the homogeneous part of Maxwell equations
 (\ref{maxw1}) and (\ref{maxw2}).]
\vskip .5cm
This fact strongly motivates to introduce the 2-form
\begin{equation}
       \label{definitionofF}
       \boxed{F:= dt \wedge \bold E \cdot d\bold r - \bold B \cdot d\bold S}
           \hskip 1cm \text{\emph{2-form of the electromagnetic field}}
\end{equation}
In terms of this form there is an extremely simple way to write down half of Maxwell equations:
\begin{equation}
       \label{dFiszero}
       dF= 0
           \hskip 1cm
           \text{\emph{homogeneous Maxwell equations}}
\end{equation}
And what about the second half?

Notice that the expressions on the l.h.s. of inhomogeneous Maxwell equations (\ref{maxw1}) and (\ref{maxw2})
may be obtained from those on the l.h.s. of the homogeneous equations by a simple replacement
\begin{equation}
       \label{replacement1}
       (\bold E, \bold B) \mapsto (-\bold B, \bold E)
\end{equation}
Now, in terms of the 2-form of the electromagnetic field $F$, the same effect is achieved
by the replacement of the basis 2-forms
\begin{equation}
       \label{replacement2}
       (dt \wedge d\bold r, d\bold S ) \mapsto (- d\bold S ,  dt \wedge d\bold r )
\end{equation}
What happens in Eq. (\ref{replacement2}) may be regarded as the action of a \emph{liner operator}
on the space of 2-forms in Minkowski space. It turns out that it is nothing but the particular case
of the \emph{Hodge star} operator $*$ (see, e.g., \$5.8 in \cite{fecko2006}):
\begin{equation}
       \label{hodgeonF}
       F\mapsto *F
       \hskip .5cm \leftrightarrow \hskip .5cm
       (\bold E, \bold B) \mapsto (-\bold B, \bold E)
\end{equation}
This means, taking into account Eq. (\ref{dontwoformF}), that the exterior derivative of $*F$
(rather than $F$ itself) produces
\begin{equation}
       \label{dontwoformstarF}
       d*F =
       dt\wedge (-\partial_t \bold E +\curl \bold B ) \cdot d\bold S -
                               (\Div \bold E)dV
\end{equation}
Now this is not to vanish by inhomogeneous Maxwell equations.
Rather, Eqs. (\ref{maxw1}) and (\ref{maxw2}) motivate us to introduce the 3-form
\begin{equation}
       \label{3formofcurrentJ}
       \boxed{J:= dt \wedge (- \bold j) \cdot d\bold S + \rho dV}
           \hskip 1cm \text{\emph{3-form of the current}}
\end{equation}
Then, in terms of $F$ and $J$, the remaining half of Maxwell equations becomes
\begin{equation}
       \label{dstarFisJ}
       d*F= -J
           \hskip 1cm
           \text{\emph{inhomogeneous Maxwell equations}}
\end{equation}
So, the whole system of Maxwell equations (\ref{maxw1}) and (\ref{maxw2})
turns, when expressed in terms of differential forms in Minkowski space,
into the following neat system:
\begin{equation}
       \label{maxwellintotal}
       \boxed{d*F= -J
           \hskip 1cm
       dF =0}
           \hskip 1cm
           \text{{\color{red}\emph{Maxwell equations}}}
\end{equation}
\vskip .5cm
\noindent
[In Eqs. (\ref{maxwellintotal}), the Hodge operator $*$ refers to the \emph{Minkowski} (``flat'')
 metric tensor $\eta$ from Eq. (\ref{minkowskimetric}). In general relativity, one studies
 space-times with more general (pseudo-Euclidean) metric tensors $g$.
 Then, one has to identify ``new Maxwell equations'', which govern electro-magnetic theory in this,
 more general, setting (in the presence of a gravi\-tational field).
 A natural choice consists in postulating that the only thing one has to do,
 in comparison with the Minkowski case, is to replace $\eta \mapsto g$ \emph{in the Hodge operator},
 $*_\eta \mapsto *_g$. Physically, this is a natural choice because of the celebrated \emph{Equivalence principle}.
 Mathematically, notice that the homogeneous equation \emph{does not} contain metric tensor at all
 (it is ``purely geometrical''),
 so that one has no reason to modify it in any way because of the change $\eta \mapsto g$.
 The \emph{only} place where $\eta$ enters the inhomogeneous equation is the Hodge star operator,
 so the replacement $*_\eta \mapsto *_g$ seems to be natural.]
\vskip .5cm

\setcounter{equation}{0}
\subsection{From 4 back to (very special) 3+1}
\label{4tospecial3plus1}

In Sec. \ref{maxwelleq} we learned how electric and magnetic field vectors,
$\bold E$ and $\bold B$, may be placed into a new object,
$2$-form of the electromagnetic field $F$, see Eq. (\ref{definitionofF}).

Let us write the result in an even slightly more succinct form
\begin{equation}
       \label{definitionofF2}
       F= dt \wedge \hat E - \hat B
\end{equation}
where we introduced
\begin{eqnarray} %{rcl}
      \label{defofhatE}
                 \hat E
             &=& \bold E \cdot d\bold r
                                       \hskip 1.2cm \text{\emph{1-form of electric field}}    \\
      \label{defofhatB}
                 \hat B
             &=& \bold B \cdot d\bold S
                                       \hskip 1.1cm \text{\emph{2-form of magnetic field}}
\end{eqnarray}
So, the direction $(\bold E, \bold B) \mapsto F$ is performed in two steps:

- first insert the fields $\bold E$ and $\bold B$ into expressions given by Eqs. (\ref{defofhatE}) and (\ref{defofhatB})

- then compose $F$ out of $\hat E$ and $\hat B$ according to Eq. (\ref{definitionofF2})

\noindent
And what about the opposite direction,  $F\mapsto (\bold E, \bold B)$?
Well, in principle one can proceed according to the explanation of the decomposition \`{a} la Eq. (\ref{decomposition1}),
i.e. express $F$ in coordinates $(t,x,y,z)$, collect all terms containing $dt$ once (you get $dt \wedge \hat E$)
and all terms which do not contain $dt$ at all (you get $-\hat B$).

There is also a ``computation'' leading to the desired result. Namely, consider the vector field
$\partial_t$ and its ``metric dual'' covector field $g(\partial_t, \ . \ ) = dt$
(see Eq. (\ref{minkowskimetric})). Then (see Eq. (\ref{definitionofF2})),

- applying $i_{\partial_t}$ on $F$ you get $\hat E$

- subtracting $dt \wedge \hat E$ from $F$ (with $\hat E$ already known from step 1) you get $-\hat B$

\begin{figure} %[h]
\begin{center}
\scalebox{0.15}{\includegraphics[clip,scale=0.9]{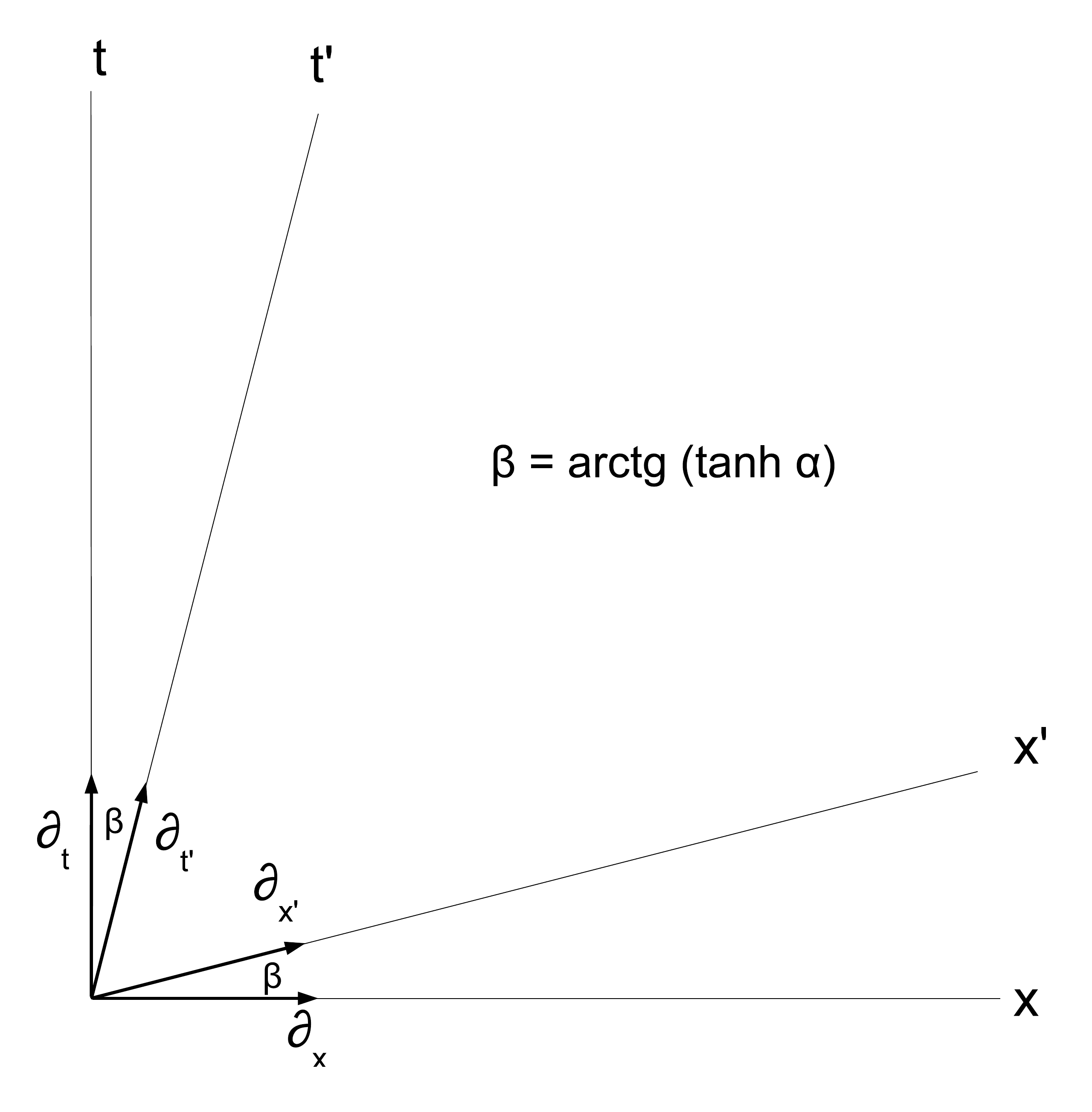}}
\hskip 2.00cm
\scalebox{0.15}{\includegraphics[clip,scale=0.9]{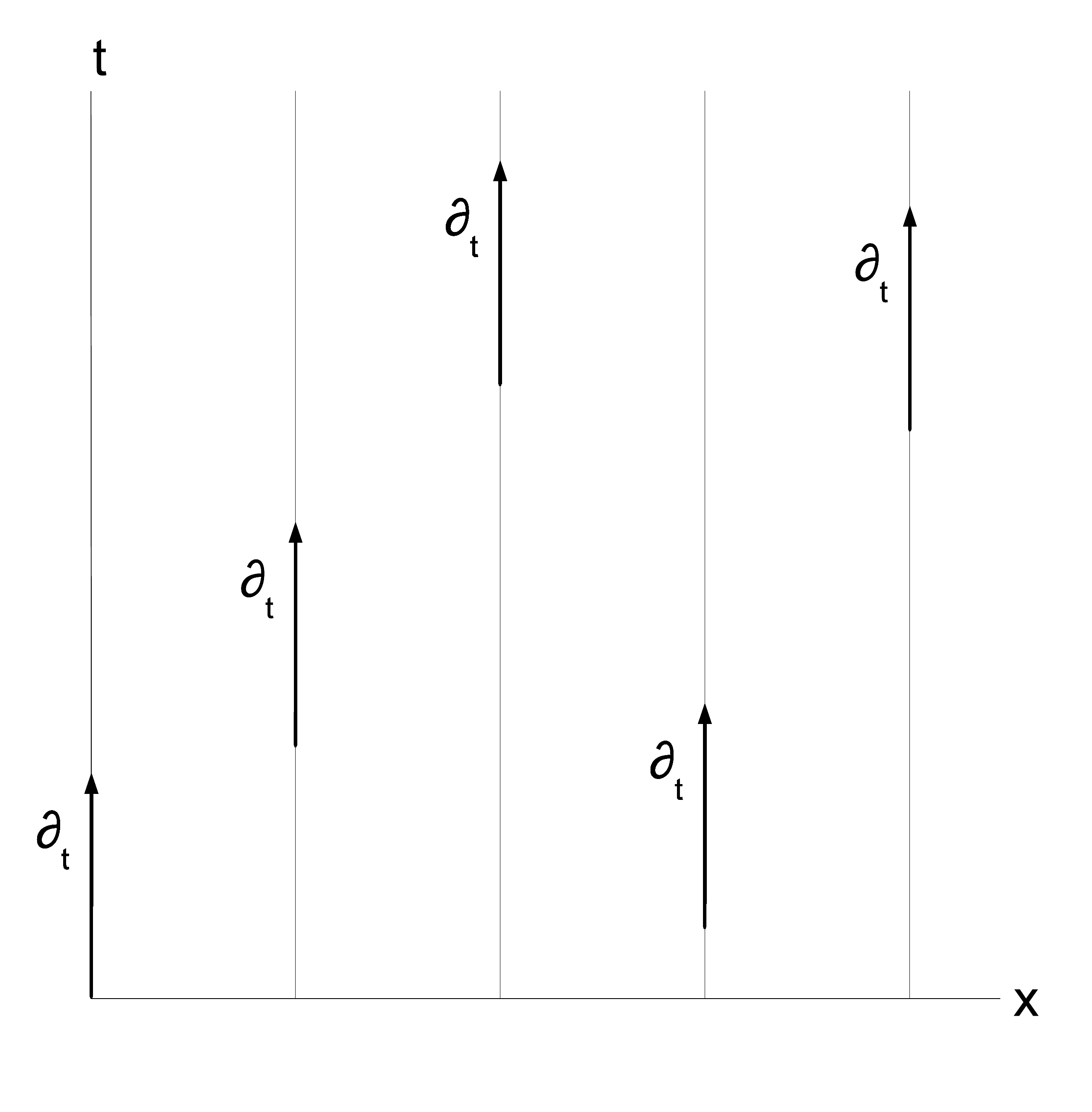}}
\caption{Left: unprimed and primed axes corresponding to the transformation given by
                                        Eqs. (\ref{twoinertialcoord0}), (\ref{twoinertialcoord1})
                                        and the corresponding tangent vectors mentioned in Eq. (\ref{partialtprime}).
                                  Right: Vector field $\partial_t$ defines, at each point of Minkowski space-time,
the time direction at this point and consequently, as the orthogonal complement, also the 3-space at the same point.
Integral curves of the field represent world-lines of observers at rest w.r.t. the reference frame using
coordinate system $(t,x,y,z)$.}
\label{fig:lorentztransf}
\end{center}
\end{figure}

\noindent
Explicitly
\begin{equation}
       \label{hatEhatBoutofF}
       \hat E = i_{\partial_t}F \hskip 1cm
      -\hat B = F - g(\partial_t, \ . \ ) \wedge (i_{\partial_t} F)
\end{equation}
What is nicely visible in formulas Eq. (\ref{hatEhatBoutofF}) is that as soon as you make the choice
what the vector field $\partial_t$ is, you automatically make the choice how $F$ ``splits''
into its electric and magnetic parts.
\vskip .2cm
\noindent
[Physically: contrary to the case of extended phase space in, say, Hamiltonian mechanics,
             time is an \emph{observer-depended} concept in relativity theory.
 So the ``direction of time axis'' within the space-time continuum, i.e. locally of $\partial_t$,
 depends on choice of reference frame.
 Consequently, the splitting of $F$ into its electric and magnetic parts
 depends on choice of reference frame as well.]
\vskip .4cm
Notice that we, so far, do \emph{not} make the \emph{further}
splitting of $\hat E$ and $\hat B$ into spatial components
(say $\hat E = E_xdx +E_ydy +E_zdz$).
We just discriminate between electric and magnetic parts of $F$.
\vskip .5cm
\noindent
{\bf Example \ref{4tospecial3plus1}.1}:
Consider, in Cartesian coordinates $(t,x,y,z)$ of an inertial frame $\mathcal O$,
the following $2$-form $F$:
\begin{equation} \label{example1formF}
                 F = dt \wedge adx - bdx \wedge dy
                        \hskip 1cm
                        a,b = \ \text{const.}
\end{equation}
Comparison with Eqs. (\ref{definitionofF2}), (\ref{defofhatE}) and (\ref{defofhatB})
shows that
\begin{eqnarray} %{rcl}
      \label{hatEinF}
                 \hat E
             &=& adx
                 \hskip 2.1cm
                 \bold E = (a,0,0)   \\
      \label{hatBinF}
                 \hat B
             &=& bdx \wedge dy
                 \hskip 1.4cm
                 \bold B = (0,0,b)
\end{eqnarray}
(we also used $dS_z = dx \wedge dy$).
Now consider another inertial frame, say $\mathcal O'$, with Cartesian coordinates $(t',x',y',z')$.
Let the primed inertial frame uniformly move along the $x$ axis w.r.t. the unprimed one
with velocity
\begin{equation} \label{vviaalpha}
                 v=\tanh \alpha
\end{equation}
Then the connection between the primed and unprimed coordinates
may be explicitly written as
\begin{eqnarray} %{rcl}
      \label{twoinertialcoord0}
                 t
             &=&  t'\cosh \alpha +x'\sinh \alpha      \\
      \label{twoinertialcoord1}
                 x
             &=&  t'\sinh \alpha +x'\cosh \alpha      \\
      \label{twoinertialcoord2}
                 y
             &=&  y'      \\
      \label{twoinertialcoord3}
                 z
             &=&  z'
\end{eqnarray}
(see the left drawing on Fig. \ref{fig:lorentztransf}). Expressing $F$ in primed coordinates, we get
\begin{eqnarray} %{rcl}
      \label{example1formF2}
                 F
             &\equiv&  dt \wedge adx - bdx \wedge dy    \\
      \label{example1formF3}
             &=&  dt' \wedge (adx' - b \sinh \alpha dy') - (b\cosh \alpha ) dx' \wedge dy'
\end{eqnarray}
This means that
\begin{eqnarray} %{rcl}
      \label{hatEprimedinF}
                 \hat E'
             &=& adx' - (b \cosh \alpha) dy'
                 \hskip 1.15cm
                 \bold E' = (a,- b \sinh \alpha,0)   \\
      \label{hatBprimedinF}
                 \hat B'
             &=& (b\cosh \alpha) dx' \wedge dy'
                 \hskip 1.4cm
                 \bold B' = (0,0,b\cosh \alpha)
\end{eqnarray}
Take, for example, the case $a=0$. Then
\begin{eqnarray} %{rcl}
      \label{hatEprimedinFazero}
                 \hat E
             &=& 0 \hskip 2.5cm \hat E' =  - (b \sinh \alpha) dy'   \\
      \label{hatBprimedinFazero}
                 \hat B
             &=&  b dx \wedge dy
                 \hskip 1.35cm \hat B' = (b\cosh \alpha) dx' \wedge dy'
\end{eqnarray}
\vskip .2cm
\noindent
[Physically: in the primed reference frame, physicists see both electric and magnetic fields
 in the situation, where physicists in unprimed frame swear to see magnetic field alone.]
\vskip .4cm
Important thing to notice is that $\hat E'$ is \emph{not} just $\hat E$
rewritten to primed coordinates (and the same is true for $\hat B'$ versus $\hat B$).
It is a \emph{genuinely different} form.
\vskip .4cm
\noindent
[For $a=0$, say, $\hat E'$ is non-zero whereas $\hat E$ vanishes.
This cannot happen by just rewriting a form to different coordinates.]
\vskip .4cm
Now, let's see how the algorithm for obtaining $\hat E'$, described in Eq. (\ref{hatEhatBoutofF}),
works.
From Eqs. (\ref{twoinertialcoord0}) and (\ref{twoinertialcoord1}) we have
\begin{equation} \label{partialtprime}
                 \partial_{t'} = \cosh \alpha \ \partial_t + \sinh \alpha \ \partial_x
                 \hskip .8cm \text{and so} \hskip .8cm
                 i_{\partial_{t'}} = \cosh \alpha \ i_{\partial_t} + \sinh \alpha \ i_{\partial_x}
\end{equation}
Therefore we get
\begin{equation} \label{hatEprime}
                 \hat E' = i_{\partial_{t'}}F = a(\cosh \alpha \ dx - \sinh \alpha \ dt) - b\sinh \alpha \ dy
\end{equation}
One can check, by expressing the form in primed coordinates,
that this \emph{is the same} $\hat E'$ we know from Eq. (\ref{hatEprimedinF}).
(For the case $a=0$ it is clear without any computation.)

Notice, that we come by this method to correct \emph{primed} electric field $1$-\emph{form}
expressed in terms of \emph{unprimed coordinates}.
But remember that it is the electric field $1$-\emph{form} that matters,
not the choice of coordinates used for its expression. So the method
computes what we are \emph{really} interested in.
The end of Example \ref{4tospecial3plus1}.1.
\vskip .3cm
Let us stress (repeat) what is so magical about the vector field $\partial_t$
that it has the power to decide, via Eq. (\ref{hatEhatBoutofF}),
which part of $F$ is electric and which one magnetic.

Well, notice that integral curves of $\partial_t$ read (see the right drawing on Fig. \ref{fig:lorentztransf})
\begin{equation} \label{intcurves}
                 t(\tau) = t_0 +\tau \hskip 1cm
                 x(\tau) = x_0       \hskip 1cm
                 y(\tau) = y_0       \hskip 1cm
                 z(\tau) = z_0
\end{equation}
So they are nothing but \emph{world-lines of observers at rest} w.r.t.
the reference frame using coordinate system $(t,x,y,z)$.
Vector field $\partial_t$ defines, at each point of Minkowski space-time,
the \emph{time direction} at this point and, consequently
(as the \emph{orthogonal complement}), also the 3-space at the same point.
So it defines, at each point, what is called $3+1$ \emph{split} of the 4-dimensional space-time.
The split depends on observer (on direction of her time axis) and it uniquely induces the
corresponding split of $F$ into two parts, $dt \wedge \hat E$ and $-\hat B$,
regarded by the observer as electric and magnetic part of the total electromagnetic field.

We will generalize this technique substantially in what follows
and then show, just for fun, how to use it in order to compute relevant quantities
from the point of view of the Faraday's law.

\subsubsection{Still more about how to go back to (very special) 3+1}
\label{4tospecial3plus1more}

In Eq. (\ref{hatEhatBoutofF}) we learned, on particular example of $2$-form of electromagnetic
field $F$, how the forms $\hat s$ and $\hat r$
in the decomposition Eq. (\ref{decompositionminkowski}) may be easily computed:
\begin{eqnarray} %{rcl}
      \label{computehats}
                 \hat s
             &=&  i_{\partial_t} \alpha   \\
      \label{computehatr}
                 \hat r
             &=&   \alpha - g(\partial_t, \ . \ ) \wedge i_{\partial_t} \alpha
\end{eqnarray}
Introduce, in addition to the interior product operator $i_{\partial_t}$, another linear operator
on forms defined as follows:
\begin{equation} \label{defjpartialt}
                 j_{\partial_t} \alpha = g(\partial_t, \ . \ ) \wedge \alpha \equiv dt \wedge \alpha
\end{equation}
Then notice that the following identity holds
\begin{equation} \label{identityitjt}
                 i_{\partial_t}j_{\partial_t} + j_{\partial_t}i_{\partial_t} = \hat 1
\end{equation}
(where $\hat 1$ is unit operator on forms) and, moreover, that our blue-eyed decomposition
of forms $\alpha = dt \wedge \hat s +\hat r$ may be regarded as application of the identity
on $\alpha$.
\newline \newline \noindent
\noindent $\blacktriangledown \hskip 0.5cm$
Indeed,
$$
\begin{array} {rcl}
  i_{\partial_t}j_{\partial_t}\alpha
  &=& i_{\partial_t}(dt \wedge \alpha) \\
  &=& (i_{\partial_t}dt) \wedge \alpha - dt \wedge i_{\partial_t}\alpha \\
  &=& \alpha - j_{\partial_t} i_{\partial_t}\alpha
\end{array}
$$
So the identity Eq. (\ref{identityitjt}) holds. Moreover,
$$
\begin{array} {rcl}
  i_{\partial_t}j_{\partial_t}\alpha
  &=& i_{\partial_t}j_{\partial_t}(dt \wedge \hat s +\hat r) \\
  &=& i_{\partial_t}(dt \wedge (dt \wedge \hat s +\hat r)) \\
  &=& i_{\partial_t}(dt \wedge \hat r) \\
  &=& \hat r
\end{array}
$$
Therefore, combining with the previous result (or computing directly),
$$
  j_{\partial_t}i_{\partial_t}\alpha = dt \wedge \hat s
$$
so, indeed, that the decomposition $\alpha = dt \wedge \hat s +\hat r$ may be regarded as application
of the identity $i_{\partial_t}j_{\partial_t} + j_{\partial_t}i_{\partial_t} = \hat 1$ on $\alpha$.
\hfill $\blacktriangle$ \par
\vskip .2cm \indent
The identity Eq. (\ref{identityitjt}) may be treated in more formal way. Introduce two linear
operators on forms
\begin{equation} \label{operatorsPQ}
                 \mathcal P := i_{\partial_t}j_{\partial_t}
                 \hskip 1.5cm
                 \mathcal Q := j_{\partial_t}i_{\partial_t}
\end{equation}
They enjoy the following properties:
\begin{eqnarray} %{rcl}
      \label{property1}
                 \mathcal P \mathcal P
             &=& \mathcal P
                 \hskip 1cm
                 \mathcal Q \mathcal Q \ = \ \mathcal Q                       \\
      \label{property2}
                 \mathcal P \mathcal Q
             &=& 0 \hskip 1.1cm \mathcal Q \mathcal P \ = \ 0                      \\
      \label{property3}
                 \mathcal P + \mathcal Q
             &=& \hat 1
\end{eqnarray}
\newline \noindent
\noindent $\blacktriangledown \hskip 0.5cm$
First notice that both $i_{\partial_t}$ and $i_{\partial_t}$ are nilpotent operators
(i.e. $i_{\partial_t}i_{\partial_t} = j_{\partial_t}j_{\partial_t}=0$).
Therefore
$$
\begin{array} {rcl}
  \mathcal P \mathcal Q
  &=& i_{\partial_t}(j_{\partial_t} j_{\partial_t})i_{\partial_t} = 0 \\
  \mathcal Q \mathcal P
  &=& j_{\partial_t}(i_{\partial_t} i_{\partial_t})j_{\partial_t} = 0
\end{array}
$$
Finally,
$$
  \mathcal P \mathcal P = \mathcal P (\hat 1 - \mathcal Q)
                        = \mathcal P  - \mathcal P \mathcal Q
                        = \mathcal P
$$
and similarly for $\mathcal Q$.
\hfill $\blacktriangle$ \par
\vskip .2cm \indent
Eqs. (\ref{property1}) express the fact that both $\mathcal P$ and $\mathcal Q$
are \emph{projectors}.
So they define, as their image spaces, two subspaces of the (infinite dimensional) space of forms.
Then Eqs. (\ref{property2}) and (\ref{property3}) say that the subspaces have no intersection (except for the zero vector)
and add to the whole space.
So, $\mathcal P$ and $\mathcal Q$ define a decomposition of the space of differential forms
into the \emph{direct sum} of exactly \emph{two subspaces}:
\begin{equation} \label{twosubspaces}
                 \alpha = (\mathcal Q + \mathcal P)\alpha
                        = \mathcal Q \alpha + \mathcal P \alpha \equiv dt \wedge \hat s + \hat r
\end{equation}
Notice that the subspace generated by the operator $\mathcal P$
(i.e. the image space Im $\mathcal P$ of $\mathcal P$) is nothing but the space of \emph{spatial} forms in
Minkowski space,
\begin{equation}
       \label{ImPisspatial}
                 \alpha = \ \text{spatial}
                 \hskip .5cm \Leftrightarrow \hskip .5cm
                 \alpha = \hat r
                 \hskip .5cm \Leftrightarrow \hskip .5cm
                 \alpha = \mathcal P \alpha
                 \hskip .5cm \Leftrightarrow \hskip .5cm
                 \mathcal Q \alpha = 0
\end{equation}

\setcounter{equation}{0}
\subsection{From 4 back to (more general) 3+1}
\label{4to3plus1}

          The decomposition of forms may be significantly generalized in the following way.
\footnote{Our approach here is based on \cite{fecko1997} (available also in Arxiv);
          see also \cite{cattaneo1963}, \cite{massa1974c} or \cite{maglevannyj1978}.
          For other applications and/or approaches, see e.g. \cite{hehlobukhov2003}, \cite{kurz2006},
          \cite{kurz2009}, \cite{raumonen2009}, \cite{raumonen2011}.}
          Rather than $\partial_t$ in Minkowski space-time,
consider a general \emph{future-oriented} vector field $V$ on a space-time $(M,g)$ satisfying
\begin{equation}
       \label{observerfield}
                 g(V,V) \equiv ||V||^2 = 1
                 \hskip 1cm
                 \text{{\color{red}observer field}}
\end{equation}
\vskip .2cm
\noindent
[Since $g(\partial_t, \partial_t)=g_{00} = 1$ for Minkowski metric tensor Eq. (\ref{matrixofmetric}),
 the field $\partial_t$ is an observer field in Minkowski space-time.
 Notice that $\partial_{t'}$ from Eq. (\ref{partialtprime}) is an observer field, too.
 (It is produced from $\partial_t$ by means of a \emph{proper Lorentz} transformation, i.e. both the
 norm and future-orientation are preserved).]
 \vskip .2cm
 Integral curves of any observer field $V$ provide a congruence of proper-time parametrized world-lines
 of observers.
\vskip .2cm
\noindent
[If $\gamma (s)$ is an integral curve of $V$, i.e. $\dot \gamma = V$, then it may be regarded as
 a world-line of an observer,  since $V\equiv \dot \gamma$ is future oriented and it has positive square.
 The normalization of $V$ to unity then guaranties that the ``length'' of the segment of $\gamma$
 between $\gamma (s_1)$ and $\gamma (s_2)$ is just $s_2 - s_1$. Therefore $s$ is proper time,
 i.e. the time measured by (ideal) clock transported along the world line $\gamma$.
 If $V$ were not normalized to unity, the proper time of the segment between $\gamma (\tau_1)$
 and $\gamma (\tau_2)$ should be calculated as $\int_{\tau_1}^{\tau_2}d\tau \sqrt{g(\dot \gamma,\dot \gamma)}$.]
 \vskip .2cm
At a given point of $M$, $V$ is directed along the local time axis of the observer at that point.
The 3-dimensional \emph{orthogonal complement} to $V$ is regarded by him as ``\emph{the space}''.
So $V$, at the point, provides \emph{space plus time} (3+1) \emph{splitting} of the 4-dimensional tangent space.
And in consequence, as we will see, it induces the corresponding decomposition of differential forms
and operations on them.

\subsubsection{Decomposition of forms}
\label{decompforms}

\begin{figure} %[h]
\vspace*{-0.4cm}
\begin{center}
\hskip 0.60cm
\scalebox{0.15}{\includegraphics[clip,scale=1.40]{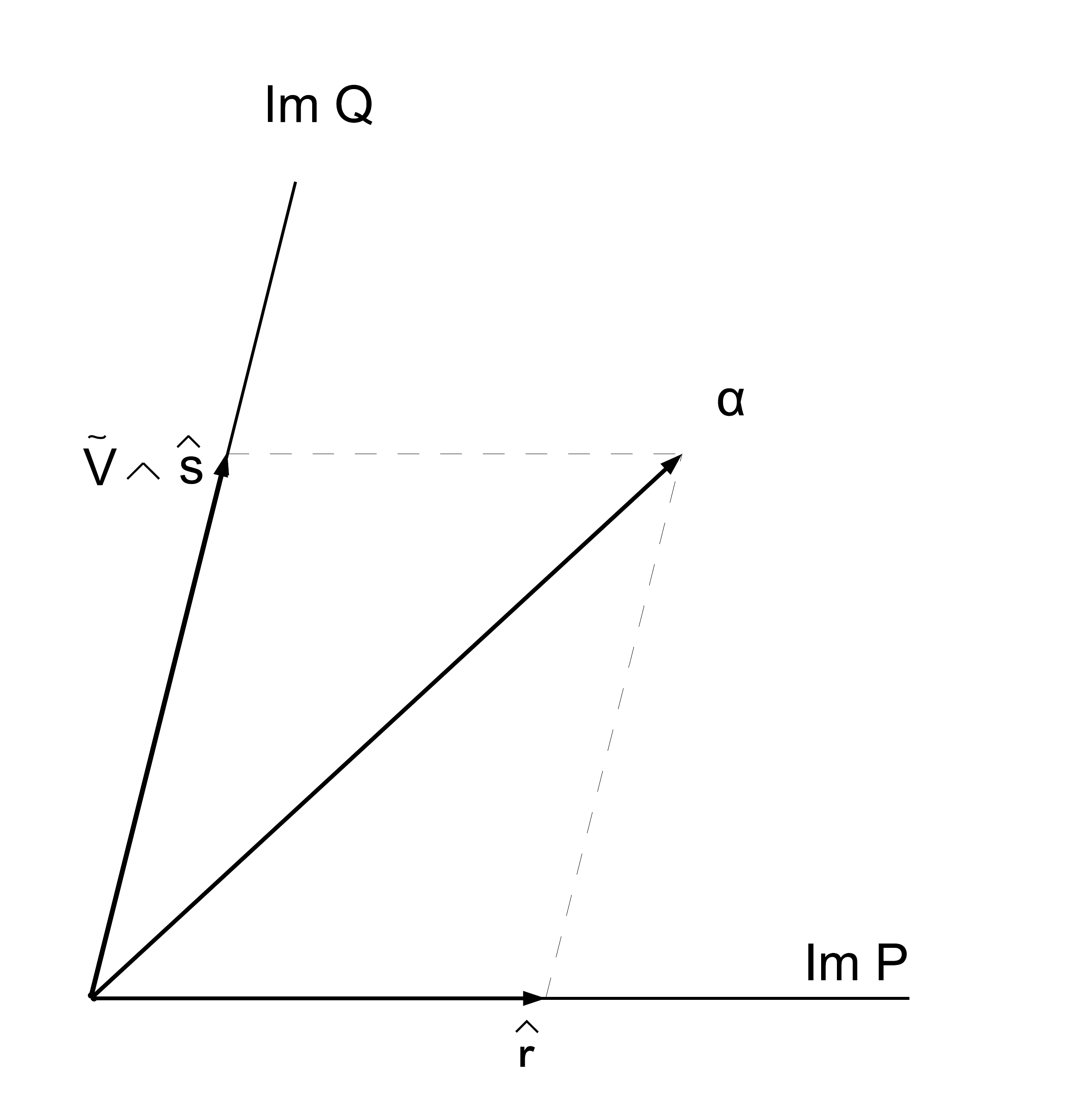}}
\scalebox{0.15}{\includegraphics[clip,scale=1.40]{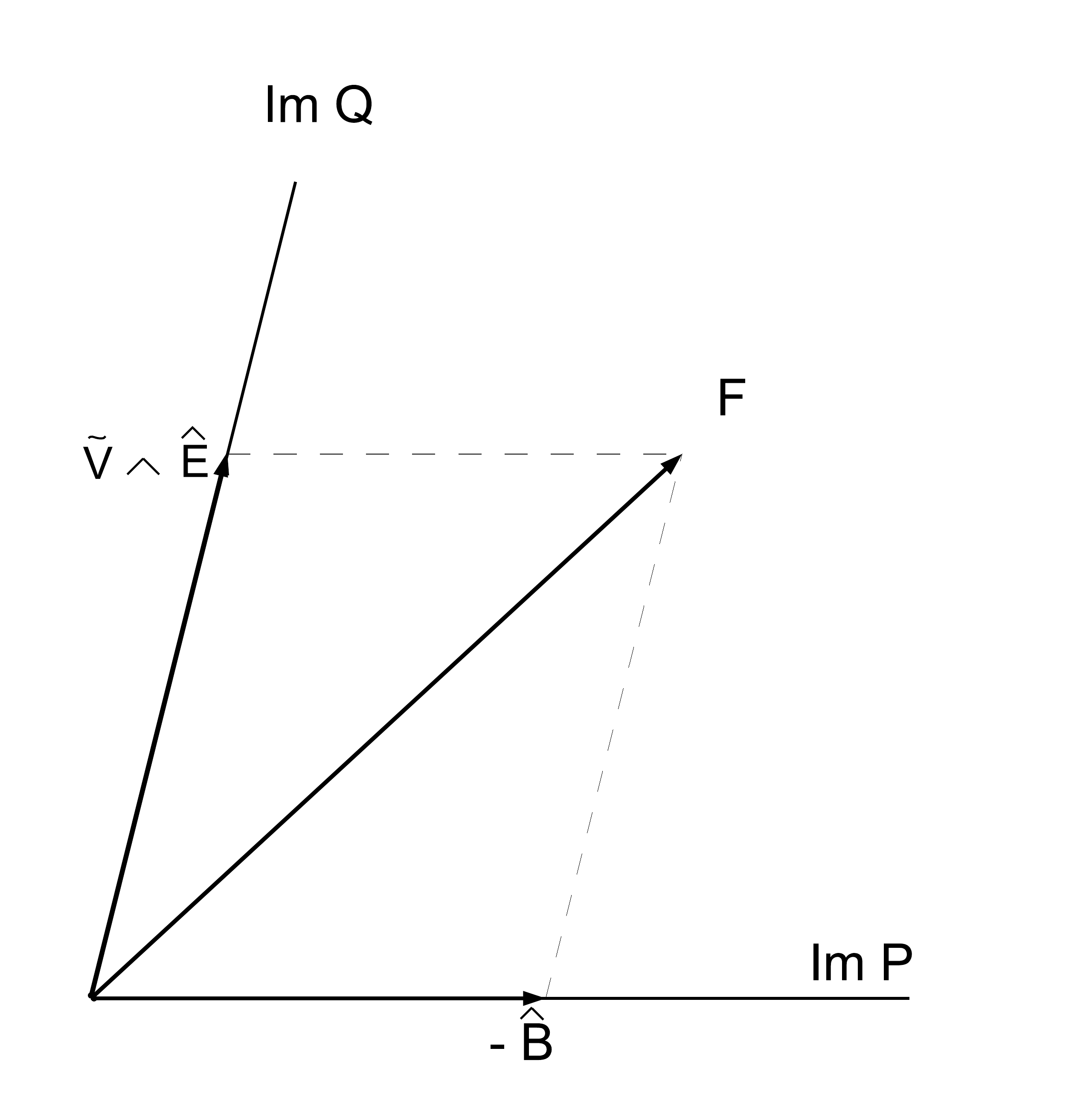}}
\vspace*{-0.2cm}
\caption{Left: decomposition of a general form $\alpha$ by means of projector operators $\mathcal Q$ and $\mathcal P$,
               see Eqs. (\ref{operatorsPQfromV}) and (\ref{decompalphawrtV}).
         Right: particular case given by the 2-form of the electromagnetic field $F$.
         The 2-form decomposes in this way into its electric and magnetic parts respectively.}
\label{fig:decomposition}
\end{center}
\end{figure}

Here we repeat the tricks from Sec. \ref{4tospecial3plus1more} with the replacement
$\partial_t \mapsto V$ (where $V$ is an observer field).
First, notice that if $j_V\alpha := \tilde V \wedge \alpha$ (where $\tilde V \equiv g(V, \ . \ )$),
the identity
\begin{equation}
       \label{identityiVjV}
                 i_Vj_V + j_Vi_V = g(V,V) \hat 1 \ \equiv \ \hat 1
\end{equation}
holds (the proof just repeats the proof of Eq. (\ref{identityitjt})).
Then, defining operators
\begin{equation} \label{operatorsPQfromV}
                 \mathcal P := i_Vj_V
                 \hskip 1.5cm
                 \mathcal Q := j_Vi_V
\end{equation}
\`{a} la Eq. (\ref{operatorsPQ}) we check that they still enjoy the properties
displayed in Eqs. (\ref{property1}) - (\ref{property3}).
Therefore, they are projectors onto two mutually complementary subspaces.
So, we have a unique decomposition of forms $\alpha = \mathcal Q \alpha + \mathcal P \alpha$
or, explicitly,
\begin{equation} \label{decompalphawrtV}
                 \alpha = \tilde V \wedge \hat s + \hat r
\end{equation}
where
\begin{eqnarray} %{rcl}
      \label{defhatsV}
                 \hat s
             &=& i_V \alpha       \\
      \label{defhatrV}
                 \hat r
             &=& i_V(j_V \alpha)
\end{eqnarray}
(see the left drawing at Fig. \ref{fig:decomposition}).
The hats on $\hat s$ and $\hat r$ mean that the two forms are \emph{spatial} (or \emph{horizontal}).
By this we mean, roughly, that they only take seriously ``spatial arguments''
(they are annihilated by temporal arguments) and so, in a sense, they live as if in the space
(rather than space-time).
\vskip .2cm
\noindent
[Precisely this means the following. Consider \emph{any} form $\beta$ of the structure
 $\beta = i_V\tau$. Decompose general vector $U$ as its temporal plus spatial parts,
 $U=U_\parallel +U_\perp \equiv \lambda V + U_\perp$, $g(U_\parallel, U_\perp)=0$. Then
$$
\begin{array} {rcl}
      \beta (U,\dots ,W)
  &=& \beta (\lambda V + U_\perp,\dots ,\mu V + W_\perp) \\
  &=& \tau (V,\lambda V + U_\perp,\dots ,\mu V + W_\perp) \\
  &=& \tau (V, U_\perp,\dots , W_\perp) \\
  &=& \beta (U_\perp,\dots , W_\perp)
\end{array}
$$
So $\beta$ completely ignores temporal (vertical) parts of arguments and only takes seriously
spatial (horizontal) parts if inserting vectors.

Notice also that if we evaluate $\alpha$ on strictly spatial arguments $(U_\perp,\dots , W_\perp)$,
only its part $\hat r$ matters since $\tilde V(U_\perp) = g(V,U_\perp) = 0$.
Therefore we call $\hat r$ the \emph{horizontal} (spatial) \emph{part} of $\alpha$.
Also the projector $\mathcal P$ onto the horizontal part of a form is denoted as ``hor''
\begin{equation} \label{operatorhor1}
                 \hor \alpha = \hor (\tilde V \wedge \hat s + \hat r) = \hat r \equiv \mathcal P \alpha
\end{equation}
An alternative definition is (check)
\begin{equation} \label{operatorhor2}
                 (\hor \alpha) (U,\dots , W) := \alpha (U_\perp,\dots , W_\perp)
\end{equation}
A form satisfying
\begin{equation} \label{horizontalform}
                 \hor \alpha = \alpha
\end{equation}
is called \emph{horizontal} (its decomposition looks $\alpha = \hat r$, i.e. it has $\hat s =0$).
Remember that a form which is horizontal w.r.t. some observer field $V$ may not be horizontal
w.r.t. other observer field $V'$.]
\vskip .2cm
Thus a general form $\alpha$ may be decomposed, w.r.t. observer field $V$,
according to Eq. (\ref{decompalphawrtV}), providing us with two spatial forms $\hat s$ and $\hat r$.
(The triple $(V,\hat s,\hat r)$ carries the same information as $(\alpha,V)$.)
Changing of $V$ causes changing of ``dividing the $\alpha$
between $\hat s$ and $\hat r$ parts''.

In particular (see the right drawing at Fig. \ref{fig:decomposition}),
we can decompose the form of our main interest, the 2-form of electromagnetic field.
We get
\begin{equation}
       \label{decompositionofF}
       \boxed{F= \tilde V \wedge \hat E - \hat B}
\end{equation}
where
\begin{eqnarray} %{rcl}
      \label{defofhatEwrtV}
                 \hat E
             &=& i_V F
                 \hskip 3.3cm
                 \text{\emph{1-form of electric field w.r.t. $V$}}    \\
      \label{defofhatBwrtV}
                 \hat B
             &=& i_V j_V F \equiv \tilde V \wedge \hat E - F
                 \hskip .75cm
                 \text{\emph{2-form of magnetic field w.r.t. $V$}}
\end{eqnarray}
If we needed, by the way, to know orthonormal components of the the two fields
w.r.t. an \emph{orthonormal spatial} frame $e_a \equiv (e_1,e_2,e_3)$
(and the dual co-frame $e^a \equiv (e^1,e^2,e^3)$), we could use standard formulas
\begin{eqnarray} %{rcl}
      \label{componentsofE}
                 \hat E
             &=& E_be^b = E^a{\hat h}_{ab}e^b \hskip 1cm {\hat h}_{ab} \equiv \delta_{ab}    \\
      \label{componentsofB}
                 \hat B
             &=& B^adS_a \equiv B^a \left( \frac 12 \epsilon_{abc} e^b \wedge e^c \right)
\end{eqnarray}
What we use here, however, are the ``complete'' forms $\hat E$ and $\hat B$ alone.

\subsubsection{Decomposition of operations on forms}
\label{decompoperforms}

Two operations on forms are to be learned in more detail for the case when our forms are decomposed
according to Eq. (\ref{decompositionofF}). Namely the \emph{Hodge star} $*$ and the
\emph{exterior derivative} $d$.

Let us begin with the Hodge star operator.

The horizontal subspace of the tangent space at each point inherits natural (``spatial'')
metric tensor $\hat h$
(with signature + + + by definition, i.e., $g=\tilde V \otimes \tilde V - \hat h$) and orientation
(a spatial frame $(e_1 ,e_2 ,e_3)$ is declared to be right-handed if the frame $(V\equiv e_0, e_1 ,e_2 ,e_3)$
is right-handed). These data are just enough for the unique \emph{spatial Hodge operator} $\hat *$
(it only acts on spatial forms and in this case it acts exactly as we know it in good old Euclidean space).
We already discussed the relation of the ``complete'' Hodge star $*$ and the spatial one $\hat *$
in Sec. \ref{formsinminkowski} (for the particular case $V=\partial_t$, see Eq. (\ref{starongeneralmink})).
Here we get the following formula (see \cite{fecko1997})
\begin{equation}
       \label{decompositionof*}
       \boxed{*(\tilde V \wedge \hat s + \hat r) = \tilde V \wedge (\hat * \hat r) + \hat * \hat \eta \hat s}
\end{equation}
So, in comparison with Eq. (\ref{starongeneralmink}), one just has to replace $dt$ by more general
object $\tilde V$.

The case of the exterior derivative is a bit more involved
(not so much technically, but rather conceptually). We are to take into account two issues.

First, if we act by $d$ on $\tilde V \wedge \hat s + \hat r$, we get
\begin{equation}
       \label{dondecomposedalpha1}
       d\alpha = d(\tilde V \wedge \hat s + \hat r)
               = d\tilde V \wedge \hat s - \tilde V \wedge d\hat s + d\hat r
\end{equation}
What is new here in comparison with Eq. (\ref{dongeneralmink}), i.e. with the case when $\tilde V=dt$,
is the fact that the first term \emph{does not vanish}, in general.
Rather, $d\tilde V$ may be decomposed, as it is true for each form,  \`{a} la Eq. (\ref{decompalphawrtV}).
In this way we get
\begin{equation}
       \label{dVdecomp}
       d\tilde V = \tilde V \wedge \hat a + \hat y
\end{equation}
where $\hat a$ and $\hat y$ are spatial $1$-form and $2$-form, respectively. They represent
\emph{kinematical characteristics} of the observer field $V$.

It may be shown (see Appendix C of the paper \cite{fecko1997}) that $\hat a$ encodes the
\emph{acceleration} of the observer field
\begin{equation}
       \label{hataisacceleration}
       \hat a = g(\nabla_V V, \ . \ ) = g(a, \ . \ )
\end{equation}
An observer field is called \emph{geodesic} if its acceleration $1$-form vanishes.
(Each observer's world-line is then geodesic.)
\vskip .2cm
\noindent
[Vector $a\equiv \nabla_V V$ is perpendicular to $V$ since $V$ has unit length (check).
 Therefore the corresponding $1$-form, $g(a, \ . \ )$, is spatial (check), i.e.
 it deserves the hat.]
\vskip .2cm

Similarly, it may be shown that the
\emph{vorticity $2$-form} $\hat y$ encodes the fact whether or not, in physical terms,
\emph{time synchronization} is possible for the system of observers under consideration.
(Non-zero $\hat y$ means that it is not possible).
Another way to say the same is, in more mathematical parlance, that $\hat y$ is the measure of
\emph{integrability} of the
\emph{spatial distribution}, i.e. it encodes whether or not the instantaneous 3-spaces
(perpendicular to $V$) mesh together to form (locally) a spatial 3-domain
(integral sub-manifold) $\mathcal D$.
(For a concrete example, see Sec. \ref{integrabilityandspatial}.)
Observer field $V$ is said to be
\emph{proper-time-synchronizable} if both $\hat a$ and $\hat y$ vanish.
Then, in adapted coordinates, $V=\partial_t$ and $\hat V=dt$.

\vskip .2cm
\vskip .5cm
\noindent
{\bf Example \ref{decompoperforms}.1}:
Consider observers in Minkowski space-time moving uniformly along $x$ with (constant) velocity $v$.
Their spatial trajectories are
$$(x(t),y(t),z(t)) = (x_0+vt,y_0,z_0)
$$
so that their world-lines read
$$\tau \mapsto (t(\tau),x(\tau),y(\tau),z(\tau)) = (t_0+\tau,x_0+v\tau ,y_0,z_0)
$$
The world-lines are integral curves of the vector field
\begin{equation}
       \label{fieldW}
       W = \partial_t +v\partial_x
\end{equation}
Its norm is not unity
$$||W||^2 = g(W,W) = 1-v^2 \hskip .5cm (= \ \text{const.})
$$
but the \emph{normalized} vector field may already play the role of $V$
\begin{equation}
       \label{fieldVfromW}
       V := \frac{W}{||W||} = \gamma (\partial_t +v\partial_x) \hskip 1cm \gamma := \frac{1}{\sqrt{1-v^2}}
\end{equation}
Then the $1$-form $\tilde V = g(V, \ . \ )$ reads
\begin{equation}
       \label{tildeVfromW}
       \tilde V = \gamma (dt - vdx) =d(\gamma (t-vx)) =: dt'
\end{equation}
So $d\tilde V=0$ and therefore

- $\hat a = 0 = \hat y$ for our observer field $V$ (it is proper-time-synchronizable)

- any constant multiple of $(t-vx)$ may serve as a new (synchronized) time $t''$

- $t''=$ const. are then $3$-space sub-manifolds

- $Vt'' = 1$ chooses the constant multiple so that $t''$ becomes \emph{proper} time $t'$

- it gives $t' = \gamma (t-vx)$

- this is proper time common for \emph{all} world-lines of the congruence

- the time $t'$ realizes the proper-time-synchronization program

\noindent
The end of Example \ref{decompoperforms}.1.
\vskip .5cm
Plugging Eq. (\ref{dVdecomp}) into Eq. (\ref{dondecomposedalpha1})
we get
\begin{equation}
       \label{dondecomposedalpha2}
       d\alpha = d(\tilde V \wedge \hat s + \hat r)
               = \tilde V \wedge (- d\hat s +\hat a \wedge \hat s ) + d\hat r +\hat y
\end{equation}
The computation is not yet finished, however, since exterior derivatives of spatial
forms ($d\hat s$ and $d\hat r$) are still to be determined
(expressed in terms of spatial exterior derivative $\hat d$).

In order to motivate proper choice of what exactly is to be defined to be $\hat d$,
one should look at \emph{Stokes theorem}
(where the exterior derivative plays prominent role) in situation where spatial objects
alone are present.

So, consider a spatial form $\hat b$ and a spatial domain $\hat D$
(i.e. the domain of any possible dimension with the property that any vector tangent
to it is spatial).
Then
\begin{eqnarray} %{rcl}
      \label{Stokes1}
                 \int_{\hat D}d\hat b
             &=& \int_{\partial \hat D}\hat b
                 \hskip 2.05cm \text{due to Stokes' theorem} \\
      \label{Stokes2}
             &=& \int_{\hat D}\hor d\hat b
                 \hskip 1.5cm \text{since $\hat D$ is spatial}
\end{eqnarray}
Combining the two expressions at the r.h.s. we get
\begin{equation}
       \label{spatialStokes}
       \boxed{\int_{\hat D}\hat d\hat b  = \int_{\partial \hat D}\hat b}
                                    \hskip 2.1cm
                                    \text{{\color{red}\emph{spatial Stokes' theorem}}}
\end{equation}
where
\begin{equation}
       \label{spatialdwrtV}
            \boxed{\hat d  := \hor d \equiv i_Vj_V d}
                                    \hskip 1.4cm
                                    \text{{\color{red}\emph{spatial exterior derivative}}}
\end{equation}
In terms of (this particular) $\hat d$ and also the Lie derivative $\mathcal L_V$,
the exterior derivative $d$ may be expressed as follows
(see Appendix D of the paper \cite{fecko1997}):
\begin{equation}
       \label{dondecomposedalpha}
       \boxed{d\alpha = d(\tilde V \wedge \hat s + \hat r)
               = \tilde V \wedge (- \hat d\hat s +\mathcal L_V \hat r +\hat a \wedge \hat s )
               + \hat d\hat r +\hat y}
\end{equation}
\vskip .2cm
\noindent
[It is worth noticing that the spatial exterior derivative $\hat d$ is, contrary to $d$,
 no longer nilpotent (i.e. $\hat d \hat d\neq 0$). Rather, it holds
\begin{equation}
       \label{hatdhatd}
       \hat d \hat d = - \hat y \wedge \mathcal L_V
\end{equation}
This may seem to be, because of $\partial \partial = 0$, in conflict with Eq. (\ref{spatialStokes}),
but a closer inspection shows it is actually not the case.
For more details, including derivation of Eqs. (\ref{dondecomposedalpha}) and (\ref{hatdhatd})
see Appendix D and G of the paper \cite{fecko1997}.]
\vskip .5cm
\noindent
{\bf Example \ref{decompoperforms}.2}:
Consider, as the simplest example, the case $V=\partial_t$ in Minkowski space
(the case we already know from Sec. \ref{formsinminkowski}).
Then $\tilde V = g(\partial_t,\partial_\mu)dx^\mu = dt$.
Therefore $0=ddt = d\tilde V = dt\wedge \hat a +\hat y$, i.e. $\hat s =0 = \hat y$.
From Eq. (\ref{dondecomposedalpha}) we get
$$d\alpha = d(dt \wedge \hat s + \hat r)
               = dt \wedge (- \hat d\hat s +\mathcal L_{\partial_t} \hat r)
               + \hat d\hat r
$$
which is exactly Eq. (\ref{dongeneralmink}), provided that the two $\hat d$-s coincide.
Well, the $\hat d$ in Eq. (\ref{dongeneralmink}) is defined in Eq. (\ref{donspatial}) by
$$
  d\hat r = dt \wedge \mathcal L_{\partial_t} \hat r + \hat d \hat r
$$
Acting on the l.h.s. by hor $= i_Vj_V=i_{\partial_t}j_{\partial_t}$ we get
$\hat d \hat r$ (with the \emph{new} $\hat d$). The same on the r.h.s. gives
$\hat d \hat r$ as well, but with the \emph{old} $\hat d$, now.
So, the two $\hat d$-s indeed coincide.
The end of Example \ref{decompoperforms}.2.

\subsubsection{How to compute easily $\hat d$ and $\hat *$}
\label{howhat*andhatd}

In practical manipulations with forms we know that actual computation of the exterior derivative
$d$ is extremely simple \emph{in coordinates}: a $k$-form on $n$-dimensional manifold is a sum of terms
of the structure
$$f(x^1,\dots, x^n)dx^i\wedge \dots \wedge dx^j
$$
and the result of the action of $d$ on this particular term is simply
\begin{equation}
       \label{resultofd}
       d(fdx^i\wedge \dots \wedge dx^j) = df\wedge dx^i\wedge \dots \wedge dx^j
\end{equation}
where
\begin{equation}
       \label{resultofdf}
       df = f_{,k}dx^k
\end{equation}
and actually one has to perform only those partial derivatives whose differentials $dx^k$ are not
already present in the original form
(i.e. for such $k$ that $dx^k$ \emph{differs} from any $dx^i$,\dots, $dx^j$).
Similar statements hold for interior product $i_V$, operator $j_V$ or $*$.

If we want to compute \emph{spatial} versions of these operators, say $\hat d$,
going to coordinates may be highly inefficient.

First, sometimes the 3-\emph{space} itself simply does not exist. Locally
(strictly speaking in the tangent space), it is defined as the orthogonal complement of $V$,
i.e. vector $u$ is spatial if $g(V,u) \equiv i_u\tilde V = 0$.
So spatial vectors annihilate $\tilde V$.
However, the distribution defined in this way may be \emph{non-integrable}.
If this is the case then, although the 3-dimensional sub-space in each tangent space exists,
the 3-space in the sense of a \emph{sub-manifold} of events with constant time does not.
(The \emph{local} ``instantaneous 3-spaces'' do not mesh together to form \emph{the} instantaneous
 3-space ``unifying'' local 3-spaces.)
And introducing coordinates in non-existing sub-manifold might be a fairly hard task.
\vskip .2cm
\noindent
[As we already mentioned above, the symptom of non-integrability of the particular distribution
 given by $\tilde V$ is \emph{non-vanishing} $\hat y$ in the decomposition Eq. (\ref{dVdecomp}).
 This is a simple consequence of Frobenius' integrability theorem
 (see Sec. \ref{integrabilityandspatial} here and/or \$19.3 in \cite{fecko2006}).]
\vskip .2cm
But even if the 3-space does exist, explicit introducing of local coordinates might be difficult
and if it is possible to compute what we are interested in without doing the difficult work
we should embrace the opportunity.

Well, let us plug $\hat s = 0$ into formulas Eqs. (\ref{dondecomposedalpha}) and (\ref{decompositionof*}).
(This means, compute $d$ and $*$ of genuinely spatial forms.) We get
\begin{eqnarray} %{rcl}
      \label{donspatialV}
                 d\hat r
             &=& \tilde V \wedge \mathcal L_V \hat r +\hat d \hat r               \\
      \label{*onspatial}
                 *\hat r
             &=& \tilde V \wedge \hat * \hat r
\end{eqnarray}
But from these equations clearly follows
\begin{equation} %{rcl}
      \label{hatdonspatial}
                 \boxed{\hat d\hat r
             = d\hat r  - \tilde V \wedge \mathcal L_V \hat r}
\end{equation}
\begin{equation}
      \label{hat*onspatial}
                 \boxed{\hat *\hat r
             = i_V * \hat r}
\end{equation}
That's all. On the r.h.s., all operations are easily performed since they do not need
coordinates in 3-space.

\setcounter{equation}{0}
\subsection{Rotating frame and electric and magnetic fields}
\label{rotframe}

Here we apply our formalism to computations in uniformly rotating frame.
First, in Sec. \ref{rotframekinematics} we compute the necessary objects for this particular kinematics.
Then, in Sec. \ref{rotframeelfield} we apply the formulas from Sec. \ref{4to3plus1}
for computation what $\hat E$ and $\hat d \hat E$ an observer in rotating frame sees
if the observer in non-rotating frame sees just a static magnetic field.
This is then in turn used for computation of the induced voltage for a wire rim
uniformly rotating in a static magnetic field (in Sec. \ref{inducedvoltage}).

\subsubsection{Basic kinematics}
\label{rotframekinematics}

We can proceed similarly as we did in Example \ref{decompoperforms}.1.

So, consider observers in Minkowski space-time rotating uniformly about the $z$-axis
with (constant) angular velocity $\omega$. Their spatial trajectories are,
in \emph{cylindrical} coordinates,
$$(r(t),\varphi (t),z(t)) = (r_0,\varphi_0 +\omega t,z_0)
$$
so that their world-lines read
$$\tau \mapsto (t(\tau),r(\tau),\varphi (\tau),z(\tau)) = (t_0+\tau,r_0 ,\varphi_0 +\omega \tau,z_0)
$$
Thus the world-lines are integral curves of vector field
\begin{eqnarray} %{rcl}
      \label{fieldWcylindr}
                 W
             &=& \partial_t +\omega \partial_\varphi
                 \hskip 2.15cm
                 \text{(in cylindrical coordinates)}            \\
      \label{fieldWcartes}
             &=& \partial_t +\omega (x\partial_y - y\partial_x)
                 \hskip .7cm
                 \text{(in Cartesian coordinates)}
\end{eqnarray}
Its norm is not unity,
$$||W||^2 = g(W,W) = 1-(\omega r)^2
$$
So, the vector field $V$ which we look for is just the \emph{normalized} vector
\begin{equation}
       \label{fieldVfromWrot}
       V := \frac{W}{||W||} = \gamma (\partial_t +\omega \partial_\varphi )
                              \hskip 1cm
                              \gamma (r) := \frac{1}{\sqrt{1-(\omega r)^2}}
\end{equation}
Thus normalization just adds overall relativistic factor $\gamma$.
\vskip .2cm
\noindent
[This makes perfect sense, when compared with a similar result in Example \ref{decompoperforms}.1.
 Just replace $v\partial_x$ with $(\omega r)e_\varphi$ in Eq. (\ref{fieldVfromW}),
 where $e_\varphi = (1/r)\partial_\varphi$ is the \emph{unit} vector directed along the motion,
 the counterpart of $e_x = \partial_x$ there.
 Notice, in particular, that the relativistic factor $\gamma$ depends on position in the space, now,
 since the magnitude of the velocity at point with coordinates $(r,\varphi,z)$ is just $v=\omega r$.]
\vskip .2cm
So, contrary to the situation discussed in Example \ref{decompoperforms}.1.,
$\gamma$ is a \emph{position dependent} quantity, here.
This, however, makes further computations a bit more complex.

Now realize that taking $\gamma$ seriously means, as we know from elementary special relativity
theory, performing computation with full (special) relativistic accuracy.
In particular it means, from the physics point of view, that we expect the velocity $v=\omega r$
being large enough on the scale of unity (which is $c=1$, the velocity of light, of course).
And this is \emph{clearly not so} for the \emph{rotating wire rim}, the case of our main interest.

Therefore it would be much wiser to make suitable approximation.
Namely, since our rim is small and it moves (very) slowly, we will in what follows content ourselves
with expressions up to terms \emph{linear in} $\omega$.
(Exact computations with all $\gamma$'s may be found, just to see they are fully tractable,
 in Appendix \ref{app:rotatingexact}.)

Now, the power series expansion of $\gamma (r)$ in $\omega r$ is
 \begin{equation}
       \label{gammaexpansion}
       \gamma (r) \equiv \frac{1}{\sqrt{1-(\omega r)^2}}
                  = 1 + \frac 12 (\omega r)^2 + \dots
\end{equation}
Therefore, within the approximation mentioned above,
we can safely put
 \begin{equation}
       \label{gammaapprox}
       \gamma \doteq 1
\end{equation}
and, consequently, $V$ reduces back to $W$:
\begin{eqnarray} %{rcl}
      \label{fieldWapproxcylindr}
                 V
             &\doteq& \partial_t +\omega \partial_\varphi
                 \hskip 2.15cm
                 \text{(in cylindrical coordinates)}            \\
      \label{fieldWapproxcartes}
             &\doteq& \partial_t +\omega (x\partial_y - y\partial_x)
                 \hskip .7cm
                 \text{(in Cartesian coordinates)}
\end{eqnarray}
Since the metric tensor Eq. (\ref{minkowskimetric}) reads
\begin{eqnarray} %{rcl}
      \label{etaincylindr}
                 g\equiv \eta = dt \otimes dt - \hat g
             &=& dt \otimes dt - (dr \otimes dr + r^2d\varphi  \otimes d\varphi + dz \otimes dz)
                        \\
      \label{etaincartes}
             &=& dt \otimes dt - (dx \otimes dx + dy \otimes dy + dz \otimes dz)
\end{eqnarray}
in cylindrical coordinates $(t,r,\varphi, z)$ and Cartesian coordinates $(t,x,y,z)$,
 respectively, the corresponding $1$-form $\tilde V = g(V, \ . \ )$ is
\begin{eqnarray} %{rcl}
      \label{tildeVincylindr}
                 \tilde V
             &\doteq& dt -\omega r^2 d\varphi
                        \\
      \label{tildeVincartes}
             &\doteq& dt -\omega (xdy - ydx)
\end{eqnarray}
Then, direct computation leads to
\begin{eqnarray} %{rcl}
      \label{dtildeVincylindrapprox1}
                 d\tilde V
             = \tilde V \wedge \hat a + \hat y \hskip 1cm \hat a &\doteq& 0     \\
      \label{dtildeVincylindrapprox2}
                 \hat y &\doteq& -2\omega dr \wedge rd\varphi \equiv -2\omega dS_z
\end{eqnarray}
\vskip .2cm
\noindent
[More exact expression for the acceleration $1$-form, up to \emph{quadratic} term in $\omega$,
 is $\hat a = \omega^2 rdr$. (Exact result is $\hat a = \gamma^2 \omega^2 rdr$,
 see Appendix \ref{app:rotatingexact}.)
 Since $\hat a = g(a, \ . \ )$, we get $a= - \omega^2 r\partial_r$.
 This is good old \emph{centripetal} acceleration of the uniform circular motion.
 It is a high-school piece of wisdom that centripetal acceleration is \emph{quadratic} in
 angular velocity $\omega$ and therefore we get $\hat a \doteq 0$
 in Eq. (\ref{dtildeVincylindrapprox1}) where just \emph{linear} term is taken into account.]
\vskip .2cm

\subsubsection{How to extract $\hat E$ and $\hat d \hat E$ from $F$}
\label{rotframeelfield}

Now we come to the situation relevant for the wire rim.

\begin{figure} %[h]
\begin{center}
\hskip -1.60cm
\scalebox{0.15}{\includegraphics[clip,scale=1.70]{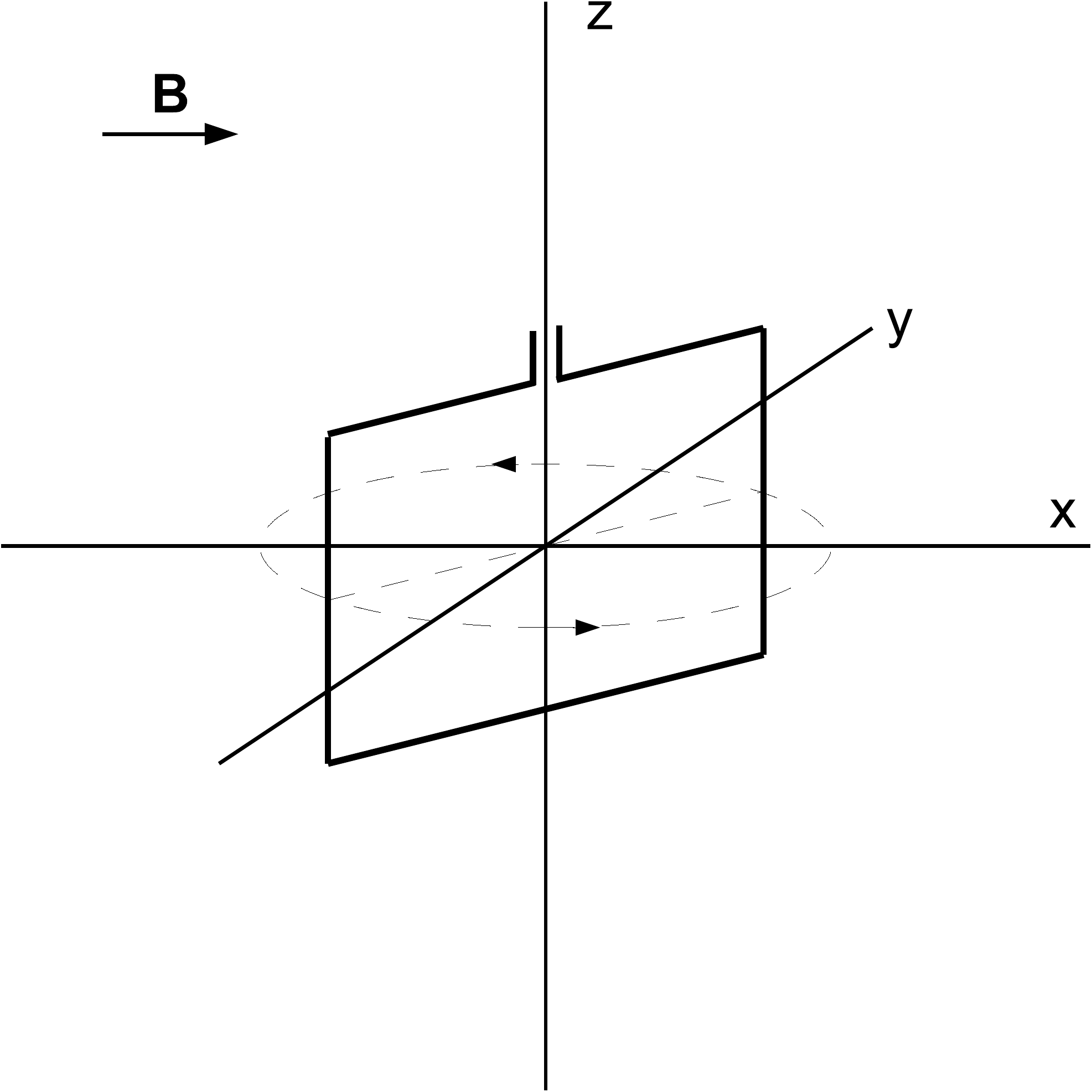}}
\scalebox{0.15}{\includegraphics[clip,scale=1.70]{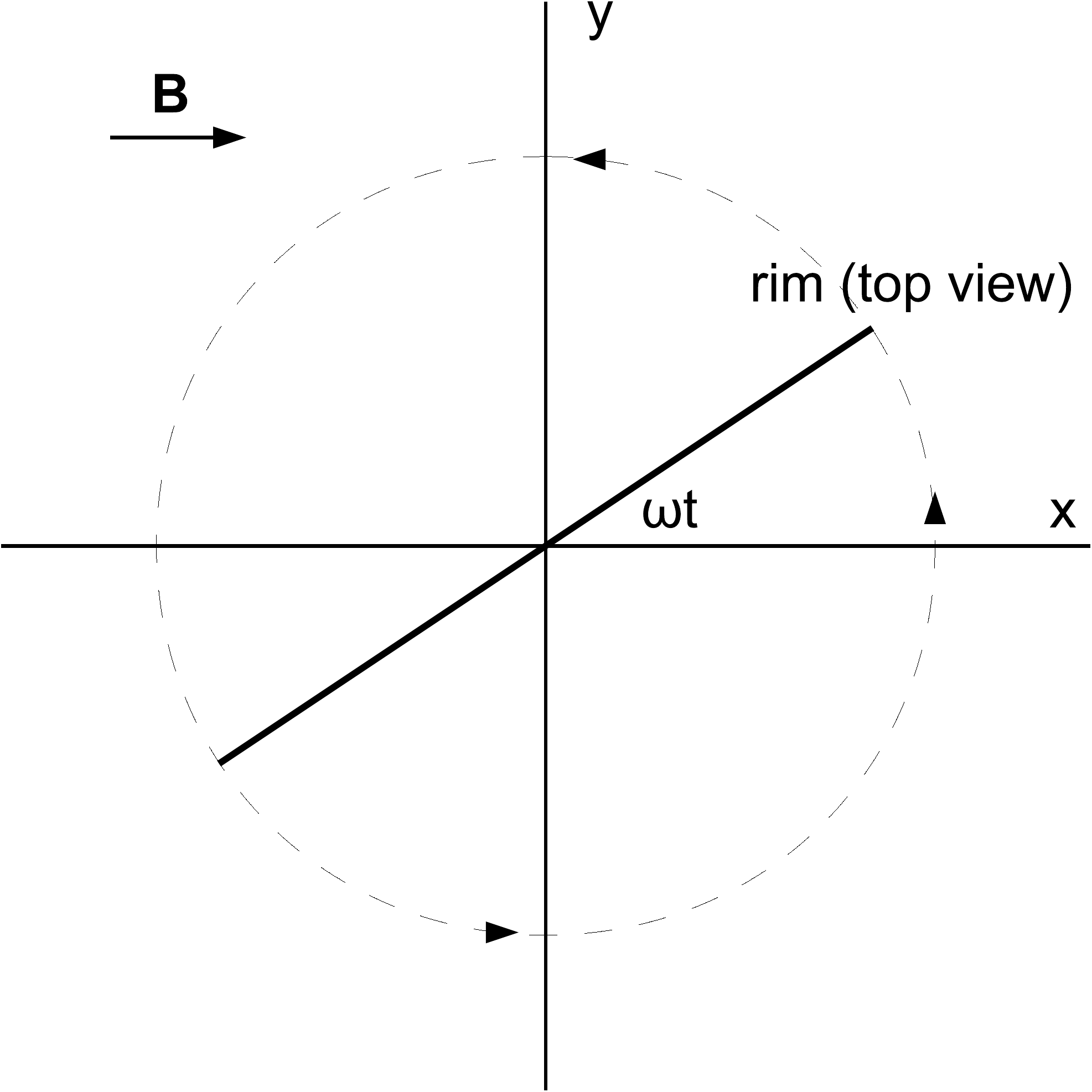}}
\caption{Wire rim rotates about $z$-axis in a static homogeneous magnetic field directed along $x$-axis.
         Left: 3D view of the situation. Right: 2D view from above.}
\label{fig:rimforfaraday}
\end{center}
\end{figure}

Consider, in the original reference frame, static homogeneous magnetic field along $x$-axis
and let the rim rotate about $z$-axis, see Fig \ref{fig:rimforfaraday}.
(So the projection of the area of the rim to $x$-axis, and consequently the flux of $\bold B$
through the rim, oscillates sinusoidally.)
Such field is described, according to Eq. (\ref{definitionofF}), by $2$-form of electromagnetic field
\begin{equation}
       \label{Ffortherim}
       F = -BdS_x = -Bdy \wedge dz
                         \hskip 1.5cm
                         B = \ \text{const.}
\end{equation}
This $F$ is observer-independent.
(The fact that it is expressed in coordinates best suited to some particular reference frame is irrelevant.)
What is observer-dependent is the splitting of this $F$ into electric and magnetic parts.
In order to compute the induced voltage we need $\hat E$
(and perhaps $\hat d\hat E$ in order to convert line integral to a surface one). We know how to find $\hat E$
from Eq. (\ref{defofhatEwrtV}). Here, it gives
\begin{equation}
       \label{hatEfortherim}
       \hat E = i_VF = i_{\partial_t +\omega (x\partial_y - y\partial_x)}(-Bdy \wedge dz)= -\omega B x dz
\end{equation}
So, as we anticipated, rotating observer sees \emph{non-zero electric} field
(in addition to magnetic field, which could be computed using Eq. (\ref{defofhatBwrtV});
 we do not need it, however).
\vskip .3cm
\noindent
[If we compare this expression with the general structure of Eq. (\ref{defofhatE}),
 we see that the electric field may be written, using coordinates $(x,y,z)$ warmly loved by
 non-rotating observer, as
 $$\bold E = (0,0,-\omega Bx)$$
 It is static $z$-directed and $x$-dependent field.
 Remember, however, that non-rotating observer sees, as a matter of fact, \emph{no} electric
 field at all.
 So, if Bob is non-rotating observer and Alice the rotating one, what Eq. (\ref{hatEfortherim})
 displays is \emph{Bob's description} of what electric field \emph{Alice sees}.
 (In more detail: Bob sees no electric field. Alice, on the contrary, sees also electric field
  and she sends a message to Bob, saying what she sees. Bob \emph{translates} her result into his
  notations and, \emph{then}, displays it in Eq. (\ref{hatEfortherim}).)
 \footnote{Presence of Alice and Bob gently suggests that the paper deals with hot stuff
           of modern science.}]
\vskip .3cm
And what $\hat d \hat E$ does she see?
Just use formula Eq. (\ref{hatdonspatial}):
\begin{equation}
       \label{hatdhatEfortherim}
       \hat d \hat E = (d - \tilde V \wedge \mathcal L_V)\hat E \doteq -\omega B dx\wedge dz
                     = \omega B dS_y
\end{equation}
\vskip .3cm
\noindent
[Bob's translation into his notations of what Alice sees is, here,
 $$\boldsymbol \nabla \times \bold E = (0,\omega B,0)
 $$
 since in general $\hat d \hat E = (\boldsymbol \nabla \times \bold E)\cdot d\bold S$,
 see Eq. (\ref{curl}).]
\vskip .3cm
\noindent $\blacktriangledown \hskip 0.5cm$
The first term is
$$d\hat E = d(-\omega B x dz) = -\omega B dx \wedge dz
$$
Concerning the second term,
$$\mathcal L_V\hat E = ... = \omega^2Bydz \doteq 0
$$
So the second term ``vanishes'' because is is at least quadratic in $\omega$.
\hfill $\blacktriangle$ \par
\vskip .2cm \indent
The most remarkable fact about Eq. (\ref{hatdhatEfortherim}) is that $\hat d \hat E$
\emph{does not vanish}. This, due to spatial Stokes theorem Eq. (\ref{spatialStokes}),
signalizes that the circulation of $\hat E$ around the rim is non-zero.
And this, in turn, is exactly induced voltage we are interesting in.

\subsubsection{Faraday's law - how to compute the induced voltage for a rim}
\label{inducedvoltage}

According to spatial Stokes theorem Eq. (\ref{spatialStokes}), we can write
\begin{equation}
       \label{spatialStokesrim}
       U = \oint_{\hat c} \hat E \equiv \oint_{\partial \hat S} \hat E = \int_{\hat S}\hat d \hat E
                                    \hskip 1cm
                                    \partial \hat S = \hat c = \ \text{the rim}
\end{equation}
where $U$ is the voltage induced by rotation of the rim.
So what we need to evaluate is the surface integral
\begin{equation}
       \label{weneedsurfaceintegral}
       \int_{\hat S} \hat d \hat E =
       \omega B \int_{\hat S} dz\wedge dx \equiv
       \omega B \int_{\hat S} dS_y
\end{equation}
over any $2$-dimensional spatial surface $\hat S$ whose boundary is the rim $\hat c$.

Remember that the rim does not move w.r.t. rotating system of observers.
So it is natural to choose the surface with the same property.

Namely, let's choose the surface such that at time $t=0$ it lies in the plane $xz$
of the non-rotating coordinate system. Call this particular surface $\hat S_0$.
Then, at further moments, $\hat S$ will be simply $\Phi_t$-image of the $\hat S_0$,
where $\Phi_t \leftrightarrow V$. So, we get
\begin{equation}
       \label{Uequals}
       U = \omega B \int_{\hat S} dS_y
         = \omega B \int_{\Phi_t(\hat S_0)} dS_y
         = \omega B \int_{\hat S_0} \Phi_t^*(dS_y)
\end{equation}
But
\begin{equation}
       \label{Phit*dSy}
       \Phi_t^*(dS_y) = dS_y \cos \omega t + dS_x \sin \omega t
\end{equation}
and, consequently,
\begin{equation}
       \label{resultingU}
       \boxed{U = \omega B \mathcal S \cos \omega t}
           \hskip 1cm
           \mathcal S = \ \text{area of} \ \hat S_0
\end{equation}
\newline \noindent
\noindent $\blacktriangledown \hskip 0.5cm$
Eq. (\ref{fieldWapproxcylindr}) implies the following flow $\Phi_t \leftrightarrow V$
$$
  \Phi_t: \ \ (t_0,r,\varphi ,z) \mapsto (t_0+t,r,\varphi +\omega t ,z)
$$
Then,
$$
\begin{array} {rcl}
      \Phi_t^*(dS_y)
  &=& \Phi_t^*(dz\wedge dx) \\
  &=& d\Phi_t^*z \wedge d\Phi_t^* x \\
  &=& dz \wedge d\Phi_t^* (r\cos \varphi ) \\
  &=& dz \wedge d (r\cos (\varphi +\omega t)) \\
  &=& dz \wedge d (r\cos \varphi \cos \omega t - r\sin \varphi \sin \omega t) \\
  &=& \cos \omega t \ dz \wedge dx   - \sin \omega t \ dz \wedge dy  \\
\end{array}
$$
i.e.
$$
  \Phi_t^*(dS_y) = \cos \omega t \ dS_y   + \sin \omega t \ dS_x
$$
Plugging this into Eq. (\ref{Uequals}) gives
$$
  U = \omega B \left( \cos \omega t \int_{\hat S_0} dS_y + \sin \omega t \int_{\hat S_0} dS_x \right)
$$
Now remembering that $\hat S_0$ lies in $xz$-plane, the second integral vanishes.
The first one is nothing but the area of $\hat S_0$, which is just $\mathcal S$.
\hfill $\blacktriangle$ \par
\vskip .2cm \indent
Eq. (\ref{resultingU}) gives the same result as we could obtain by elementary means,
i.e. compute the magnetic flux for the rim rotated by angle $\omega t$
(it is $-B\mathcal S \sin \omega t$), differentiate it with respect to $t$
and take minus of the result.
\vskip .5cm
\noindent
[Still another way to get the voltage formally
(as minus time derivative of magnetic flux through the moving surface of the rim,
 see \$13.5 of \cite{nearing})
provides $k=2$ version of \emph{Reynolds transport theorem} mentioned in Sec. (\ref{reynoldstransport}).
It reads, for a general vector field $\bold B(\bold r,t)$ (see Eq. (\ref{transport2}))
$$\frac{d}{dt}\int_{S(t)} \bold B \cdot d\bold S
                  =
                  \int_{S(t)} (\partial_t\bold B + (\boldsymbol \nabla \cdot \bold B) \bold v) \cdot d\bold S
                  +
                  \oint_{\partial S(t)} (\bold B \times \bold v) \cdot d\bold r
$$
For \emph{any} magnetic field $\boldsymbol \nabla \cdot \bold B$ vanishes and for \emph{our}
(static) field the same also holds for $\partial_t\bold B$.
So, the first integral (over surface $S(t)$) drops out and the formula simplifies to
$$\frac{d}{dt}\int_{S(t)} \bold B \cdot d\bold S
                  = \oint_{\partial S(t)} (\bold B \times \bold v) \cdot d\bold r
$$
Here, $\bold v$ is the velocity of the element of the rim. Since the rim rotates with
constant angular velocity $\boldsymbol \omega = (0,0,\omega)$, we have
$\bold v = \boldsymbol \omega \times \bold r$ and combining this with $\bold B = (B,0,0)$
we get
$$\frac{d}{dt}\int_{S(t)} \bold B \cdot d\bold S
                  = B\omega \oint_{\partial S(t)} xdz
                  = - B\omega \int_{S(t)} dz \wedge dx
                  = - B\omega \int_{S(t)} dS_y
$$
so that the induced voltage, which is according to Faraday's law \emph{minus} the l.h.s.
of the equation, is
$$
  U = B\omega \int_{S(t)} dS_y = B\mathcal S \omega \cos \omega t
$$
in concordance with Eq. (\ref{resultingU}).]
\vskip .3cm

\subsubsection{A note on integrability and spatial Stokes theorem}
\label{integrabilityandspatial}

In Eq. (\ref{spatialStokesrim}), we used \emph{spatial Stokes} theorem (see Eq. (\ref{spatialStokes}))
for computing the line integral $\oint \hat E$. This step perhaps deserves to pause a bit,
since the important concept of \emph{integrability} of a distribution is tacitly hidden behind
and it might be a good place to discuss this stuff on our concrete example.

Recall that \emph{spatial domain} $\hat D$ plays central role in spatial Stokes theorem
\begin{equation}
       \label{spatialStokes2}
       \int_{\hat D}\hat d\hat b  = \int_{\partial \hat D}\hat b
\end{equation}
Since we speak of $4$-dimensional space-time, there are, in principle,
$4$ possible degrees for spatial forms $\hat b$
($0,1,2$ and $3$-forms). However, $\hat d \hat b$
has to be spatial as well and the possibilities reduce to $0,1$ and $2$-forms alone.
Therefore, the l.h.s. of Eq. (\ref{spatialStokes2}) assumes the existence
of $1,2$ and $3$-dimensional spatial domains $\hat D$.
And this innocent-looking assumption may turn out to be more problematic than it looks at first sight.

Let's start with $3$-dimensional spatial domains $\hat D$.

Existence of $3$-dimensional spatial domains is equivalent to the statement, that the
\emph{spatial distribution is integrable}. What does it mean?

By definition, spatial vectors are vectors orthogonal to $V$ (to the local \emph{time direction}),
i.e. $g(V,w)\equiv \tilde V(w)=0$ for spatial $w$.
So we can also characterize them as vectors \emph{annihilating} the $1$-form $\tilde V$
\begin{equation}
       \label{defofspatial}
       w \ \text{is spatial}
       \hskip 1cm \Leftrightarrow \hskip 1cm
       \tilde V(w) = 0
\end{equation}
At each point of space-time, spatial vectors form a $3$-dimensional subspace of the
$4$-dimensional tangent space. This collection of all subspaces is called spatial \emph{distribution}.

\emph{Integrability} of the distribution means that the space-time (locally) slices to
 $3$-dimen\-sio\-nal sub-manifolds, \emph{integral sub-manifolds} of the spatial distribution,
 such that each of them is spatial. That is, at each point, the collection of vectors tangent
 to the $3$-di\-men\-sional sub-manifold passing through the point coincides with the collection of vectors
 orthogonal to $V$ at this point (i.e. the collection of \emph{spatial} vectors at this point).

 If the spatial distribution in the space-time is \emph{non}-integrable
 (and, as we will see in a moment, this is exactly the case for \emph{our} distribution, here),
 the very concept of the \emph{space} is only well-defined at the level of the \emph{tangent space}
 at \emph{each} point of the space-time, but it is \emph{not} possible to glue (``integrate'')
 the picture together and to form $3$-dimensional spatial sub-manifolds.
 There are\emph{ no} $3$-dimensional spatial domains $\hat D$ in this case.
 So spatial Stokes theorem \emph{for} \emph{$2$-forms} $\hat b$ is of no help in this case,
 since the needed $\hat D$ does not exist.

 The technical tool for deciding whether a general distribution is or is not integrable
 is given by \emph{Frobenius integrability theorem} (see \$19.3 in \cite{fecko2006}).
Its ``vector fields version'' says:
\begin{equation}
       \label{Frobenius}
       \text{integrability}
       \hskip .7cm \Leftrightarrow \hskip .7cm
       \{ \ u,w \ \text{spatial}
       \hskip .2cm \Rightarrow \hskip .2cm
       [u,w] \ \text{also spatial} \ \}
\end{equation}
This may be restated in a form which is more convenient for us.

In order to do that consider $(u,w)$, \emph{two spatial} vectors. So $\tilde V(u) = \tilde V(w) = 0$.
Then, due to Cartan's formula for computation of the exterior derivative $d$ (see 6.2.13 in \cite{fecko2006}),
$$(d\tilde V)(u,w) = u\tilde V(w) - w\tilde V(u) - \tilde V([u,w]) = -\tilde V([u,w])
$$
At the same time, from the decomposition $d\tilde V = \tilde V\wedge \hat a +\hat y$ we have
$(d\tilde V)(u,w) = \hat y(u,w)$. So,
\begin{equation}
       \label{yonuw}
       \tilde V([u,w]) = - \hat y(u,w)
\end{equation}
Combination of Eqs. (\ref{defofspatial}), (\ref{Frobenius}) and (\ref{yonuw})
(and the fact that $u,w$ are \emph{arbitrary spatial} vectors) immediately leads to the
following consequence:
\begin{equation}
       \label{spatialintegrable}
       \hat y = 0
       \hskip 1cm \Leftrightarrow \hskip 1cm
       \text{spatial distribution integrable}
\end{equation}
So, technically speaking, complete information about integrability of spatial distribution resides
in the $2$-form $\hat y$.

Now look at Eqs. (\ref{dtildeVincylindrapprox1}) and (\ref{dtildeVincylindrapprox2}).
We see that for $\omega \neq 0$ (i.e. whenever there is a real rotation), $d\tilde V\neq 0$.
Moreover, contrary to $\hat a$, there is non-vanishing $\hat y$ already at the lowest order
approximation, so in this reference frame (system of observers given by our particular $V$)
the \emph{$3$-space} distribution is \emph{non-integrable}.
\vskip .4cm
\noindent
[This is often expressed differently, namely as that \emph{time-synchronization} is \emph{impossible}
 in this case. Why?

 If the spatial distribution is integrable, i.e. if the slicing in terms of spatial $3$-dimensional
 sub-manifolds exists, we can identify any particular sub-manifold with ``the space''
 and introduce a smooth function, \emph{time}, which is constant at each particular sub-manifold
 and has different values on different sub-manifolds.
 This is, in physical terms, time-synchronization: one can imagine a clock sitting at each point
 of the sub-manifold and all the clocks show the same time.
 So we can (locally) introduce a common value of time at the $3$-space sub-manifold,
 i.e. set up the same reading at each clock in ``the space''.]
\vskip .4cm
So, in our particular case of $V$ corresponding to rotating observers, one \emph{cannot} use spatial
Stokes theorem for $\hat b$ being a $2$-form, since there are
\emph{no spatial} $3$-\emph{dimensional domains} $\hat D$ available.

Hopefully we are more fortunate when we apply spatial Stokes theorem in the case of
\emph{$1$-form} $\hat E$ (i.e. for $2$-dimensional $\hat D$).
Well, let us check.

Consider a \emph{$2$-dimensional spatial} distribution
given by some particular spatial vector fields $(u,w)$,
i.e. the $2$-dimensional subspace of tangent space at each point of space-time,
spanned by the values of the (point-wise linearly independent) spatial fields $u$ and $w$.
In order for the needed spatial $2$-dimensional surface $\hat D$ to exist,
the distribution must be integrable.
Notice that we used the spatial Stokes theorem for the surface ``inside'' the rotating rim
and this surface is ``built'' on the (spatial! - check) pair $u=\partial_r$ and $w= \partial_z$
(the rim is ``in the $rz$-plane'').
Now these two fields \emph{commute}. So the distribution \emph{is} integrable due to
Frobenius criterion.
Therefore \emph{our } application of spatial Stokes theorem
for converting the line integral of the electric field $1$-form $\hat E$ over the rim
to surface integral of $\hat d\hat E$ over the surface ``inside'' the rim \emph{is legal}.

\vskip .4cm
\noindent
[In Appendix \ref{app:rotatingexact} ortho-normal spatial frame
 $(\mathcal E_1, \mathcal E_2, \mathcal E_3)$
 is computed exactly
 (i.e. with no approximations in powers of $\omega$; see Eqs. (\ref{E_0} - \ref{E_3}))
 by ``Lorentz-rotating'' the cylindrical frame $(e_r,e_\varphi,e_z)$:
\begin{equation}
       \label{EEE}
       \mathcal E_1   = \partial_r
       \hskip 1cm
       \mathcal E_2   = \gamma (\omega r \partial_t + (1/r) \partial_\varphi)
       \hskip 1cm
       \mathcal E_3   = \partial_z
\end{equation}
(we recover the original frame when $\omega =0$).

From this expression we see that $2$-dimensional spatial distributions spanned by
$(\mathcal E_1, \mathcal E_3)$ (this corresponds to our ``rim surface'') as well as
$(\mathcal E_2, \mathcal E_3)$ (corresponding to the surface of a cylinder)
are ``safe'' (in both cases the two fields commute)
whereas the distribution spanned by $(\mathcal E_1, \mathcal E_2)$
(say, the ``floor surface'') is ``bad''
(the commutator $[\mathcal E_1,\mathcal E_2]$ cannot be expressed in terms of the two fields themselves;
 actually it is outside the span of the \emph{whole triple} $(\mathcal E_1, \mathcal E_2, \mathcal E_3)$,
 so it is exactly this pair which ``spoils'' integrability of the \emph{complete} $3$-dimensional spatial
 distribution).

In terms of time-synchronization this means the following:

i) It \emph{is} possible within ``rim-like surfaces'' (based on $(\mathcal E_1, \mathcal E_3)$).

ii) It \emph{is} possible within ``cylinders'' (based on $(\mathcal E_2, \mathcal E_3)$).

iii) It \emph{is not} possible for the pair $(\mathcal E_1, \mathcal E_2)$ (there is no such surface).

Notice also, that the term proportional to $\partial_t$ in the explicit expression of
$\mathcal E_2$ (contrary to $\mathcal E_1$ and $\mathcal E_3$) means,
that although the surface mentioned as ii) is regarded as \emph{spatial} by rotating observers
(i.e. it only consists of points with the same value of time), non-rotating observers
(those using $\partial_t$ as $V$)
claim that the surface ``mixes'' points of space-time with different (their) time values.]
\vskip .2cm
And finally, since \emph{any} $1$-dimensional distribution \emph{is} integrable
(just think of Frobenius theorem), there are no problems with spatial Stokes theorem
for $1$-dimensional $\hat D$.

\setcounter{equation}{0}
\subsection{Maxwell equations with respect to general $V$}
\label{maxwellforgeneralV}

As we know from Sec. \ref{maxwelleq}, Maxwell equations read, in $4$-dimensional language,
\begin{equation}
       \label{maxwellin4}
       d*F= -J
           \hskip 1cm
       dF =0
           \hskip 1cm
           \text{\emph{Maxwell equations}}
\end{equation}
With respect to an observer field $V$, each form $\alpha$ decomposes
as $\alpha = \tilde V \wedge \hat s + \hat r$ and also basic operations, $d$ and $*$,
may be expressed in terms of their spatial counterparts.

So, one can

- perform the decomposition of both $F$ and $J$

- apply decomposed versions of $d$ and $*$ on the decomposed forms from above

- decompose the \emph{two} equations (\ref{maxwellin4}) themselves

- get a set of \emph{four spatial} equations in this way

\noindent
(one should equate independently $\hat s$-parts as well as $\hat r$-parts
 of both sides of the equations).

 For general $V$, one gets more complex expressions than those known from (\ref{maxw1}) and (\ref{maxw2})
 in the resulting set of equations.
 Namely (see e.g. \cite{fecko1997}), the following set of spatial {\color{red}\emph{Maxwell equations w.r.t. $V$}}
 drops out:
$$\hskip -2cm
  \text{\emph{inhomogeneous}}
  \hskip 4cm
  \text{\emph{homogeneous}}
$$
\begin{eqnarray} %{rcl}
            \label{maxw1wrtV}
            \hat d \hat * \hat E + \hat y \wedge \hat * \hat B
      &=& \rho \hat \Omega \hskip 1.6cm \hat d \hat E  + \mathcal L_V \hat B - \hat a \wedge \hat E \ = \ 0 \\
            \label{maxw2wrtV}
            \hat d \hat * \hat B -\mathcal L_V \hat * \hat E - \hat a \wedge \hat * \hat B
      &=& \hat J \hskip 3cm  \hat d \hat B - \hat y \wedge \hat E  \ = \ 0
\end{eqnarray}
Here,
$$
\begin{array} {rcl}
      F &=& \tilde V \wedge \hat E - \hat B   \\
      J &=& \tilde V \wedge (-\hat J) +\rho \hat \Omega  \\
 \Omega &=& \tilde V \wedge \hat \Omega
\end{array}
$$
where $\Omega$ is the volume $4$-form, $\hat \Omega$ is the spatial volume $3$-form,
$\hat J$ is the spatial electric current $2$-form and $\rho$ is the electric charge density
($0$-form). All spatial quantities are meant w.r.t. $V$.

Clearly, for $V=\partial_t$ we should recover in this way (and indeed do recover)
 original Maxwell equations
 (\ref{maxw1}) and (\ref{maxw2}):
$$\hskip -2cm
  \text{\emph{inhomogeneous}}
  \hskip 4cm
  \text{\emph{homogeneous}}
$$
\begin{eqnarray} %{rcl}
            \label{maxw1wrtpartialt}
            \Div \bold E
      &=& \rho  \hskip 3cm \curl \bold E +\partial_t \bold B = 0 \\
            \label{maxw2wrtpartialt}
            \curl \bold B -\partial_t \bold E
      &=& \bold j \hskip 4.2cm  \Div \bold B   = 0
\end{eqnarray}
\vskip .5cm
\noindent
[Here,
$$
\begin{array} {rcl}
      F &=& dt \wedge \bold E \cdot d\bold r  - \bold B \cdot d\bold S   \\
      J &=& dt \wedge (- \bold j \cdot d\bold S)  + \rho dV   \\
 \Omega &=& dt \wedge dV
\end{array}
$$
where $dV = dx\wedge dy \wedge dz$ is the usual volume $3$-form (the letter $V$ is not to be confused
with observer field $V$!).]
\vskip .5cm
The terms in Eqs. (\ref{maxw1wrtV}) and (\ref{maxw2wrtV}) containing $\hat a$ or $\hat y$ have
\emph{no counterpart} in good old Maxwell equations. They are ``genuinely new'',
in comparison with (\ref{maxw1wrtpartialt}) and (\ref{maxw2wrtpartialt}),
resulting from possible ``kinematic complications'' connected with particular reference system given
by general $V$.

\subsubsection{Approximate Maxwell equations for our $V$}
\label{maxwellforourV}

In this section we, first, write down Maxwell equations for \emph{our} $V$ corresponding to rotating observers
(well, only \emph{up to first order} approximation in $\omega$), Then, we explicitly check that
\emph{our} $\hat E$ and $\hat B$ indeed \emph{satisfy} the equations.
(This is just to check that ``strange terms'' entering general equations are to be really there.)

In order to make the equations more transparent, let us write each quantity of interest
as a $1$-st order polynomial in $\omega$, i.e. as $(\text{something})_0 +\omega (\text{something})_1$.

For example recall that, up to linear terms in $\omega$, we obtained
\begin{equation}
       \label{hataandhaty}
       \hat a \doteq 0 \equiv \hat a_0 + \omega \hat a_1
           \hskip 1cm
       \hat y \doteq - 2\omega dr \wedge rd\varphi \equiv \hat y_0 + \omega \hat y_1
\end{equation}
(see Eqs. (\ref{dtildeVincylindrapprox1}), (\ref{dtildeVincylindrapprox2})), i.e.
\begin{equation}
       \label{hataandhaty0and1}
       \hat a_0 = \hat a_1 = 0
           \hskip .8cm
       \hat y_0 = 0
           \hskip .8cm
       \hat y_1 = - 2 dr \wedge rd\varphi \equiv - 2 dx\wedge dy
\end{equation}
Next, recall that the field $F=-BdS_x \equiv -Bdy \wedge dz$ from Eq. (\ref{Ffortherim}),
i.e. the static magnetic field, $\bold E = (0,0,0)$ and $\bold B = (0,0,B)$,
satisfies Maxwell equations  (\ref{maxw1}) and (\ref{maxw2}) with \emph{no sources},
$\rho =0, \bold j = \bold 0$. So the $3$-form $J$ vanishes, i.e. $\rho =0=\hat J$ w.r.t.
\emph{any} $V$.

Let's go on to fields $\hat E$ and $\hat B$.

We know from Eq. (\ref{hatEfortherim})
that $\hat E = -\omega Bxdz \equiv \hat E_0+\omega \hat E_1$ with
$\hat E_0 =0$ and $\hat E_1 = -Bxdz$.

Magnetic $2$-form $\hat B$ may be easily computed from $\hat B = \tilde V \wedge \hat E - F$
and the result is
\begin{equation}
       \label{hatBforrim1}
       \hat B = B(dy - \omega xdt) \wedge dz
              \equiv \hat B_0+\omega \hat B_1
\end{equation}
with
\begin{equation}
       \label{hatBforrim2}
       \hat B_0 = Bdy \wedge dz
           \hskip .4cm
       \hat B_1 = -Bxdt \wedge dz
\end{equation}
Finally, the observer field is $V=\partial_t +\omega \partial_\varphi$, i.e.
$V_0=\partial_t$ and $V_1 = \partial_\varphi \equiv -y\partial_x +x\partial_y$.

So, we can summarize all quantities of interest as follows:
\begin{eqnarray} %{rcl}
            \label{hatE01}
            \hat E &=& \hat E_0+\omega \hat E_1
            \hskip .8cm
            \hat E_0 \ = \ 0 \hskip 2.25cm \hat E_1 \ = \ -Bxdz      \\
            \label{hatB01}
            \hat B &=& \hat B_0+\omega \hat B_1
            \hskip .8cm
            \hat B_0 \ = \ Bdy \wedge dz \hskip 1cm \hat B_1 \ = \ dt \wedge \hat E_1      \\
            \label{hatV01}
            \hat V &=& \hat V_0+\omega \hat V_1
            \hskip 1cm
            \hat V_0 \ = \ \partial_t \hskip 2.1cm \hat V_1 \ = \ -y\partial_x +x\partial_y      \\
            \label{hata01}
            \hat a &=& \hat a_0+\omega \hat a_1
            \hskip 1.05cm
            \hat a_0 \ = \ 0 \hskip 2.3cm \hat a_1 \ = \ 0      \\
            \label{haty01}
            \hat y &=& \hat y_0+\omega \hat y_1
            \hskip 1.1cm
            \hat y_0 \ = \ 0 \hskip 2.3cm \hat y_1 \ = \ -2dx \wedge dy
\end{eqnarray}
Now we are to check that these expression do satisfy source-less Maxwell equations
(\ref{maxw1wrtV}) and (\ref{maxw2wrtV})
\begin{eqnarray} %{rcl}
            \label{maxw1wrtVour}
            \hat d \hat * \hat E + \hat y \wedge \hat * \hat B
      &=& 0 \hskip 1.6cm \hat d \hat E  + \mathcal L_V \hat B - \hat a \wedge \hat E \ = \ 0 \\
            \label{maxw2wrtVour}
            \hat d \hat * \hat B -\mathcal L_V \hat * \hat E - \hat a \wedge \hat * \hat B
      &=& 0 \hskip 2.8cm  \hat d \hat B - \hat y \wedge \hat E  \ = \ 0
\end{eqnarray}
up to first order accuracy in $\omega$. Remember that we should use Eqs. (\ref{hatdonspatial})
and (\ref{hat*onspatial}) in order to compute $\hat *$ and $\hat d$.
After a small exercise we come to conclusion that all 4 equations are satisfied.

\section*{Acknowledgement}
\addcontentsline{toc}{section}{Acknowledgement}

         I acknowledge the invitation of Peter Marko\v{s} to write this paper.
         He, after some effort, finally succeeded to convince me that the writing makes sense.
         Please do not blame him for this if you think he was actually wrong.
         %I also acknowledge the willingness of Stanislav Fecko, my son, who created the figures.

\newpage
\section*{Appendices}
\renewcommand{\theequation}{\thesection.\arabic{equation}}
\appendix

\setcounter{equation}{0} \setcounter{figure}{0} \setcounter{table}{0}
\section{\label{app:vectoranalysis}Vector analysis and differential forms}

When our manifold is standard \emph{3-dimensional} \emph{Euclidean} space,
formulas computed and displayed in terms of differential forms may also be computed and expressed
in terms of classical \emph{vector analysis}.
Since a major part of literature on topics treated in this text uses vector analysis
as its principal tool, it is very useful to be able to convert formulas
from one language to another.

Here we collect the most frequent results of this type.
A systematic exposition may be found, e.g., in Sections \$8.5 and \$16.1 of \cite{fecko2006}.

First, since the space is $3$-dimensional, the possible (non-vanishing) forms are $0,1,2$ and $3$-forms.
They may be parametrized in terms of scalar and vector fields (nothing else is needed!) as follows:
\begin{equation}
        \label{possibleformsine3}
        f
        \hskip 1cm
        \bold A \cdot d\bold r
        \hskip 1cm
        \bold A \cdot d\bold S
        \hskip 1cm
        f dV
\end{equation}
There are three basic \emph{linear algebra} operations on forms.

First, let us mention the \emph{exterior product} of forms.
Here, the interesting results are the following:
\begin{eqnarray} %{rcl}
      \label{crossproduct}
                 (\bold A \cdot d\bold r) \wedge  (\bold B \cdot d\bold r)
             &=& (\bold A \times \bold B) \cdot d\bold S
                                       \hskip 1.2cm \text{cross product}                        \\
      \label{dotproduct}
                 (\bold A \cdot d\bold r) \wedge  (\bold B \cdot d\bold S)
             &=& (\bold A \cdot \bold B) dV
                                       \hskip 1.6cm \text{dot product}
\end{eqnarray}
Then, there is the \emph{interior product} of a vector field and a form.
Non-trivial results:
\begin{eqnarray} %{rcl}
      \label{dotproductagain}
                 i_{\bold A} (\bold B \cdot d\bold r)
             &=& \bold A \cdot \bold B
                                       \hskip 1.6cm \text{dot product again}                        \\
      \label{crossproductagain}
                 i_{\bold A} (\bold B \cdot d\bold S)
             &=& (\bold B \times \bold A) \cdot d\bold r
                                       \hskip .6cm \text{cross product again}                        \\
      \label{multiplication}
                 i_{\bold A} (fdV)
             &=& f\bold A \cdot d\bold S
                                       \hskip 1.3cm \text{just a multiplication}
\end{eqnarray}
And finally, there is the \emph{Hodge star} operator. Here, it just interchanges
$$d\bold r \leftrightarrow d\bold S
  \hskip 1cm \text{and} \hskip 1cm 1 \leftrightarrow dV
$$
(so, when applied twice, it just gives the identity), i.e., in full,
\begin{equation}
        \label{hodgestar}
        *f = fdV
        \hskip .5cm
        *(\bold A \cdot d\bold r) = \bold A \cdot d\bold S
        \hskip .5cm
        *(\bold A \cdot d\bold S) = \bold A \cdot d\bold r
        \hskip .5cm
        *(f dV)=f
\end{equation}
Now, we come to main \emph{differential} operator on forms,
the \emph{exterior derivative} $d$. Here, in $E^3$, it produces the three well-known
differential operations from vector analysis:
\begin{eqnarray} %{rcl}
      \label{gradient}
             df                    &=& \boldsymbol \nabla f \cdot d\bold r
                                       \hskip 1.6cm \text{gradient}                        \\
      \label{curl}
        d (\bold A \cdot d\bold r) &=& (\boldsymbol \nabla \times \bold A) \cdot d\bold S
                                       \hskip .7cm \text{curl}                             \\
      \label{divergence}
        d (\bold A \cdot d\bold S) &=& (\boldsymbol \nabla \cdot \bold A) dV
                                       \hskip 1.1cm \text{divergence}                      \\
      \label{nothing}
        d (fdV)                    &=& 0
\end{eqnarray}
From the general property $dd=0$ we immediately get the two notorious identities
\begin{equation}
        \label{divrotarotgradje0}
        \text{div rot} = 0
        \hskip 1.5cm
        \text{rot grad} = 0
\end{equation}
If the \emph{Lie derivative} of forms is needed, we can remember the general Cartan's formula
\begin{equation}
        \label{cartanlvjeivdplusdiv}
        \mathcal L_{\bold A} = i_{\bold A}d+di_{\bold A}
\end{equation}
and use the formulas given above. In this way, we get
\begin{eqnarray} %{rcl}
      \label{LAondeg0}
             \mathcal L_{\bold A}f                    &=&
                      \bold A \cdot \boldsymbol \nabla f                                     \\
      \label{LAondeg1}
        \mathcal L_{\bold A} (\bold B \cdot d\bold r) &=&
                    ((\text{curl} \ \bold B )\times \bold A
                   +\boldsymbol \nabla (\bold A \cdot \bold B)) \cdot d\bold r               \\
      \label{LAondeg2}
        \mathcal L_{\bold A} (\bold B \cdot d\bold S) &=&
        ((\text{div} \bold B) \ \bold A + \text{curl} \ (\bold B \times \bold A)) \cdot d\bold S  \\
      \label{LAondeg3}
        \mathcal L_{\bold A} (fdV)                    &=& \text{div}(f\bold A)dV
\end{eqnarray}
And, finally, in \emph{integral} calculus of forms, the key result is given by (general)
\emph{Stokes theorem}
\begin{equation}
        \label{generalStokes}
        \int_{D} d\alpha = \int_{\partial D} \alpha
\end{equation}
Since there are three relevant realizations of $d$ here, Eqs. (\ref{gradient}), (\ref{curl})
and (\ref{divergence}), we get as many as three particular versions of the theorem in $E^3$,
namely
\begin{eqnarray} %{rcl}
      \label{gradienttheorem}
             \int_C\boldsymbol \nabla f \cdot d\bold r               &=&  f(B) - f(A)
                                       \hskip .8cm \text{gradient theorem}                        \\
      \label{stokestheorem}
        \int_S (\text{curl} \ \bold A) \cdot d\bold S &=& \oint_{\partial S} \bold A \cdot d\bold r
                                       \hskip 1.2cm \text{Stokes theorem}                             \\
      \label{gausstheorem}
        \int_V (\text{div} \ \bold A) dV &=& \oint_{\partial V} \bold A \cdot d\bold S
                                       \hskip 1.1cm \text{Gauss theorem}
\end{eqnarray}
Here $C$ is a path between points $A$ and $B$, $S$ is a surface with boundary $\partial S$ and
$V$ is a volume with boundary $\partial V$.

If we switch to the \emph{extended} space $E^3 \times \mathbb R$,
differential forms may be decomposed (see Sec. \ref{formsonmtimesr}) as
\begin{equation}
      \alpha = dt \wedge \hat s + \hat r
      \label{decomposition1app}
\end{equation}
Here spatial forms $\hat s$ and $\hat r$ are \emph{time-dependent} versions of expressions
displayed in Eq. (\ref{possibleformsine3}). So, possible $0,1,2,3$ and $4$-forms look as follows:
\begin{equation}
        \label{possibleformsine3r}
        f
        \hskip .6cm
        fdt + \bold a \cdot d\bold r
        \hskip .6cm
        dt \wedge \bold a \cdot d\bold r + \bold b \cdot d\bold S
        \hskip .6cm
        dt\wedge \bold a \cdot d\bold S + f dV
        \hskip .6cm
        fdt\wedge dV
\end{equation}
with both scalar and vector fields $f,\bold a, \bold b$ depending, in general, on $(\bold r,t)$.

\setcounter{equation}{0}
\section{\label{app:cannotend}Why vortex filament cannot end in a fluid}

According to second Helmholtz theorem (see Section \ref{euler_stationary_helmholtz_related}),
a \emph{vortex filament} (vortex tube with very small cross-section area) cannot end in a fluid.
Why this happens?

Actually, similar statement is well known to be true for \emph{magnetic} filaments, so our discussion
covers both situations.

Let's start with observation that both vector fields
(vorticity field $\boldsymbol \omega$ as well as magnetic field $\bold B$)
are \emph{divergence-free} (or \emph{solenoidal}):
\begin{eqnarray} %{rcl}
      \label{vorticitydivfree}
            0  &=&  dd(\bold v \cdot d\bold r)
                =  d(\boldsymbol \omega \cdot d\bold S)
                =  (\Div \boldsymbol \omega) dV
                \hskip .7cm \text{so} \hskip .7cm
                \boxed{\Div \boldsymbol \omega = 0}\\
      \label{magneticdivfree}
            0  &=&  dd(\bold A \cdot d\bold r)
                =  d(\bold B \cdot d\bold S)
                =  (\Div \bold B) dV
                \hskip .6cm \text{so} \hskip .65cm
                \boxed{\Div \bold B = 0}
\end{eqnarray}
\begin{figure}[tb]
\begin{center}
\includegraphics[clip,scale=0.30]{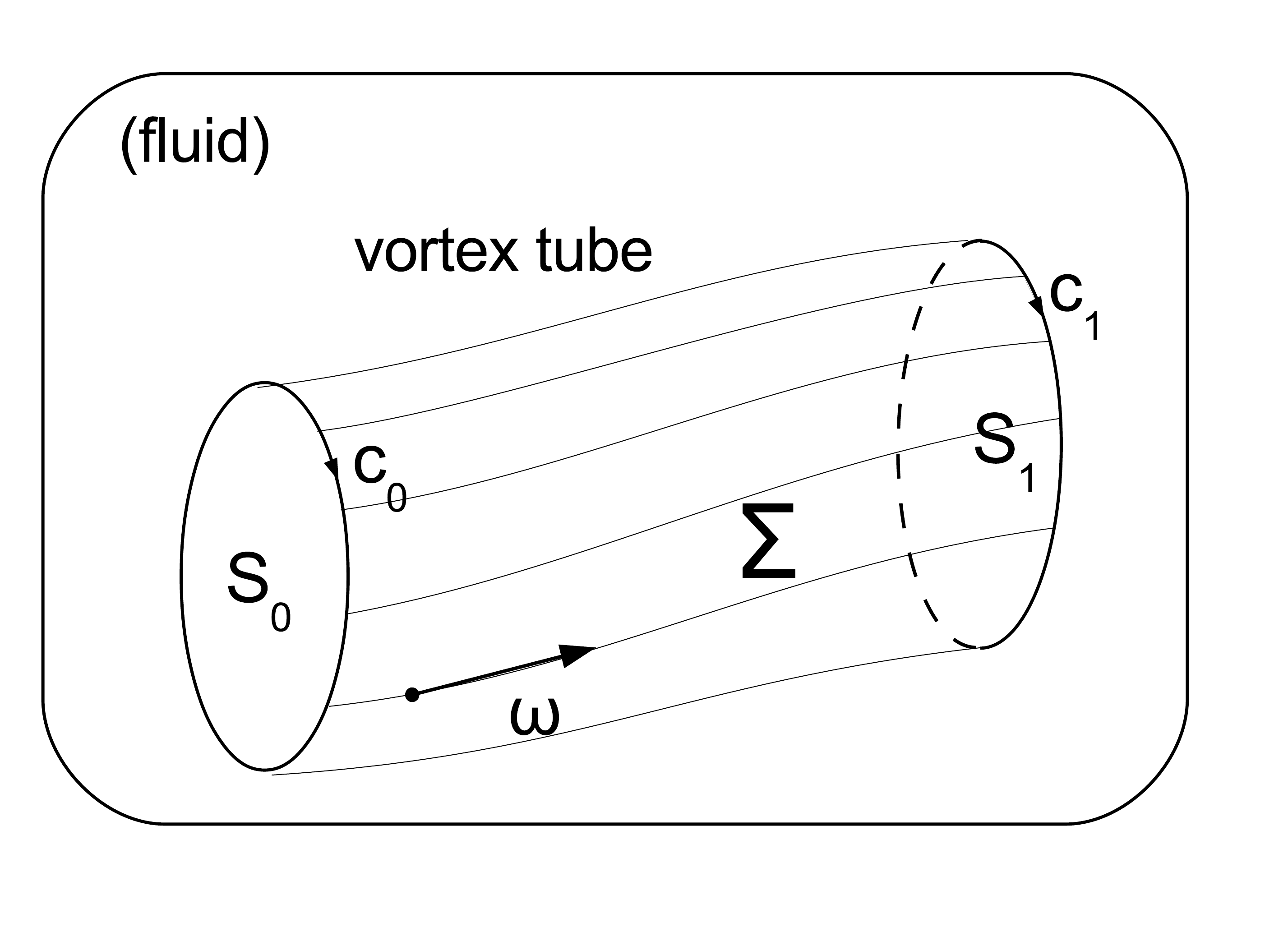}
\caption{A solid vortex tube $V$ (the tube \emph{inside}) is made of vortex lines emanating from the left cap $S_1$
         and entering the right cap $S_2$.
         Boundary $\partial V$ of the solid cylinder $V$ consists of 3 parts, $\partial V = \Sigma +S_1-S_0$,
         hollow cylinder $\Sigma$ (``side'' of the solid cylinder) and the two caps, $S_0$ and $S_1$.
         We have $0=\partial \partial V = \partial \Sigma +\partial S_1- \partial S_0$,
         so $\partial \Sigma = \partial S_0- \partial S_1 = c_0 - c_1$.
         The cycles $c_0$ and $c_1$ encircle the same tube of vortex lines.}
\label{fig:vortextube2}
\end{center}
\end{figure}
Now, consider $V$, the \emph{solid} vortex tube. By this we mean the $3$-dimensional domain enclosed
by the \emph{hollow} $2$-dimensional vortex tube $\Sigma$ and two $2$-dimensional ``cross section''
surfaces (left cap and right cap) $S_0$ and $S_1$, see Fig \ref{fig:vortextube2}.
So $\partial V = \Sigma + S_1 - S_0$ (and also $0=\partial \partial V = \partial \Sigma + c_1 - c_0$).
Then
\begin{equation} \label{solidtubeagain}
                 0 = \int_{V}dd(\bold v \cdot d\bold r)
                   = \int_{\partial V}\boldsymbol \omega \cdot d\bold S
                   = \int_{\Sigma}\boldsymbol \omega \cdot d\bold S
                   + \int_{S_1}\boldsymbol \omega \cdot d\bold S
                   - \int_{S_0}\boldsymbol \omega \cdot d\bold S
\end{equation}
But the integral over $\Sigma$ vanishes due to the argument mentioned in Eq. (\ref{ofthestructure}),
since vector $\boldsymbol \omega$ (which is, by construction, tangent to $\Sigma$) annihilates
the form under integral sign:
$$i_{\boldsymbol \omega} (\boldsymbol \omega \cdot d\bold S)
                        = (\boldsymbol \omega \times \boldsymbol \omega) \cdot d\bold r
                        = 0
$$
and we get
\begin{equation} \label{vortexflux}
        \oint_{S_0}\boldsymbol \omega \cdot d\bold S  = \oint_{S_1}\boldsymbol \omega \cdot d\bold S
\end{equation}
if $S_0$ and $S_1$ enclose a common \emph{solid vortex tube}.
So the \emph{vortex flux} is constant along vortex tube.
(We already know this from Eq. (\ref{indepofsection}).
 Recall that the constant is known as the\emph{ strength} of the tube.)

In the same way we get, clearly, the corresponding ``magnetic'' statement:
\begin{equation} \label{magneticflux}
        \oint_{S_0}\bold B \cdot d\bold S  = \oint_{S_1}\bold B \cdot d\bold S
\end{equation}
if $S_0$ and $S_1$ enclose a common \emph{solid magnetic tube}.
So the \emph{magnetic flux} is constant along magnetic tube.

Notice that the only property of the vector fields $\boldsymbol \omega$ and $\bold B$
which we needed was that both of them are \emph{divergence-free}.

The following intuitive rough picture might be instructive to understand the two equations.
Let the cross-section surface $S_0$ be perpendicular to the tube and let $\mathcal S_0$ be its area.
Then the l.h.s. of Eq. (\ref{magneticflux}) may be written as $\mathcal S_0B_0$, where
$B_0$ is the \emph{average} value of the component of $\bold B$ along the tube.
Then Eq. (\ref{magneticflux}) may be rewritten as
\begin{equation} \label{magneticflux2}
       \mathcal S_0B_0  = \mathcal S_1B_1
\end{equation}
So, the narrower is the tube, the greater average value of the corresponding component
of the magnetic field is necessary to keep the constant value of the flux.

The same holds, of course, for the vorticity case:
\begin{equation} \label{vortexflux2}
       \mathcal S_0\omega_0  = \mathcal S_1\omega_1
\end{equation}
The narrower is the tube, the greater average value of the corresponding component
of the vorticity vector is necessary to keep the constant value of the flux
(the strength of the tube).

And now the argument may go as follows.

Imagine a vortex tube. It is characterized by its strength.
It means that at \emph{any} cross-section surface $S$ we have \emph{nonzero} average value of $|\boldsymbol \omega|$.
So, we \emph{can} expand the tube \emph{further}. It does not end at (any) cross-section $S$.
That's all.

Notice, however, that the argument says \emph{nothing} about individual vortex \emph{lines}.
The statement holds for tubes or filaments but it does not exclude the possibility that
there is an ``odd fish'', exceptional line which, flagrantly ignoring the folklore wisdom,
\emph{does end} in the fluid.
\vskip .5cm
\noindent
[The value of the resulting flux is not influenced by the contribution from an individual line.
 So if some misbehaving line decides to end in the fluid,
 the flux of the whole tube still remains the same thanks to
 enough lines with high standards of decorum in the neighborhood of the misbehaving one.
 Consequently, the rest of the tube does not end there.]
\vskip .5cm
The ``odd fish'' mentioned above is not just a possibility.
One can easily find a smooth divergence-free vector field whose (exceptional) lines
\emph{really do end} within the region where the field is ``perfectly ok''.
(See an explicit example in Appendix \ref{app:linesandtubes}.)

Is this phenomenon known in literature? Yes, it is.

What Helmholtz writes in his original paper \cite{helmholtz1858} is:

``... ein Wirbelfaden nirgends innerhalb der Fl\"{u}{\ss}igkeit aufh\"{o}ren d\"{u}rfe ...''

And in Sec. I.10 of \cite{truesdell} the following warning may be found (p.17):

``This statement becomes incorrect if ``vortex-line'' rather than ``vortex-filament'' be substituted
for ``Wirbelfaden''''.

          Truesdell
\footnote{I thank Peter Guba for turning my attention to this valuable source.}
          refers to Chapter 2, \$6 of the book \cite{kellogg1929}. There, we can read:
\newline \indent
 ``Since Newtonian fields are solenoidal in free space, ceasing to be so
only at points where masses are situated, it is customary to say that
\emph{lines of force originate and terminate only at points of the acting masses}.
But this should be understood in terms of tubes of force. For an individual
line may fail to keep its continuity of direction, and even its identity
throughout free space. As X, Y and Z are continuous, this may
happen only when they vanish simultaneously, that is, at a point of
equilibrium.''
\vskip .3cm
\noindent
[Kellog also mentions the following example: Newtonian field of gravity of two equal point masses.
 It has an equilibrium (i.e. it vanishes) in the mid-point between the two masses and it is
 clear that two distinct lines \emph{start} in this mid-point going in opposite directions,
 each one to its own point mass. So, in the mid-point, some field lines start in spite of the
 fact that there is \emph{no mass} sitting there.]
\vskip .3cm

In spite of this, overwhelming majority of textbooks on electromagnetism or fluid dynamics,
including the best ones, ignore subtleties of this sort and they explicitly speak of \emph{lines},
both in magnetic and in vortex context.

 For example, we can read in \$ 13-4 of \cite{feynman1964} the following sentences:

 ``There are no magnetic charges from which lines of $\bold B$ can emerge.
 If we think in terms of ``lines'' of the vector field $\bold B$, they can never start and they never stop.''

 Variation on the same theme in \$ 5.3.4 of \cite{griffiths1999}:

``Electric field lines originate on positive charges and terminate on negative ones;
  magnetic field lines do not begin or end anywhere - to do so would require a nonzero divergence.
  They either form closed loops or extend out to infinity. ... This is the physical content
  of the statement $\boldsymbol \nabla \cdot \bold B=0$.''

 Concerning vorticity, in \$ 40-5 of the same \cite{feynman1964} we read:

 ``So vortex lines are like lines of $\bold B$ - they never start or stop, and will tend to go in closed loops.''

\setcounter{equation}{0}
\section{\label{app:linesandtubes}Lines and tubes of solenoidal fields - an instructive example}

In Appendix \ref{app:cannotend} we mentioned that one can easily write down
a smooth and \emph{divergence-free} vector field such that it possesses lines
which end (and/or start) somewhere (not in infinity).

Well, consider the ``magnetic'' field expressed, in cartesian coordinates $(x,y,z)$, as follows:
\begin{equation} \label{thefield}
       \bold B  = x\partial_x - y\partial_y
                  \hskip 1cm \text{i.e.} \hskip 1cm
                   \bold B = (x,-y,0)
\end{equation}
It can be generated from vector potential
\begin{equation} \label{thevectorpotential}
       \bold A  = xy\partial_z
                  \hskip 1cm \text{i.e.} \hskip 1cm
                   \bold A = (0,0,xy)
\end{equation}
In terms of differential forms: we have the 1-form
\begin{equation} \label{oneformA}
       \tilde A = \bold A \cdot d\bold r = xydz
\end{equation}
and its exterior derivative, the 2-form
\begin{equation} \label{twoformB}
       \tilde B = d\tilde A = d(\bold A \cdot d\bold r)
                = \bold B \cdot d\bold S
                = xdy\wedge dz +ydx\wedge dz
                = xdS_x - ydS_y
\end{equation}
Therfore
\begin{equation} \label{ddAiszerosodivB0}
       (\Div \bold B) dV = d(\bold B \cdot d\bold S) = d\tilde B = dd\tilde A = 0
\end{equation}

\begin{figure} %[h]
\begin{center}
\scalebox{0.15}{\includegraphics[clip,scale=1.10]{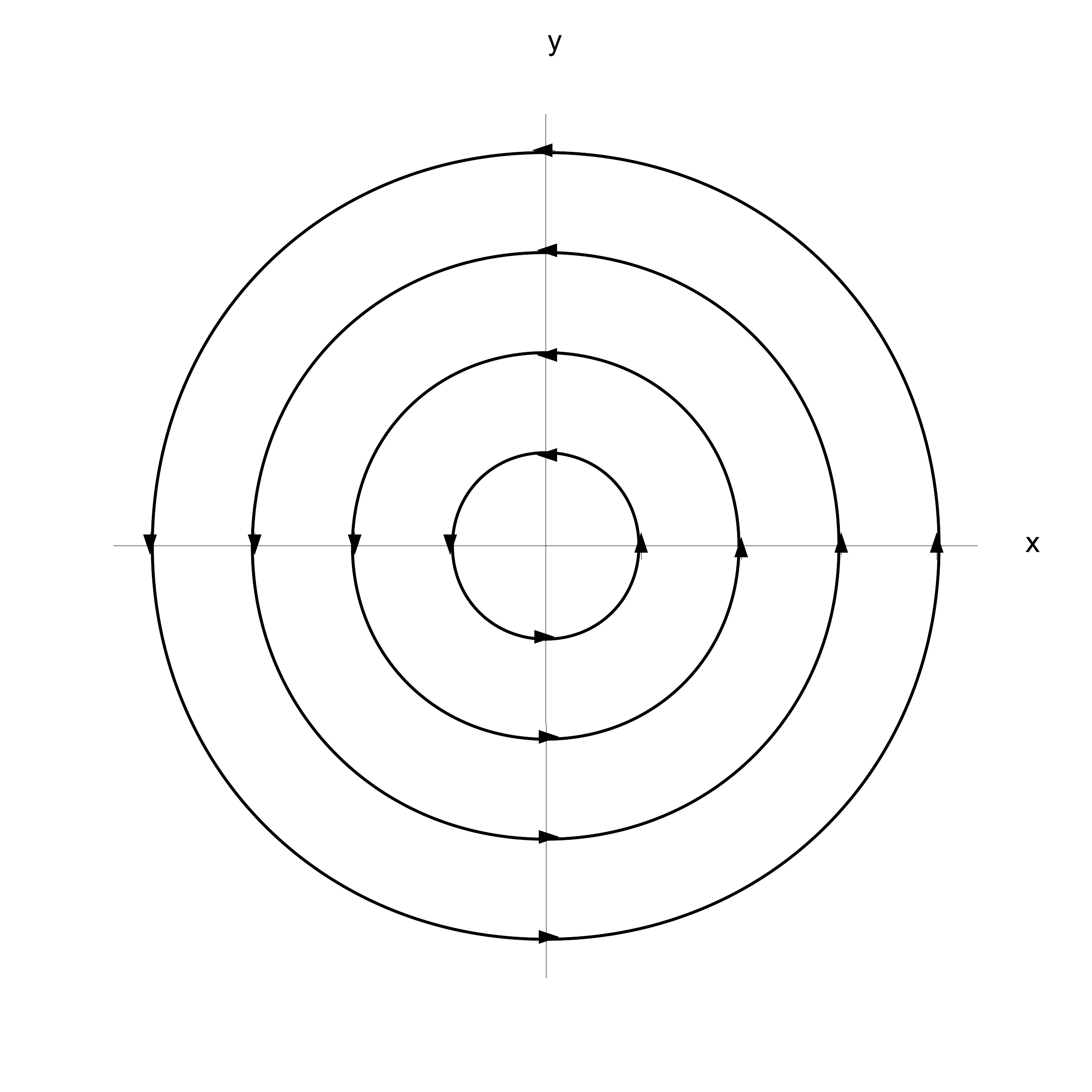}}
\scalebox{0.15}{\includegraphics[clip,scale=1.10]{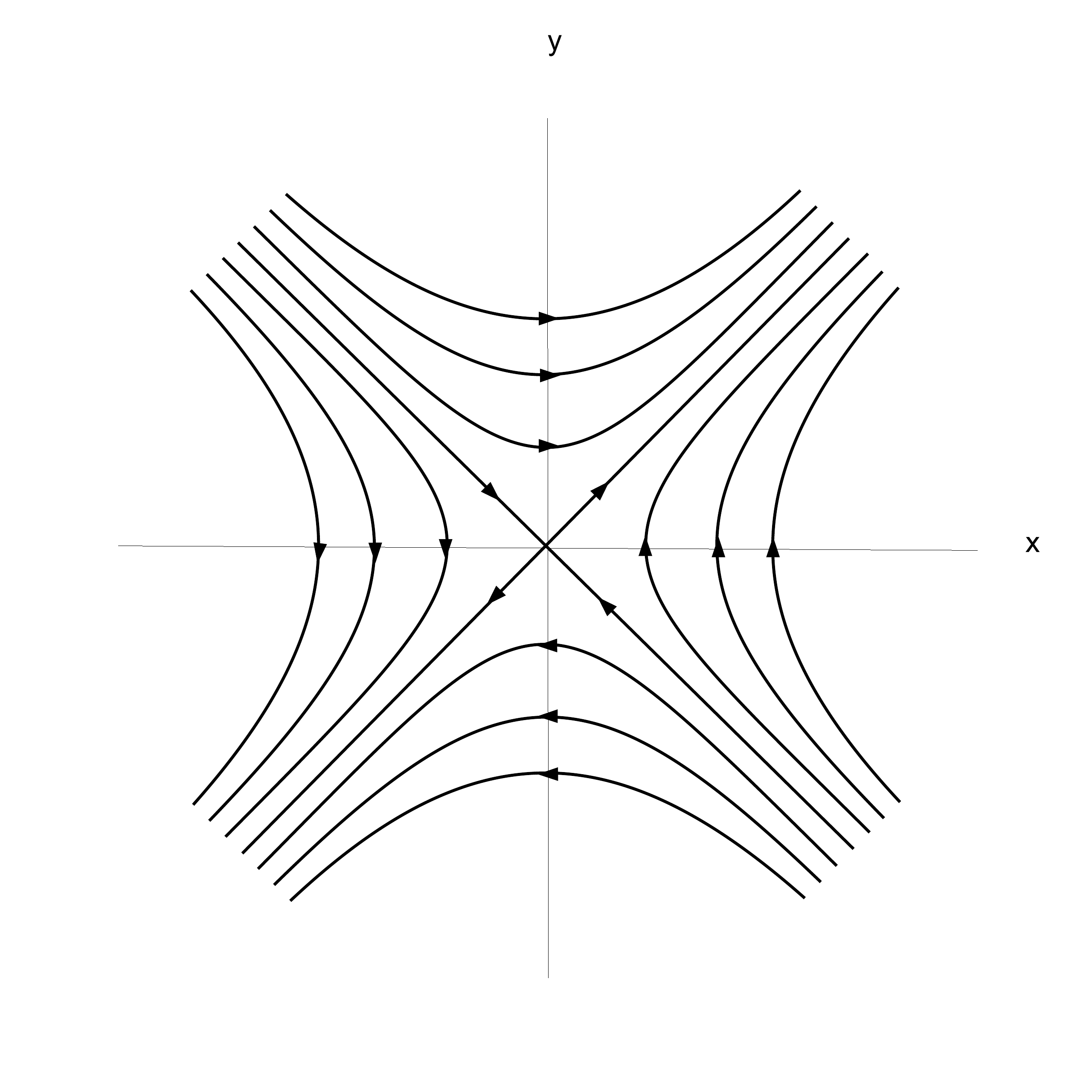}}
\scalebox{0.15}{\includegraphics[clip,scale=1.10]{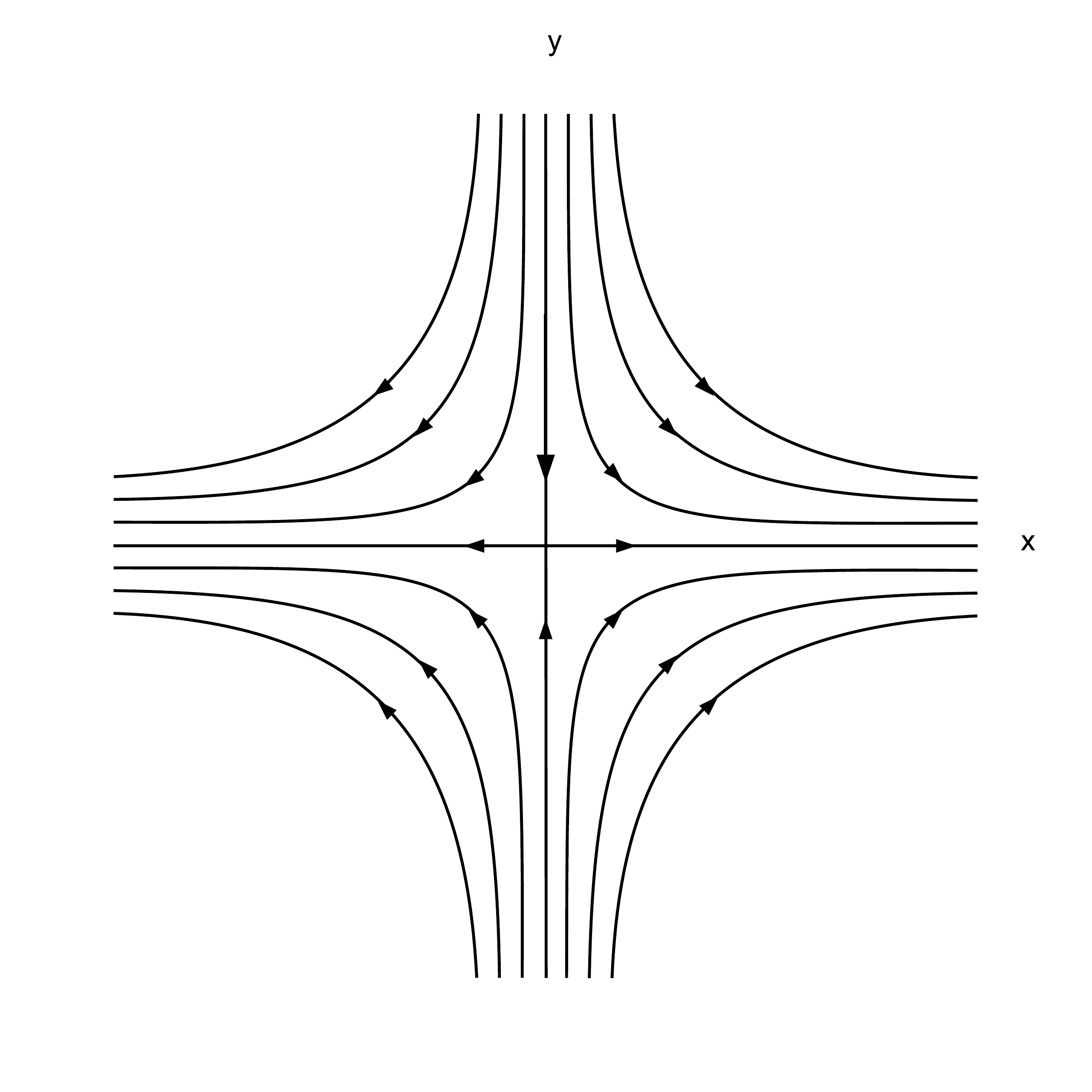}}
\caption{Three examples of lines of solenoidal (divergence-free) vector fields in the plane.
                                     Left: lines of the field $y\partial_x - x\partial_y$; it generates clock-wise \emph{rotations}.
                                     Middle: lines of the field $y\partial_x + x\partial_y$; this field generates \emph{hyperbolic} rotations
                                             (pseudo-rotations; also \emph{Lorentz transformations}, when axes are labeled $(t,x)$).
                                     Right: lines of the field $x\partial_x - y\partial_y$; this may be regarded as the
                                             $\pi/4$-rotated version of the previous example.}
\label{fig:threefields}
\end{center}
\end{figure}

Clearly, our $\bold B$ is well-defined smooth and divergence-free vector field
in whole space $E^3$, it has no singularity anywhere.
Notice, however, that each point of the $z$-axis
(and no other points) serves as \emph{equilibrium} (i.e. the field vanishes there).

Since the $z$-component of the field $\bold B$ vanishes and $x$ and $y$ components do not depend
on $z$ as well, the field lines lie in planes parallel to the $xy$-plane
and, moreover, repeat the pattern in each such plane.
So, everything interesting actually ``happens'' in any ``representative'' plane of this type,
say, in the $xy$-plane (the plane $z=0$) itself.
          Therefore we restrict our discussion to the 2-dimensional (planar) vector field
\footnote{I thank Pavol \v{S}evera for turning my attention to this particular vector field.}
\begin{equation} \label{thefieldplanar}
       \bold B  = x\partial_x - y\partial_y
                  \hskip 1cm \text{i.e.} \hskip 1cm
                   \bold B = (x,-y)
\end{equation}
So, we have just a single equilibrium, here, the origin of the coordinate system, $P=(0,0)$.
The flow induced by the field reads
\begin{equation} \label{thefieldplanarflow}
       x\partial_x - y\partial_y
       \hskip 1cm \leftrightarrow \hskip 1cm
       (x,y) \mapsto (e^tx,e^{-t}y)
\end{equation}
So, as we also see from the right drawing in Fig. \ref{fig:threefields},
there are

- \emph{two lines ending} at the origin (along $\pm$ of the $y$-axis)

- \emph{two lines starting} at the origin (along $\pm$ of the $x$-axis)

(There are also similar four lines on the central drawing in Fig. \ref{fig:threefields}.)

These four lines are examples for the ``odd fish'' behavior. All other lines are
``well-educated lines of a divergence-free field'',
namely they start at infinity and also end at infinity
(so, no line is an example of closed loop, here; notice, however, that \emph{all} lines form loops
 in the case of the \emph{left} drawing in Fig. \ref{fig:threefields}).

Notice also that all lines may be written (when ignoring their orientation) as $xy=$ const.,
so they form \emph{hyperbolas} (therefore ``hyperbolic rotations''). For the constant being zero
the hypobolas \emph{degenerate} to \emph{lines} and these degenerate hyperbolas are of interest for us.

\begin{figure} %[h]
\begin{center}
\hskip -1.60cm
\scalebox{0.15}{\includegraphics[clip,scale=1.70]{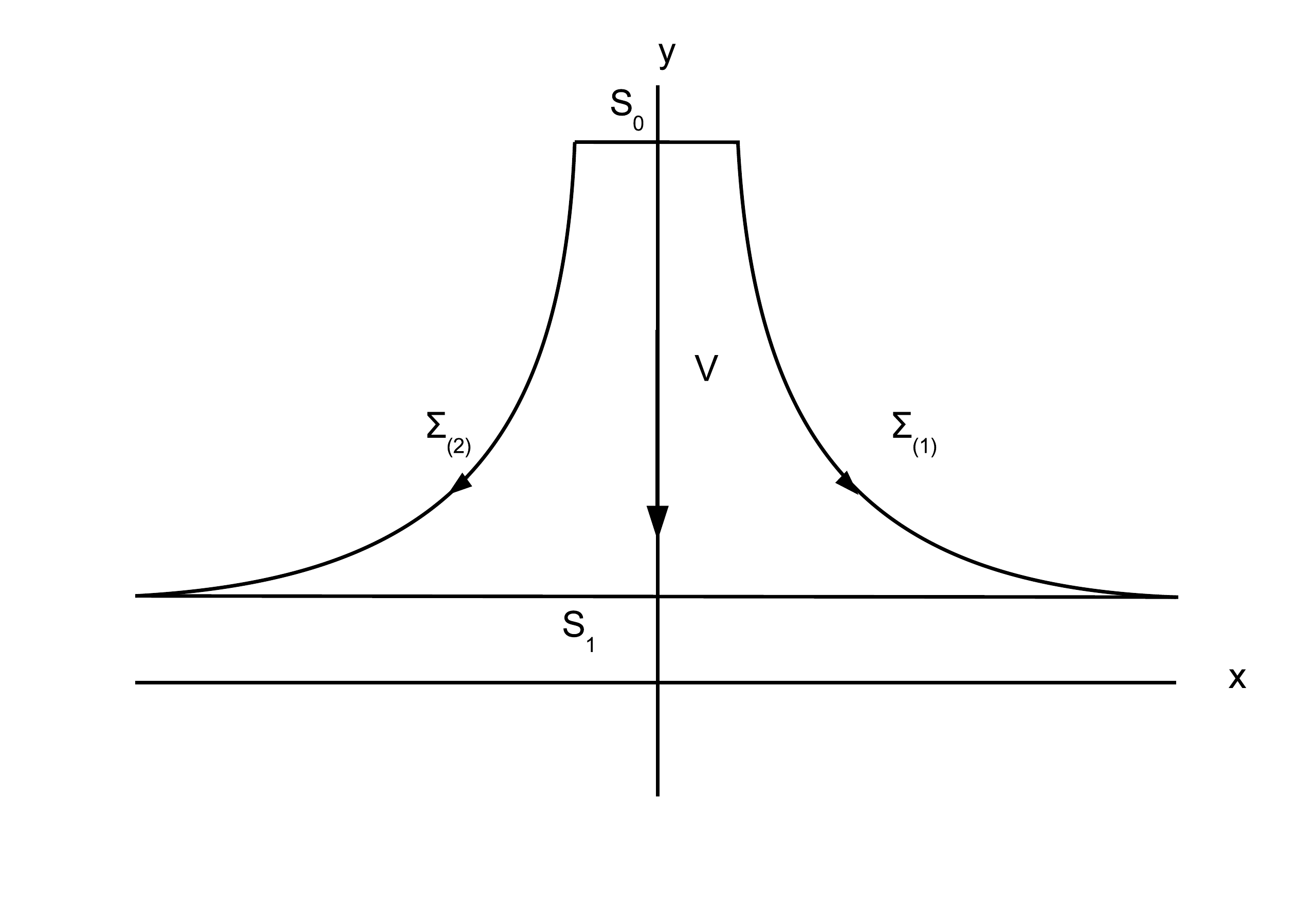}}
\scalebox{0.15}{\includegraphics[clip,scale=1.70]{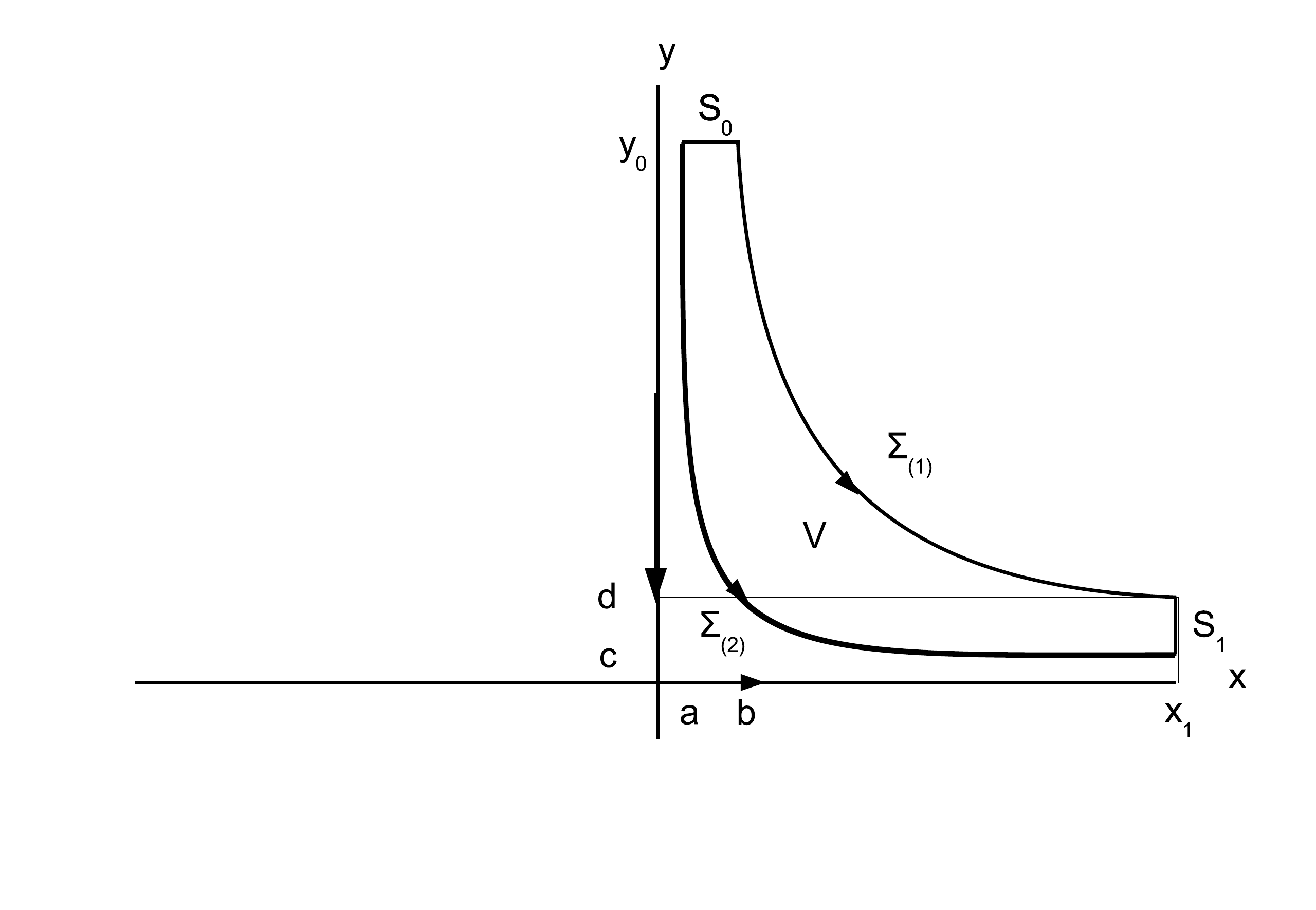}}
\caption{Two examples of solid tubes corresponding to the field $x\partial_x - y\partial_y$ in the plane
         (see the right drawing on Fig. \ref{fig:threefields}).
         In both cases, the sections (``caps'') $S_0$ and $S_1$ as well as the ``hollow tube'' $\Sigma$ reduce to lines.
         $\Sigma$ is not connected, it is made of two pieces, $\Sigma_{(1)}$ and $\Sigma_{(2)}$.}
\label{fig:tubes2d}
\end{center}
\end{figure}

\vskip .3cm
\noindent
[The two lines ending at the origin are given by
 $$ (0,y_0) \mapsto (0,e^{-t}y_0)
 $$
 for $y_0$ positive and negative. The two lines starting at the origin are given by
 $$ (x_0,0) \mapsto (e^tx_0,0)
 $$
 for $x_0 = \epsilon$ very small, positive and negative. So, they do not literally start at the origin
 (the origin flows as $(0,0) \mapsto (e^t 0,e^{-t} 0) \equiv (0,0)$, so it does not move at all),
 but rather they ``start'' at a point ``very close'' to origin. More meaningful formulation is,
 that the \emph{reversed} line (the line with reversed orientation) \emph{ends} in origin.]
\vskip .3cm
Now let us go on to \emph{solid tubes}. On Fig. \ref{fig:tubes2d} two possible (solid) tubes are displayed.
Notice that a solid tube is only 2-dimensional object, here, so its boundary, the \emph{hollow} tube
$\Sigma$ and the two ``caps'' $S_0$ and $S_1$, are 1-dimensional
(the hollow tube is even not connected, here, it is made from two side lines).
Recall the general ($n$-dimensional) \emph{Gauss} formula
\begin{equation} \label{generalGauss}
       \int_V (\Div \bold B)dV = \oint_{\partial V} \bold B \cdot d\bold S
                          \equiv \oint_{\partial V} B^id\Sigma_i
\end{equation}
where
\begin{equation} \label{generalGaussagain}
      dV = \sqrt{|g|}dx^1\wedge \dots \wedge dx^n
       \hskip 1cm
       d\Sigma_i = \frac{1}{(n-1)!}\sqrt{|g|}\epsilon_{ij\dots k} dx^j\wedge \dots \wedge dx^k
\end{equation}
(see 8.2.7 in \cite{fecko2006}). We used it, for $n=3$, in Eq. (\ref{solidtubeagain}).
Here, for $n=2$, it reduces to
\begin{equation} \label{Gaussfor2d}
       \int_V (B^x_{ \ ,x}+B^y_{ \ ,y})dx dy = \oint_{\partial V} B^xdy - B^ydx
\end{equation}
(i.e., actually, to \emph{Green's} formula). So the vanishing ``surface'' integral
over $\Sigma +S_1-S_0$ in \ref{solidtubeagain} becomes, here,
\begin{equation} \label{surfaceisline1}
       0= \oint_{\partial V} xdy + ydx
        = \int_{\Sigma} xdy + ydx
        + \int_{S_1} xdy + ydx
        - \int_{S_0} xdy + ydx
\end{equation}
and, finally (since integral over $\Sigma$ vanishes), we get
\begin{equation} \label{surfaceisline2}
        \int_{S_0} xdy + ydx  = \int_{S_1} xdy + ydx
\end{equation}
\noindent $\blacktriangledown \hskip 0.5cm$
From Eq. (\ref{generalGaussagain}) we get, here, $dV=dx dy$ and
$$\bold B \cdot d\bold S = B^i d\Sigma_i = B^i\epsilon_{ij}dx^j
                         = B^1dx^2 - B^2dx^1
                         = B^xdy - B^ydx
$$
For our $\bold B$, Eq. (\ref{thefieldplanar}), we have then
$$
\begin{array} {rcl}
  i_{\bold B} (\bold B \cdot d\bold S)
  &=& i_{(x\partial_x - y\partial_y)} (xdy + ydx)  \\
  &=& xy - xy                           \\
  &=& 0
\end{array}
$$
and therefore the integral over $\Sigma$ vanishes.
The two cross-sections $S_0$ and $S_1$ are arbitrary lines crossing the (solid) tube.
\hfill $\blacktriangle$ \par
\vskip .2cm \indent
Let us check Eq. (\ref{surfaceisline2}). Take the tube from the right picture displayed at Fig. \ref{fig:tubes2d}.
Then $S_0$ is the horizontal line which may be parametrized as follows:
\begin{equation} \label{S0parametrize}
        S_0: s\mapsto (x(s),y(s)) = (s,y_0)
              \hskip 1cm s\in \langle a,b \rangle
\end{equation}
Similarly, $S_1$ is the vertical line
\begin{equation} \label{S1parametrize}
        S_1: s\mapsto (x(s),y(s)) = (x_1,s)
              \hskip 1cm s\in \langle c,d \rangle
\end{equation}
In order that we speak of a fixed solid tube, however, the parameters are to be related
as follows:
\begin{equation} \label{arerelated}
               cx_1 = a y_0
               \hskip 1cm
               dx_1 = b y_0
\end{equation}
Now
$$
\begin{array} {rcl}
        \int_{S_0} \bold B \cdot d\bold S
       &\equiv& \int_{S_0} (xdy + ydx) =  \int_a^b(sdy_0+y_0ds)
                                       = y_0\int_a^bds = y_0(b-a)  \\
        \int_{S_1} \bold B \cdot d\bold S
       &\equiv& \int_{S_1} (xdy + ydx)
                                       =  \int_c^d(x_1ds+sdx_1)
                                       = x_1 \int_c^dds = x_1(d-c)
\end{array}
$$
Because of Eq. (\ref{arerelated}), the two fluxes are equal.

When $a= 0$ (and also $c = 0$) and $b$ (so that also $d$) is small, we get what we called
a \emph{filament}. Its entering flux is $y_0b$.
This is the limit ($\epsilon \to 0$) of the flux of ``a bit narrower'' filament
with flux $y_0(b-\epsilon)$. The whole narrower filament turns to the right (from our perspective)
and results in the final flux $x_1(d-\epsilon y_0/x_1) = x_1d-\epsilon y_0$.
Its limit is $x_1d$.

So nothing bad happens, from the point of view of fluxes,
after inclusion of our strange line (ending at origin)
as the ``right bank of the river'', it has no influence on the resulting flux;
only the lines turning
to the right count for the value $x_1d$ (we have $y_0b = x_1d$, now).

\setcounter{equation}{0}
\section{\label{app:rotatingexact}Kinematics of rotating observers - exact expressions}

In Section \ref{rotframe} we mostly use \emph{approximate} expressions for $V$, $\tilde V$ etc.
Here we show exact results.

First, normalizing $W$ given in Eq. (\ref{fieldWcylindr}) we get (see (\ref{fieldVfromWrot}))
\begin{eqnarray} %{rcl}
      \label{Vincylindr}
                    V
             &=& \gamma (\partial_t +\omega \partial_\varphi)
                        \\
      \label{tildeVincylindrex}
             \tilde V &=& \gamma (dt -\omega r^2 d\varphi)
\end{eqnarray}
where
\begin{equation}
       \label{gammadependentonr}
             \gamma (r) := \frac{1}{\sqrt{1-(\omega r)^2}}
\end{equation}
Direct computation (using Eqs. (\ref{defhatsV}) and (\ref{defhatrV})) then leads to
\begin{eqnarray} %{rcl}
      \label{dtildeVincylindr}
                 d\tilde V
             = \tilde V \wedge \hat a + \hat y \hskip 1cm \hat a &=& \gamma^2 \omega^2 rdr     \\
      \label{dtildeVincartes}
                 \hat y &=& -2\gamma^3(\omega r) dr \wedge (d\varphi - \omega dt)
\end{eqnarray}
From this we can, in retrospection, confirm and better understand approximate expressions given in
Eqs. (\ref{dtildeVincylindrapprox1}) and (\ref{dtildeVincylindrapprox2}).
\begin{figure} %[h]
\vspace*{-0.2cm}
\begin{center}
\scalebox{0.15}{\includegraphics[clip,scale=1.70]{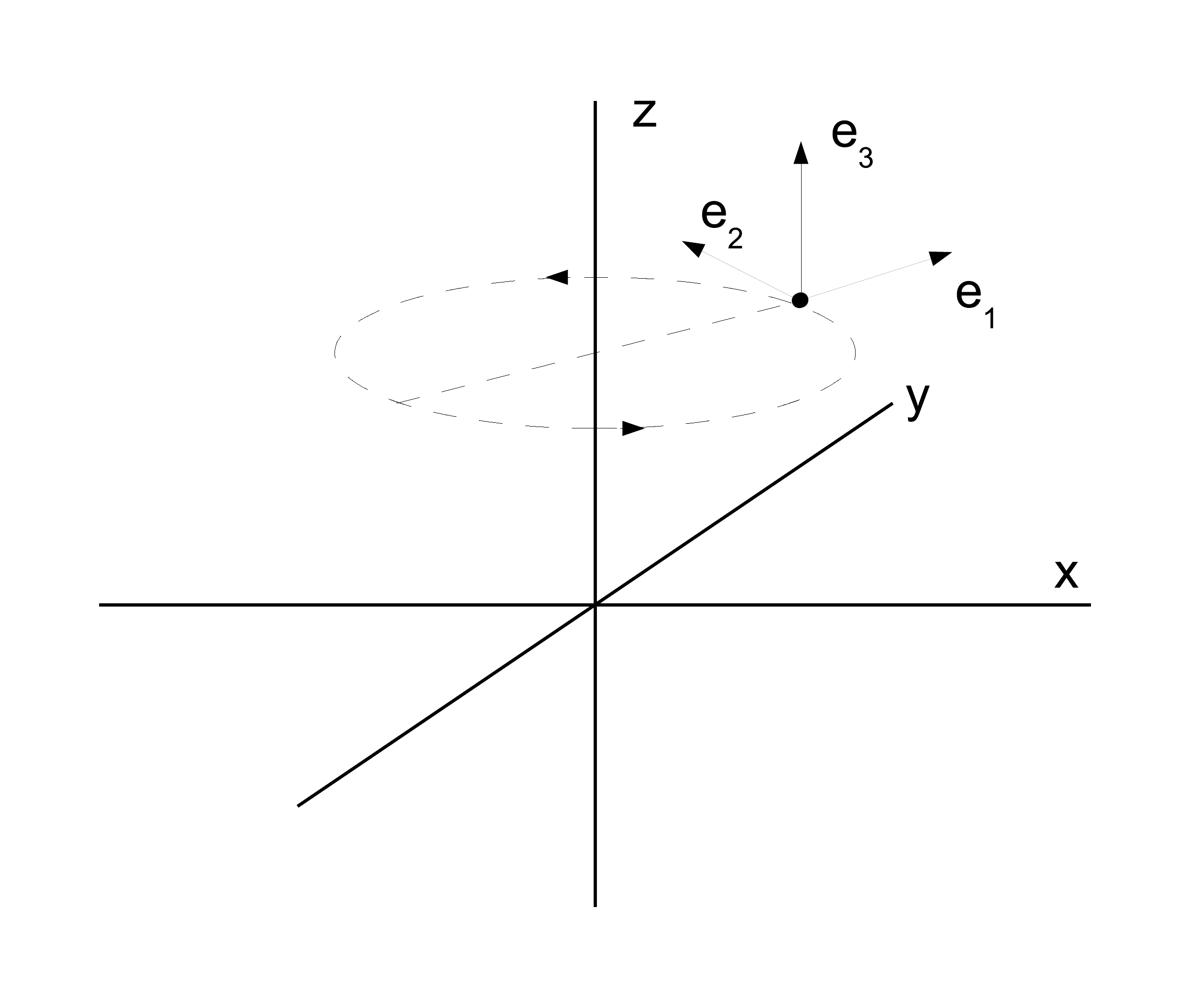}}
\scalebox{0.15}{\includegraphics[clip,scale=1.70]{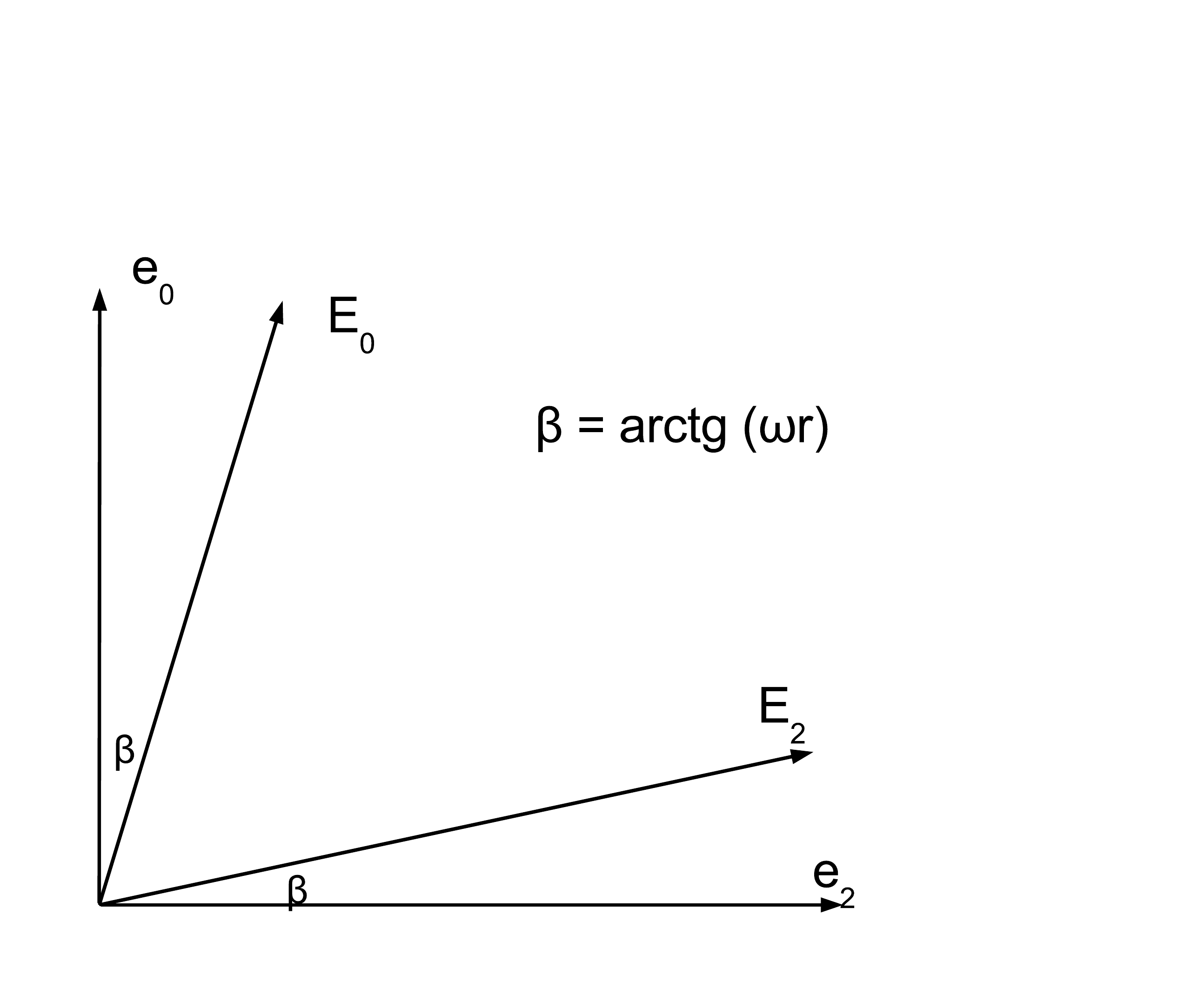}}
\vspace*{-0.2cm}
\caption{Left: standard cylindrical orthonormal frame $(e_1,e_2,e_3) \equiv (e_r,e_{\varphi},e_z)$ in $E^3$
         given by formulas Eqs. (\ref{e_1}) - (\ref{e_3}).
         Right: the frame $(\mathcal E_0,\mathcal E_1,\mathcal E_2,\mathcal E_3)$
         displayed in Eqs. (\ref{E_0}) - (\ref{E_3}) is obtained from
         the standard orthonormal cylindrical frame $(e_0, e_1, e_2, e_3)$
         in Minkowski space by local hyperbolic rotation (local Lorentz transformation) in the plane spanned by
         $(e_0, e_2)$. Notice that the corresponding parameter $\beta$ is indeed position dependent,
         namely it depends on $r$.}
\label{fig:spatialframe}
\end{center}
\vspace*{-0.3cm}
\end{figure}
We can also express spatial forms $\hat a$ and $\hat y$ differently, using appropriate
\emph{spatial frame}.

Recall, that the original (non-rotating) orthonormal ``cylindrical'' frame $e_a$ and co-frame $e^a$
(where $a=0,1,2,3$) read
\begin{eqnarray} %{rcl}
      \label{e_0}
             e_0   &=& \partial_t  \hskip 1.85cm    e^0 \ = \ dt     \\
      \label{e_1}
             e_1   &=& \partial_r  \hskip 1.85cm    e^1 \ = \ dr     \\
      \label{e_2}
             e_2   &=& (1/r)\partial_\varphi  \hskip 1cm    e^2 \ = \ rd\varphi     \\
      \label{e_3}
             e_3   &=& \partial_z  \hskip 1.85cm    e^3 \ = \ dz
\end{eqnarray}
Now, since $V$ is future-oriented normalized vector, it may serve as $\mathcal E_0$
of some other orthonormal frame $\mathcal E_a$.
Each such frame (coframe) may be obtained from the original frame (coframe)
by scrambling by a \emph{Lorentzian matrix} $\Lambda$,
\begin{equation} \label{Lorentz_frame}
        \mathcal E_a = e_b\Lambda^b_{ \ a}
        \hskip 1cm
        \mathcal E^a = {(\Lambda^{-1})}^a_{ \ b} e^b
\end{equation}
Rewriting Eq. (\ref{Vincylindr}) as
\begin{equation} \label{VasLorentzframe}
        \mathcal E_0 = \gamma \ e_0 +\gamma \omega r \ e_2   \equiv e_b\Lambda^b_{ \ 0}
\end{equation}
we see that the particular Lorentzian matrix realizes hyperbolic rotation in $02$-plane:
\begin{equation} \label{matrixLambda}
  \Lambda^a_{ \ b} =
  \left( \begin{matrix}
                        \gamma  &  0 &  \gamma \omega r &  0\\
                        0 & 1 &  0 &  0\\
                        \gamma \omega r &  0 & \gamma &  0\\
                        0 &  0 &  0 & 1
                        \end{matrix} \right)
  \hskip .7cm
  {(\Lambda^{-1})}^a_{ \ b} =
  \left( \begin{matrix}
                        \gamma  &  0 &  -\gamma \omega r &  0\\
                        0 & 1 &  0 &  0\\
                        -\gamma \omega r &  0 & \gamma &  0\\
                        0 &  0 &  0 & 1
                        \end{matrix} \right)
\end{equation}
(Notice that the matrices are \emph{position-dependent}. It should be so since the particular
hyperbolic rotation depends on velocity and velocity in turn depends od $r$.)
Plugging these particular matrices into Eq. (\ref{Lorentz_frame}) we get
\begin{eqnarray} %{rcl}
      \label{E_0}
       V \equiv \mathcal E_0 &=& \gamma (\partial_t +\omega \partial_\varphi)
         \hskip 1.5cm
         \tilde V \equiv \mathcal E^0 \ = \ \gamma (dt -\omega r^2 d\varphi)     \\
      \label{E_1}
             \mathcal E_1   &=& \partial_r  \hskip 3.75cm    \mathcal E^1 \ = \ dr     \\
      \label{E_2}
             \mathcal E_2   &=& \gamma (\omega r \partial_t + (1/r) \partial_\varphi)
                                \hskip 1.3cm    \mathcal E^2 \ = \ \gamma (rd\varphi - \omega r dt) \\
      \label{E_3}
             \mathcal E_3   &=& \partial_z  \hskip 3.75cm    \mathcal E^3 \ = \ dz
\end{eqnarray}
Clearly, for $\omega \to 0$,
\begin{equation} \label{omegato0}
        \Lambda^a_{ \ b} \to \delta^a_b
        \hskip 1cm
        \mathcal E_a \to e_a
        \hskip 1cm
        \mathcal E^a \to e^a
\end{equation}
and first-order (in $\omega$) approximations are obtained simply by putting $\gamma \doteq 1$
in the exact expressions.

From the duality property $\langle \mathcal E^a, \mathcal E_b \rangle = \delta^a_b$
(and the fact that $i_V$(spatial form)$=0$) it is clear that \emph{any spatial} form may be decomposed
w.r.t. \emph{basis spatial} $1$-forms
$\mathcal E^1$, $\mathcal E^2$, $\mathcal E^3$. And indeed, we see from Eq. (\ref{dtildeVincylindr})
that
\begin{equation} \label{aandy}
        \hat a = \gamma^2 \omega^2 r \mathcal E^1
        \hskip 1cm
        \hat y = -2\gamma^2 \omega \mathcal E^1 \wedge \mathcal E^2
\end{equation}

\newpage
\fancyhead[LO]{References}


\addcontentsline{toc}{section}{References}
\begin{thebibliography}{999}




\bibitem[Arnold 1989]{arnold1989} Arnold V., Mathematical Methods of Classical Mechanics, Springer-Verlag, (1989)

\bibitem[Batchelor 2002]{batchelor} Batchelor, G.K., An Introduction to Fluid Dynamics, Cambridge University Press (2002)

\bibitem[Cartan 1922]{cartan1922} Cartan \'{E}., Le\c{c}ons sur les invariants int\'egraux, Hermann, Paris (1922)

\bibitem[Cattaneo Gasparini 1963]{cattaneo1963} Cattaneo Gasparini I., Proiezioni dei tensori di curvatura
                                 di una variet\`{a} riemanniana a metrica iperbolica normale,
                                 Rendiconti di Matematica -(V)  \textbf{22}, Fascicolo 1-2, 127-146 (1963)

\bibitem[Chorin, Marsden 1993]{chormarsden1993} Chorin, A.J., Marsden, J.E., A Mathematical Introduction to Fluid Mechanics, Springer (Third Edition) (1993)

\bibitem[Ertel 1942]{ertel1942} Ertel H., Ein neuer hydrodynamischer Erhaltungssatz,
                                      Meteorologische Zeitschrift \textbf{59}, 277–281 (1942)

\bibitem[Fecko 1997]{fecko1997} Fecko M., On 3+1 decompositions with respect to an observer field via differential forms,
                                      J.Math.Phys. \textbf{38} (1997) 4542-4560
                                       (online at http://arxiv.org/abs/gr-qc/9701066)

\bibitem[Fecko 2006]{fecko2006} Fecko, M., Differential Geometry and Lie Groups for Physicists, Cambridge University Press (2006)

\bibitem[Fecko 2007]{fecko2007} Fecko, M., Differential Geometry in Physics. An introductory exposition for true non-experts
                               (lectures from Regensburg, 23pp);
                                available online at http://sophia.dtp.fmph.uniba.sk/\~{}fecko/referaty/regensburg\_2007.pdf


\bibitem[Fecko 2013a]{fecko2013a} Fecko M., On p-form vortex-lines equations on extended phase space,
                                          arXiv:1305.3167 [math-ph] (2013)

\bibitem[Fecko 2013b]{fecko2013b} Fecko M., On symmetries and conserved quantities in Nambu mechanics,
                                      J.Math.Phys. \textbf{54}, 102901 (2013); doi: 10.1063/1.4824684


\bibitem[Feynman 1964]{feynman1964} Feynman, R.P., Leighton, R.B., Sands, M.,
                                             The Feynman Lectures on Physics, Vol. 2, Addison Wesley Publ. Comp. (1964)

\bibitem[Fitzpatrick 2006]{fitzpatrick2006} Fitzpatrick, R., Thermodynamics and Statistical Mechanics: An intermediate level course;
                                online at
                                http://farside.ph.utexas.edu/teaching/sm1/statmech.pdf


\bibitem[Friedlander 1995]{friedlander1995} Friedlander S., Lectures on stability and instability of an ideal fluid,
                          In Hyperbolic equations and frequency interactions (Park City, UT, 1995), pages 227--304.
                          Amer. Math. Soc., Providence, RI, 1999. MR1662831

\bibitem[Griffiths 1999]{griffiths1999} Griffiths, D.J., Introduction to Electrodynamics, 3-rd Ed., Prentice Hall (1999)


\bibitem[Hehl, Obukhov 2003]{hehlobukhov2003} Hehl, F.W., Obukhov, Y.N., Foundations of Classical Electrodynamics, Birkh\"auser (2003)

\bibitem[Helmholtz 1858]{helmholtz1858} Helmholtz, H.,
                        \"Uber Integrale der hydrodynamischen Gleichungen, welcher der Wirbelbewegungen entsprechen'',
                        Journal f\"ur die reine und angewandte Mathematik (in German) \textbf{55}: 25–55, (1858)

\bibitem[Kellogg 1929]{kellogg1929} Kellogg, O.D., Foundations of Potential Theory, Verlag von Julius Springer, Berlin (1929)

\bibitem[Kelvin 1869]{kelvin1869} Kelvin, Lord, On vortex motion,
                                      Trans.Roy.Soc.Edinburgh, \textbf{25} (Math. and Phys.Papers \textbf{4}),(1869)

\bibitem[Kurz et al. 2006]{kurz2006} Kurz S., Flemisch B., Wohlmuth B.,
                              A framework for Maxwell's equations in non-inertial frames based on differential forms,
                              Proceedings of ICAP 2006 Chamonix, France (2006)

\bibitem[Kurz et al. 2009]{kurz2009} Kurz S., Auchmann B., Flemisch B.,
                              Dimensional reduction of field problems in a differential-forms framework,
                              COMPEL, \textbf{28}, Issue 4, 907-921, (2009)

\bibitem[Landau, Lifshitz 1994]{lanlif94} Landau, L.D., Lifshitz, E.M., The Classical Theory of Fields, Butterworth-Heinemann (Fourth Edition) (1994)

\bibitem[Landau, Lifshitz 1987]{lanlif87} Landau, L.D., Lifshitz, E.M., Fluid Mechanics, Pergamon Press (Second Edition) (1987)

\bibitem[Maglevannyj 1978]{maglevannyj1978} Maglevannyj I.I., Formalizm vneshnikh form i neinertsialnye sistemy odscheta,
                                      Izv.vyssh.ucheb.zav., Fizika, \textbf{7} 91-96 (1978)

\bibitem[Massa 1974a]{massa1974a} Massa E., Space Tensors in General Relativity I. Spatial Tensor Algebra and Analysis,
                                      Gen. Rel. Grav. \textbf{5} (1974) 555

\bibitem[Massa 1974b]{massa1974b} Massa E., Space Tensors in General Relativity II. Physical Applications,
                                      Gen. Rel. Grav. \textbf{5} (1974) 573

\bibitem[Massa 1974c]{massa1974c} Massa E., Space Tensors in General Relativity III. The Structural Equations,
                                      Gen. Rel. Grav. \textbf{5} (1974) 715-836

\bibitem[Nearing 2010]{nearing} Nearing, J., Mathematical Tools for Physics, Dover Publications (2010)

\bibitem[Poincar\'{e} 1890]{Poincare1} Poincar\'e H., Sur le probl\`{e}m des trois corps et les \'equations de la dynamique,
\newline \indent
                                Acta Mathematica, Vol. XIII, 1890
\bibitem[Poincar\'{e} 1899]{Poincare2} Poincar\'e H., Les M\'ethodes nouvelles de la M\'ecanique C\'eleste,
\newline \indent
                                (III, Invatiants int\'egraux), Gauthier-Villars et fils, 1899

\bibitem[Raumonen 2009]{raumonen2009} Raumonen P.,
                              Mathematical Structures for Dimensional Reduction end Equivalence Classification
                              of Electromagnetic Boundary Value Problems,
                              PhD.Thesis, Tampere University of Technology (2009),
                              online at http://dspace.cc.tut.fi/dpub/bitstream/handle/123456789/6032/raumonen.pdf?sequence=3

\bibitem[Raumonen et al. 2011]{raumonen2011} Raumonen P., Suuriniemi S., Kettunen L.,
                              Dimensional reduction of electromagnetic boundary value problems,
                              Boundary Value Problems, Article Number: 9,   DOI: 10.1186/1687-2770-2011-9   (2011)





\bibitem[Reynolds 1903]{reynolds1903} Reynolds O., Papers on mechanical and physical subjects-the sub-mechanics of the Universe,
                                      Collected Work, Volume III, Cambridge University Press, (1903)

\bibitem[Saffman 1993]{saffmanf} Saffman, P.G., Vortex Dynamics, Cambridge University Press (1993)

\bibitem[Schutz 1980]{schutz} Schutz, B.F., Geometrical Methods of Mathematical Physics, Cambridge University Press (1980)

\bibitem[Truesdell 1954]{truesdell} Truesdell, C., The Kinematics of Vorticity, Indiana University Press (1954)


\end{thebibliography}
\end{document}